\documentclass[useAMS,usenatbib]{mn2e}
\usepackage{times}
\usepackage{amsmath}
\usepackage{amssymb}
\usepackage{longtable}
\usepackage{natbib}
\usepackage{epsfig}

\title[The Optical Recombination Spectrum of NGC\,7009]
{Very deep spectroscopy of the bright Saturn Nebula NGC\,7009 -- II.
Analysis of the rich optical recombination spectrum}

\author[X. Fang and X.-W. Liu]{X. Fang$^1$
        \thanks{E-mail: fangx@pku.edu.cn} and X.-W. Liu$^{1,2}$\\
        $^1$Department of Astronomy, School of Physics, Peking University,
        Beijing 100871, P.~R. China\\
        $^2$Kavli Institute for Astronomy and Astrophysics at Peking
        University, Beijing 100871, P.~R. China}

\begin{document}

\date{Accepted . Received }

\pagerange{\pageref{firstpage}--\pageref{lastpage}} \pubyear{2011}

\maketitle

\label{firstpage}

\begin{abstract}
In Paper~I, we presented deep, long-slit spectrum of the bright Saturn nebula
NGC\,7009. Numerous permitted lines emitted by the C$^{+}$, N$^{+}$, O$^{+}$
and Ne$^{+}$ ions were detected. Gaussian profile fitting to the spectrum
yielded more than 1000 lines, the majority of which are optical recombination
lines (ORLs) of heavy element ions. In the current paper, we present a
critical analysis of the rich optical recombination spectrum of NGC\,7009,
in the context of the bi-abundance nebular model proposed by Liu et al.
Transitions from individual multiplets are checked carefully for potential
blended lines. The observed relative intensities are compared with the
theoretical predictions based on high quality effective recombination
coefficients, now available for the recombination line spectrum of a number
of heavy element ions.

The possibility of plasma diagnostics using the ORLs of various heavy element
ions is discussed in detail. The line ratios that can be used to determine
electron temperature are presented for each ion, although there is still a
lack of adequate atomic data and some of the lines are still not detected in
the spectrum of NGC\,7009 due to weakness and/or line blending.
Plasma diagnostics based on the N~{\sc ii} and O~{\sc ii} recombination
spectra both yield electron temperatures close to 1000~K, which is lower
than those derived from the collisionally excited line (CEL) ratios (e.g.,
the [O~{\sc iii}] and [N~{\sc ii}] nebular-to-auroral line ratios; see
Paper~I for details) by nearly one order of magnitude. The very low
temperatures yielded by the O~{\sc ii} and N~{\sc ii} ORLs indicate that
they originate from very cold regions. The C$^{2+}$/H$^+$, N$^{2+}$/H$^+$,
O$^{2+}$/H$^+$ and Ne$^{2+}$/H$^+$ ionic abundance ratios derived from ORLs
are consistently higher, by about a factor of 5, than the corresponding
values derived from CELs. In calculating the ORL ionic abundance ratios, we
have used the newly available high quality effective recombination
coefficients, and adopted an electron temperature of $\sim$1000~K, as given
by the ORL diagnostics and as a consequence presumably representing the
physical conditions prevailing in the regions where the heavy element ORLs
arise. Measurements of the ultraviolet (UV) and infrared (IR) CELs from the
literature are used to calculate CEL ionic abundance ratios when optical
data are not available for the ionic species. A comparison of results of
plasma diagnostics and abundance determinations for NGC\,7009 points to the
existence of ``cold", metal-rich (i.e., H-deficient) inclusions embedded in
the hot, diffuse ionized gas, first postulated by Liu et al.

At electron temperatures yielded by the N~{\sc ii} and O~{\sc ii} ORLs, the
predicted relative intensities of ORLs agree well with the observed values,
indicating that the current quantum calculations of the recombination spectra
of those two ionic species well represent the recombination processes under
nebular conditions. Deviations from the {\it LS}\,coupling, noticed in an
earlier quantitative spectroscopy by Liu et al. for the same object, are again
confirmed, especially for recombination lines of the 4f\,--\,3d transition
array. For N~{\sc ii}, as well as for O~{\sc ii}, the ionic abundances derived
from different $J$-resolved transitions within a multiplet, or from the
transitions belonging to different multiplets, agree with each other. This
is another evidence that the new effective recombination coefficients are
reliable. New calculations of the effective recombination coefficients for
the Ne~{\sc ii} lines at nebular temperatures and densities are needed.
\end{abstract}


\section{\label{introduction}
Introduction}

The bright Saturn Nebula NGC\,7009 is known for its rich and prominent optical
recombination lines (ORLs) of heavy element ions, especially those of O~{\sc
ii}, ever since the spectrophotographic observations of Wyse \cite{wyse1942},
who published and analyzed deep spectra of the Orion Nebula and nine planetary
nebulae (PNe), including NGC\,7009. He identified and measured several dozen
O~{\sc ii} permitted lines in NGC\,$7009$ in the wavelength range 3700 --
6750\,{\AA}, although accurate measurements of many of those O~{\sc ii} lines
were hampered by line blending. At the end of this paper Wyse \cite{wyse1942}
expressed the desire of having more accurate measurements of the O~{\sc ii}
permitted lines.
Aller \& Kaler \cite{ak1964} identified more than 100 O~{\sc ii}
permitted lines in the spectrum of NGC\,7009. Large numbers of permitted lines
of other ionic species, such as C~{\sc ii}, N~{\sc ii}, N~{\sc iii}, O~{\sc
iii}, Ne~{\sc ii}, were also detected. The majority of these permitted lines
are mainly excited by recombination. Other possible excitation mechanisms,
such as the dielectronic recombination, radiative charge transfer, and
resonance fluorescence by starlight or by some other prominent nebular
emission lines, are all by their nature selective, which means that they
tend to excite lines from specific spectral terms of certain parity and
multiplicity only (e.g., Grandi \citealt{grandi1976}; Liu \& Danziger
\citealt{ld1993a}; Liu, Danziger \& Murdin \citealt{ldm93}). With high
signal-to-noise ratio, high spectral resolution and wide wavelength-coverage
spectra of PNe now available, more and more ORLs of fainter intensities from
heavy element ions that arise from many different multiplets are observed and
provide an opportunity to study the radiative and dielectronic recombination
processes and test the accuracy of the recombination theories for
non-hydrogenic ions. The first systematic study of the ORLs in NGC\,7009 was
carried out by Liu et al. (\citealt{liu1995}, hereafter LSBC), who analyzed
dozens of O~{\sc ii} ORLs, using effective recombination coefficients
calculated in the intermediate coupling scheme for transitions from the
3d\,--\,3p and 4f\,--\,3d arrays, and coefficients calculated in the {\it
LS}\,coupling scheme for transitions from the 3p -- 3s array. LSBC found
clear deviations from the {\it LS}\,coupling in the 3d\,--\,3p and 4f\,--\,3d
transitions. Luo, Liu \& Barlow (\citealt{luo2001}, hereafter LLB01) presented
high-quality observations of several dozens Ne~{\sc ii} ORLs in NGC\,7009,
and derived the Ne$^{2+}$/H$^+$ abundance ratios from them.

Along with the advance of observational techniques that have enabled the
detections of many faint ORLs of heavy element ions in photoionized gaseous
nebulae, the recombination theories of heavy element ions, such as C~{\sc ii},
N~{\sc ii}, O~{\sc ii}, and Ne~{\sc ii}, have seen steady improvements since
early 1980s (e.g. Storey \citealt{storey1981}; Nussbaumer \& Storey
\citealt{ns1983}, \citealt{ns1984}, \citealt{ns1986}, \citealt{ns1987};
Escalante \& Victor \citealt{ev1990}; P\'{e}quignot, Petitjean \& Boisson
\citealt{ppb1991}; Storey \citealt{storey1994}; LSBC;
Kisielius et al. \citealt{kisielius1998}; Davey, Storey \& Kisielius
\citealt{davey2000}; Kisielius \& Storey \citealt{ks1999}, \citealt{ks2002};
Fang, Storey \& Liu \citealt{fsl2011}). The high-quality atomic data have
been widely used to reveal the physical conditions (electron temperatures and
densities) under which the ORLs of heavy element ions arise, and to determine
ionic and elemental abundances from them (e.g. Liu et al. \citealt{liu2000}).

In nebular astrophysics there has been a long-standing dichotomy whereby the
ionic and elemental abundances of C, N, O and Ne relative to hydrogen
determined from ORLs (e.g. C~{\sc ii} M6 $\lambda$4267, N~{\sc ii} M39b
$\lambda$4041, O~{\sc ii} M1 $\lambda$4649 and M48a $\lambda$4089, Ne~{\sc ii}
M55e $\lambda$4392) are systematically higher than those derived from the much
brighter collisionally excited lines (CELs, often referred to as forbidden
lines). With high-quality optical spectra now available, detailed studies of
this problem have been carried out for several archetypal PNe (LSBC and LLB01
for NGC\,7009; Liu et al. \citealt{liu2000} for NGC\,6153; Liu et al.
\citealt{liu2001b} for M\,1-42 and M\,2-36; Liu et al. \citealt{lbzbs06} for
Hf\,2-2; Garnett \& Dinerstein \citealt{gd2001} for NGC\,6720). Several deep
optical spectroscopic surveys of PNe, which allow for the analyses of
nebulae based on ORLs, have been carried out during the past decade (Tsamis
et al. \citealt{tsamis2003}, \citealt{tsamis2004}; Liu et al.
\citealt{liu2004a},\,b; Robertson-Tessi \& Garnett \citealt{rtg2005}; Wesson,
Liu \& Barlow \citealt{wlb2005}; Wang \& Liu \citealt{wl2007}). The abundance
discrepancy factors (ADFs), defined as the ratio of the abundance derived
from ORLs to that deduced from CELs, typically lie in the range 1\,--\,3. But
for a significant number of PNe, ADF values exceeding 5, or even 10, are seen.
The highest ADF value ($\sim$70) of all PNe is found in Hf\,2-2 (Liu et al.
\citealt{lbzbs06}).  Another dichotomy that is closely related to the problem
of abundance discrepancy is that nebular electron temperatures derived from
the traditional diagnostic [O~{\sc iii}] nebular-to-auroral line ratio are
generally higher than those derived from the Balmer jump (BJ) of hydrogen
recombination spectrum (e.g. Peimbert \citealt{peimbert1971}; Liu \& Danziger
\citealt{ld1993b}). A number of postulations have been raised to explain
these problems (e.g. Peimbert \citealt{peimbert1967}; Rubin
\citealt{rubin1989}; Viegas \& Clegg \citealt{vc1994}), but all failed to
provide a consistent interpretation of all the available observations.
Recenltly, Nicholls, Dopita \& Sutherland \cite{nicholls12} explored the
possibility that electrons in H~{\sc ii} regions and PNe depart from a
Maxwell-Boltzmann equilibrium energy distribution and suggested that a
``$\kappa$-distribution" for the electron energies, which are widely found
in solar system plasmas, can explain the temperature and abundance
discrepancies in H~{\sc ii} regions and PNe. The bi-abundance nebular model
proposed by Liu et al. \cite{liu2000}, who postulated that PNe (probably also
H~{\sc ii} regions) contain H-deficient inclusions, provides a better and
natural explanation of the dichotomy. In this model, the faint ORLs of heavy
element ions originate mainly from the ``cold", H-deficient inclusions, while
the stronger CELs are emitted from the warmer ambient plasma with `normal'
chemical composition. Deep spectroscopic surveys and recombination line
analysis of individual nebulae in the past decade has yielded strong evidence
for the existence of such a ``cold" component (see recent reviews by Liu
\citealt{liu2003}, \citealt{liu2006a}, \citealt{liu2011}).

This is the second of the two papers devoted to very deep spectroscopy of
NGC\,7009. In the previous paper (Fang \& Liu \citealt{fl2011}, hereafter
Paper~I), we presented high-quality spectra of NGC\,7009 and tabulation of
all detected lines, including their observed and dereddened intensities, many
of which were obtained via careful deblending using the technique of
multi-Gaussian profile fitting. We also carried out plasma diagnostics using
the CEL ratios, the H~{\sc i} recombination spectrum (including the Balmer
and Paschen decrements of the line spectrum, and the Balmer and Paschen jumps
of the continuum spectrum), and the He~{\sc i} and He~{\sc ii} recombination
spectrum (including the He~{\sc i} recombination line ratios, and
discontinuities of the He~{\sc i} and He~{\sc ii} recombination continua).
The average electron temperature yielded by CELs, $T_\mathrm{e}$(CELs), is
higher than that from the H~{\sc i} Balmer jump, $T_\mathrm{e}$(H~{\sc i}~BJ),
which in turn is higher than the temperature derived from the He~{\sc i}
recombination line ratios, $T_\mathrm{e}$(He~{\sc i}). The current paper
focuses on analyses of the optical recombination spectra of heavy element
ions detected in the spectrum of NGC\,7009. New effective recombination
coefficients, including those for the N~{\sc ii} and O~{\sc ii} recombination
spectrum that were calculated in the intermediate coupling scheme, are now
available and are utilized in the analyses. Plasma diagnostics based on the
ORLs of heavy element ions are carried out in Section\,2, and the electron
temperatures derived from the N~{\sc ii} and O~{\sc ii} ORL ratios agree
with each other and are both close to 1200~K. Thus the general pattern of
electron temperatures, $T_\mathrm{e}$(CELs) $\gtrsim$ $T_\mathrm{e}$(H~{\sc
i}~BJ) $\gtrsim$ $T_\mathrm{e}$(He~{\sc i}) $\gtrsim$ $T_\mathrm{e}$(N~{\sc
ii},\,O~{\sc ii}~ORLs), which was predicted by the bi-abundance nebular model
(Liu \citealt{liu2003}) and has been seen in many PNe, is confirmed in the
current analysis of NGC\,7009. A comprehensive analysis of individual
multiplets of the C~{\sc ii}, N~{\sc ii}, O~{\sc ii}, and Ne~{\sc ii}
recombination spectra are presented in Section\,3. The lines are critically
examined for potential blending effects. Comparison is made for the observed
and predicted relative intensities of the best observed transitions, using the
latest effective recombination coefficients. Ionic and elemental abundances
are derived in Section\,4, where ADFs for the C, N, O, and Ne ionic
abundances are calculated. The results are discussed in Section\,5, followed
by a summary in Section\,6.

\section{\label{diagnose} Plasma diagnostics based on the ORLs of heavy
element ions}

\subsection{\label{diagnose:data}
Effective recombination coefficients}

Reliable atomic data, most importantly the effective recombination coefficients
of abundant heavy element ions such as C~{\sc ii}, N~{\sc ii}, O~{\sc ii}, and
Ne~{\sc ii}, are key to the spectroscopic analysis of photoionized gaseous
nebulae. Most of the {\it ab~initio} calculations of heavy element ions aimed
for astrophysical applications hitherto were carried out in the {\it
LS}\,coupling scheme. This approximation tacitly assumes a statistical
distribution in the population of the fine-structure levels of the recombining
ions (i.e., 1\,:\,2 for the N$^{2+}$ $^{2}$P$^{\rm o}_{1/2}$ and $^{2}$P$^{\rm
o}_{3/2}$ levels in the case of N~{\sc ii}; 1\,:\,3\,:\,5 for the O$^{2+}$
$^{3}$P$_{0}$, $^{3}$P$_{1}$ and $^{3}$P$_{2}$ levels in the case of O~{\sc
ii}). The assumption of {\it LS}\,coupling may give satisfactory results for
some of the low-lying transitions such as those belonging to the 3p\,--\,3s
configuration, but not for many of the transitions from the higher 3d\,--\,3p
or 4f\,--\,3d configurations. In low-density objects such as H~{\sc ii}
regions and evolved PNe, the relative populations of the ground-term
fine-structure levels of the recombining ion actually have density-dependence
and deviate from the statistical distribution, and so do the relative
emissivities of resultant recombination lines. A better treatment of the
recombination and the following cascading in a proper coupling scheme is vital
for probing the physical conditions in gaseous nebulae.

\begin{figure*}
\begin{center}
\includegraphics[width=9.5cm,angle=-90]{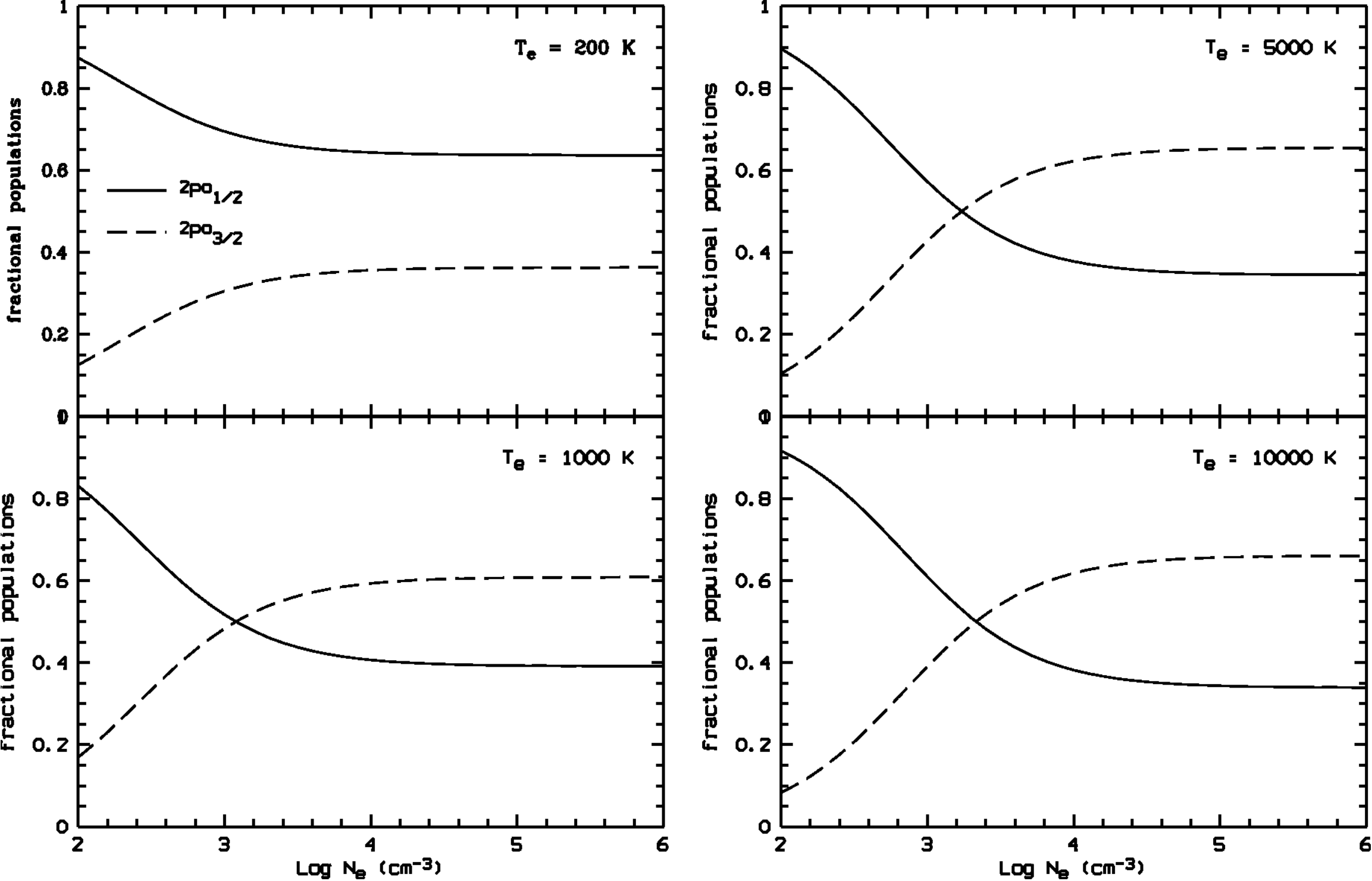}
\caption{Fractional populations of the N$^{2+}$ $^{2}$P$^{\rm o}_{1/2}$ and
$^{2}$P$^{\rm o}_{3/2}$ fine-structure levels. Four temperature cases, 200,
1000, 5000 and 10\,000~K, are shown. This figure is obtained by solving
level population equations for a five-level atomic model.}
\label{niii:frac}
\end{center}
\end{figure*}

\begin{figure*}
\begin{center}
\includegraphics[width=10cm,angle=-90]{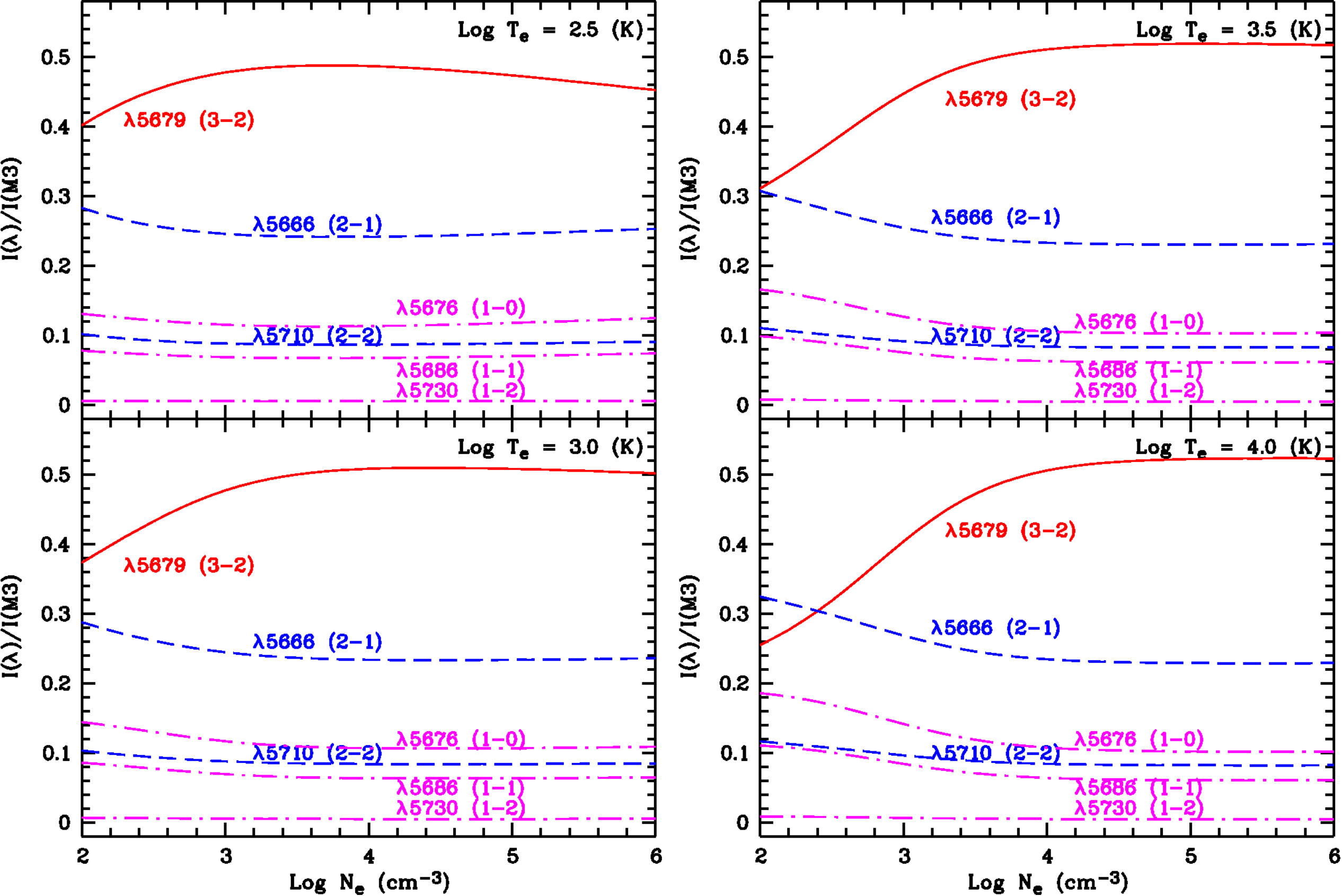}
\caption{Fractional intensities of the N~{\sc ii} M3
2p3p\,$^3$D\,--\,2p3s\,$^3$P$^{\rm o}$ $\lambda$5679 multiplet as a function
of electron density. The numbers in the brackets ($J_2\,-\,J_1$) following
the wavelength labels are the total angular momentum quantum numbers of the
upper and lower levels, respectively. Transitions from the upper levels with
the same angular momentum quantum number $J_2$ are represented by curves of
the same color and line type. Four temperature cases, $\log{T_\mathrm{e}}$~[K]
= 2.5, 3.0, 3.5, and 4.0, are presented. The calculations were based on the
effective recombination coefficients of FSL11.}
\label{nii:v3}
\end{center}
\end{figure*}

\begin{figure*}
\begin{center}
\includegraphics[width=10cm,angle=-90]{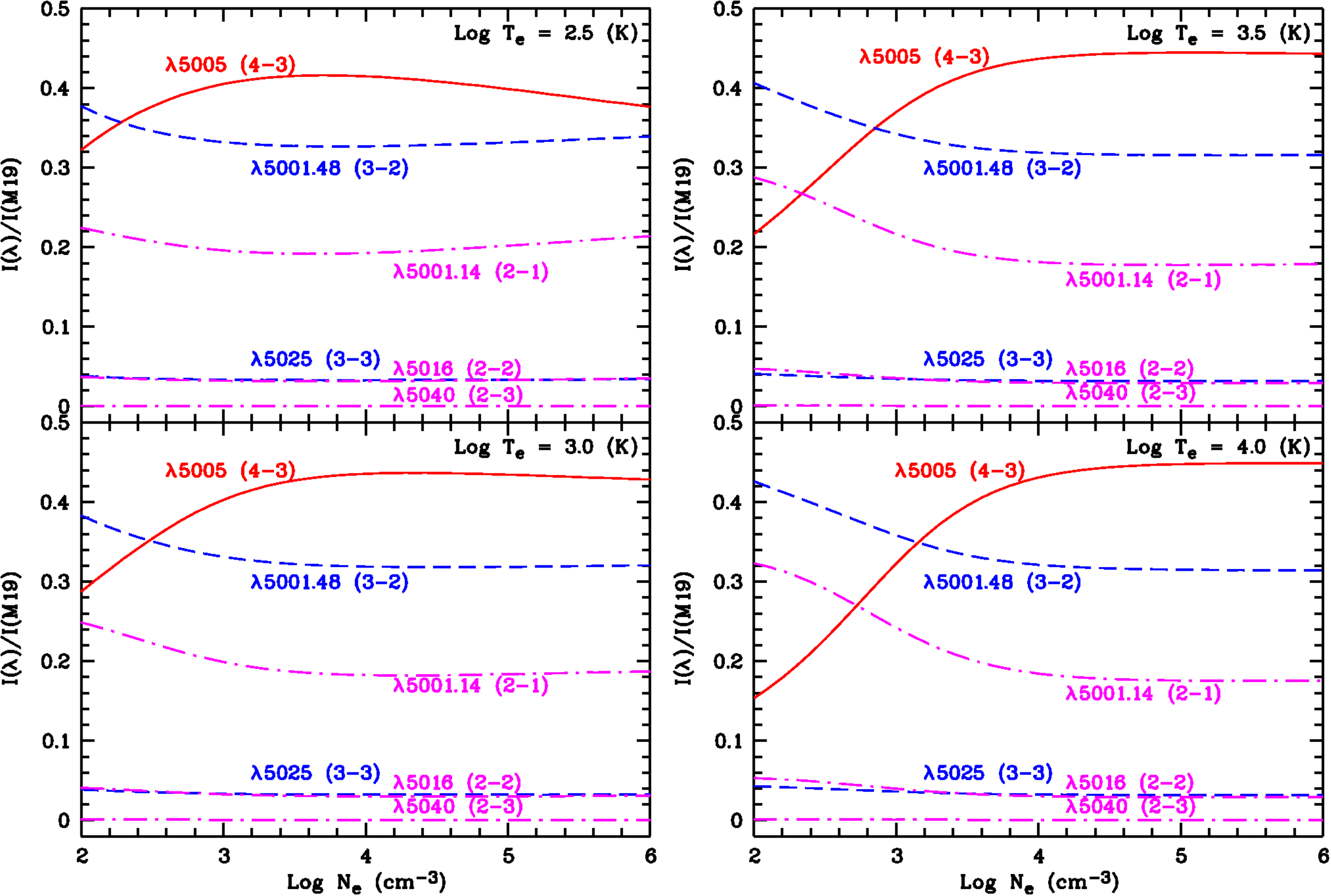}
\caption{Same as Fig.\,\ref{nii:v3} but for the fractional intensities of
the N~{\sc ii} M19 2p3d\,$^{3}$F$^{\rm o}$\,--\,2p3p\,$^{3}$D $\lambda$5004
multiplet.}
\label{nii:v19}
\end{center}
\end{figure*}

\begin{figure*}
\begin{center}
\includegraphics[width=10cm,angle=-90]{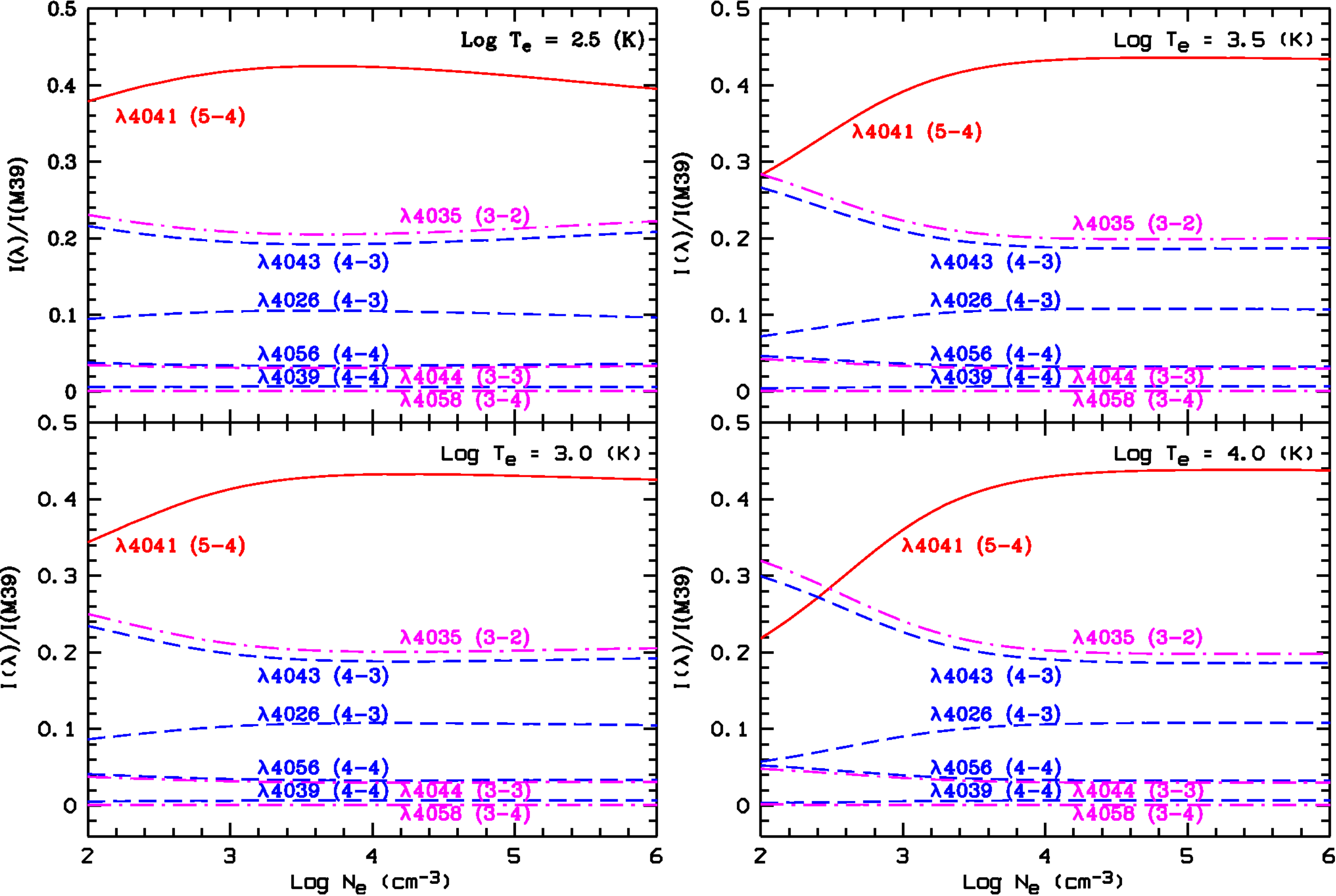}
\caption{Same as Fig.\,\ref{nii:v3} but for the fractional intensities of
the N~{\sc ii} M39 2p4f\,G[7/2,9/2]\,--\,2p3d\,$^{3}$F$^{\rm o}$ $\lambda$4041
multiplet.}
\label{nii:v39}
\end{center}
\end{figure*}

New {\it ab initio} calculation of the effective recombination coefficients
for the N~{\sc ii} recombination spectrum was presented by Fang, Storey \&
Liu (\citealt{fsl2011}, hereafter FSL11)\footnote{Dr. Daniel P\'{e}quignot
found some anomalies in the published data of FSL11, which were due to
mislabeling of five bound-state energy levels of N~{\sc ii}. The labeling has
recently been corrected and the effective recombination coefficients for the
N~{\sc ii} lines were re-calculated. A corrigendum has been in preparation.
Figs.\,\ref{nii:v3}, \ref{nii:v19} and \ref{nii:v39} in the current paper are
based on the revised effective recombination coefficients of N~{\sc ii}.},
who took into account the density dependence of effective recombination
coefficients arising from the density-dependence of relative populations of
the ground fine-structure levels of the recombining ion (i.e. N$^{2+}$
$^2$P$^{\rm o}_{1/2}$ and $^2$P$^{\rm o}_{3/2}$), an elaboration that has not
been attempted before for this ion. The availability of such data opens up the
possibility of electron density determination via recombination line
analysis. Fig.\,\ref{niii:frac} (also Fig.\,$3$ in FSL11) shows the relative
populations of the N$^{2+}$ $^{2}$P$^{\rm o}_{1/2}$ and $^{2}$P$^{\rm
o}_{3/2}$ fine-structure levels as a function of electron density under
typical nebular conditions. Photoionization cross-sections, bound state
energies, and oscillator strengths of N~{\sc ii} with $n\leq{11}$ and
$l\leq{4}$ were obtained using the close-coupling R-matrix method in the
intermediate coupling scheme. Photoionization data were computed using an
energy mesh which accurately map out the near-threshold resonances, and were
used to derive recombination coefficients, including radiative and
dielectronic recombination. Also new is the inclusion in the calculations of
the effects of dielectronic recombination via high-$n$ resonances lying
between the $^2$P$^{\rm o}_{1/2}$ and $^2$P$^{\rm o}_{3/2}$ thresholds. The
calculated coefficients are valid for temperatures down to an unprecedentedly
low level ($\sim$100~K). Figs.\,\ref{nii:v3}, \ref{nii:v19} and \ref{nii:v39}
(also Figs.\,$5$, $6$ and $7$ in FSL11) show the theoretical relative
intensities of the fine-structure components of the M3
2p3p\,$^3$D\,--\,2p3s\,$^3$P$^{\rm o}$, M19 2p3d\,$^{3}$F$^{\rm
o}$\,--\,2p3p\,$^{3}$D and M39 2p4f\,G[7/2,9/2]\,--\,2p3d\,$^{3}$F$^{\rm o}$
multiplets of N~{\sc ii}, respectively, as a function of electron density.

\begin{figure*}
\begin{center}
\includegraphics[width=9.5cm,angle=-90]{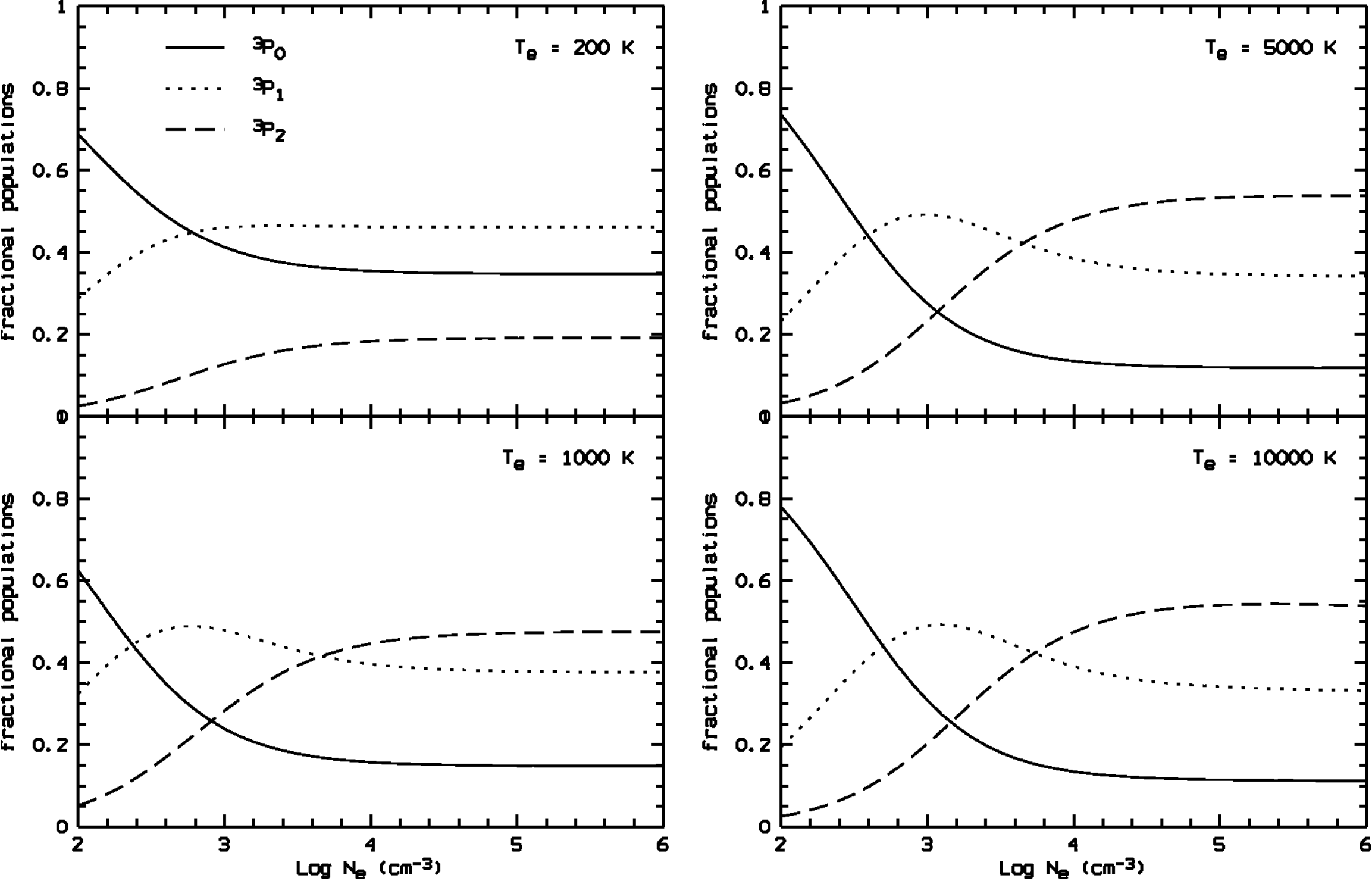}
\caption{Fractional populations of the O$^{2+}$ $^{3}$P$_{0}$, $^{3}$P$_{1}$
and $^{3}$P$_{2}$ fine-structure levels. Four temperature cases, 200, 1000,
5000 and 10\,000~K, are shown. This figure is obtained by solving the level
population equations for a five-level atomic model.}
\label{oiii:frac}
\end{center}
\end{figure*}

\begin{figure*}
\begin{center}
\includegraphics[width=10cm,angle=-90]{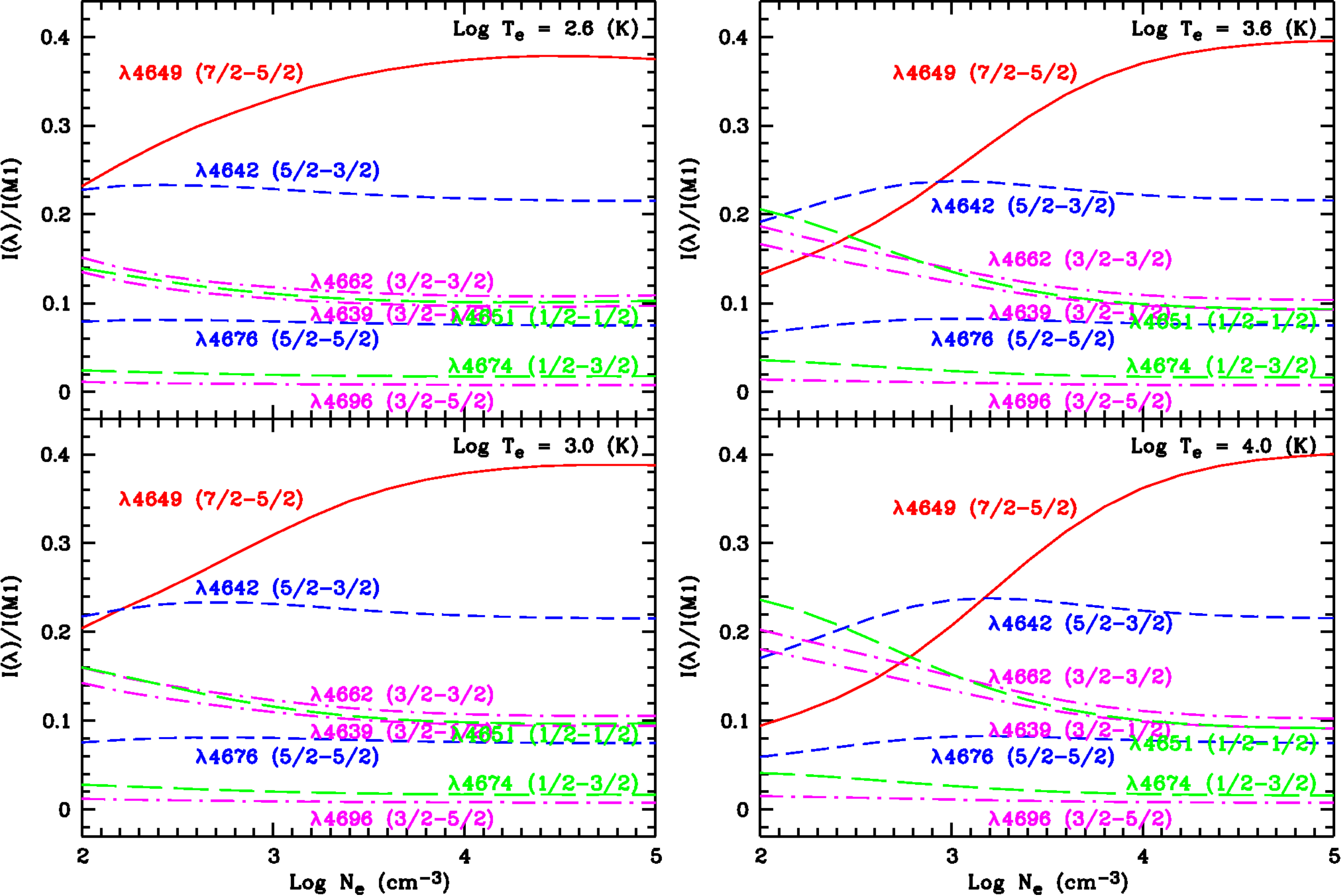}
\caption{Same as Fig.\,\ref{nii:v3} but for the fractional intensities
of the O~{\sc ii} M1 2p$^{2}$3p\,$^{4}$D$^{\rm o}$\,--\,2p$^{2}$3s\,$^{4}$P
$\lambda$4652 multiplet. The calculations were based on the unpublished
effective recombination coefficients of PJS.}
\label{oii:v1}
\end{center}
\end{figure*}

\begin{figure*}
\begin{center}
\includegraphics[width=10cm,angle=-90]{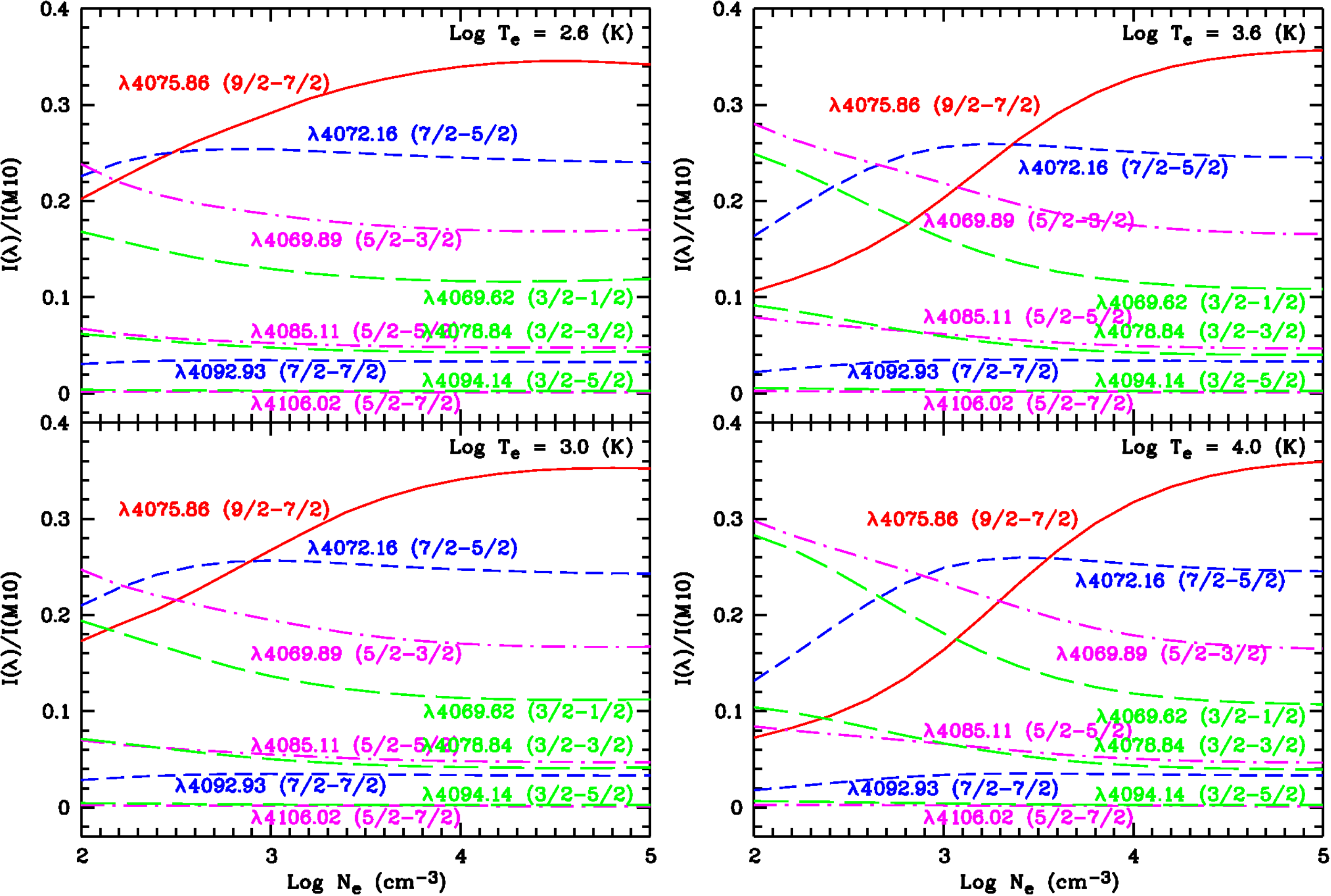}
\caption{Same as Fig.\,\ref{oii:v1} but for the fractional intensities of
the O~{\sc ii} M10 2p$^{2}$3d\,$^{4}$F\,--\,2p$^{2}$3p\,$^{4}$D$^{\rm o}$
$\lambda$4075 multiplet.}
\label{oii:v10}
\end{center}
\end{figure*}

\begin{figure*}
\begin{center}
\includegraphics[width=10cm,angle=-90]{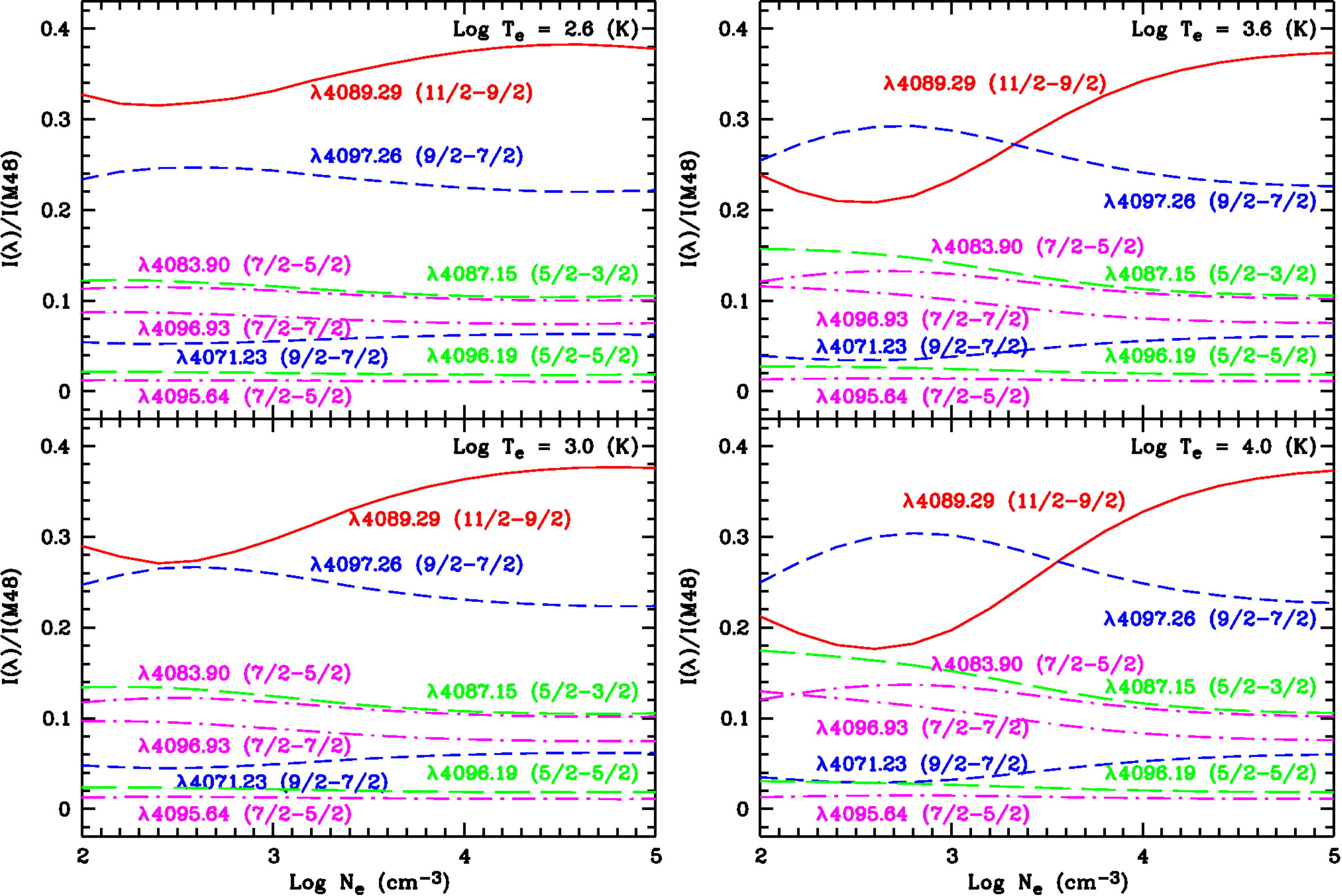}
\caption{Same as Fig.\,\ref{oii:v1} but for the fractional intensities of the
O~{\sc ii} M48 4f\,G[5,4,3]$^{\rm o}$\,--\,3d\,$^4$F $\lambda$4089 multiplet.}
\label{oii:v48}
\end{center}
\end{figure*}

So far, most calculations of the O~{\sc ii} effective recombination
coefficients have been in the {\it LS}\,coupling assumption. The first
comprehensive treatment of the O~{\sc ii} recombination at nebular
temperatures and densities was by Storey \cite{storey1994}, who adopted the
bound-bound and bound-free radiative data of O~{\sc ii} from the Opacity
Project data base (Cunto et al. \citealt{cunto1993}) and took into account
cascading as well as the effects of collisions. LSBC
presented partial treatment of intermediate coupling effects in transitions
between the ($^3$P)4f, ($^3$P)3d, and ($^3$P)3p electron configurations. The
most recent calculations of effective recombination coefficients for the
O~{\sc ii} recombination spectrum was carried out by P.~J. Storey (private
communication, hereafter PJS) in the intermediate coupling scheme. Density
dependence of the relative populations of the ground-term fine-structure
levels of the recombining ion was considered in the level population
calculations. Fig.\,\ref{oiii:frac} shows the fractional populations of
the recombining ion O$^{2+}$ $^{3}$P$_{0}$, $^{3}$P$_{1}$ and $^{3}$P$_{2}$
fine-structure levels as a function of electron density. The new O~{\sc ii}
recombination coefficients were calculated down to a temperature of 400~K.
Figs.\,\ref{oii:v1}, \ref{oii:v10} and \ref{oii:v48} show the theoretical
relative intensities of the fine-structure components of the O~{\sc ii} M1
2p$^{2}$3p\,$^{4}$D$^{\rm o}$\,--\,2p$^{2}$3s\,$^{4}$P, M10
2p$^{2}$3d\,$^{4}$F\,--\,2p$^{2}$3p\,$^{4}$D$^{\rm o}$ and M48
4f\,G[5,4,3]$^{\rm o}$\,--\,3d\,$^4$F multiplets, respectively, as a
function of electron density.

The new effective recombination coefficients for the N~{\sc ii} and O~{\sc
ii} recombination spectra provide an opportunity to construct nebular plasma
diagnostics based on the ORLs of heavy element ions. With those new atomic
data, we have determined electron temperatures and densities for over 100
Galactic PNe and 40 Galactic and extragalactic H~{\sc ii} regions (McNabb et
al. \citealt{mfls2011}). By comparing our results of plasma diagnostics based
on the N~{\sc ii} and O~{\sc ii} ORLs with the electron temperatures given in
literature ($T_\mathrm{e}$'s derived from CELs, H~{\sc i} Balmer jump and the
He~{\sc i} recombination lines), we find a temperature sequence for about 50
PNe, $T_\mathrm{e}$([O~{\sc iii}]) $\gtrsim$ $T_\mathrm{e}$(H~{\sc i} BJ)
$\gtrsim$ $T_\mathrm{e}$(He~{\sc i}) $\gtrsim$ $T_\mathrm{e}$(N~{\sc
ii}~\&~ O~{\sc ii}~ORLs), which is consistent with predictions from the
bi-abundance nebular model postulated by Liu et al. \cite{liu2000}.

\begin{figure}
\begin{center}
\includegraphics[width=10cm,angle=-90]{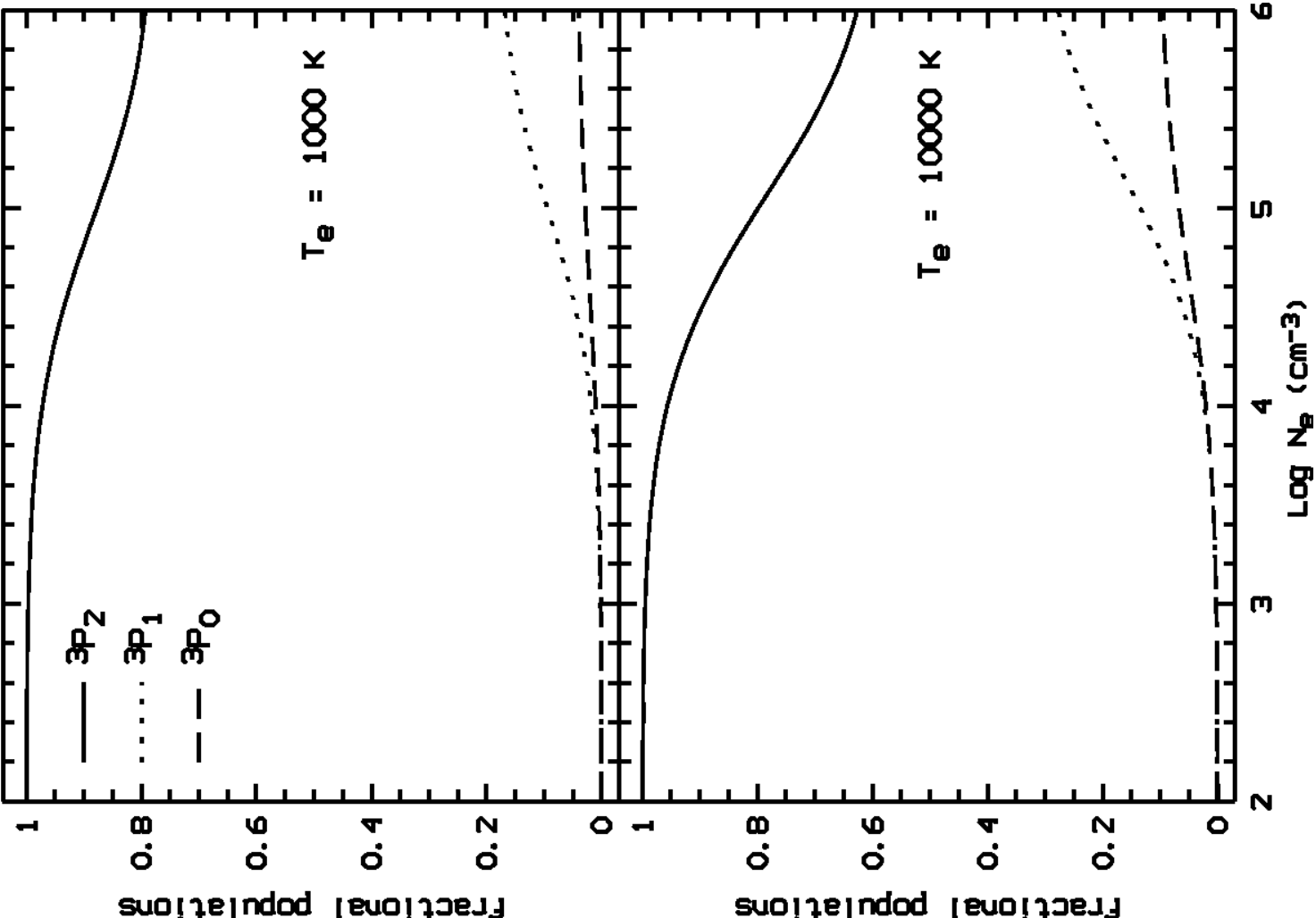}
\caption{Fractional populations of the Ne$^{2+}$ $^{3}$P$_{2}$, $^{3}$P$_{1}$
and $^{3}$P$_{0}$ fine-structure levels. Two temperature cases, 1000 and
10\,000~K, are shown. This figure is obtained by solving the level population
equations for a five-level atomic model.}
\label{neiii:frac}
\end{center}
\end{figure}

Kisielius et al. \cite{kisielius1998} published the Ne~{\sc ii} effective
recombination coefficients that were calculated in the {\it LS}\,coupling
scheme. Only transitions between states with $l\leq{2}$ were presented.
Preliminary effective recombination coefficients for a few selected lines
from the 4f\,--\,3d configuration are available (P.~J. Storey, private
communication), but only for a single temperature and density case. All the
previous calculations of the Ne~{\sc ii} recombination spectrum assumed that
the three ground-term fine-structure levels of the recombining ion Ne$^{2+}$,
$^{3}$P$_{2}$, $^{3}$P$_{1}$ and $^{3}$P$_{0}$, are thermalized, i.e. they
are populated according to the statistical weights. However, the $^{3}$P$_{1}$
and $^{3}$P$_{0}$ levels have relatively large critical densities:
2.0$\times$10$^5$~cm$^{-3}$ for $^{3}$P$_{1}$ and 2.9$\times$10$^4$~cm$^{-3}$
for $^{3}$P$_{0}$ at 10\,000~K, and these values drop to about half when the
electron temperature decreases to 1000~K. At physical conditions lower than
the critical densities, the $^{3}$P$_{1}$ and $^{3}$P$_{0}$ levels are
underpopulated compared to the values under thermal equilibrium.
Fig.\,\ref{neiii:frac} shows the fractional populations of the three Ne~{\sc
iii} levels as a function of electron density. The effects of the
non-equilibrium level populations of Ne~{\sc iii} on the effective
recombination coefficients for the 4f -- 3d transitions are not clear and may
vary from line to line. For the strongest 4f\,--\,3d lines that form
exclusively from recombination of target $^{3}$P$_{2}$ plus cascades, their
effective recombination coefficients will be underestimated if a thermal
equilibrium of the Ne~{\sc iii} ground levels is assumed, and that will cause
a corresponding overestimation of the derived Ne$^{2+}$/H$^+$.

Many Ne~{\sc ii} recombination lines from different multiplets have been
observed in deep spectra of PNe and H~{\sc ii} regions and ionic abundances
derived (e.g. LLB01). However, a proper
analysis of those data requires new calculations in an appropriate coupling
scheme for the strongest Ne~{\sc ii} recombination lines, especially those
belonging to the 3d\,--\,3p and 4f\,--\,3d transition arrays.

\subsection{\label{diagnose:cii}
Electron temperature from the C~{\sc ii} recombination lines}

\begin{figure}
\begin{center}
\includegraphics[width=7.0cm,angle=-90]{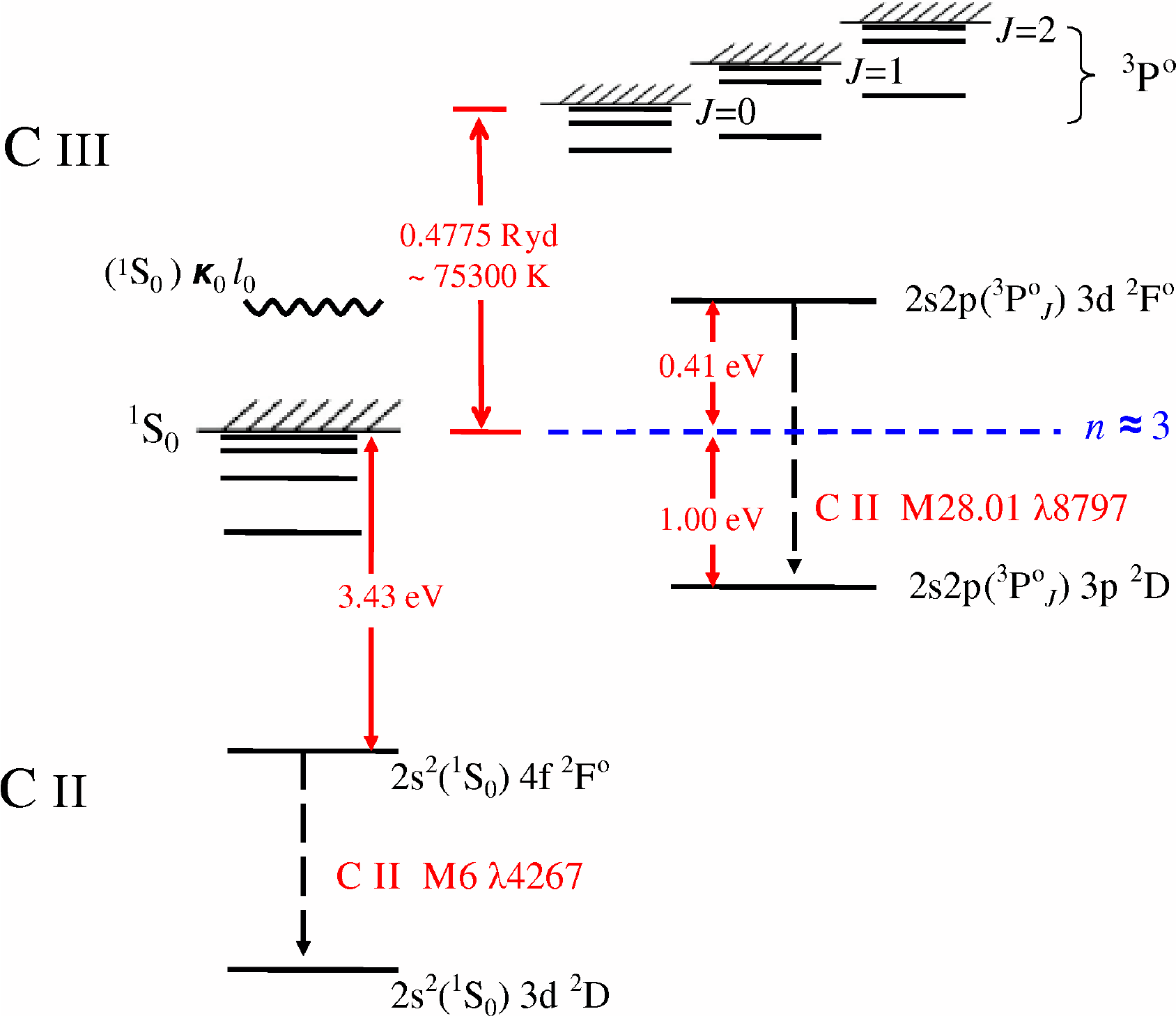}
\caption{Schematic figure showing the dielectronic recombination of C~{\sc
ii} through the autoionizing state between the ionization thresholds
2s$^{2}$\,$^{1}$S$_{0}$ and 2s2p\,$^{3}$P$^{\rm o}_{J}$ of C~{\sc iii}. The
electrons captured to the 2s2p($^{3}$P$^{\rm o}_{J}$)3d\,$^{2}$F$^{\rm o}$
autoionizing state either go back to a true continuum state
2s$^{2}$($^{1}$S$_{0}$)$\kappa_{0}\l_{0}$ through autoionization, or decay to
the 2s2p($^{3}$P$^{\rm o}_{J}$)3p\,$^{2}$D bound state through the C~{\sc ii}
M28.01 $\lambda$8797 transition. Also shown is the C~{\sc ii} M6 $\lambda$4267
radiative recombination transition between the 4f\,$^{2}$F$^{\rm o}$ and the
3d\,$^{2}$D bound states.}
\label{ciii_diele}
\end{center}
\end{figure}

Most C~{\sc ii} lines detected in the spectrum of NGC\,7009 are mainly
excited by radiative recombination, except for a few for which dielectronic
recombination dominates. Examples of the latter include the C~{\sc ii} M28.01
3d$^{\prime}$\,$^2$F$^{\rm o}$\,--\,3p$^{\prime}$\,$^2$D $\lambda$8797
multiplet, which originates from dielectronic capture of an electron to the
2s2p($^{3}$P$^{\rm o}_{J}$)3d\,$^{2}$F$^{\rm o}$ autoionization state that
lies 0.41~eV (Moore \citealt{moore1993}) above the first ionization threshold
2s$^{2}$\,$^1$S$_{0}$ and the subsequent decay to the 2s2p($^{3}$P$^{\rm
o}_{J}$)3p\,$^{2}$D bound state that lies about 1.00~eV below the ionization
threshold. Fig.\,\ref{ciii_diele} is a schematic diagram that shows the
dielectronic and radiative recombination of C~{\sc ii}.

\begin{figure}
\begin{center}
\includegraphics[width=7.0cm,angle=-90]{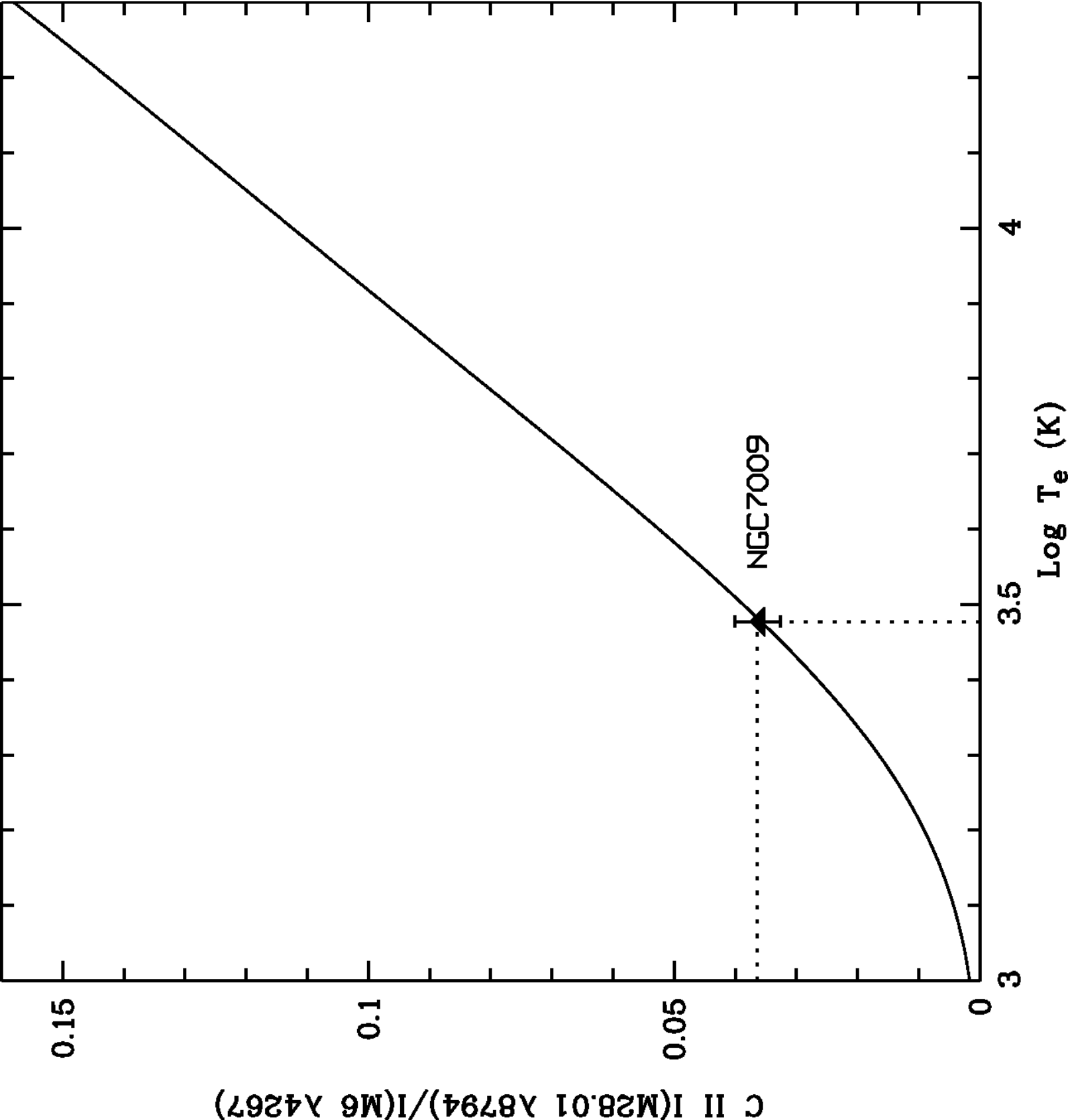}
\caption{The C~{\sc ii} $\lambda$8794/$\lambda$4267 ratio as a function
of electron temperature. The plot is based on the effective dielectronic and
radiative recombination coefficients calculated by Nussbaumer \& Storey
\citet{ns1984} and P\'{e}quignot, Petitjean \& Boisson \citet{ppb1991},
respectively. The observed C~{\sc ii} $\lambda$8794/$\lambda$4267 ratio in
NGC\,7009 yields an electron temperature of $\sim$3000$\pm$250~K. The error
bar is calculated from measurement uncertainties of the two lines.}
\label{cii_te}
\end{center}
\end{figure}

The electron on an autoionizing state either decays to another autoionizing
or bound state with the emission of radiation, or autoionizes to a true
continuum state leaving an ion and a free electron with no emission of
radiation. The latter process usually dominates, and the population of
autoionization states is close to that given by Saha and Boltzmann equations
as in the case of the local thermodynamic equilibrium (LTE). The emissivity of
a dielectronic recombination line is sensitive to electron temperature through
the Boltzmann factor $\exp{(-E/kT_\mathrm{e})}$, where $E$ is the excitation
energy of the upper state relative to the ionization threshold. By comparing
the strength of a dielectronic recombination line to that of an ordinary (i.e.
radiative recombination dominated) recombination line, whose emissivity has a
relatively weak power-law dependence on electron temperature
($\sim\,T_\mathrm{e}^{\alpha}$, where $\alpha\sim$1), one can determine the
electron temperature. The C~{\sc ii} dielectronic lines have been used to
determine electron temperatures in stellar winds of PNe (e.g. De~Marco et al.
\citealt{dsb1998}). The strongest C~{\sc ii} recombination line detected in
the spectra of nebulae is the M6 4f\,$^{2}$F$^{\rm o}$\,--\,3d\,$^{2}$D
$\lambda$4267 line, which is excited by radiative recombination only. The
upper state of the $\lambda$4267 line lies about 3.4~eV below the ionization
threshold 2s$^{2}$\,$^{1}$S$_{0}$ (see Fig.\,\ref{ciii_diele}), and its
population is far from LTE, and thus has a very different
temperature-dependence from that of the upper state of the M28.01
$\lambda$8797 transition (i.e. 3d$^{\prime}$\,$^2$F$^{\rm o}$). We use the
intensity ratio of the $\lambda$8793.80 (3d$^{\prime}$\,$^2$F$^{\rm
o}_{7/2}$\,--\,3p$^{\prime}$\,$^2$D$_{5/2}$) line, the stronger fine-structure
component of the C~{\sc ii} M28.01 multiplet, and the $\lambda$4267 line to
determine electron temperature. In NGC\,7009, this line ratio yields a
temperature of 3000~K, as shown in Fig.\,\ref{cii_te}. The atomic data used
here are the effective dielectronic and radiative recombination coefficients
of Nussbaumer \& Storey \cite{ns1984} and P\'{e}quignot, Petitjean \& Boisson
\cite{ppb1991}, respectively. Measurements of the C~{\sc ii} M28.01 lines are
presented in Section\,\ref{orls:cii:v28.01}.

\subsection{\label{diagnose:niioii}
Electron temperatures and densities from the N~{\sc ii} and O~{\sc ii}
recombination lines}

In the low-density conditions in nebulae, the relative populations of the
ground-term fine-structure levels of a recombining ion (e.g., N$^{2+}$
$^{2}$P$^{\rm o}_{1/2}$ and $^{2}$P$^{\rm o}_{3/2}$ in the case of N~{\sc ii})
vary with electron density, and this is reflected in the density dependence
of the resultant emissivities (i.e. the effective recombination coefficients)
of different recombination lines within a multiplet of the
recombined ion. Thus by comparing the intensities of two ORLs belonging to
the same multiplet but formed from different parent levels one can determine
electron density. At typical nebular conditions, emissivities of heavy element
recombination lines have only a weak, power-law dependence on electron
temperature, $\epsilon\propto{T_\mathrm{e}^{-\alpha}}$ ($\alpha\sim{1}$), and
in general, the line ratios depend very little on temperature. However, the
temperature sensitivity still differs for recombination lines decaying from
levels of different orbital angular momentum quantum number $l$, and this
difference becomes more profound if two lines of very different $l$ are
compared. Thus the intensity ratio of two lines from multiplets of different
$l$ can be used to determine electron temperature, provided that the
measurements are precise enough (e.g. Liu \citealt{liu2003}; FSL11).

\begin{figure}
\begin{center}
\includegraphics[width=7.5cm,angle=-90]{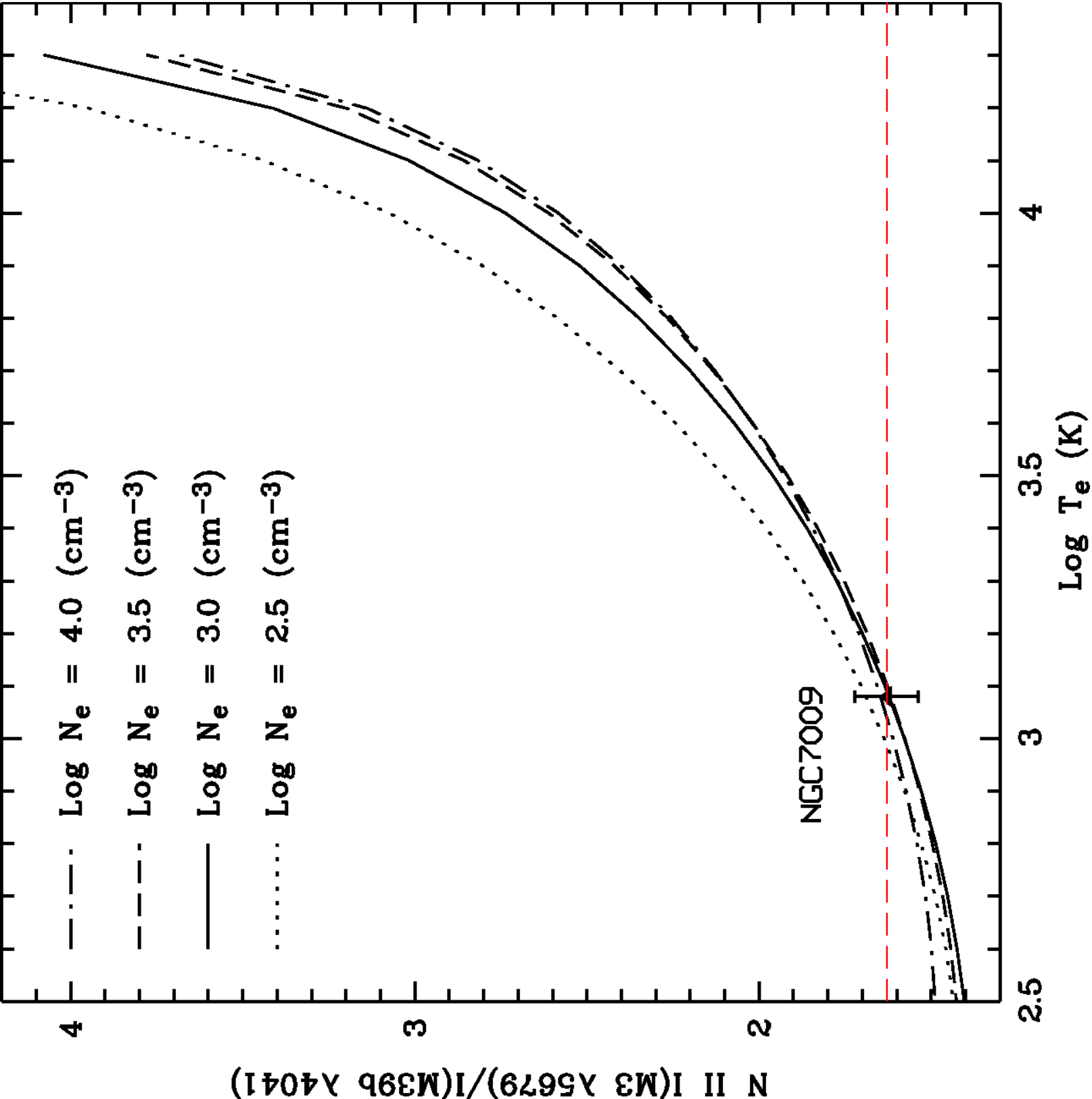}
\caption{The N~{\sc ii} $\lambda$5679/$\lambda$4041 ratio as a function of
electron temperature. Different curves represent different density cases,
and are based on the calculation of FSL11. The observed N~{\sc ii}
$\lambda$5679/$\lambda$4041 ratio in NGC\,7009 yields an electron temperature
of about 1200$\pm$200~K. The error bar was calculated from measurement
uncertainties of the two lines.}
\label{nii_te:v3v39}
\end{center}
\end{figure}

\begin{figure}
\begin{center}
\includegraphics[width=7.5cm,angle=-90]{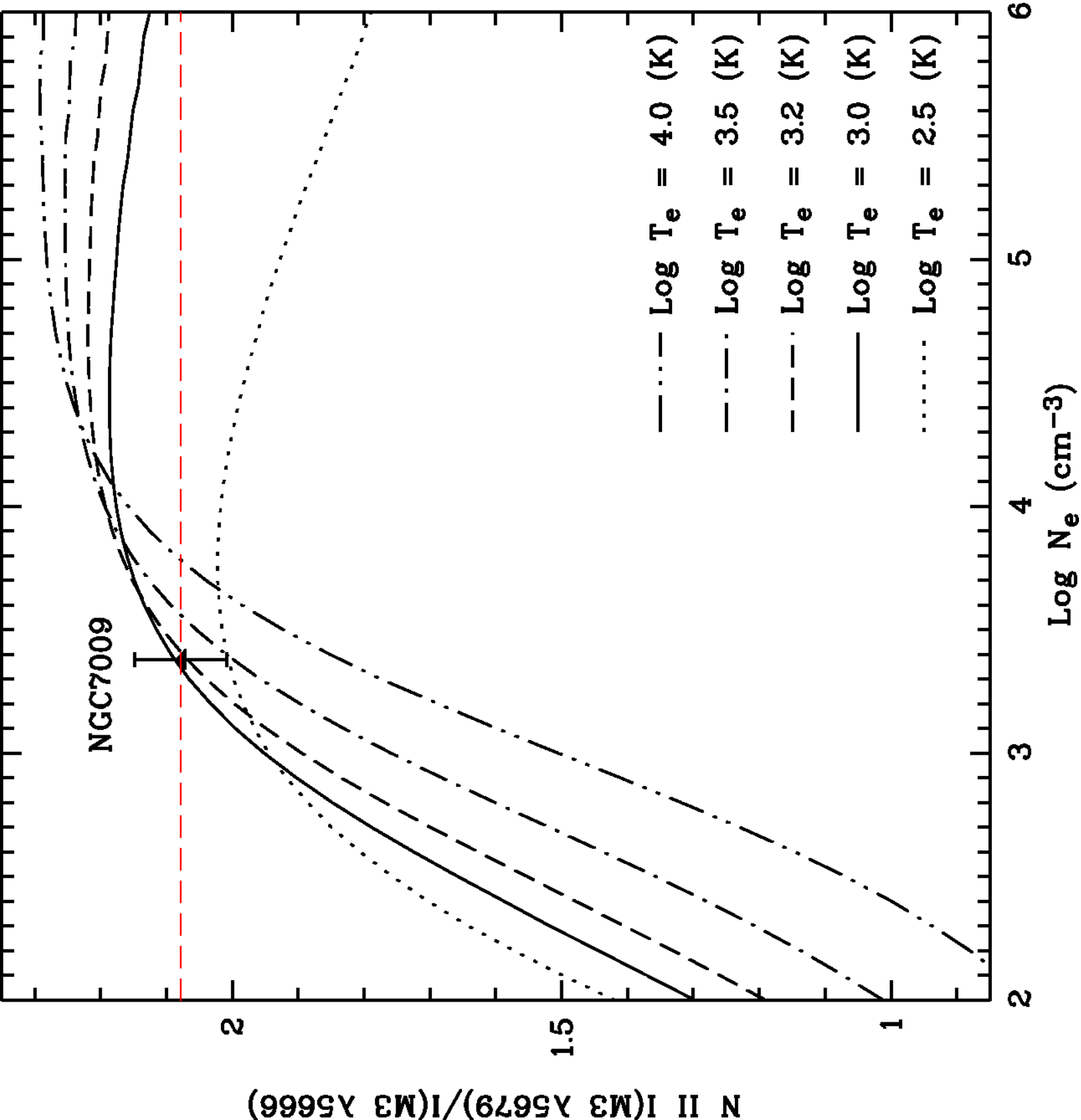}
\caption{The N~{\sc ii} $\lambda$5679/$\lambda$5666 ratio as a function of
electron density. Different curves represent different temperature cases,
and are based on the calculation of FSL11. The observed N~{\sc ii}
$\lambda$5679/$\lambda$5666 ratio in NGC\,7009 yields an electron density of
2000\,--\,3000~cm$^{-3}$. The error bar was calculated from measurement
uncertainties of the two lines.}
\label{nii_ne:v3}
\end{center}
\end{figure}

\begin{figure}
\begin{center}
\includegraphics[width=7.5cm,angle=-90]{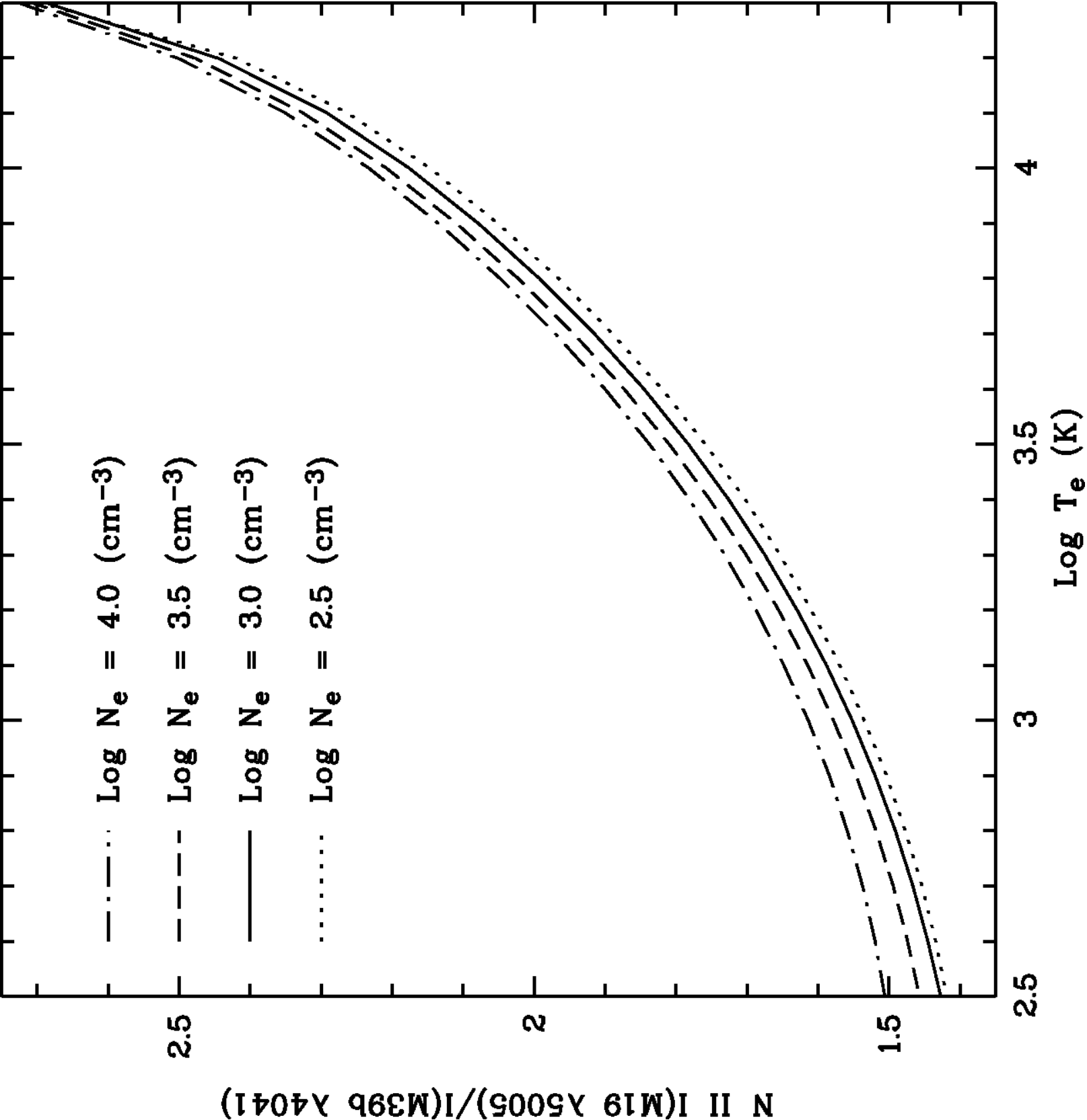}
\caption{The N~{\sc ii} $\lambda$5005/$\lambda$4041 ratio as a function of
electron temperature. Different curves represent different density cases,
and are based on the calculation of FSL11.}
\label{nii_te:v19v39}
\end{center}
\end{figure}

\begin{figure}
\begin{center}
\includegraphics[width=7.5cm,angle=-90]{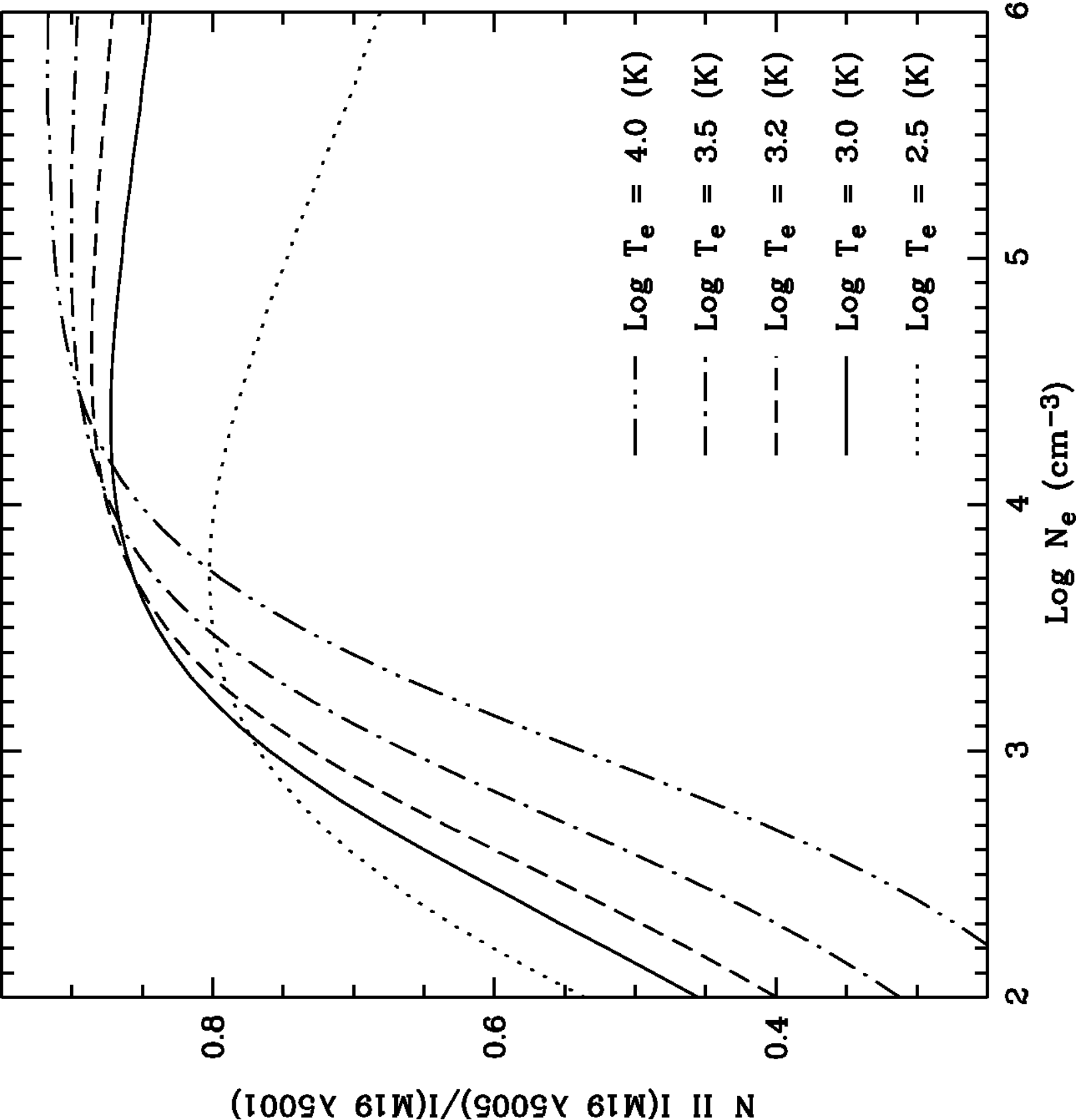}
\caption{The N~{\sc ii} $\lambda$5005/$\lambda$5001 ratio as a function of
electron density. Here the intensity of the $\lambda$5001 line is a sum of the
$\lambda$5001.48 (M19 3d\,$^{3}$F$^{\rm o}_{3}$\,--\,3p\,$^{3}$D$_{2}$) and
$\lambda$5001.14 (M19 3d\,$^{3}$F$^{\rm o}_{2}$\,--\,3p\,$^{3}$D$_{1}$) lines.
Different curves represent different temperature cases, and are based on the
calculation of FSL11.}
\label{nii_ne:v19}
\end{center}
\end{figure}

Figs.\,\ref{nii_te:v3v39} and \ref{nii_ne:v3} show that the N~{\sc ii} line
ratio $\lambda$5679/$\lambda$4041 observed in NGC\,7009 yields an electron
temperature of $\sim$1200$\pm$200~K, whereas the $\lambda$5679/$\lambda$5666
ratio yields a density of 2000\,--\,3000~cm$^{-3}$. The $\lambda$5679.56 line
is the strongest fine-structure component of the N~{\sc ii} M3
3p\,$^{3}$D\,--\,3s\,$^{3}$P$^{\rm o}$ $\lambda$5679 multiplet, and forms
exclusively from the $^2$P$^{\rm o}_{3/2}$ core capturing an electron plus
cascades from higher levels, while the second strongest line $\lambda$5666.63
can form, in addition, from recombination of the $^2$P$^{\rm o}_{1/2}$ core.
For the target N$^{2+}$, the population of the fine-structure level
$^2$P$^{\rm o}_{3/2}$ relative to $^2$P$^{\rm o}_{1/2}$ increases with
electron density due to collisional excitation, and this results in an
increase of the $\lambda$5679.56 intensity relative to the $\lambda$5666.63
line with density, as shown in Fig.\,\ref{nii:v3}. Thus the
$\lambda$5679/$\lambda$5666 ratio can be used as a density diagnostic. The
$\lambda$4041.31 line belongs to the N~{\sc ii} M39b
4f\,G[9/2]\,--\,3d\,$^3$F$^{\rm o}$ multiplet and is the strongest among the
N~{\sc ii} 4f\,--\,3d array. It forms from recombination of the $^2$P$^{\rm
o}_{3/2}$ core plus cascades from higher levels. The intensity ratio of the
$\lambda$5679.56 and $\lambda$4041.31 lines has a relatively strong temperature
dependence, and thus can serve as a temperature diagnostic. In the spectrum of
NGC\,7009, the $\lambda$5666.63 line is free of blending and is amongst the
best observed N~{\sc ii} ORLs, while the $\lambda$5679.56 and $\lambda$4041.31
lines are affected by line blending. Accurate measurements of the latter two
lines were obtained using multi-Gaussian profile fits (see
Section\,\ref{nii_orls}).
The M19 2p3d\,$^3$F$^{\rm o}$\,--\,2p3p\,$^3$D multiplet is the strongest of
the 3d\,--\,3p configuration of N~{\sc ii}. The density-dependence of the
relative emissivities of the two strongest fine-structure components of M19
is noticeable (Fig.\,\ref{nii:v19}). The intensity ratio of those components,
$\lambda$5005.15/($\lambda$5001.14\,+\,$\lambda$5001.48), may serve as
another density diagnostic. Similarly, the intensity ratio of the
$\lambda$5005.15 and M39b $\lambda$4041.31 lines may be used to determine
electron temperature. Figs.\,\ref{nii_te:v19v39} and \ref{nii_ne:v19} show
the $\lambda$5005/$\lambda$5001 and $\lambda$5005/$\lambda$4041 ratios of
N~{\sc ii} as a function of electron density and temperature, respectively.
However, accurate measurements of the N~{\sc ii} M19 lines are essentially
impossible due to the presence of the extraordinarily strong [O~{\sc iii}]
$\lambda$5007 line, which is often strongly saturated in deep spectra.

\begin{figure}
\begin{center}
\includegraphics[width=7.5cm,angle=-90]{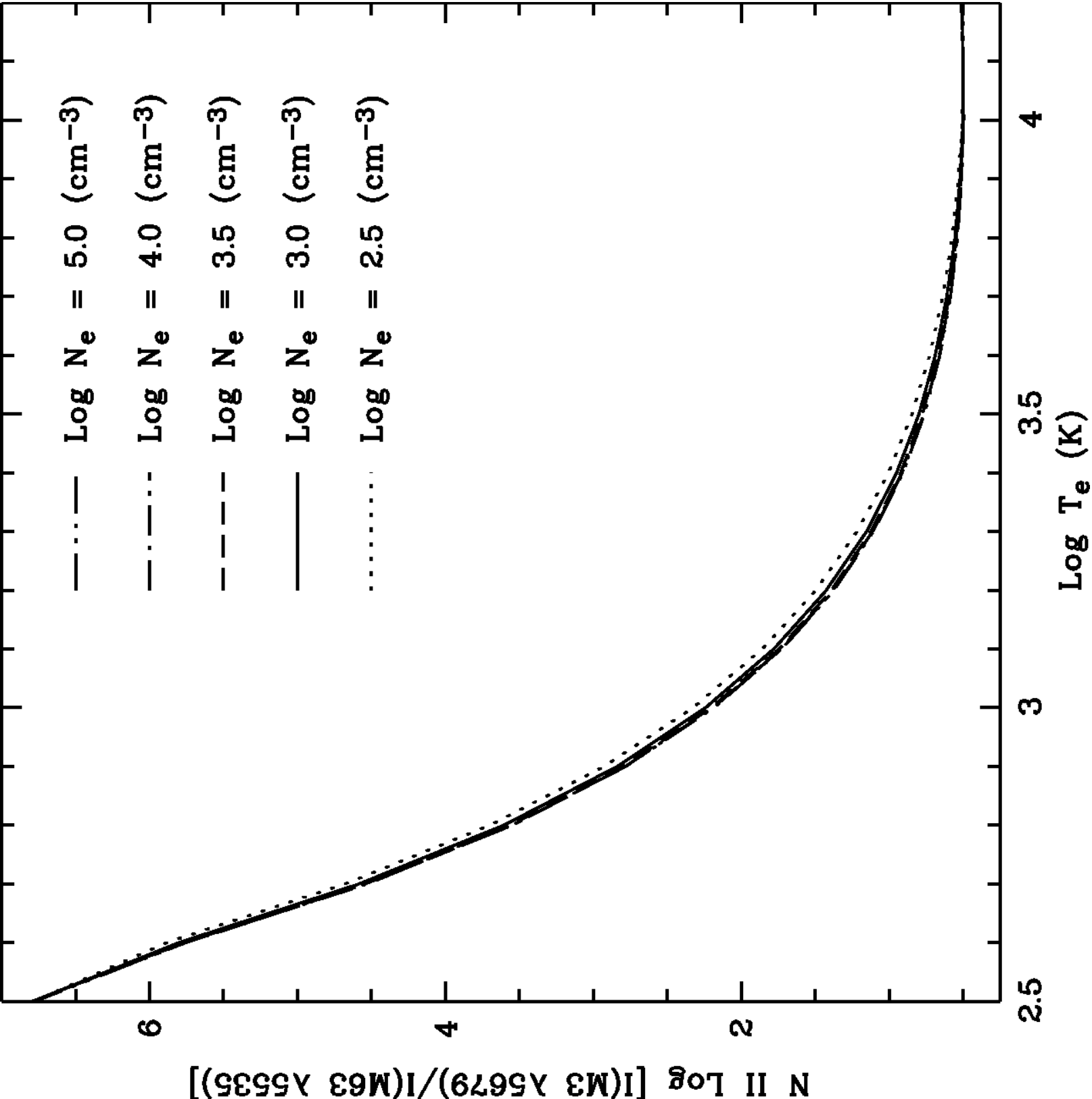}
\caption{The N~{\sc ii} $\lambda$5679/$\lambda$5535 ratio as a function of
electron temperature. Here the intensity of the $\lambda$5535 line is a sum
of the $\lambda$5535.36 (M63
3p$^{\prime}$\,$^{5}$D$^{\rm o}_{4}$\,--\,3s$^{\prime}$\,$^{5}$P$_{3}$) and
$\lambda$5535.36 (M63
3p$^{\prime}$\,$^{5}$D$^{\rm o}_{1}$\,--\,3s$^{\prime}$\,$^{1}$P$_{3}$) lines.
Different curves represent different density cases, and are based on the
calculation of FSL11.}
\label{nii_te:extra}
\end{center}
\end{figure}

Some N~{\sc ii} states of parentage other than $^{2}$P$^{\rm o}$ have energies
even higher than the 2p($^{2}$P$^{\rm o}$)4f\,G[9/2] spectral term, which is
the upper state of the M39b $\lambda$4041.31 line. The intensity ratio of an
N~{\sc ii} recombination line that originates from one of those high-energy
states to the M3 $\lambda$5679.56 line can also be used as a temperature
diagnostic. Possible candidates in the optical waveband for such application
are, e.g. the M63 3p$^{\prime}$\,$^{5}$D$^{\rm o}$\,--\,3s$^{\prime}$\,$^{5}$P,
M66 3d$^{\prime}$\,$^{5}$F\,--\,3p$^{\prime}$\,$^{5}$D$^{\rm o}$ and M72
4f$^{\prime}$\,$^{5}$G$^{\rm o}$\,--\,3d$^{\prime}$\,$^{5}$F multiplets.
According to the experimental data given by NIST\footnote{The NIST Spectra
Database\\ $http://physics.nist.gov/PhysRefData/ASD/levels_form.html$}, the
upper state of the M63 multiplet is about 1.85~eV below the ionization
threshold N~{\sc iii} $^{2}$P$^{\rm o}_{1/2}$, while the upper states of the
M66 and M72 multiplets are 0.53 and 3.67~eV, respectively, above this
threshold. The R-matrix calculation of the bound-state energy levels of
N~{\sc ii} in FSL11 only extends to about
0.45~eV (corresponding to $n\,=\,11$ in the principal series of N~{\sc ii})
below the ionization threshold. Thus only the energy levels of the
2s2p$^{2}$($^{4}$P)\,3s and 2s2p$^{2}$($^{4}$P)\,3p configurations (i.e., the
levels of the $^{5}$P, $^{3}$P, $^{3}$S$^{\rm o}$, $^{5}$D$^{\rm o}$,
$^{5}$P$^{\rm o}$, $^{3}$D$^{\rm o}$, $^{5}$S$^{\rm o}$ and $^{3}$P$^{\rm o}$
spectral terms, in the energy order given by NIST) are included in the
R-matrix calculation and the N~{\sc ii} recombination lines that originate
from those levels are precisely calculated. In principle, the intensity ratio
of the $\lambda$5679.56 and the $\lambda$5535.36 lines, the strongest
fine-structure components of the M3 3p\,$^{3}$D\,--\,3s\,$^{3}$P$^{\rm o}$
and the M63 3p$^{\prime}$\,$^{5}$D$^{\rm o}$\,--\,3s$^{\prime}$\,$^{5}$P
multiplets of N~{\sc ii}, respectively, can be used to determine electron
temperature. Fig.\,\ref{nii_te:extra} shows the $\lambda$5679/$\lambda$5535
ratio as a function of electron temperature, and this relation is quite
insensitive to electron density in the logarithmic scale. However, accurate
measurements of the $\lambda$5535.36 line is difficult due to weakness (about
10$^{4}$ times weaker than H$\beta$). We have not detected any N~{\sc ii}
lines of the parentage other than $^{2}$P$^{\rm o}$ in the deep spectrum of
NGC\,7009.

\begin{figure}
\begin{center}
\includegraphics[width=7.5cm,angle=-90]{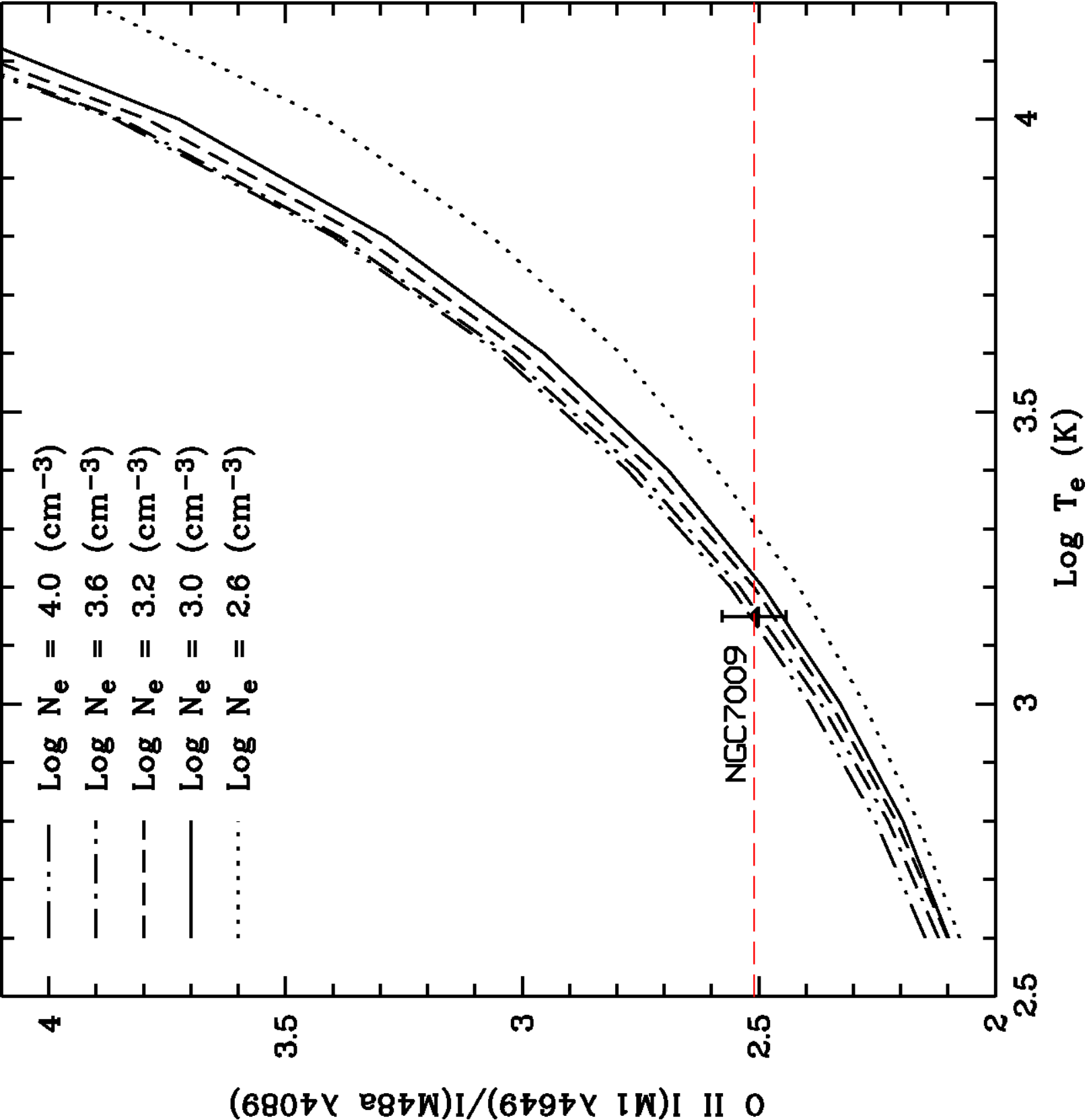}
\caption{The O~{\sc ii} $\lambda$4649/$\lambda$4089 ratio as a function of
electron temperature. Different curves represent different density cases,
and are based on the unpublished calculation of PJS. The observed O~{\sc ii}
$\lambda$4649/$\lambda$4089 ratio in NGC\,7009 yields an electron temperature
of about 1400$\pm$300~K. The error bar was calculated from measurement
uncertainties of the two lines.}
\label{oii_te:v1v48}
\end{center}
\end{figure}

\begin{figure}
\begin{center}
\includegraphics[width=7.5cm,angle=-90]{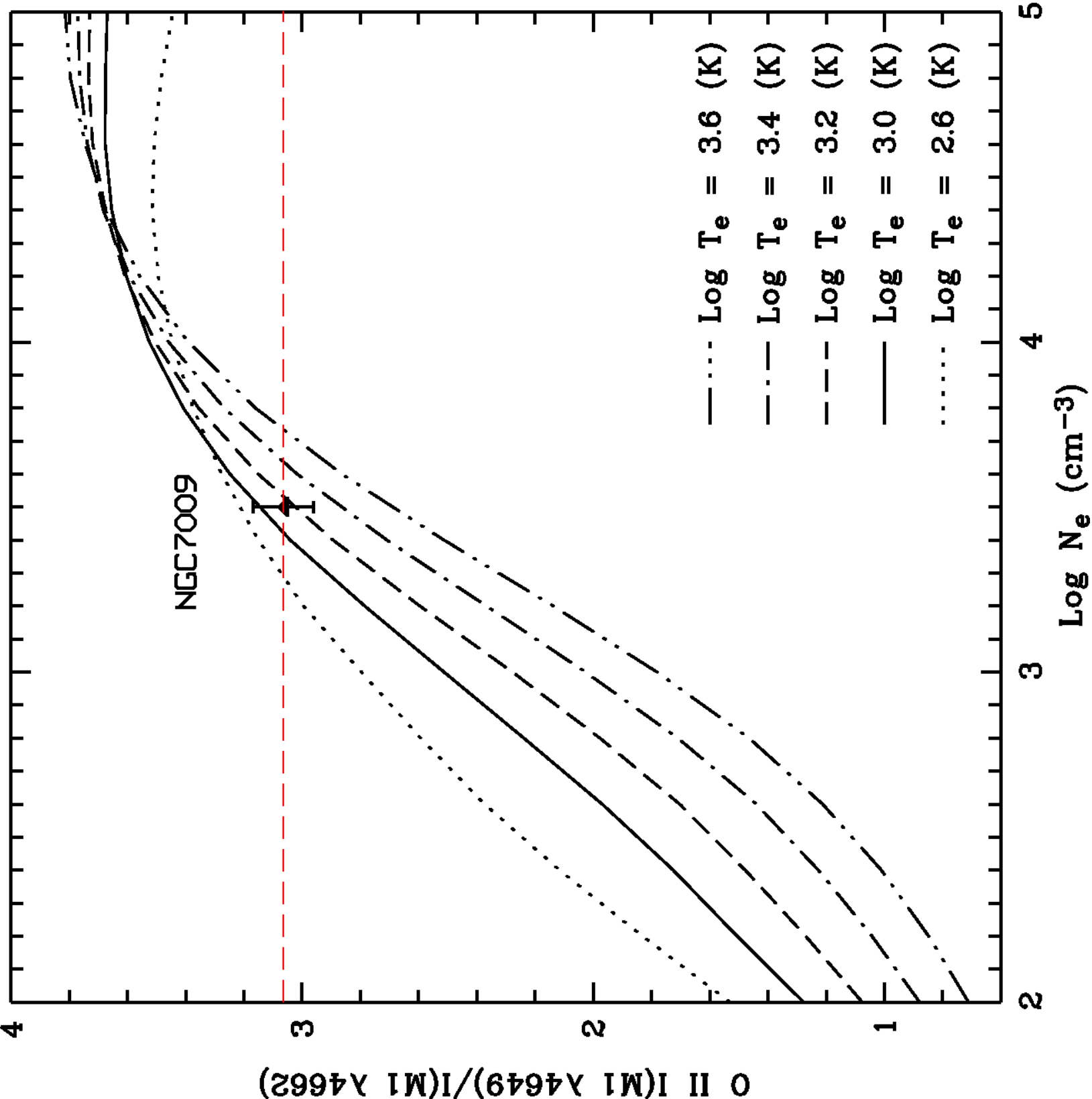}
\caption{The O~{\sc ii} $\lambda$4649/$\lambda$4662 ratio as a function of
electron density. Different curves represent different temperature cases,
and are based on the unpublished calculation of PJS. The observed O~{\sc ii}
$\lambda$4649/$\lambda$4662 ratio in NGC\,7009 yields an electron density of
2500\,--\,4000~cm$^{-3}$. The error bar was calculated from measurement
uncertainties of the two lines.}
\label{oii_ne:v1}
\end{center}
\end{figure}

The $\lambda$4649.13 line is the strongest of the O~{\sc ii} M1
3p\,$^{4}$D$^{\rm o}$\,--\,3s\,$^{4}$P multiplet, and forms only from
recombination of the $^3$P$_{2}$ core plus cascades from higher energy levels,
while another O~{\sc ii} M1 line $\lambda$4661.63 can form, in addition, from
recombination of the $^3$P$_{0}$ and $^3$P$_{1}$ cores. For the recombining
ion O$^{2+}$, the population of the fine-structure level $^3$P$_{2}$ relative
to $^3$P$_{0}$ and $^3$P$_{1}$ increases with electron density due to
collisional excitation, and so does the resultant emissivity of the
$\lambda$4649.13 line relative to the $\lambda$4661.63 line, as is shown in
Fig.\,\ref{oii:v1}. Thus the intensity ratio $\lambda$4649/$\lambda$4662 can
serve as a density diagnostic. The $\lambda$4089.29 line (M48a 4f\,G[5]$^{\rm
o}_{11/2}$\,--\,3d\,$^{4}$F$_{9/2}$) is the strongest amongst the O~{\sc ii}
4f\,--\,3d array, and forms from recombination of the $^3$P$_{2}$ core. The
intensity ratio of the $\lambda$4089.29 and the $\lambda$4649.13 lines has a
strong temperature dependence, and can be used to determine electron
temperature. Figs.\,\ref{oii_te:v1v48} and \ref{oii_ne:v1} show that the
observed O~{\sc ii} line ratios $\lambda$4649/$\lambda$4089 and
$\lambda$4649/$\lambda4662$ in NGC\,7009 yield an electron temperature of
$\sim$1400$\pm$300~K and a density of 2500\,--\,4000~cm$^{-3}$, respectively.
Although the $\lambda$4661.63 line is the third strongest in the O~{\sc ii}
M1 multiplet, it is free from line blending and thus best observed, while the
$\lambda$4649.13 and $\lambda$4089.29 lines both suffer from line blending:
the $\lambda$4649.13 line is blended with another O~{\sc ii} M1 line
$\lambda$4650.84 and the three C~{\sc iii} M1 lines $\lambda\lambda$4647.42,
4650.25 and 4651.47; the $\lambda$4089.29 line is contaminated by the Si~{\sc
iv} M1 $\lambda$4088.86 (4p\,$^{2}$P$^{\rm o}_{3/2}$\,--\,4s\,$^{2}$S$_{1/2}$)
line. Multi-Gaussian fitting was carried out to derive the intensities of the
two O~{\sc ii} ORLs, and both intensities are accurate to within 20 per cent.
Details of spectral fits are given in Section\,\ref{oii_orls}.

\begin{figure}
\begin{center}
\includegraphics[width=7.5cm,angle=-90]{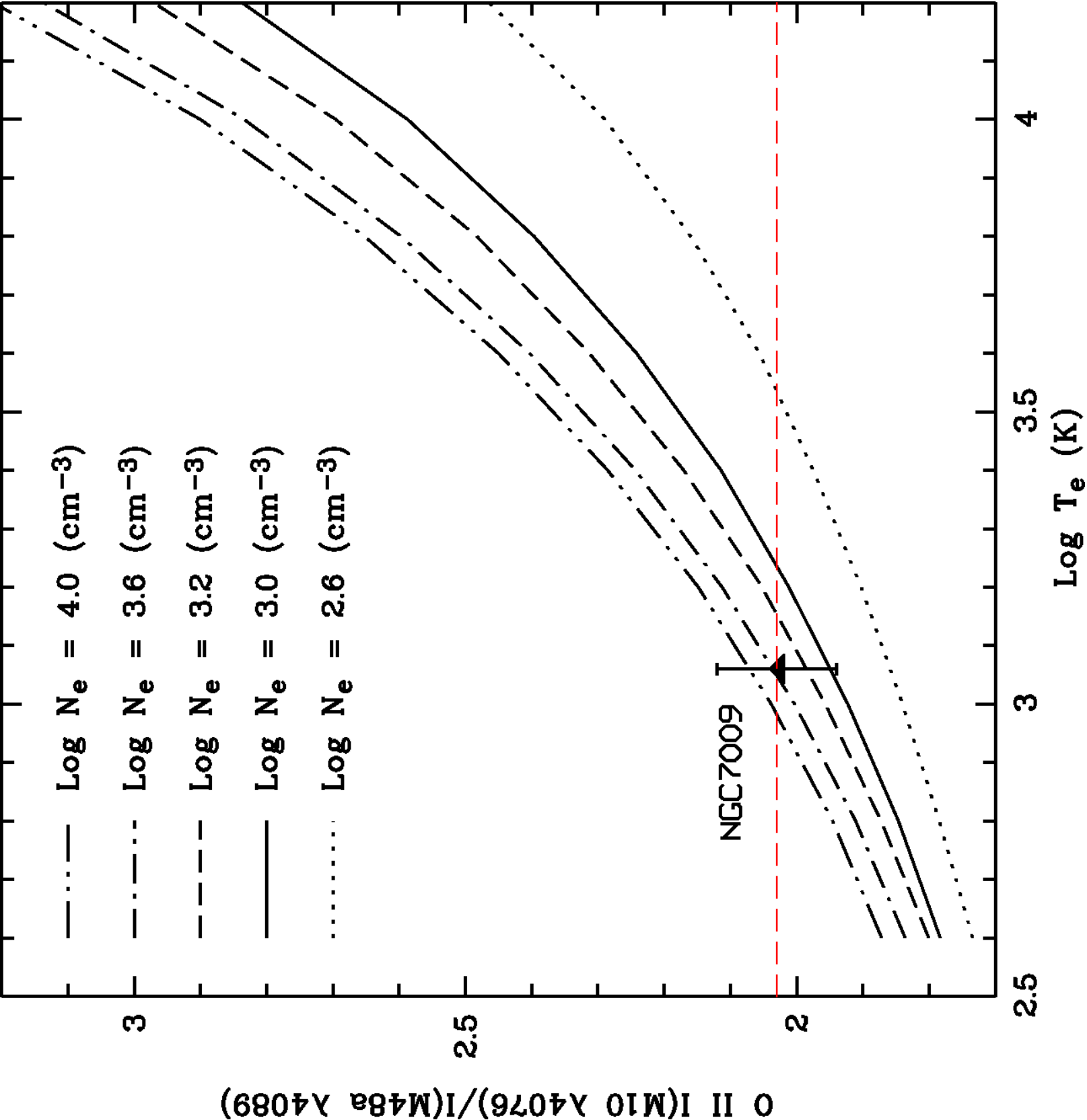}
\caption{Same as Fig.\,\ref{oii_te:v1v48} but for the O~{\sc ii}
$\lambda$4076/$\lambda$4089 ratio as a function of electron temperature.}
\label{oii_te:v10v48}
\end{center}
\end{figure}

\begin{figure}
\begin{center}
\includegraphics[width=7.5cm,angle=-90]{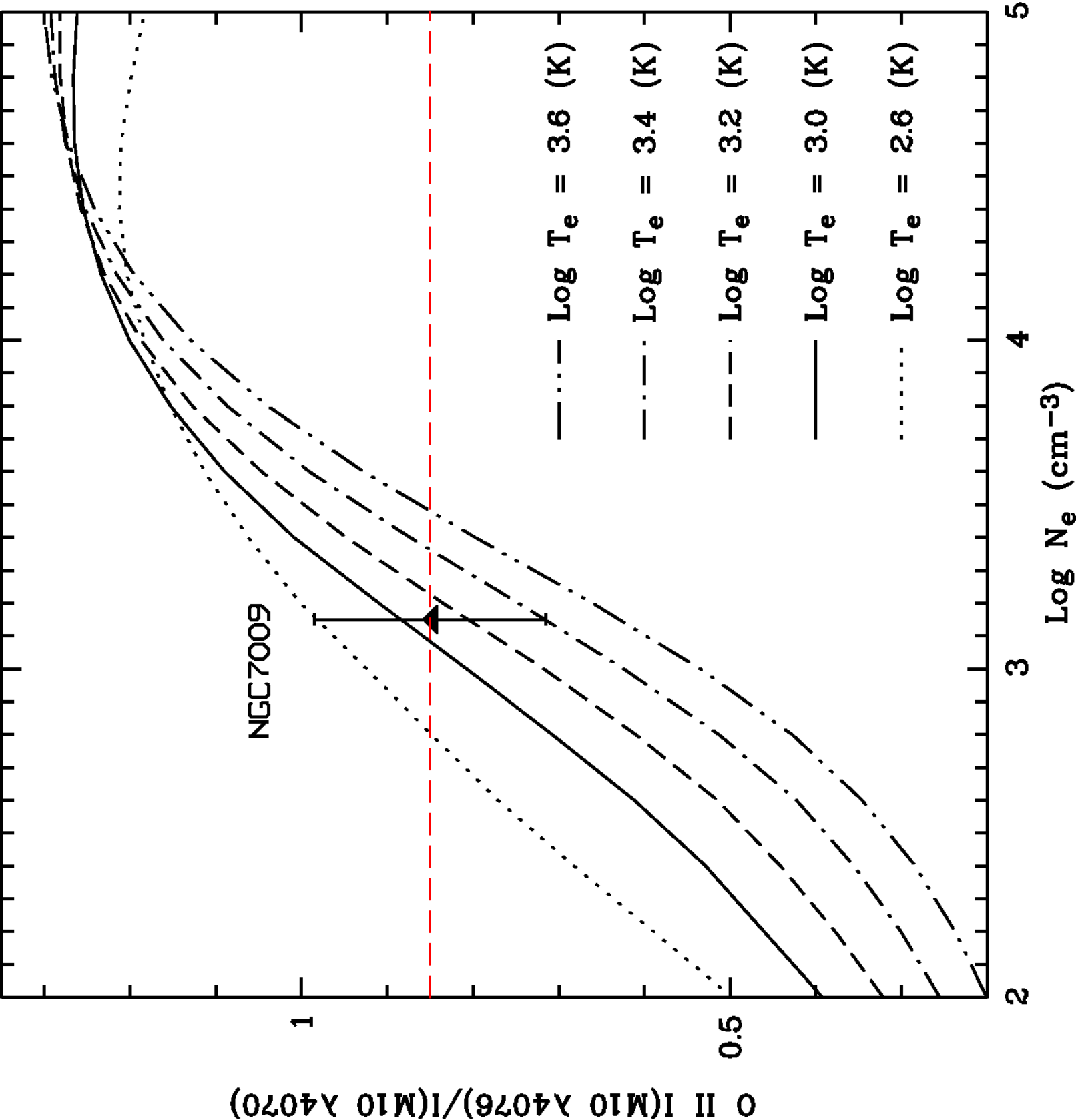}
\caption{Same as Fig.\,\ref{oii_ne:v1} but for the O~{\sc ii}
$\lambda$4076/$\lambda$4070 ratio as a function of electron density. Here
the intensity of the $\lambda$4070 line is a sum of the $\lambda$4069.89
(M10 3d\,$^{4}$F$_{5/2}$\,--\,3p\,$^{4}$D$^{\rm o}_{3/2}$) and
$\lambda$4069.62 (M10 3d\,$^{4}$F$_{3/2}$\,--\,3p\,$^{4}$D$^{\rm o}_{1/2}$)
lines.}
\label{oii_ne:v10}
\end{center}
\end{figure}

The M10 3d\,$^4$F\,--\,3p\,$^4$D$^{\rm o}$ $\lambda$4075 multiplet is the
strongest transition of the 3d\,--\,3p configuration of O~{\sc ii}. Given
the opposite trends of the fractional intensities of the $\lambda$4075.86
and $\lambda$4069.89,62 lines, the three fine-structure components of M10,
as a function of electron density, as shown in Fig.\,\ref{oii:v10}, the
intensity ratio of the two lines may serve as another density diagnostic.
Here the $\lambda$4075.86 line is the strongest component of the M10
multiplet. The intensity ratio of the $\lambda$4075.86 line and the
$\lambda$4089.29 line, the strongest fine-structure component of the M48a
4f\,G[5]$^{\rm o}$\,--\,3d\,$^{4}$F multiplet of O~{\sc ii}, can be used as
another temperature diagnostic. Figs.\,\ref{oii_te:v10v48} and \ref{oii_ne:v10}
show the $\lambda$4076/$\lambda$4089 and $\lambda$4076/$\lambda$4070 as a
function of electron temperature and density, respectively. Here the
intensity of the $\lambda$4070 line is a sum of the $\lambda$4069.89 (M10
3d\,$^{4}$F$_{5/2}$\,--\,3p\,$^{4}$D$^{\rm o}_{3/2}$) and $\lambda$4069.62
(M10 3d\,$^{4}$F$_{3/2}$\,--\,3p\,$^{4}$D$^{\rm o}_{1/2}$) lines. If we assume
a density of about 4300~cm$^{-3}$, as derived from CEL ratios (Paper~I), the
electron temperature deduced from the O~{\sc ii} $\lambda$4076/$\lambda$4089
ratio is 1150$\pm$300~K for NGC\,7009. The electron density derived from the
O~{\sc ii} $\lambda$4076/$\lambda$4070 ratio is of large uncertainty, due to
the relatively large measurement uncertainties of the two lines. The
$\lambda$4075.86 line is blended with [S~{\sc ii}] $\lambda$4076.35
(3p$^{3}$\,$^{2}$P$^{\rm o}_{1/2}$\,--\,$^{4}$S$^{\rm o}_{3/2}$) line, while
the $\lambda$4069.89,62 line is blended with the [S~{\sc ii}]
$\lambda$4068.60 (3p$^{3}$\,$^{2}$P$^{\rm o}_{3/2}$\,--\,$^{4}$S$^{\rm
o}_{3/2}$) line and the three C~{\sc iii} M16
5g\,$^{3}$G\,--\,4f\,$^{3}$F$^{\rm o}$ lines $\lambda\lambda$4067.94, 4068.92
and 4070.31. Multi-Gaussian profile fitting was carried out to obtain line
fluxes (c.f. Section\,\ref{oii_orls:v10}).

\begin{figure}
\begin{center}
\includegraphics[width=7.5cm,angle=0]{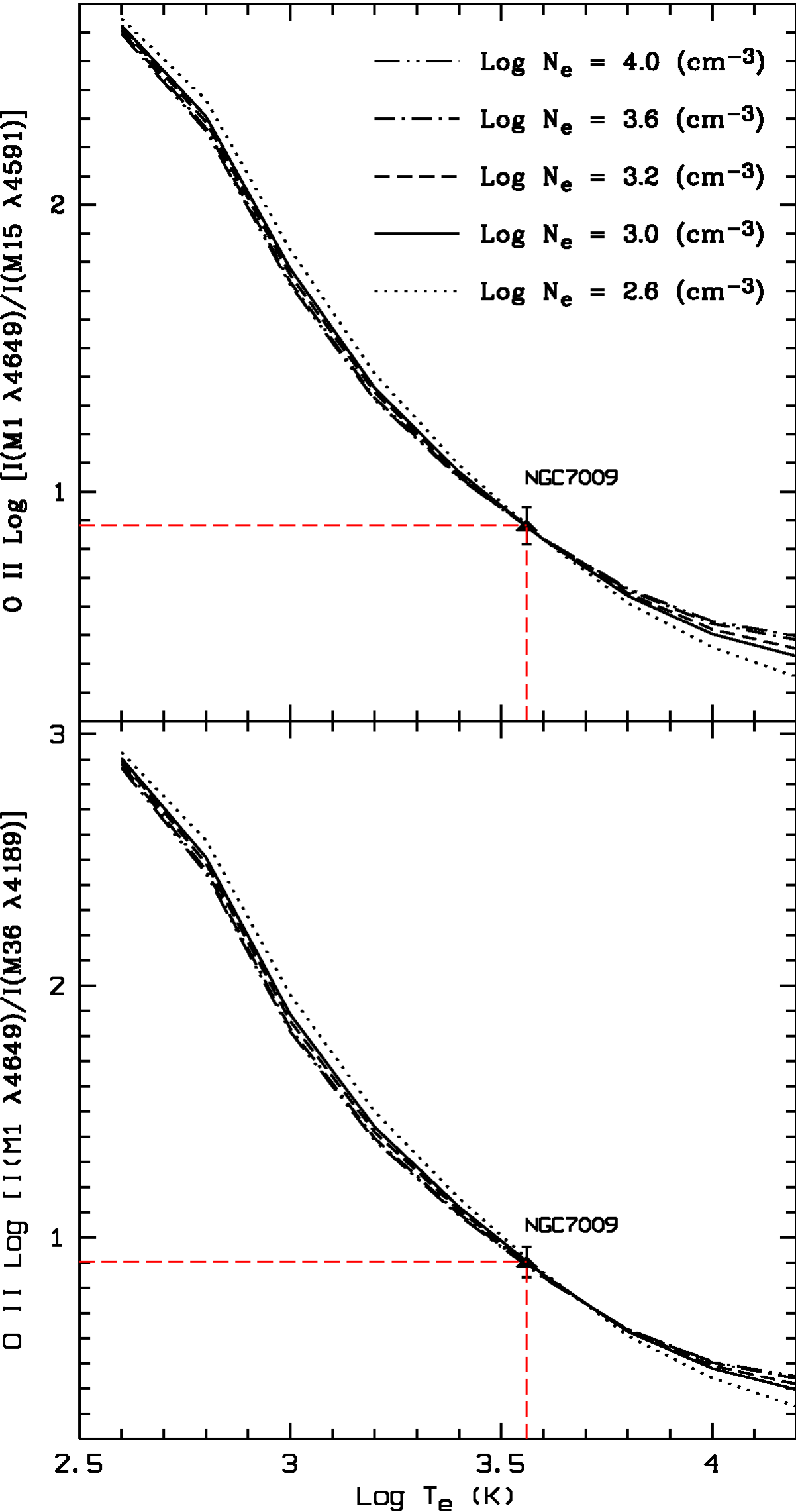}
\caption{The O~{\sc ii} $\lambda$4649/$\lambda$4591 ($upper$) and
$\lambda$4649/$\lambda$4189 ($lower$) ratios as a function of electron
temperature. Here the intensity of the $\lambda$4189 line is a sum of the
$\lambda$4189.79 (M36
3d$^{\prime}$\,$^{2}$G$_{9/2}$\,--\,3p$^{\prime}$\,$^{2}$F$^{\rm o}_{7/2}$)
and $\lambda$4189.59 (M36
3d$^{\prime}$\,$^{2}$G$_{7/2}$\,--\,3p$^{\prime}$\,$^{2}$F$^{\rm o}_{7/2}$)
lines. Different curves represent different density cases, and are based on
the unpublished calculation of PJS. Both O~{\sc ii} line ratios observed in
the spectrum of NGC\,7009 yield electron temperatures close to 3600$\pm$500~K.
The error bars were calculated from measurement uncertainties of the lines.}
\label{oii_te:extra}
\end{center}
\end{figure}

Several O~{\sc ii} recombination lines with parentage other than $^{3}$P
have been detected in the spectrum of NGC\,7009. These lines belong to the
M15 3p$^{\prime}$\,$^{2}$F$^{\rm o}$\,--\,3s$^{\prime}$\,$^{2}$D,
M36 3d$^{\prime}$\,$^{2}$G\,--\,3p$^{\prime}$\,$^{2}$F$^{\rm o}$,
M101 4f$^{\prime}$\,H[5]$^{\rm o}$\,--\,3d$^{\prime}$\,$^{2}$G, and
M105 4f$^{\prime}$\,P[1]$^{\rm o}$\,--\,3d$^{\prime}$\,$^{2}$S multiplets of
O~{\sc ii}. According to the experimental data from NIST, the upper states
of the M15 and M36 multiplets are 6.76 and 3.80~eV, respectively, below the
ionization threshold O~{\sc iii} $^{3}$P$_{0}$, while the upper states of
the M101 and M105 multiplets are about 0.89 and 0.87~eV, respectively, below
this threshold. In the most recent calculation of PJS for the O~{\sc ii}
effective recombination coefficients, only the transitions between the levels
with principal quantum number $n\leq$6 (i.e. corresponding to 1.50~eV below
the ionization threshold $^{3}$P$_{0}$) were presented. Thus only the
effective recombination coefficients of the M15 and M36 lines are available.
The strongest fine-structure components of the M15 and M36 multiplets are
$\lambda$4590.97 (3p$^{\prime}$\,$^{2}$F$^{\rm
o}_{7/2}$\,--\,3s$^{\prime}$\,$^{2}$D$_{5/2}$) and $\lambda$4189.79
(3d$^{\prime}$\,$^{2}$G$_{9/2}$\,--\,3p$^{\prime}$\,$^{2}$F$^{\rm o}_{7/2}$),
respectively, and both lines are detected in the deep spectrum of NGC\,7009,
as shown in Figs.\,\ref{4555-4625} and \ref{4176-4260}. Although the
effective recombination coefficients of the M101 and M105 multiplets are not
available, the strongest fine-structure component of the M101 multiplet, the
$\lambda$4253.90 (4f$^{\prime}$\,H[5]$^{\rm
o}_{11/2}$\,--\,3d$^{\prime}$\,$^{2}$G$_{9/2}$) line, is also detected
(Fig.\,\ref{4176-4260}). Fig.\,\ref{oii_te:extra} shows the O~{\sc ii} line
ratios $\lambda$4649.13/$\lambda$4590.97 and
$\lambda$4649.13/$\lambda$4189.79 as a function of electron temperature. The
figure also shows that the two line ratio--temperature relations are
insensitive to electron density, indicating that they are good temperature
diagnostics. Both line ratios detected in the spectrum of NGC\,7009 yield
electron temperatures close to 3600~K.

\subsection{\label{diagnose:neii}
The Ne~{\sc ii} recombination lines as potential plasma diagnostics}

\begin{figure}
\begin{center}
\includegraphics[width=7.5cm,angle=0]{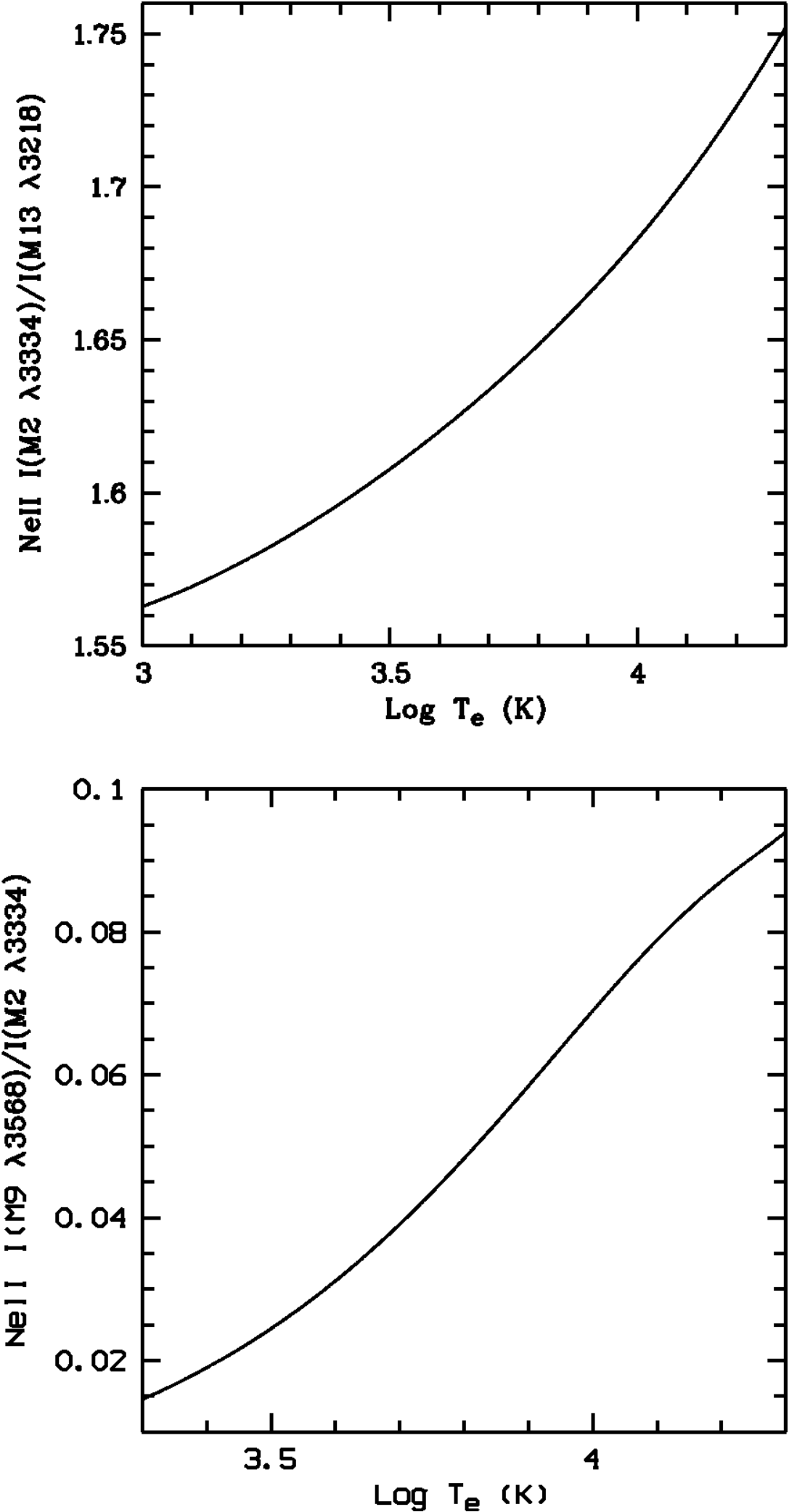}
\caption{The Ne~{\sc ii} recombination line ratios as a function of electron
temperature. $Upper$: The $\lambda$3334/$\lambda$3218 ratio. $Lower$: The
$\lambda$3568/$\lambda$3334 ratio. The plot is based on the effective
recombination coefficients of Kisielius et al. \citet{kisielius1998}. Only a
density case of 10\,000~cm$^{-3}$ is presented.}
\label{neii_te}
\end{center}
\end{figure}

So far no efforts have been attempted for plasma diagnostics based on the
Ne~{\sc ii} recombination spectrum, partly due to the lack of suitable
effective recombination coefficients. Since all Ne~{\sc ii} effective
recombination coefficients were calculated under the {\it LS}\,coupling
scheme, and relative populations of the $^{3}$P$_{2}$, $^{3}$P$_{1}$ and
$^{3}$P$_{0}$ parent levels were assumed to be proportional to the statistical
weights, no density diagnostic is possible with the current available atomic
data. However, the Ne~{\sc ii} recombination line ratios may still serve as
temperature diagnostics using the effective recombination coefficients of
Kisielius et al. \cite{kisielius1998}. The M2 3p~$^{4}$D$^{\rm
o}$\,--\,3s~$^{4}$P $\lambda$3337 multiplet is the strongest transition of
the 3p\,--\,3s configuration, and M13 3d~$^{4}$F\,--\,3p~$^{4}$D$^{\rm o}$
$\lambda$3220 is the strongest multiplet of the 3d\,--\,3p configuration of
Ne~{\sc ii}. The intensity ratio of the strongest fine-structure components
of those two multiplets, $\lambda$3334.84/$\lambda$3218.19, may be serve as
a temperature diagnostic, as shown in the upper panel of Fig.\,\ref{neii_te}.
In order to obtain reliable electron temperature, the measurement
uncertainty of the $\lambda$3334/$\lambda$3218 ratio needs to be less than
10 per cent, which is very demanding to achieve.

Recombination of the Ne$^{2+}$ $^1$D core plus cascades gives rise to another
series of Ne~{\sc ii} recombination lines, and the strongest multiplet of
this series is M9 3p$^{\prime}$\,$^2$F$^{\rm o}$\,--\,3s$^{\prime}$\,$^2$D
$\lambda$3571. The intensity ratio of the $\lambda$3568.50 line, the strongest
fine-structure component of the M9 multiplet, and the M2 $\lambda$3334.84 line
may be used as another temperature diagnostic, as shown in the lower panel of
Fig.\,\ref{neii_te}. The calculation of Kisielius et al. \cite{kisielius1998}
shows that the $\lambda$3568/$\lambda$3334 ratio is only marginally sensitive
to electron temperature. In order to derive reliable temperature, the line
ratio (especially the $\lambda$3568 line) needs to be measured to a very high
accuracy level. Although the temperature range considered in the calculation
of Kisielius et al. \cite{kisielius1998} is from 1000 to 20\,000~K, the
current analytic fit to the effective recombination coefficient for the
$\lambda$3568 line is only valid for 2000\,--20\,000~K. As a consequence, the
usage of the diagnostic curve of the $\lambda$3568/$\lambda$3334 ratio in
Fig.\,\ref{neii_te} outside this temperature range is not recommended. The
Ne~{\sc ii} $\lambda$3334/$\lambda$3218 and $\lambda$3568/$\lambda$3334
ratios observed in NGC\,7009 are 1.86 and 0.40, respectively, both falling
outside the diagnostic ranges of Fig.\,\ref{neii_te}.

\subsection{\label{diagnose:ciii}
The C~{\sc iii}, N~{\sc iii} and O~{\sc iii} recombination lines as potential
temperature diagnostics}

In this Section, we discuss the possibility of using the C~{\sc iii}, N~{\sc
iii} and O~{\sc iii} optical recombination line ratios to determine electron
temperatures. Although some of those lines are detected in the spectrum of
NGC\,7009, they are not used for plasma diagnostics in the current paper,
due to the lack of adequate atomic data. Unless otherwise specified, the
effective recombination coefficients of Nussbaumer \& Storey \cite{ns1984}
and P\'{e}quignot, Petitjean \& Boisson \cite{ppb1991} are used to create
the diagnostic curves.

\begin{figure}
\begin{center}
\includegraphics[width=7.5cm,angle=0]{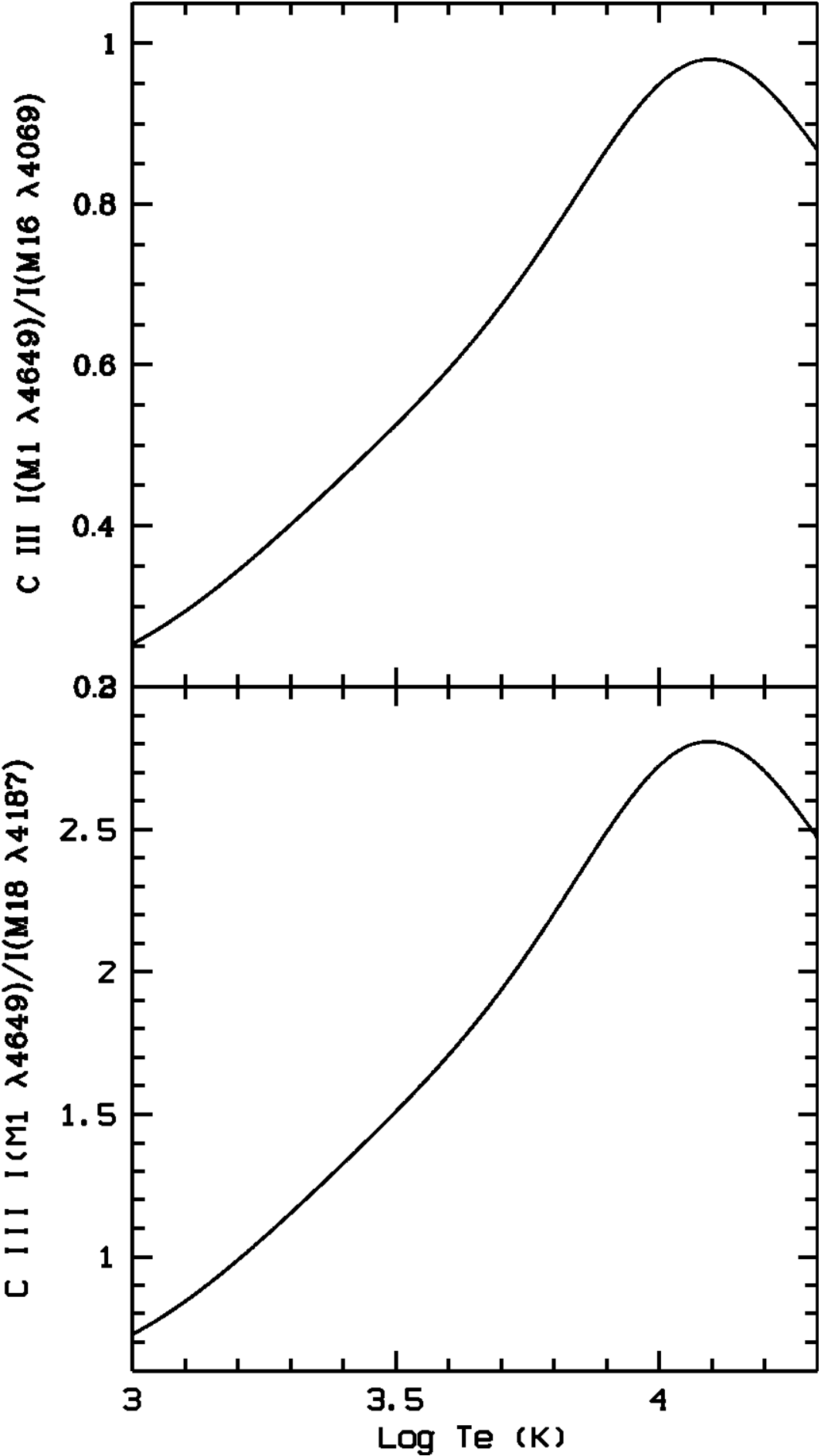}
\caption{The C~{\sc iii} recombination line ratios as a function of electron
temperature. $Upper$: The $I$(M1~$\lambda$4649)/$I$(M16~$\lambda$4069) ratio.
$Lower$: The $I$(M1~$\lambda$4649)/$I$(M18~$\lambda$4187) ratio. The figure is
based on the effective dielectronic and radiative recombination coefficients
calculated by Nussbaumer \& Storey \citet{ns1984} and P\'{e}quignot, Petitjean
\& Boisson \citet{ppb1991}, respectively.}
\label{ciii_te}
\end{center}
\end{figure}

The C~{\sc iii} lines are excited by recombination only, and the ratios of
the best observed lines can be used as temperature probes. The intensity
ratio of the M1 $\lambda$4649 (3p\,$^{3}$P$^{\rm o}$\,--\,3s\,$^{3}$S) and
M16 $\lambda$4069 (5g\,$^{3}$G\,--\,4f\,$^{3}$F$^{\rm o}$) multiplets of
C~{\sc iii} is sensitive to electron temperature, as shown in the upper panel
of Fig.\,\ref{ciii_te}. The intensity ratio of the C~{\sc iii} triplet M1
$\lambda$4649 and singlet M18 $\lambda$4187
(5g\,$^{1}$G\,--\,4f\,$^{1}$F$^{\rm o}$) can also be used to determine
electron temperature, as shown in the lower panel of Fig.\,\ref{ciii_te}.
However, those C~{\sc iii} lines all suffer from line blending. The
$\lambda$4187 line is blended with the O~{\sc ii} M36 $\lambda$4185.45
(3d$^{\prime}$\,$^{2}$G$_{7/2}$\,--\,3p$^{\prime}$\,$^{2}$F$^{\rm o}_{5/2}$)
line, but its intensity can be measured to a high accuracy using multi-Gaussian
profile fitting (Fig.\,\ref{4176-4260}). The C~{\sc iii} M1 $\lambda$4649
triplets are blended with the O~{\sc ii} M1 3p\,$^{4}$D$^{\rm
o}$\,--\,3s\,$^{4}$P lines $\lambda\lambda$4649.13 and 4650.84
(Fig.\,\ref{4625-4680}), and the C~{\sc iii} M16 $\lambda$4069 triplets are
blended with three O~{\sc ii} M10 3d\,$^{4}$F\,--\,3p\,$^{4}$D$^{\rm o}$
lines and the [S~{\sc ii}] $\lambda$4068.60 line (Fig.\,\ref{4060-4115}).
Intensities of the C~{\sc iii} M1 and M16 multiplets are obtained from
multi-Gaussian profile fitting. The intensity ratio of the fine-structure
components of each multiplet was assumed to be as in {\it LS}\,coupling
(Sections\,\ref{oii_orls:v1} and \ref{oii_orls:v10}). In NGC\,7009, the
C~{\sc iii} $I$(M1~$\lambda$4649)/$I$(M16~$\lambda$4069) and
$I$(M1~$\lambda$4649)/$I$(M18~$\lambda$4187) ratios are 1.05 and 3.41,
respectively. Both ratios are beyond the diagnostic ranges of
Fig.\,\ref{ciii_te}.

\begin{figure}
\begin{center}
\includegraphics[width=8.0cm,angle=0]{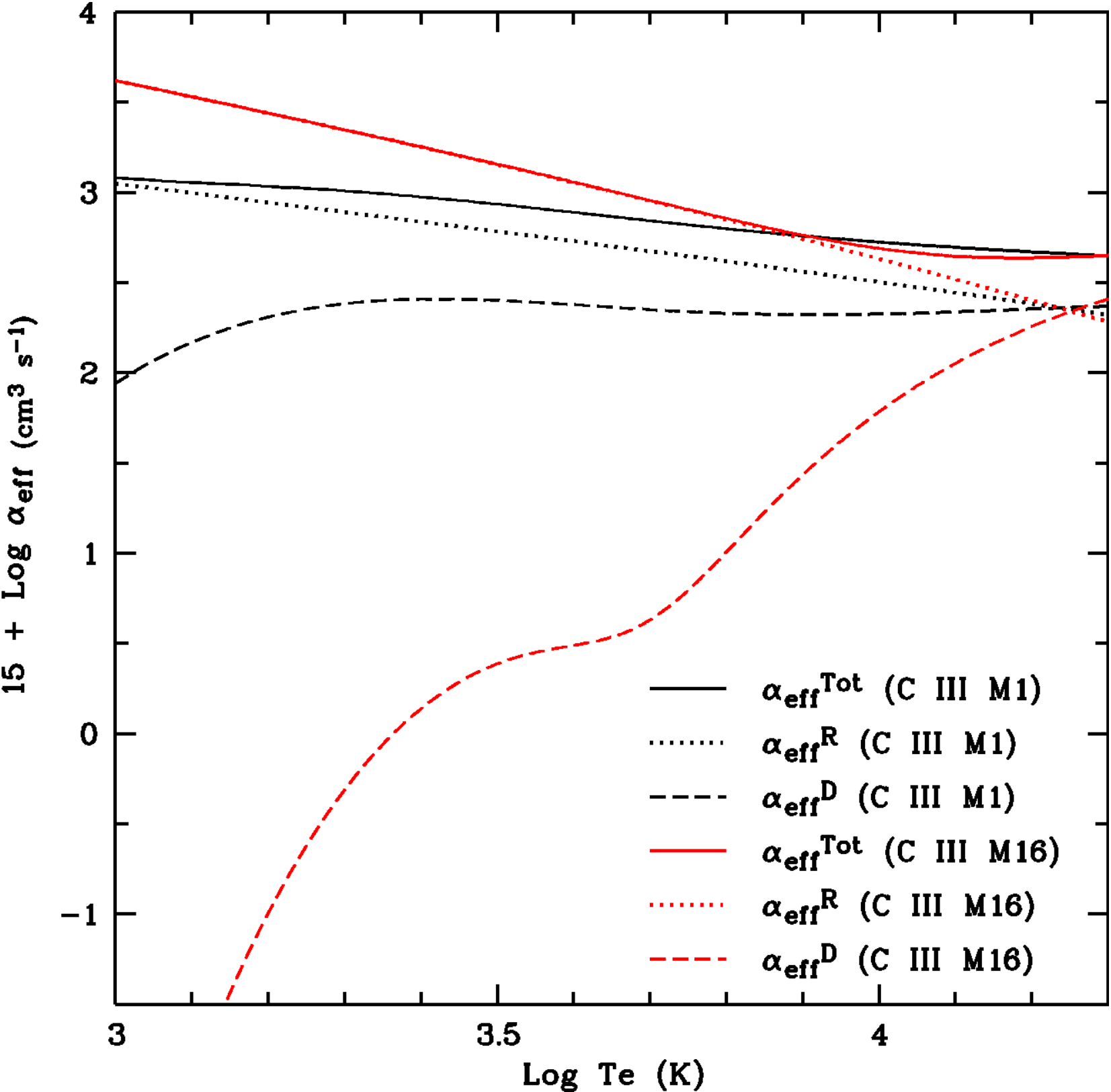}
\caption{The effective recombination coefficients of the C~{\sc iii} M1
$\lambda$4649 (black curves) and M16 $\lambda$4069 (red curves) multiplets
as a function of electron temperature. Different curves of the same color
represent different recombination coefficients: radiative recombination
coefficient $\alpha_{\rm eff}^{\rm R}$ (dotted line), dielectronic
recombination coefficient $\alpha_{\rm eff}^{\rm D}$ (dashed line), and total
effective recombination coefficient $\alpha_{\rm eff}^{\rm Tot}$ (solid line).
Data source of the plot is the same as Fig.\,\ref{ciii_te}.}
\label{ciii_coeff}
\end{center}
\end{figure}

Fig.\,\ref{ciii_te} shows that the C~{\sc iii}
$I$(M1~$\lambda$4649)/$I$(M16~$\lambda$4069) and
$I$(M1~$\lambda$4649)/$I$(M18~$\lambda$4187) ratios increase with electron
temperature below 10\,000~K, but both decrease when the temperature goes
beyond $\sim$12\,600~K ($\log{T_\mathrm{e}}\sim$\,4.1). In order to explain
those trends, the effective radiative ($\alpha_{\rm eff}^{\rm R}$) and
dielectronic ($\alpha_{\rm eff}^{\rm D}$) recombination coefficients as well
as the total effective recombination coefficients ($\alpha_{\rm eff}^{\rm
Tot}$) of the C~{\sc iii} M1 and M16 multiplets are shown in
Fig.\,\ref{ciii_coeff} as a function of electron temperature. Below 10\,000~K,
the C~{\sc iii} M16 multiplet is dominated by radiative recombination. In this
temperature regime, the total effective recombination coefficient of the
C~{\sc iii} M16 multiplet decreases much faster than that of the M1 multiplet
as temperature increases. When the temperature goes above 10\,000~K, the
decreasing rate of the M16 multiplet slows down because its monotonically
increasing dielectronic recombination coefficient becomes relatively
significant, while that of the M1 multiplet does not change much.

\begin{figure}
\begin{center}
\includegraphics[width=7.0cm,angle=-90]{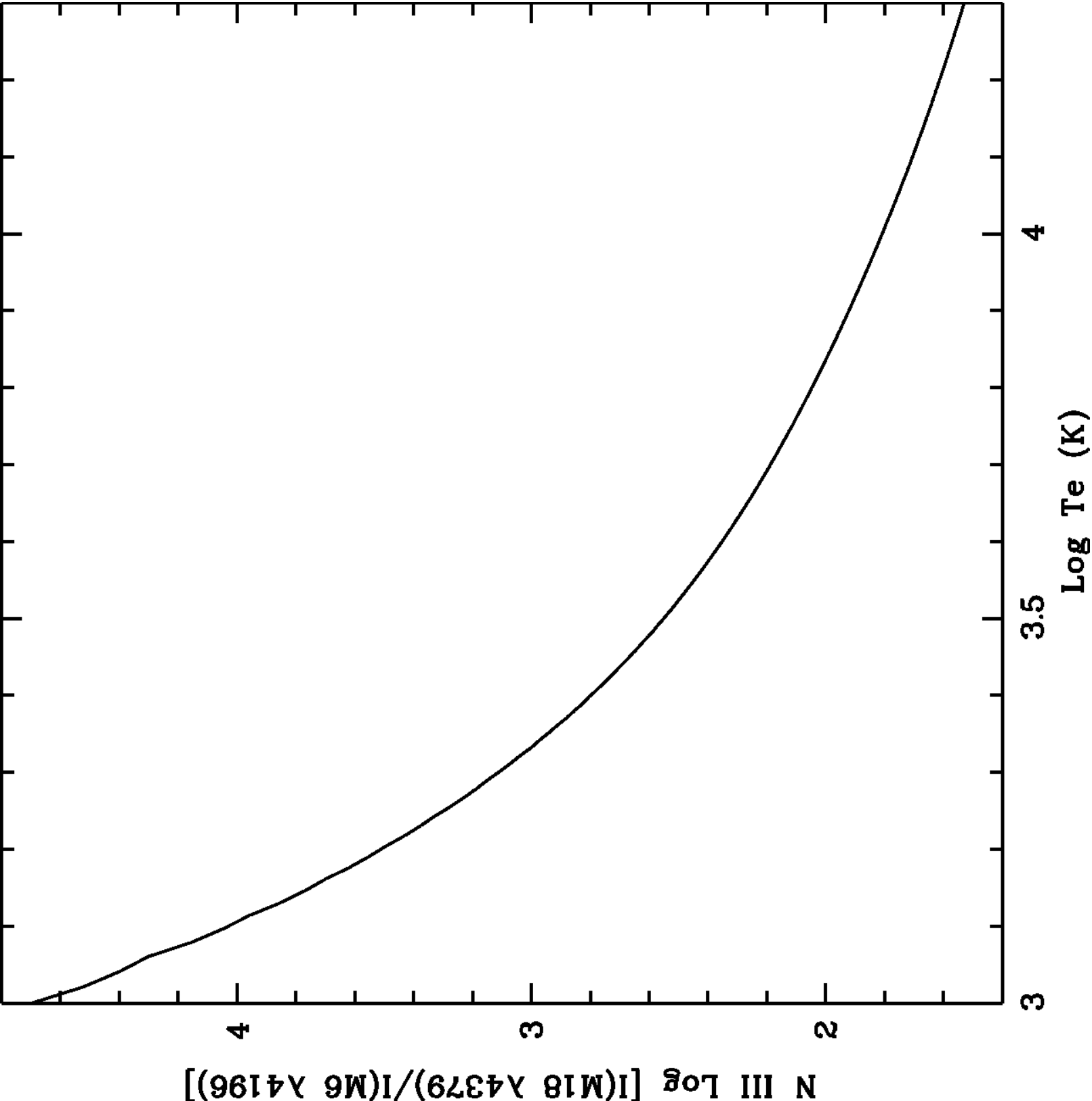}
\caption{The N~{\sc iii} $I$(M18~$\lambda$4379)/$I$(M6~$\lambda$4196) ratio
as a function of electron temperature. Data source of the figure is the same
as Fig.\,\ref{ciii_te}.}
\label{niii_te}
\end{center}
\end{figure}

The most prominent permitted transitions of N~{\sc iii} in optical, the M1
$\lambda$4100 (3p\,$^{2}$P$^{\rm o}$\,--\,3s\,$^{2}$S) and M2 $\lambda$4641
(3d\,$^{2}$D\,--\,3p\,$^{2}$P$^{\rm o}$) multiplets, are affected by the Bowen
fluorescence mechanism (e.g. Bowen \citealt{bowen1934}, \citealt{bowen1935}).
The N~{\sc iii} M18 $\lambda$4379.11 (5g\,$^{2}$G\,--\,4f\,$^{2}$F$^{\rm o}$)
line is amongst the best observed N~{\sc iii} lines in the spectrum of
NGC\,7009, which are not affected by the fluorescence processes. The
intensity ratio of the $\lambda$4379.11 line and the $\lambda$4195.76 line,
which is the second strongest fine-structure component of the N~{\sc iii} M6
3p$^{\prime}$\,$^{2}$D\,--\,3s$^{\prime}$\,$^{2}$P$^{\rm o}$ multiplet, can be
used as a temperature diagnostic. Fig.\,\ref{niii_te} shows the N~{\sc iii}
$\lambda$4379/$\lambda$4196 ratio as a function of electron temperature. The
dominant excitation mechanism of the N~{\sc iii} M18 multiplet is radiative
recombination, while the M6 multiplet is mainly excited by dielectronic
recombination. The other two fine-structure components of the N~{\sc iii} M6
multiplet, $\lambda\lambda$4200.10 and 4215.77 cannot be used: the former one
is blended with the He~{\sc ii} $\lambda$4199.83
(11g\,$^{2}$G\,--\,4f\,$^{2}$F$^{\rm o}$) line, which is more than three
times stronger, and the latter one cannot be accurately measured due to
weakness ($<$10$^{-4}$ of the H$\beta$ intensity). Another N~{\sc iii}
multiplet, M17 5f\,$^{2}$F$^{\rm o}$\,--\,4d\,$^{2}$D $\lambda$4003, when
used in pair with the N~{\sc iii} M6 $\lambda$4195.76 line, may also be a
temperature diagnostic, but its radiative recombination coefficients are
unknown. The N~{\sc iii} M17 lines are detected in the spectrum of NGC\,7009
(Fig.\,\ref{3940-4006}).

\begin{figure}
\begin{center}
\includegraphics[width=7.5cm,angle=0]{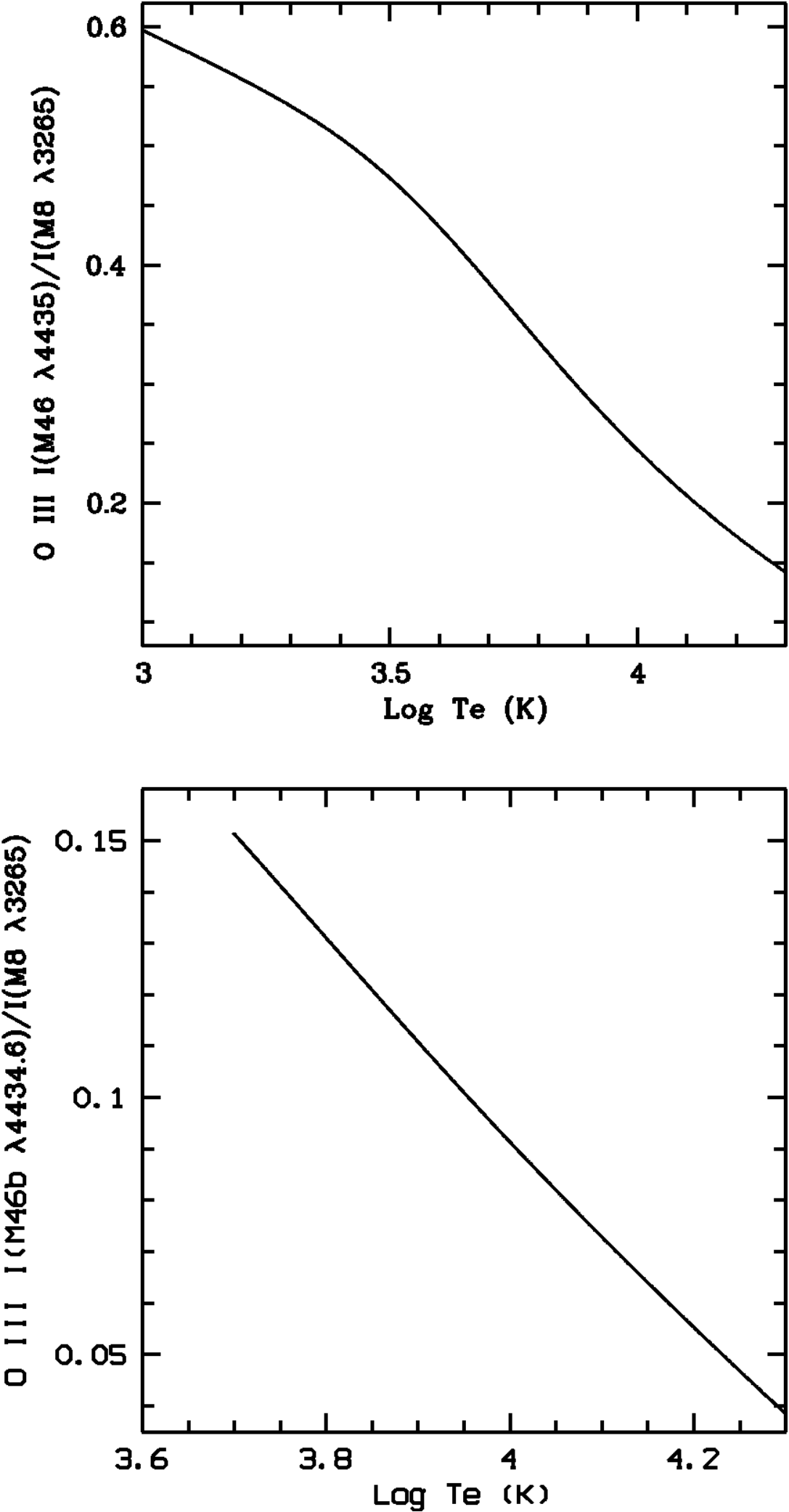}
\caption{$Upper$: The O~{\sc iii} $\lambda$4435/$\lambda$3265 ratio as a
function of electron temperature. Here the $\lambda$4435 line is the M46b
5g\,H[11/2]$^{\rm o}$\,--\,4f\,G[9/2] multiplet of O~{\sc iii}. The figure
is based on the dielectronic and radiative recombination coefficients of
Nussbaumer \& Storey \citet{ns1984} and P\'{e}quignot, Petitjean \& Boisson
\citet{ppb1991}, respectively.
$Lower$: The O~{\sc iii} $\lambda$4434.60/$\lambda$3265 ratio as a function
of electron temperature. Here the $\lambda$4434.60 line is the strongest
fine-structure component of the O~{\sc iii} M46b multiplet, and the effective
recombination coefficients of this line are from Kisielius \& Storey
\citet{ks1999}, whose calculations were carried out in the intermediate
coupling scheme and are valid from 5000 to 20\,000~K.}
\label{oiii_te}
\end{center}
\end{figure}

The majority of the O~{\sc iii} triplets of the 3d\,--\,3p and 3p\,--\,3s
configurations detected in the spectrum of NGC\,7009 are mainly excited by
the fluorescence or charge-transfer mechanism (e.g. Liu \& Danziger
\citealt{ld1993a}; Liu, Danziger \& Murdin \citealt{ldm93}). Thus those lines
are not suitable for plasma diagnostics or abundance determinations. However,
the O~{\sc iii} M8 3d\,$^{3}$F$^{\rm o}$\,--\,3p\,$^{3}$D multiplet is
unaffected by such mechanisms. The intensity of the strongest component of
the O~{\sc iii} M8 multiplet, $\lambda$3265.32 (3d\,$^{3}$F$^{\rm
o}_{4}$\,--\,3p\,$^{3}$D$_{3}$), relative to the best observed O~{\sc iii}
5g\,--\,4f line, can in principle be used as a temperature diagnostic. The
O~{\sc iii} M8 multiplet is mainly excited by radiative recombination at
temperatures below 5000~K, and the contribution of dielectronic recombination
to the total recombination rate catches up with that of the radiative
recombination at about 16\,000~K (Nussbaumer \& Storey \citealt{ns1984};
P\'{e}quignot, Petitjean \& Boisson \citealt{ppb1991}). The O~{\sc iii}
5g\,--\,4f lines are dominantly excited by radiative recombination.
The upper panel of Fig.\,\ref{oiii_te} shows the intensity ratio of the M46b
$\lambda$4435 (5g\,H[11/2]$^{\rm o}$\,--\,4f\,G[9/2]) multiplet, the strongest
transition of the 5g\,--\,4f configuration, and the M8 $\lambda$3265.32 line
as a function of electron temperature. P\'{e}quignot, Petitjean \& Boisson
\cite{ppb1991} present the radiative recombination coefficients of both the
M8 and M46b multiplets of O~{\sc iii}, while Nussbaumer \& Storey
\cite{ns1984} only give the dielectronic recombination coefficients of M8.
The lower panel of Fig.\,\ref{oiii_te} shows the intensity ratio of the
$\lambda$4434.60 (5g\,H[11/2]$^{\rm o}_{6}$\,--\,4f\,G[9/2]$_{5}$) line, the
strongest fine-structure component of the M46b multiplet, and the M8
$\lambda$3265.32 line as a function of electron temperature. The effective
recombination coefficients of the M46b $\lambda$4434.60 line are adopted
from Kisielius \& Storey \cite{ks1999}, whose calculations for the O~{\sc
iii} 5g\,--\,4f recombination spectrum were carried out in the intermediate
coupling scheme and valid from 5000 to 20\,000~K. Measurement of the
$\lambda$4434.60 line is of large uncertainty due to line blending. The other
O~{\sc iii} 5g\,--\,4f lines are not detected in the spectrum of NGC\,7009.

\subsection{\label{diagnose:summary}
Summary of the ORL diagnostics}

We have discussed about the possibility of using various recombination line
ratios of heavy element ions to determine electron temperatures and densities.
The line ratios are illustrated as a function of temperature or density.
For some cases, there is still a lack of adequate effective recombination
coefficients or the expected lines are not detected in the spectrum of
NGC\,7009 due to weakness and/or line blending. The O~{\sc iii} lines of the
5g\,--\,4f configuration are not detected. Fig.\,\ref{oiii_te} is probably
applicable once deep spectra with higher resolution are available. The
effective recombination coefficients for the C~{\sc iii}, N~{\sc iii} and
O~{\sc iii} lines quoted from Nussbaumer \& Storey \cite{ns1984} and
P\'{e}quignot, Petitjean \& Boisson \cite{ppb1991} are probably inadequate,
which can be inferred from the fact that the observed line ratios are all
outside of the diagnostic ranges of Figs.\,\ref{ciii_te} and \ref{niii_te}.

The applicability of Figs.\,\ref{nii_te:v19v39} and \ref{nii_ne:v19} is
quite small, given that the N~{\sc ii} M19 lines are close to the O~{\sc iii}
$\lambda$5007 line. The N~{\sc ii} and O~{\sc ii} recombination lines of
parentage other than the ground states of the recombining ions are good
temperature diagnostics. As shown in Figs.\,\ref{nii_te:extra} and
\ref{oii_te:extra}, those line ratios are insensitive to electron density.
The electron temperatures derived from the two O~{\sc ii} line
ratios ($\lambda$4649/$\lambda$4591 and $\lambda$4649/$\lambda$4189) in
Fig.\,\ref{oii_te:extra} are consistent with each other, and are close to
the temperature value yielded by the C~{\sc ii} $\lambda$8794/$\lambda$4267
ratio as shown in Fig.\,\ref{cii_te}.
Although no N~{\sc ii} lines of parentage other than $^{2}$P$^{\rm o}$ are
detected in the spectrum of NGC\,7009 due to weakness, they are promising
diagnostic tools in spectroscopy. The most reliable temperatures derived from
the ORLs of heavy element ions are those yielded by the N~{\sc ii} and O~{\sc
ii} lines, as shown in Figs.\,\ref{nii_te:v3v39} and \ref{oii_te:v1v48}.
Currently only the N~{\sc ii} and O~{\sc ii} recombination lines can be used
to determine electron density, since the density dependence of the population
distributions of the energetically lowest fine-structure levels of the
recombining ions have been taken into account in the recombination calculations
for those two ions (this effect was also considered by Kisielius \& Storey
\citealt{ks1999} for the calculation of the 5g\,--\,4f recombination lines
of O~{\sc iii}). Figs.\,\ref{nii_ne:v3}, \ref{oii_ne:v1} and \ref{oii_ne:v10}
all show large scatter in electron density for a given line ratio observed in
NGC\,7009. This is reasonable because the ORL ratios only have very weak
density dependence. New treatment of the Ne~{\sc ii} recombination in the
intermediate coupling scheme is needed.

\section{\label{orls}
The optical recombination spectrum of heavy elements}

In this section, we present a comprehensive analysis of the most significant
permitted transitions of C~{\sc ii}, N~{\sc ii}, O~{\sc ii}, and Ne~{\sc ii}
as well as C~{\sc iii}, N~{\sc iii} and O~{\sc iii} detected in the spectrum
of NGC\,7009 reported in Paper~I. The lines are critically examined for
potential blending effects and compared to theoretical predictions using the
latest atomic data. Unless specified otherwise, all intensities quoted
throughout the paper are corrected for interstellar extinction\footnote{In
Paper~I, we derived a mean value of 0.174 for the logarithmic extinction at
H$\beta$, $c$(H$\beta$), using the observed H~{\sc i} Balmer line ratios
H$\alpha$/H$\beta$ and H$\gamma$/H$\beta$. The predicted H~{\sc i} line
ratios in the Case~B assumption were adopted from Storey \& Hummer
\cite{sh1995}, with $T_\mathrm{e}$ = 10\,000~K and $N_\mathrm{e}$ =
10\,000~cm$^{-3}$.} and in units of $I$(H$\beta$) = 100, and the theoretical
intensities/ratios are predicted assuming an electron temperature of 1000~K
as given by the N~{\sc ii} and O~{\sc ii} ORL diagnostics
(Figs.\,\ref{nii_te:v3v39} and \ref{oii_te:v1v48}). The wavelengths of
atomic transitions are adopted from the compilation of laboratory and
theoretical values of Hirata \& Horaguchi \cite{hh1995}. The
extinction-corrected flux of H$\beta$, $I$(H$\beta$), of NGC\,7009 is derived
using $\log{I({\rm H}\beta)}$\,=\,$\log{F({\rm H}\beta)}$\,+\,$c$(H$\beta$),
where $F$(H$\beta$) is the observed H$\beta$ flux ($-9.80$ in logarithm),
which is adopted from Cahn, Kaler \& Stanghellini \cite{cks92}, and
$c$(H$\beta$) is the logarithmic extinction at H$\beta$, which was derived
from the H~{\sc i} Balmer line ratios H$\alpha$/H$\beta$ and H$\gamma$/H$\beta$
(Paper~I). The value of $c$(H$\beta$) we derived for NGC\,7009 is 0.174, which
agrees with the value (0.17) given by Cahn, Kaler \& Stanghellini \cite{cks92},
who used the radio/H$\beta$ flux ratio. Thus in NGC\,7009 we have $I$(H$\beta$)
= 10$^{-9.63}$~erg\,cm$^{-2}$\,s$^{-1}$.

\subsection{\label{orls:cii}
The C~{\sc ii} optical recombination spectrum}

Several dozen emission lines were identified as the permitted transitions of
C~{\sc ii}, with 41 being solid identifications (Paper~I). The strongest
transitions are presented in the current paper. As an example, we give
principles of fits for the multiplets M6 and M28.01 in this section. The
other multiplets are presented in Appendix\,\ref{appendix:a}. The effective
recombination coefficients of Davey, Storey \& Kisielius \cite{davey2000}
are used for ORL analysis.

\subsubsection{\label{orls:cii:v6}
Multiplet 6, 4f $^2$F$^{\rm o}$ -- 3d $^2$D}

C~{\sc ii} M6 $\lambda$4267 is the strongest C~{\sc ii} multiplet observed in
NGC\,7009 (see Fig.\,\ref{4260-4310}). The three fine-structure components of
this multiplet have close wavelengths: 4267.00, 4267.26 and 4267.26\,{\AA}.
Single Gaussian profile fitting to the emission feature gives an intensity of
0.880 (normalized to a scale where H$\beta$ = 100), with an uncertainty of
less than 5 per cent. Here the contributions from the O~{\sc ii} M53c
4f\,D[1]$^{\rm o}_{3/2}$\,--\,3d\,$^4$P$_{5/2}$ $\lambda$4263.27 and Ne~{\sc
ii} M57c 4f~1[3]$^{\rm o}_{7/2}$\,--\,3d~$^4$F$_{9/2}$ $\lambda$4267.38 lines
are negligible. This intensity value agrees with LSBC,
whose observation yields a value of 0.838. The calculation of Bastin
(\citealt{bastin2006}, hereafter B06) shows that the Case~A effective
recombination coefficient for the C~{\sc ii} M6 $\lambda4267$ line differs
from that in Case~B by 1.5 per cent, and a similar difference is given by
Davey, Storey \& Kisielius \cite{davey2000}, indicating that this transition
is case insensitive.

\subsubsection{\label{orls:cii:v28.01}
Multiplet 28.01, 3d$^{\prime}$ $^2$F$^{\rm o}$ -- 3p$^{\prime}$ $^2$D}

\begin{figure}
\begin{center}
\includegraphics[width=7.5cm,angle=-90]{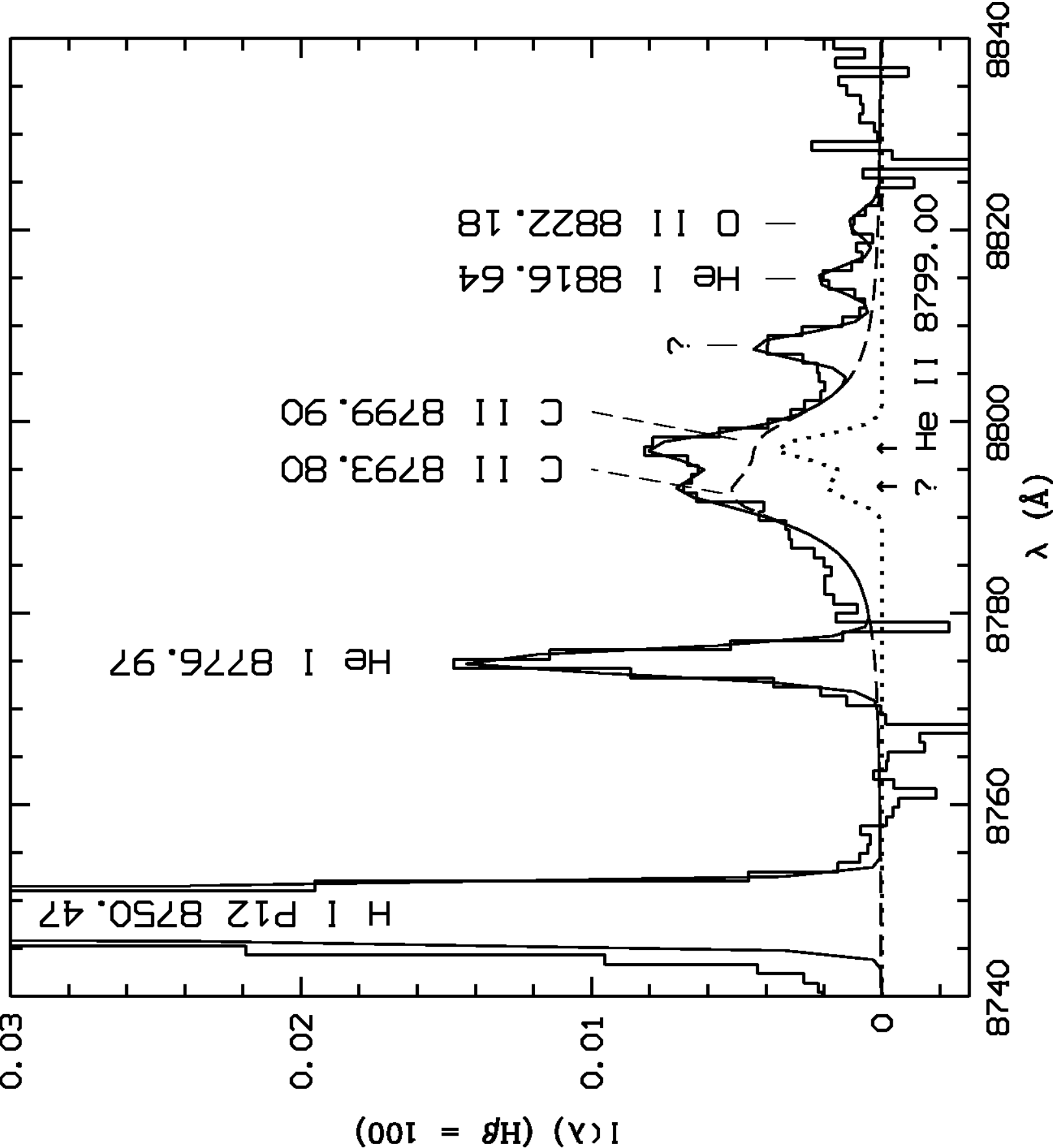}
\caption{Spectrum of NGC\,7009 from 8740 to 8840\,{\AA} showing the two C~{\sc
ii} M28.01 autoionization lines $\lambda\lambda$8793.80 and 8799.90. The
dashed curve is the sum of Voigt profile fits to the two C~{\sc ii} lines,
which probably blend with two weaker features, He~{\sc ii} $\lambda$8799.00
and an unknown component, whose profiles are assumed to be Gaussian and are
represented by the dotted curve. The solid continuous curve is the sum of all
fits. Here a Voigt profile is the convolution of a Gaussian ($\sim$3.0\,{\AA}
FWHM) and a Lorentz profile ($\sim$6.86\,{\AA} FWHM). Continuum has been
subtracted and the spectrum has been normalized such that H$\beta$ has an
integrated flux of 100. Extinction has not been corrected for.}
\label{8740-8840}
\end{center}
\end{figure}

This multiplet is a dielectronic transition, which is a result of cascading
from the autoionization state 2s2p($^3$P$^{\rm o}$)\,3d\,$^2$F$^{\rm o}$ that
lies about 0.41~eV above the first ionization threshold to the
2s2p($^3$P$^{\rm o}$)\,3p\,$^2$D state that lies 1.00~eV below the ionization
threshold (Moore \citealt{moore1993}). The features of the two fine-structure
components, $\lambda$8793.80 (3d$^{\prime}$\,$^2$F$^{\rm
o}_{7/2}$\,--\,3p$^{\prime}$\,$^2$D$_{5/2}$) and $\lambda$8799.90
(3d$^{\prime}$\,$^2$F$^{\rm o}_{5/2}$\,--\,3p$^{\prime}$\,$^2$D$_{3/2}$) are
very broad (Fig.\,\ref{8740-8840}). The wings of the two lines obviously
affect the weaker emission features nearby. Detailed analysis of the complex
indicates that at least two more emission lines are blended with the two
C~{\sc ii} lines: one is the He~{\sc ii} 23p\,$^2$P$^{\rm o}$\,--\,6s\,$^2$S
$\lambda$8799.0 line, while the other is unknown. The results of fitting to
the features are shown in Fig.\,\ref{8740-8840}.

For each of the two C~{\sc ii} M28.01 lines, we used a simulated Lorentz
profile with an intrinsic width of 6.86\,{\AA} convolved with a Gaussian
instrumental profile with a full width at half-maximum (FWHM) of 3.00\,{\AA}
to fit the observed feature. The convolution of the Lorentz profile and
Gaussian gives a Voigt profile with a width of 8.50\,{\AA}, which fits the
observed features quite well (Fig.\,\ref{8740-8840}). The intensity
contribution of the blended He~{\sc
ii} $\lambda$8799 line was estimated from the He~{\sc ii} 4f\,$^2$F$^{\rm
o}$\,--\,3d\,$^2$D $\lambda$4686 line and the hydrogenic theory of Storey \&
Hummer \cite{sh1995}. an electron temperature of 10\,000~K and a density of
10\,000~cm$^{-3}$ were assumed. After correcting for the contribution from
the He~{\sc ii} line, the intensity of the $\lambda$8799.90 line is
0.022$\pm$0.003. With the assumption that the relative intensity of the two
C~{\sc ii} M28.01 lines are as in the {\it LS}\,coupling, i.e. 1.0\,:\,0.7,
the intensity of the $\lambda$8793.80 line is 0.032, which is lower than the
total intensity of the broad feature at $\lambda$8793 (see
Fig.\,\ref{8740-8840}), indicating that there is probably unknown blending.
Multiple trial fitting to the profile shows that an emission feature with an
observed wavelength of 8793.20\,{\AA} best fits the feature, and its intensity
is $<$0.01. The {\sc emili}\footnote{{\sc emili} is developed by Dr. B.
Sharpee et al. and is designed to aid in the identification of weak emission
lines, particularly the weak recombination lines seen in high dispersion and
high signal-to-noise (S/N) spectra. URL:
$http://www.pa.msu.edu/astro/software/emili$} code (Sharpee et al.
\citealt{sharpee03}) identified this probably blended weak line as a [Cr~{\sc
ii}] line with a laboratory wavelength of 8795.17\,{\AA}. More efforts are
needed to verify this identification.

The intensity ratio of the C~{\sc ii} $\lambda$8793.80 line and the C~{\sc ii}
M6 $\lambda$4267 multiplet is 0.036, and that yields an electron temperature
of $\sim\,3000$~K (Section\,\ref{diagnose:cii} and Fig.\,\ref{cii_te}). As
discussed in Section\,\ref{diagnose:cii}, this temperature is questionable,
due to different excitation mechanisms of the C~{\sc ii} M28.01 $\lambda$8797
and the M6 $\lambda$4267 multiplets. Measurements of the C~{\sc ii} M28.01
lines are inaccurate unless detailed modeling of the autoionization levels of
C~{\sc ii} is carried out. Besides, too many skylines in the near-red of the
spectrum of NGC\,7009 and the relatively poor sky subtraction in this
wavelength region also makes accurate measurements of the C~{\sc ii} lines
difficult (c.f. Section\,$2.1$ in Paper~I).

\subsubsection{\label{orls:cii:summary}
Comments on the C~{\sc ii} recombination spectrum}

The current effective recombination coefficients used for analysis of the
C~{\sc ii} lines are mainly from Davey, Storey \& Kisielius \cite{davey2000}
and Bastin \cite{bastin2006}. The former calculation takes much care in the
low temperature case ($T_\mathrm{e}\,<$\,5000~K), while the latter one mainly
covers higher temperatures (5000\,--\,50\,000~K) and includes the effects of
high temperature dielectronic recombination, which rapidly becomes important
above an electron temperature of $\sim$15\,000~K. Both calculations were
carried out in the {\it LS}\,coupling assumption and only for the transitions
of the parentage $^{1}$S. In the current analysis, we adopt the calculation
of Davey, Storey \& Kisielius \cite{davey2000}, assuming a temperature of
1000~K.

The best observed (i.e. the most accurate) multiplets of C~{\sc ii} in the
spectrum of NGC\,7009 are M6 (4f\,$^2$F$^{\rm o}$\,--\,3d\,$^2$D) and some
transitions that belong to the $n$g\,--\,4f ($n\,\geq$\,5) array. For the
transitions with parentage other than $^{1}$S, only the M28.01
(3d$^{\prime}$\,$^{2}$F$^{\rm o}$\,--\,3p$^{\prime}$\,$^{2}$D) multiplet is
detected, but measurements of this multiplet could be unreliable, as mentioned
in Section\,\ref{orls:cii:v28.01}. The effective dielectronic recombination
coefficients of Nussbaumer \& Storey \cite{ns1984} and radiative recombination
coefficients of P\'{e}quignot, Petitjean \& Boisson \cite{ppb1991} were used
for the analysis of the M28.01 multiplet. A full treatment of the C~{\sc ii}
recombination in an appropriate coupling assumption (i.e. intermediate
coupling), with transitions between the autoionization levels taken into
account, is needed in the future.

\subsection{\label{nii_orls}
The N~{\sc ii} optical recombination spectrum}

In this section, we present intensities of the N~{\sc ii} ORLs detected
in the spectrum of NGC\,7009, and analyze these lines using the N~{\sc ii}
effective recombination coefficients of FSL11.
Unless otherwise specified, the theoretical relative intensities of the
N~{\sc ii} lines quoted in this section are all based on that calculation.
Comparison of the observed and predicted relative intensities of the N~{\sc
ii} lines with accurate intensities is made to assess the new atomic data.
An electron temperature of 1000~K is assumed throughout the analysis. In this
section, spectral fits and discussion of results are only given for the
M3 3p\,$^{3}$D\,--\,3s\,$^{3}$P$^{\rm o}$, M28 3d\,$^{3}$D$^{\rm
o}$\,--\,3p\,$^{3}$P multiplets and the strongest multiplets M39a,b of the
4f\,--\,3d transition array. Discussion of other multiplets of N~{\sc ii} are
given in Appendix\,\ref{appendix:b}.

\subsubsection{\label{nii_orls:v3}
Multiplet 3, 3p $^3$D -- 3s $^3$P$^{\rm o}$}

\begin{figure}
\begin{center}
\includegraphics[width=7.5cm,angle=-90]{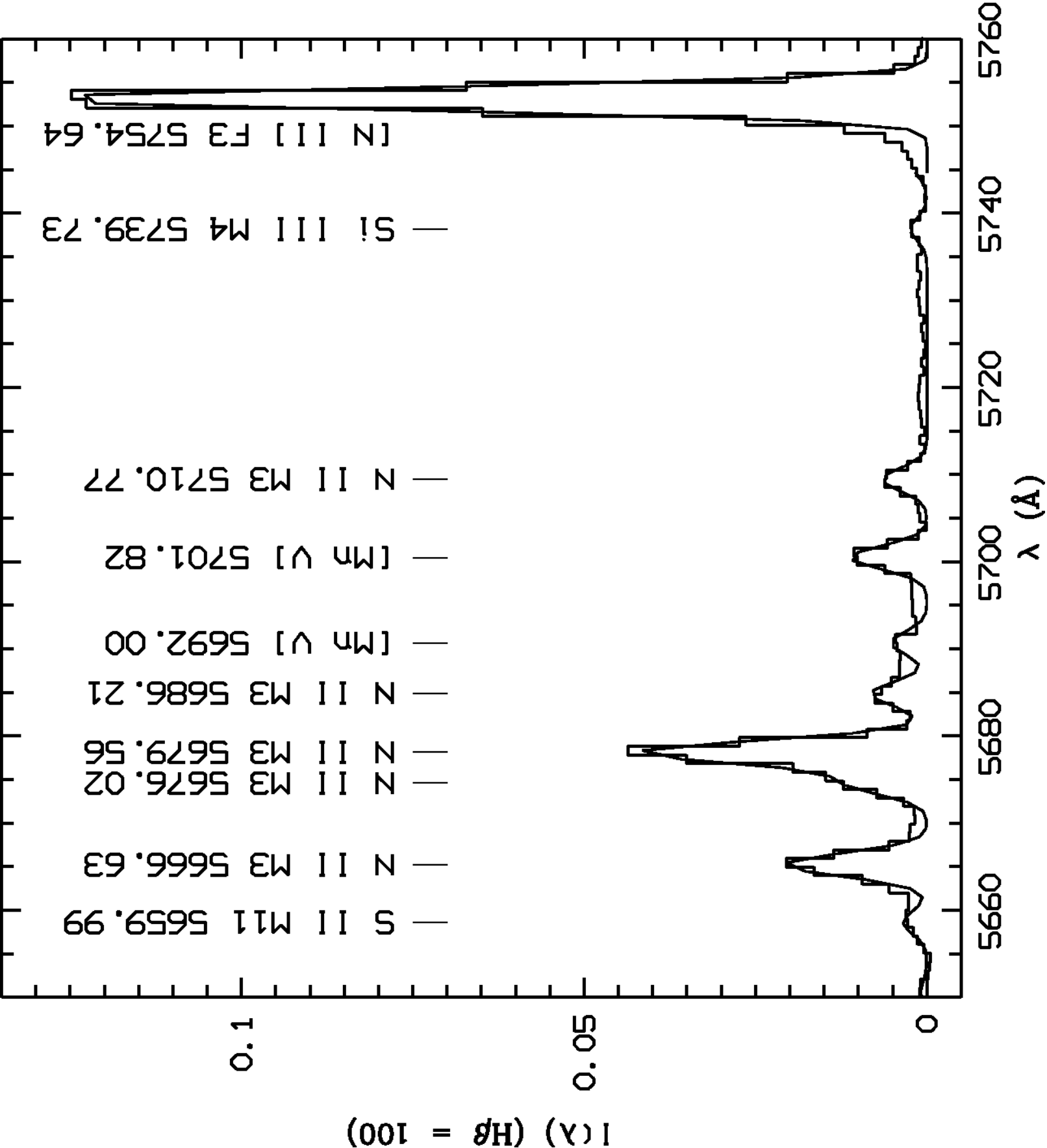}
\caption{Spectrum of NGC\,7009 from 5650 to 5760\,{\AA} showing the N~{\sc ii}
M3 lines and other emission features. The continuous curve is the sum of
Gaussian profile fits. Continuum has been subtracted and the spectrum has been
normalized such that H$\beta$ has an integrated flux of 100. Extinction has
not been corrected for.}
\label{5650-5760}
\end{center}
\end{figure}

This multiplet is the strongest of N~{\sc ii} in optical. The intensities of
the N~{\sc ii} M3 lines are presented in column\,5 of
Table\,\ref{relative:nii_v3} in unites of $I(\lambda5679.56)$ = 1.0. Also
presented are the theoretical relative intensities in the {\it LS}\,coupling
assumption (column 3) and the intermediate coupling (column 4). The theoretical
predictions in intermediate coupling are calculated from the N~{\sc ii}
effective recombination coefficients of FSL11.
Comparisons of the observed and predicted relative intensities are in columns
6 and 7. Results of multi-Gaussian profile fitting to the wavelength range
5650\,--\,5760\,{\AA} are also presented in Fig.\,\ref{5650-5760}.

The strongest M3 line, $\lambda$5679.56, is blended with $\lambda$5676.02 of
the same multiplet (Fig.\,\ref{5650-5760}). Two Gaussian profiles with the
same width were used to fit them. The intensities of $\lambda\lambda$5679.56
and 5676.02 are 0.135$\pm$0.007 and 0.035$\pm$0.004, respectively. Thus the
$\lambda$5676.02/$\lambda$5679.56 ratio is 0.273, which agrees with both
theoretical ratios within errors (Table\,\ref{relative:nii_v3}). Another two
lines, $\lambda\lambda$5666.63 and 5710.77, are free of line blending. The
intensity of the $\lambda$5710.77 line is 0.020$\pm$0.003, which agrees better
with the intermediate coupling (Table\,\ref{relative:nii_v3}). The intensity
of the $\lambda$5666.63 line is 0.065$\pm$0.007, which also agrees better with
intermediate coupling (Table\,\ref{relative:nii_v3}). Another M3 line
$\lambda$5686.21 is partially blended with a weaker feature, which was
identified as {Mn~{\sc v}} $\lambda$5692.00 (Fig.\,\ref{5650-5760}). The
fitted intensity of $\lambda$5686.21 is 0.024$\pm$0.005, which seems to agree
better with {\it LS}\,coupling (Table\,\ref{relative:nii_v3}). However, the
intensity of this line is questionable due to weakness. The other M3 line,
$\lambda$5730.65, is not observed in our spectrum.

\begin{table}
\centering
\caption{Comparison of the observed and predicted relative intensities of
the N~{\sc ii} M3 lines detected in the spectrum of NGC\,7009. $I_{\rm IC}$
is the theoretical intensity deduced from the effective recombination
coefficients of FSL11, and $I_{\rm LS}$ is the value in the {\it
LS}\,coupling assumption. The above two symbols have the same meaning in
other tables of the current paper. An electron temperature of 1000~K is
assumed for the theoretical predictions $I_{\rm IC}$.}
\label{relative:nii_v3}
\begin{tabular}{lcccccc}
\hline
Line & $J_2-J_1$ & $I_{\rm LS}$ & $I_{\rm IC}$ & $I_{\rm obs}$ &
$\frac{I_{\rm obs}}{I_{\rm LS}}$ & $\frac{I_{\rm obs}}{I_{\rm IC}}$\\
\hline
$\lambda$5666.63 & 2 -- 1 & 0.536 & 0.466 & 0.481 & 0.897 & 1.032\\
$\lambda$5676.02 & 1 -- 0 & 0.238 & 0.215 & 0.273 & 1.147 & 1.271\\
$\lambda$5679.56 & 3 -- 2 & 1.000 & 1.000 & 1.000 & 1.000 & 1.000\\
$\lambda$5686.21 & 1 -- 1 & 0.179 & 0.128 & 0.187 & 1.047 & 1.464\\
$\lambda$5710.77 & 2 -- 2 & 0.179 & 0.167 & 0.151 & 0.844 & 0.902\\
\hline
\end{tabular}
\end{table}

\subsubsection{\label{nii_orls:v28}
Multiplet 28, 3d $^3$D$^{\rm o}$ -- 3p $^3$P}

Gaussian profile fitting to M28 $\lambda$5941.65 (3d\,$^3$D$^{\rm
o}_{3}$\,--\,3p\,$^3$P$_{2}$) yields an intensity of 0.030$\pm$0.004. The
intensity contribution from the M28 $\lambda$5940.24 (3d\,$^3$D$^{\rm
o}_{1}$\,--\,3p\,$^3$P$_{1}$) line is negligible. The intensity ratio of
$\lambda$5941.65 and the N~{\sc ii} M3 $\lambda$5679.56 line is 0.228, which
is 42 per cent lower than the theoretical ratio.

Another M28 line $\lambda$5927.81 (3d\,$^3$D$^{\rm
o}_{1}$\,--\,3p\,$^3$P$_{0}$) is blended with M28 $\lambda$5931.78
(3d\,$^3$D$^{\rm o}_{2}$\,--\,3p\,$^3$P$_{1}$), which coincides in wavelength
with the He~{\sc ii} 25h\,$^2$H$^{\rm o}$\,--\,5g\,$^2$G $\lambda$5931.83
line. Two Gaussian profiles were used to fit the complex, and the intensity
of the $\lambda$5927.81 line is 0.007, with a large uncertainty. The intensity
ratio of $\lambda$5927.81 to the $\lambda$5941.65 line is 0.250, which agrees
with the predicted ratio 0.235. Assuming that the He~{\sc ii} $\lambda$5931.83
line contributes 55 per cent to the total intensity of the blend at
$\lambda$5932, as estimated from the hydrogenic theory of Storey \& Hummer
\cite{sh1995}, we obtained an intensity of the $\lambda$5931.78 line which is
much higher than the theoretical prediction. Another M28 line $\lambda$5952.39
(3d\,$^3$D$^{\rm o}_{2}$\,--\,3p\,$^3$P$_{2}$) is blended with the He~{\sc ii}
24h\,$^2$H$^{\rm o}$\,--\,5g\,$^2$G $\lambda$5952.93 line. The intensity of
the $\lambda$5952.39 line is much higher than the predicted value. The other
M28 line $\lambda$5960.90 (3d\,$^3$D$^{\rm o}_{1}$\,--\,3p\,$^3$P$_{2}$) is
not observed.

\begin{table}
\centering
\caption{Same as Table\,\ref{relative:nii_v3} but for a comparison of the
observed and predicted relative intensities of the N~{\sc ii} M28 lines
detected in the spectrum of NGC\,7009.}
\label{relative:nii_v28}
\begin{tabular}{lcccccc}
\hline
Line & $J_2-J_1$ & $I_{\rm LS}$ & $I_{\rm IC}$ & $I_{\rm obs}$ &
$\frac{I_{\rm obs}}{I_{\rm LS}}$ & $\frac{I_{\rm obs}}{I_{\rm IC}}$\\
\hline
$\lambda$5941.65$^a$ & 3 -- 2 & 1.000 & 1.000 & 1.000 & 1.000 & 1.000\\
$\lambda$5952.39$^b$ & 2 -- 2 & 0.152 & 0.122 & 0.255 & 1.686 & 2.103\\
$\lambda$5931.78$^c$ & 2 -- 1 & 0.455 & 0.412 & 0.471 & 1.037 & 1.143\\
$\lambda$5960.90$^d$ & 1 -- 2 & 0.010 & 0.009 & --    & 0.000 & 0.000\\
$\lambda$5927.81     & 1 -- 0 & 0.202 & 0.219 & 0.317 & 1.570 & 1.448\\
\hline
\end{tabular}
\begin{description}
\item [$^a$] Including the $\lambda$5940.24 (3d\,$^3$D$^{\rm
o}_{1}$\,--\,3p\,$^3$P$_{1}$) line.
\item [$^b$] Corrected for the contribution from the He~{\sc ii}
$\lambda$5952.93 (24h\,$^{2}$H$^{\rm o}$\,--\,5g\,$^{2}$G) line (74 per cent).
\item [$^c$] Corrected for the contribution from the He~{\sc ii}
$\lambda$5931.83 (25h\,$^{2}$H$^{\rm o}$\,--\,5g\,$^{2}$G) line (57 per cent).
\item [$^d$] Not detected.
\end{description}
\end{table}

\begin{figure}
\begin{center}
\includegraphics[width=7.5cm,angle=-90]{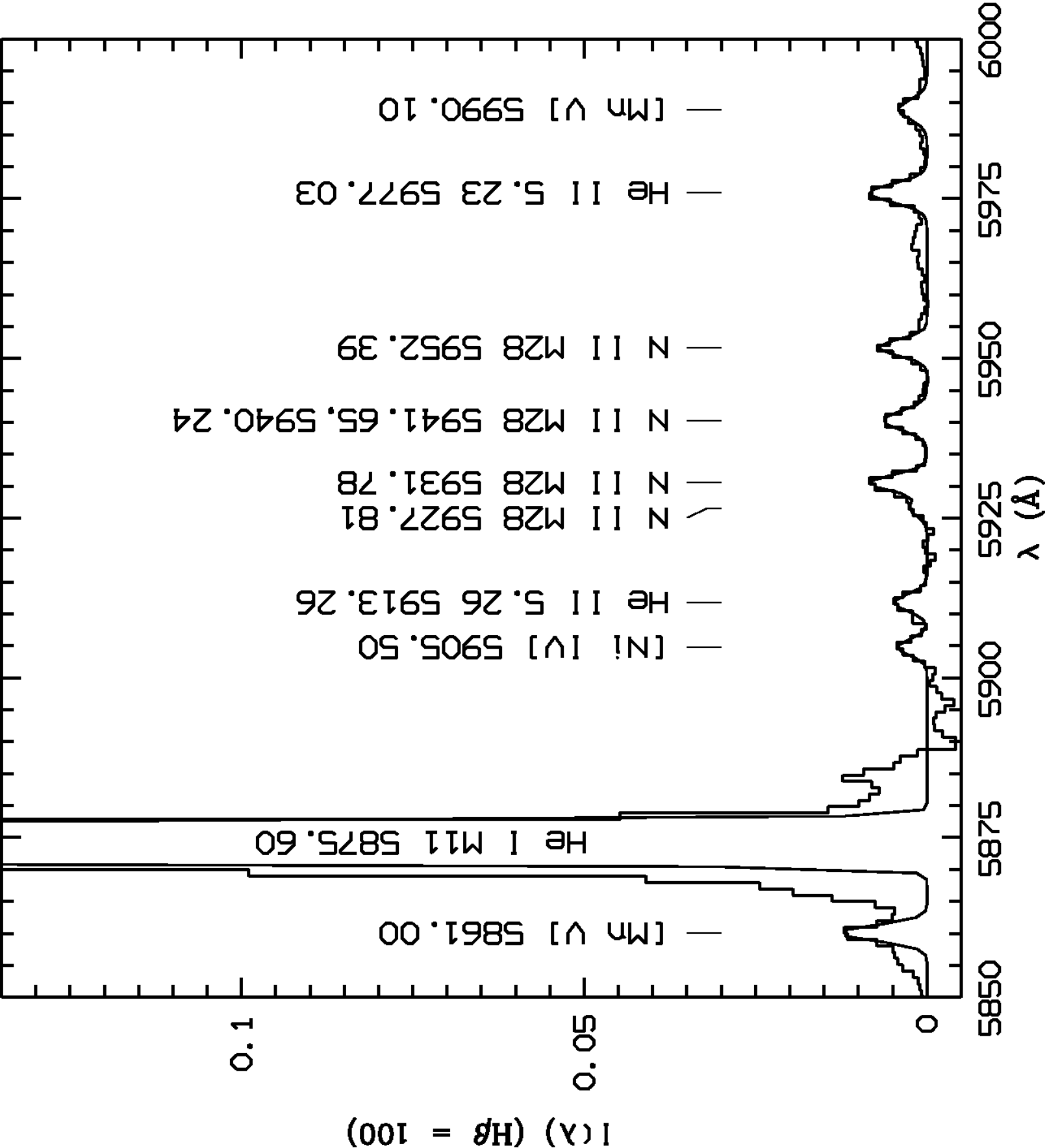}
\caption{Spectrum of NGC\,7009 from 5850 to 6000\,{\AA} showing the N~{\sc ii}
M28 lines and other emission features. The continuous curve is the sum of
Gaussian profile fits. Continuum has been subtracted and the spectrum has been
normalized such that H$\beta$ has an integrated flux of 100. Extinction has
not been corrected for.}
\label{5850-6000}
\end{center}
\end{figure}

\subsubsection{\label{nii_orls:4f-3d}
4f -- 3d transitions}

\begin{figure}
\begin{center}
\includegraphics[width=7.5cm,angle=-90]{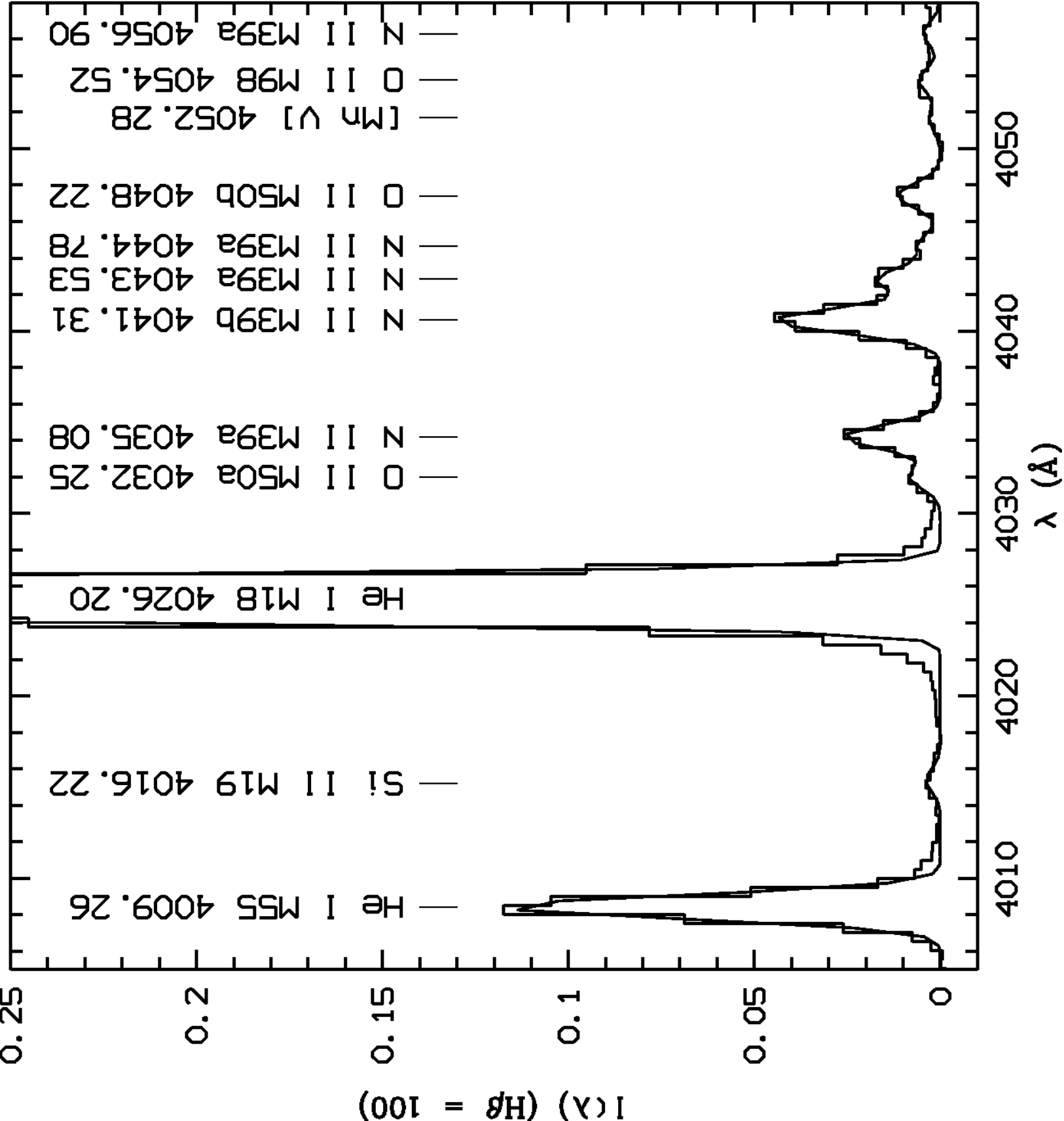}
\caption{Spectrum of NGC\,7009 from 4005 to 4058\,{\AA} showing the N~{\sc ii}
ORLs of the M39a 4f\,G[7/2]\,--\,3d\,$^3$F$^{\rm o}$ and M39b
4f\,G[9/2]\,--\,3d\,$^3$F$^{\rm o}$ multiplets. The strongest N~{\sc ii} lines
of the 4f\,--\,3d array, $\lambda$4041.31 (M39b
4f\,G[9/2]$_{5}$\,--\,3d\,$^3$F$^{\rm o}_{4}$) is observed. The continuous
curve is the sum of Gaussian profile fits. Continuum has been subtracted and
the spectrum has been normalized such that H$\beta$ has an integrated flux of
100. Extinction has not been corrected for.}
\label{4005-4058}
\end{center}
\end{figure}

\begin{figure*}
\begin{minipage}{128mm}
\begin{center}
\includegraphics[width=10cm,angle=-90]{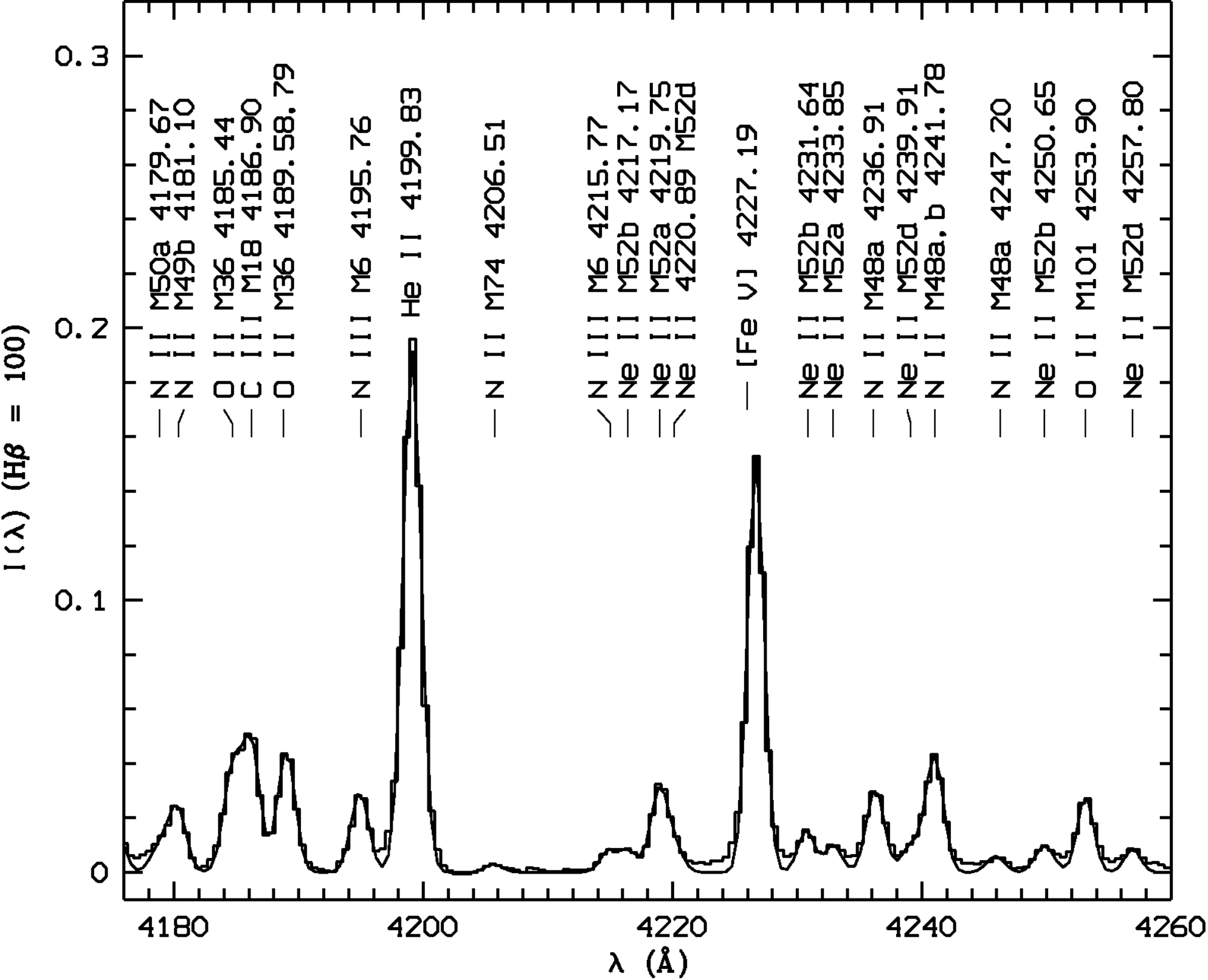}
\caption{Spectrum of NGC\,7009 from 4176 to 4260\,{\AA} showing the N~{\sc ii}
ORLs of the M48a 4f\,F[5/2]\,--\,3d\,$^3$D$^{\rm o}$, M48b
4f\,F[7/2]\,--\,3d\,$^3$D$^{\rm o}$, M49a 4f\,G[7/2]\,--\,3d\,$^3$D$^{\rm o}$,
and M49b 4f\,G[9/2]\,--\,3d\,$^3$D$^{\rm o}$ multiplets. The continuous curve
is the sum of Gaussian profile fits. Continuum has been subtracted and the
spectrum has been normalized such that H$\beta$ has an integrated flux of 100.
Extinction has not been corrected for.}
\label{4176-4260}
\end{center}
\end{minipage}
\end{figure*}

The 4f\,--\,3d transitions of N~{\sc ii} are located in the blue side of our
spectrum ($<$4500\,{\AA}) and suffer from line blending. Accurate measurements
of most of them are difficult. Table\,\ref{relative:nii_4f-3d} presents the
observed and predicted relative intensities of those 4f\,--\,3d lines with the
most reliable intensities. Figs.\,\ref{4005-4058} and \ref{4176-4260} show
some of the N~{\sc ii} lines of the 4f\,--\,3d transition array detected in
the spectrum of NGC\,7009. The results of multi-Gaussian profile fitting are
overplotted. Only the M39a and M39b multiplets are discussed here. Discussion
of other 4f\,--\,3d transitions of N~{\sc ii} are presented in
Appendix\,\ref{appendix:nii:4f-3d}.

\paragraph{\label{nii:4f-3d:v39a}
Multiplet 39a, 4f G[7/2] -- 3d $^3$F$^{\rm o}$:}

The $\lambda$4035.08 (4f\,G[7/2]$_{3}$\,--\,3d\,$^3$F$^{\rm o}_{2}$) and
$\lambda$4043.53 (4f\,G[7/2]$_{4}$\,--\,3d\,$^3$F$^{\rm o}_{3}$) lines are
shown in Fig.\,\ref{4005-4058}. The $\lambda$4035.08 line is blended with
the O~{\sc ii} M50b 4f\,F[3]$^{\rm o}_{5/2}$\,--\,3d\,$^4$F$_{5/2}$
$\lambda$4035.07 line, which contributes 15 per cent to the total intensity,
and the O~{\sc ii} M50b 4f\,F[3]$^{\rm o}_{7/2}$\,--\,3d\,$^4$F$_{5/2}$
$\lambda$4035.46 line, which is negligible. The $\lambda$4035 line has an
intensity 0.037$\pm$0.006. Here the contribution from the O~{\sc ii}
$\lambda$4035.07 line has been corrected for. This intensity agrees with the
predicted relative intensity. The $\lambda$4043.53 line is partially blended
with the N~{\sc ii} M39b 4f\,G[9/2]$_{5}$\,--\,3d\,$^3$F$^{\rm o}_{4}$
$\lambda$4041.31 line, which is more than 2 times stronger. The intensity of
the $\lambda$4043.53 line also agrees well with the predicted value. Reliable
measurements of the $\lambda$4044.78 (4f\,G[7/2]$_{3}$\,--\,3d\,$^3$F$^{\rm
o}_{3}$) line are difficult. The other two lines $\lambda$4056.90
(4f\,G[7/2]$_{4}$\,--\,3d\,$^3$F$^{\rm o}_{4}$) and $\lambda$4058.16
(4f\,G[7/2]$_{3}$\,--\,3d\,$^3$F$^{\rm o}_{4}$) are too weak.

\paragraph{\label{nii:4f-3d:v39b}
Multiplet 39b, 4f G[9/2] -- 3d $^3$F$^{\rm o}$:}

The $\lambda$4041.31 (4f\,G[9/2]$_{5}$\,--\,3d\,$^3$F$^{\rm o}_{4}$) line is
blended with the O~{\sc ii} M50c 4f\,F[2]$^{\rm o}_{5/2}$\,--\,3d~$^4$F$_{5/2}$
$\lambda$4041.28 and O~{\sc ii} M50c 4f\,F[2]$^{\rm
o}_{3/2}$\,--\,3d\,$^4$F$_{5/2}$ $\lambda$4041.95 lines. The two O~{\sc ii}
lines contribute only $\sim$7 per cent to the total intensity of the blend at
$\lambda$4041. The intensity of the $\lambda$4041.31 line is 0.082$\pm$0.008.
The intensity ratio of $\lambda$4041.31 to the N~{\sc ii} M3 $\lambda$5679.56
line is 0.604, which agrees quite well with the predicted ratio 0.598. Another
M39b line $\lambda$4026.08 (4f\,G[9/2]$_{4}$\,--\,3d\,$^3$F$^{\rm o}_{3}$) is
blended with the He~{\sc i} M18 5d\,$^3$D\,--\,2p\,$^3$P$^{\rm o}$
$\lambda$4026.20 line. The other M39b line $\lambda$4039.35
(4f\,G[9/2]$_{4}$\,--\,3d\,$^3$F$^{\rm o}_{4}$) is not observed.

\begin{table}
\centering
\caption{Same as Table\,\ref{relative:nii_v3} but for a comparison of the
observed and predicted relative intensities of the N~{\sc ii} 4f\,--\,3d
lines detected in the spectrum of NGC\,7009.}
\label{relative:nii_4f-3d}
\begin{tabular}{lcllr}
\hline
Line & $J_2-J_1$ & $I_{\rm IC}$ & $I_{\rm obs}$ &
$\frac{I_{\rm obs}}{I_{\rm IC}}$\\
\hline
M39a 4f G[7/2] -- 3d $^3$F$^{\rm o}$ & & & &\\
$\lambda$4035.08$^a$ & 3--2 & 0.477 & 0.540 & 1.132\\
$\lambda$4043.53     & 4--3 & 0.436 & 0.432 & 0.992\\
M39b 4f G[9/2] -- 3d $^3$F$^{\rm o}$ & & & &\\
$\lambda$4041.31$^b$ & 5--4 & 1.000 & 1.000 & 1.000\\
M43a 4f F[5/2] -- 3d $^1$D$^{\rm o}$ & & & &\\
$\lambda$4176.16$^c$ & 3--2 & 0.293 & 0.343 & 1.171\\
M43b 4f F[7/2] -- 3d $^1$D$^{\rm o}$ & & & &\\
$\lambda$4171.61     & 3--2 & 0.296 & 0.281 & 0.950\\
M48a 4f F[5/2] -- 3d $^3$D$^{\rm o}$ & & & &\\
$\lambda$4236.91$^d$ & 2--1 & 0.539 & 0.669 & 1.241\\
M48b 4f F[7/2] -- 3d $^3$D$^{\rm o}$ & & & &\\
$\lambda$4241.78$^e$ & 4--3 & 0.984 & 0.996 & 1.012\\
M55a 4f D[5/2] -- 3d $^3$P$^{\rm o}$ & & & &\\
$\lambda$4442.02$^f$ & 2--1 & 0.130 & 0.242 & 1.859\\
M58a 4f G[7/2] -- 3d $^1$F$^{\rm o}$ & & & &\\
$\lambda$4552.53$^g$ & 4--3 & 0.201 & 0.395 & 1.965\\
M58b 4f G[9/2] -- 3d $^1$F$^{\rm o}$ & & & &\\
$\lambda$4530.41$^h$ & 4--3 & 0.474 & 0.593 & 1.251\\
M61a 4f D[5/2] -- 3d $^1$P$^{\rm o}$ & & & &\\
$\lambda$4694.64$^i$ & 2--1 & 0.147 & 0.222 & 1.510\\
\hline
\end{tabular}
\begin{description}
\item [$^a$] Including the contribution from the O~{\sc ii} M50b
4f\,F[3]$^{\rm o}_{5/2}$\,--\,3d\,4F$_{5/2}$ $\lambda$4035.07 line.
Neglecting the O~{\sc ii} M50b 4f\,F[3]$^{\rm
o}_{7/2}$\,--\,3d\,4F$_{5/2}$ $\lambda$4035.46 line.
\item [$^b$] Neglecting the contributions from the O~{\sc ii} M50c
4f\,F[2]$^{\rm o}_{5/2}$\,--\,3d\,$^4$F$_{5/2}$ $\lambda$4041.28 and O~{\sc
ii} M50c 4f\,F[2]$^{\rm o}_{3/2}$\,--\,3d\,$^4$F$_{5/2}$ $\lambda$4041.95
lines.
\item [$^c$] Including N~{\sc ii} M43a 4f\,F[5/2]$_{2}$\,--\,3d\,$^1$D$^{\rm
o}_{2}$ $\lambda$4175.66.
\item [$^d$] Including N~{\sc ii} M48b 4f\,F[7/2]$_{3}$\,--\,3d\,$^3$D$^{\rm
o}_{2}$ $\lambda$4237.05.
\item [$^e$] Including N~{\sc ii} M48a 4f\,F[5/2]$_{3}$\,--\,3d\,$^3$D$^{\rm
o}_{2}$ $\lambda$4241.78. Neglecting N~{\sc ii} M48a
4f\,F[5/2]$_{2}$\,--\,3d\,$^3$D$^{\rm o}_{2}$ $\lambda$4241.24 and Ne~{\sc
ii} M52c 4f\,2[1]$^{\rm o}_{3/2}$\,--\,3d\,$^4$D$_{1/2}$ $\lambda$4242.04.
\item [$^f$] Including Ne~{\sc ii} M60b 4f\,1[4]$^{\rm
o}_{7/2}$\,--\,3d\,$^2$F$_{5/2}$ $\lambda$4442.69. Neglecting O~{\sc iii}
M49b 5g\,F[3]$^{\rm o}_{2,\,3}$\,--\,4f\,D[3]$_{2}$ $\lambda$4442.02.
\item [$^g$] Including Ne~{\sc ii} M55d 4f\,2[2]$^{\rm
o}_{5/2}$\,--\,3d\,$^4$F$_{3/2}$ $\lambda$4553.17 and Si~{\sc iii} M2
4p\,$^3$P$^{\rm o}_{2}$\,--\,4s\,$^3$S$_{1}$ $\lambda$4552.62.
\item [$^h$] Including N~{\sc iii} M3
3p$^{\prime}$\,$^4$D$_{1/2}$\,--\,3s$^{\prime}$\,$^4$P$^{\rm o}_{3/2}$
$\lambda$4530.86.
\item [$^i$] Overestimated.
\end{description}
\end{table}

%

\subsubsection{\label{nii_orls:summary}
Comments on the N~{\sc ii} recombination spectrum}

The effective recombination coefficients used for the analysis of the N~{\sc
ii} recombination spectrum are from FSL11, which
is dedicated to low temperatures ($T_\mathrm{e}\,<$\,10\,000~K) and is an
improvement over all previous calculations for this ion, as described
Section\,\ref{diagnose:data}. The best observed N~{\sc ii} lines in our
spectrum are M3 (3p\,$^{3}$D\,--\,3s\,$^{3}$P$^{\rm o}$), M12
(3p\,$^{1}$D\,--\,3s\,$^{1}$P$^{\rm o}$), and the strongest lines of the
4f\,--\,3d array, e.g. $\lambda$4041.31 (M39b
4f\,G[9/2]$_{5}$\,--\,3d\,$^{3}$F$^{\rm o}_{4}$), $\lambda$4043.53 (M39a
4f\,G[7/2]$_{4}$\,--\,3d\,$^{3}$F$^{\rm o}_{4}$). Those N~{\sc ii} lines
have been used for plasma diagnostics (Section\,\ref{diagnose:niioii}). The
fine-structure components of the N~{\sc ii} multiplets M5
(3p\,$^{3}$P\,--\,3s\,$^{3}$P$^{\rm o}$), M20 (3d\,$^{3}$D$^{\rm
o}$\,--\,3p\,$^{3}$D), M28 (3d\,$^{3}$D$^{\rm o}$--\,3p\,$^{3}$P), and M29
(3d\,$^{3}$P$^{\rm o}$\,--\,3p\,$^{3}$P) observed in our spectrum are
incomplete due to line blending. Those that are detected are also blended with
other lines. Although multi-Gaussian profile fitting has been carried out
and effective recombination coefficients used to correct for the blended line
of other ionic species, the derived intensities of those N~{\sc ii} lines
could still be questionable. Grandi \cite{grandi1976} shows that the strongest
components of the M3, M5 and M28 multiplets are affected by the fluorescence
mechanism in the Orion nebula. However, such effects is probably insignificant
in NGC\,7009 (C. Morisset, private communication). The 4f\,--\,3d transitions
are almost certainly free of fluorescence enhancement because they require
such high excitation energy photons that the central star of NGC\,7009 cannot
afford.

\subsection{\label{oii_orls}
The O~{\sc ii} optical recombination spectrum}

LSBC observed eight multiplets of the 3\,--\,3 transition
arrays and a few dozen 4f\,--\,3d lines of O~{\sc ii}. The effective radiative
recombination coefficients for the O~{\sc ii} 3d\,--\,3p and 4f\,--\,3d
transitions were calculated under the intermediate coupling scheme, and were
used for spectral analysis. For the 3p,--\,3s transitions, the effective
recombination coefficients of Storey \cite{storey1994}, whose calculations
were carried out in the {\it LS}\,coupling scheme, were utilized. LSBC
confirmed the breakdown of {\it LS}\,coupling in the O~{\sc ii} transitions,
especially those of the 4f\,--\,3d configuration. In Paper~I, we presented
very deep spectrum of NGC\,7009. The data quality is higher than that in LSBC.
In the current paper, we present the intensities of the O~{\sc ii} ORLs, and
analyze the O~{\sc ii} recombination spectrum using the new O~{\sc ii}
effective recombination coefficients of PJS. Unless otherwise specified, the
theoretical relative intensities of the O~{\sc ii} lines quoted in this
section are all based on that calculation. Comparison of the observed and
predicted relative intensities of the O~{\sc ii} lines with accurate
intensities is made to assess the new atomic data. An electron temperature
of 1000~K is assumed throughout the analysis. In this section, spectral fits
and discussion of the results are only given for the M1 3p\,$^{4}$D$^{\rm
o}$\,--\,3s\,$^{4}$P, M10 3d\,$^{4}$F\,--\,3p\,$^{4}$D$^{\rm o}$ multiplets
and the strongest multiplets M39a,b of the 4f\,--\,3d transition array.
Discussion of other multiplets of O~{\sc ii} are given in
Appendix\,\ref{appendix:c}.

\subsubsection{\label{oii_orls:v1}
Multiplet 1, 3p $^4$D$^{\rm o}$ -- 3s $^4$P}

\begin{figure*}
\begin{minipage}{128mm}
\begin{center}
\includegraphics[width=10cm,angle=-90]{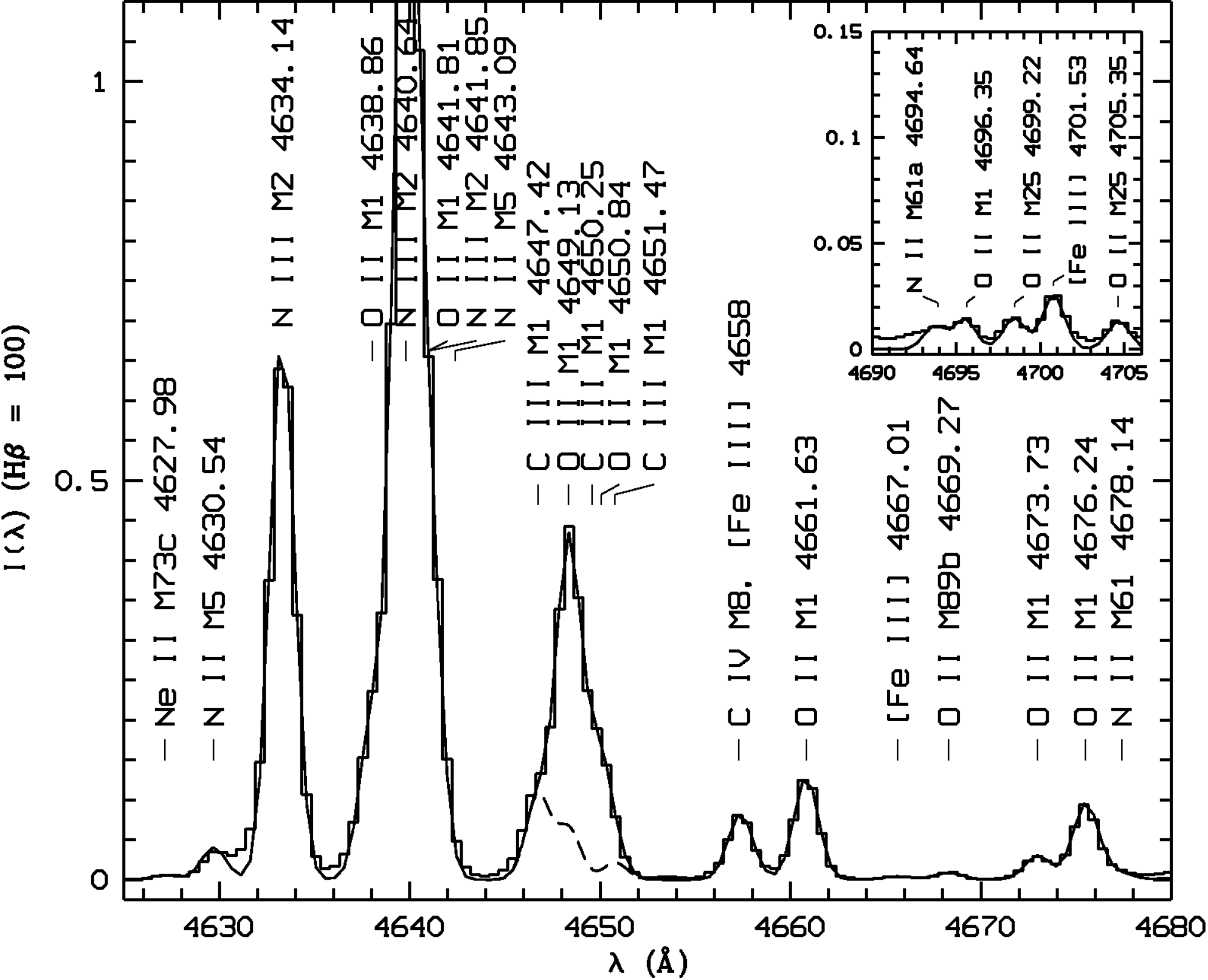}
\caption{Spectrum of NGC\,7009 from 4625 to 4680\,{\AA} showing the O~{\sc
ii} M1 lines and other emission features. The solid continuous curve is the
sum of Gaussian profile fits. The dashed curve is the sum of the Gaussian
profiles of the three C~{\sc iii} M1 (3p\,$^{3}$P$^{\rm o}$\,--\,3s\,$^{3}$S)
lines $\lambda\lambda$4647.42, 4650.25 and 4651.47, whose intensity ratio is
fixed to be as in {\it LS}\,coupling, i.e. 1\,:\,3\,:\,5. The inset shows the
very weak O~{\sc ii} M1 $\lambda$4696.35 (3p\,$^{4}$D$^{\rm
o}_{3/2}$\,--\,3s\,$^{4}$P$_{5/2}$) line, which is located in a different
wavelength region. Continuum has been subtracted and the spectrum has been
normalized such that H$\beta$ has an integrated flux of 100. Extinction has
not been corrected for.}
\label{4625-4680}
\end{center}
\end{minipage}
\end{figure*}

The M1 multiplet is the strongest amongst all the O~{\sc ii} permitted
transitions, and is one of the best observed multiplets. Comparisons of the
observed and predicted relative intensities of the M1 fine-structure
components are presented in Table\,\ref{relative:oii_v1}. The results of
multi-Gaussian profile fitting are shown in Fig.\,\ref{4625-4680}.

The strongest M1 line $\lambda$4649.13 is blended with $\lambda$4650.84 of the
same multiplet; also blended are the three lines of C~{\sc iii} M1
3p~$^3$P$^{\rm o}$ -- 3s~$^3$S: $\lambda\lambda$4647.42, 4650.25 and 4651.47.
Five Gaussian profiles of the same FWHM were used to fit the complex, with the
laboratory wavelength differences utilized. The relative intensities of the
three C~{\sc iii} M1 lines were assumed to be as in {\it LS}\,coupling, i.e.
5\,:\,3\,:\,1, but the relative intensities of the two O~{\sc ii} lines were
not constrained. The intensity of the $\lambda$4649.13 line is 0.667$\pm$0.030.
The intensity of the $\lambda$4650.84 line is 0.169$\pm$0.008. The intensity
ratio of the $\lambda$4650.84 and $\lambda$4649.13 lines agrees with the
theoretical ratios predicted in the intermediate coupling, but is slightly
higher than the {\it LS}\,coupling value (Table\,\ref{relative:oii_v1}).
Another three M1 lines, $\lambda\lambda$4661.63, 4673.73 and 4676.24, are free
of line blending. The fitted intensities of the three lines agree
with the predicted values, except for $\lambda$4673.73, whose measurement is
obviously higher than both predicted values (Table\,\ref{relative:oii_v1}). The
measurement uncertainties of the three lines are all less than 10 per cent.
Such large difference between the observed and predicted intensity of
$\lambda$4673.73 cannot be explained explicitly. $\lambda$4673.73 coincides
in wavelength with C~{\sc iii} M5 3p$^{\prime}$~$^3$P$_{1}$ --
3s$^{\prime}$~$^3$P$^{\rm o}_{2}$ $\lambda$4673.95. However, as discussed in
LSBC, a significant contribution from the C~{\sc iii}
$\lambda$4673.95 line was unlikely because another C~{\sc iii} M5 line
$\lambda$4665.86, which is expected to be much stronger than $\lambda$4673.95,
is not observed. $\lambda$4676.24 is blended with O~{\sc ii} M91 4f~G[4]$^{\rm
o}_{7/2}$ -- 3d~$^2$D$_{5/2}$ $\lambda$4677.07 and N~{\sc ii} M61b
4f~D[3/2]$_{2}$ -- 3d~$^1$P$^{\rm o}_{1}$ $\lambda$4678.14, but the
contributions of these two lines are probably insignificant, as estimated from
the new effective recombination coefficients.

Another two M1 lines, $\lambda\lambda$4638.86 and 4641.81, are both blended
with two N~{\sc iii} M2 lines $\lambda\lambda$4640.64 and 4641.85, which are
excited by the Bowen fluorescence mechanism. Also blended with this feature
is N~{\sc ii} M5 3p~$^3$P$_{1}$ -- 3s~$^3$P$^{\rm o}_{2}$ $\lambda$4643.09.
Taking into account another N~{\sc iii} M2 line $\lambda$4634.14 and N~{\sc ii}
M5 3p~$^3$P$_{2}$ -- 3s~$^3$P$^{\rm o}_{2}$ $\lambda$4630.54, we used six
Gaussian profiles (the O~{\sc ii} $\lambda$4641.81 line coincides in wavelength
with N~{\sc iii} $\lambda$4641.85, thus they were treated as a single
component) to fit the complex, assuming that all the six components had the
same FWHM. The results of the fitting are plotted in Fig.\,\ref{4625-4680}.
Here the contribution of N~{\sc iii} M2 $\lambda$4641.85 to the total
intensity of the blend at $\lambda$4642 was estimated from N~{\sc iii} M2
$\lambda$4634.14, which is free of line blending. The intensity ratio of the
two N~{\sc iii} lines was assumed to be as in pure {\it LS}\,coupling, i.e.
1\,:\,5, considering the fact that the two lines decay from the same upper
level, thus their intensity ratio depends only on the coupling scheme instead
of the excitation mechanism. The relative intensity of the two O~{\sc ii} M1
lines were not constrained. The intensity of $\lambda$4641.81 thus obtained
is 0.437$\pm$0.085. This measurement agrees well with the predicted value in
the intermediate coupling (Table\,\ref{relative:oii_v1}). The resultant
intensity of the $\lambda$4638.86 line is higher than the theoretical ratios
(Table\,\ref{relative:oii_v1}), but its intensity could be unreliable due to
the strength of the N~{\sc iii} M2 $\lambda$4640.64 line, which is more than
10 times stronger. The C~{\sc ii} M12.01 6d\,$^2$D -- 4p\,$^2$P$^{\rm o}$
lines, $\lambda\lambda$4637.63, 4638.91 and 4639.07, may be also blended in
the $\lambda$4638 feature, but taking them into account makes the task of
line deblending more difficult.

The other M1 line $\lambda$4696.35, which is expected to be the faintest of
O~{\sc ii} M1, is observed (the inset in Fig.\,\ref{4625-4680}). It coincides
in wavelength with O~{\sc ii} M89a 4f~D[3]$^{\rm o}_{5/2}$ -- 3d~$^2$D$_{3/2}$
$\lambda$4696.35, and is partially blended with another weak feature which was
identified as N~{\sc ii} M61a 4f~D[5/2]$_{2}$ -- 3d~$^1$P$^{\rm o}_{1}$
$\lambda$4694.64. Accurate measurements of $\lambda$4696.35 are difficult due
to weakness. Assuming that the O~{\sc ii} M89a $\lambda$4696.35 line
contributes 38 per cent to total flux of the blend at $\lambda$4696, as
estimated from the new O~{\sc ii} effective recombination coefficients, we
obtained an intensity of 0.015 for the M1 $\lambda$4696.35 line, which agrees
well with the newly predicted value (Table\,\ref{relative:oii_v1}).

\begin{table}
\centering
\caption{Same as Table\,\ref{relative:nii_v3} but for a comparison of the
observed and predicted relative intensities of O~{\sc ii} M1 lines detected
in the spectrum of NGC\,7009.}
\label{relative:oii_v1}
\begin{tabular}{lcccccc}
\hline
Line & $J_2-J_1$ & $I_{\rm LS}$ & $I_{\rm IC}$ & $I_{\rm obs}$ &
$\frac{I_{\rm obs}}{I_{\rm LS}}$ & $\frac{I_{\rm obs}}{I_{\rm IC}}$\\
\hline
$\lambda$4638.86 & 3/2--1/2 & 0.208 & 0.283 & 0.502 & 2.414 & 1.774\\
$\lambda$4641.81 & 5/2--3/2 & 0.525 & 0.632 & 0.656 & 1.249 & 1.037\\
$\lambda$4649.13 & 7/2--5/2 & 1.000 & 1.000 & 1.000 & 1.000 & 1.000\\
$\lambda$4650.84 & 1/2--1/2 & 0.208 & 0.290 & 0.263 & 1.264 & 0.907\\
$\lambda$4661.63 & 3/2--3/2 & 0.267 & 0.317 & 0.326 & 1.222 & 1.028\\
$\lambda$4673.73 & 1/2--3/2 & 0.042 & 0.051 & 0.078 & 1.857 & 1.535\\
$\lambda$4676.24 & 5/2--5/2 & 0.225 & 0.220 & 0.237 & 1.054 & 1.078\\
$\lambda$4696.35 & 3/2--5/2 & 0.025 & 0.024 & 0.023 & 0.920 & 0.962\\
\hline
\end{tabular}
\end{table}

\subsubsection{\label{oii_orls:v10}
Multiplet 10, 3d $^4$F -- 3p $^4$D$^{\rm o}$}

\begin{figure*}
\begin{minipage}{128mm}
\begin{center}
\includegraphics[width=10cm,angle=-90]{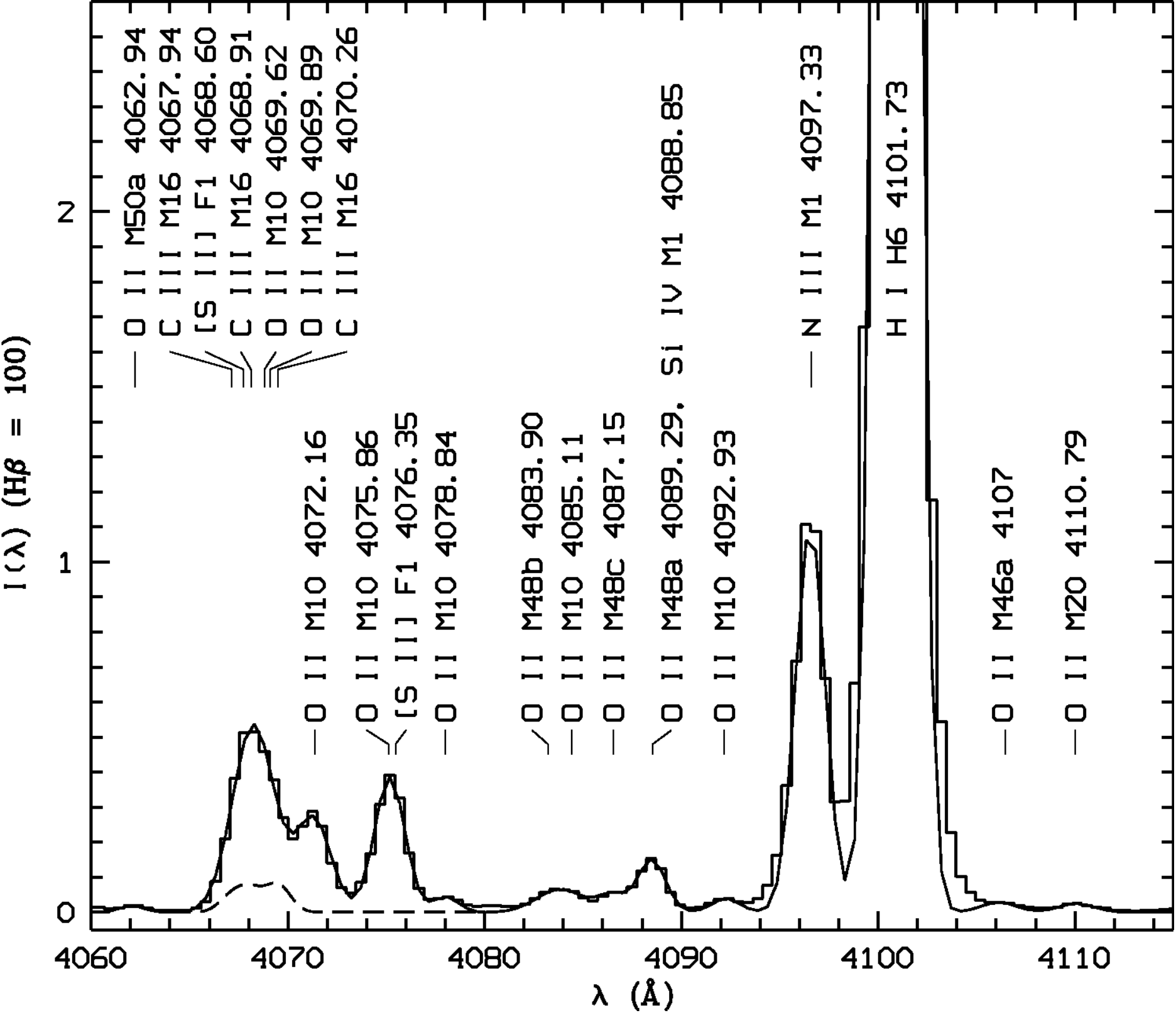}
\caption{Spectrum of NGC\,7009 from 4060 to 4115\,{\AA} showing the O~{\sc
ii} M10 lines and some O~{\sc ii} lines from the 4f\,--\,3d configuration. The
strongest O~{\sc ii} 4f\,--\,3d line $\lambda$4089.29 (M48a 4f\,G[5]$^{\rm
o}_{11/2}$\,--\,3d\,$^4$F$_{9/2}$) is observed. The solid continuous curve is
the sum of Gaussian profile fits. The dashed curve is the sum of the Gaussian
profiles of the three C~{\sc iii} M16 (5g\,$^{3}$G\,--\,4f\,$^{3}$F$^{\rm o}$)
lines $\lambda\lambda$4067.94, 4068.92 and 4070.31, whose intensity ratio is
fixed to be as in {\it LS}\,coupling, i.e. 1.00\,:\,1.31\,:\,1.71. Continuum
has been subtracted and the spectrum has been normalized such that H$\beta$
has an integrated flux of 100. Extinction has not been corrected for.}
\label{4060-4115}
\end{center}
\end{minipage}
\end{figure*}

The observed and predicted relative intensities of O~{\sc ii} M10 lines
detected in NGC\,7009 are presented in table\,\ref{relative:oii_v10}. The
emission features of the M10 lines and the results of Gaussian profile fitting
are shown in Fig.\,\ref{4060-4115} $\lambda\lambda$4069.62 and 4069.89, are
blended with [S~{\sc ii}] 3p$^{3}$~$^{2}$P$^{\rm o}_{3/2}$ --
3p$^{3}$~$^{4}$S$^{\rm o}_{3/2}$ $\lambda$4068.60 and the three C~{\sc iii}
M16 5g~$^{3}$G -- 4f~$^{3}$F$^{\rm o}$ lines: $\lambda\lambda$4067.94, 4068.91
and 4070.26. The blend feature at $\lambda$4070 is also partially resolved
from another O~{\sc ii} M10 line $\lambda$4072.16. Six Gaussian profiles of
the same FWHM were used to fit the complex ($\lambda\lambda$4069.62 and 4069.89
were treated as a single component, given their close wavelengths), assuming
that the differences of the observed wavelengths were the same as those of the
laboratory ones. Assumption was also made that the relative intensities of the
three C~{\sc iii} M16 lines were as in pure {\it LS}\,coupling, i.e.
1.00~:~1.31~:~1.71. The $\lambda$4072.16 line blends with N~{\sc ii} M38b
4f~F[7/2]$_{3}$ -- 3d~$^3$F$^{\rm o}_{2}$ $\lambda$4073.05 and O~{\sc ii} M48a
4f~G[5]$^{\rm o}_{9/2}$ -- 3d~$^{4}$F$_{7/2}$ $\lambda$4071.23. The intensity
of the N~{\sc ii} $\lambda$4073.05 line was assumed to be negligible, and the
O~{\sc ii} $\lambda$4071.23 line contributes about 9 per cent to the total
intensity of the blend at $\lambda$4072. Despite of the assumptions above, no
further constraint was set for the relative intensities of the following four
features in Gaussian profile fitting: The C~{\sc iii} M16 multiplet, [S~{\sc
ii}] $\lambda$4068.60, the two O~{\sc ii} M10 lines at $\lambda$4070, and
O~{\sc ii} $\lambda$4072.15. The resultant total intensity of
$\lambda$4069.62,\,89 is 0.635$\pm$0.060. The intensity of the
$\lambda$4072.16 line derived from the fits is 0.549$\pm$0.029. These
intensities agree well with the theoretical values predicted in the
intermediate coupling (Table\,\ref{relative:oii_v10}).

The total intensity of the C~{\sc iii} M16 $\lambda$4069 multiplet yielded by
Gaussian profile fitting is 0.288$\pm$0.058. The intensity ratio of the C~{\sc
iii} M16 multiplet to the C~{\sc iii} M1 $\lambda$4650 multiplet (the
measurements of C~{\sc iii} M1 are described in Section\,\ref{oii_orls:v1})
agrees with the predicted ratio within errors. Here the predicted C~{\sc iii}
M16/M1 ratio is derived based on the radiative and dielectronic recombination
coefficients given by P\'{e}quignot, Petitjean \& Boisson \cite{ppb1991} and
Nussbaumer \& Storey \cite{ns1984}, respectively. The C~{\sc iii} M1
$\lambda$4650 multiplet is mainly excited by dielectronic recombination, while
the C~{\sc iii} M16 $\lambda$4069 multiplet is by radiative recombination
(LSBC).

$\lambda$4075.86 is expected to be the strongest in O~{\sc ii} M10
(Table\,\ref{relative:oii_v10}). It is blended with [S~{\sc ii}]
3p$^{3}$~$^{2}$P$^{\rm o}_{1/2}$ -- 3p$^{3}$~$^{4}$S$^{\rm o}_{3/2}$
$\lambda$4076.35. We used the same technique as LSBC to
derive the intensities. The flux contribution of [S~{\sc ii}] $\lambda$4076.35
to the blend at $\lambda$4076 was estimated from the measured intensity of
[S~{\sc ii}] $\lambda$4068.60. A five-level atomic model was constructed to
calculate the level population of S$^{+}$, with an appropriate electron
temperature and density assumed. The calculated [S~{\sc ii}]
$\lambda$4068.60/$\lambda$4076.35 intensity ratio was 3.04, the same as the
ratio given by LSBC. The resultant intensity of the O~{\sc
ii} $\lambda$4075.86 line is 0.688$\pm$0.070. The intensity ratio of the
$\lambda$4075.86 line and the O~{\sc ii} M1 $\lambda$4661.63 (the measurements
of the M1 $\lambda$4661.63 line is given in Section\,\ref{oii_orls:v1}) is
3.164, in close agreement with the predicted ratio 3.004 based on the latest
O~{\sc ii} effective recombination coefficients. Here an electron temperature
N~{\sc ii} M38a 4f~F[5/2]$_{2}$ -- 3d~3F$^{\rm o}_{2}$ $\lambda$4076.91 and
N~{\sc ii} M38a 4f~F[5/2]$_{3}$ -- 3d~3F$^{\rm o}_{2}$ $\lambda$4077.40 also
blend in the $\lambda$4076 feature, but their flux contributions were assumed
to be negligible.

$\lambda$4078.84 is much weaker, and is close to the blend feature at
$\lambda$4076 (Fig.\,\ref{4060-4115}). The intensity of the $\lambda$4078.84
line derived from Gaussian profile fitting is 0.089$\pm$0.009. The intensity
ratio of $\lambda$4078.84 to the $\lambda$4075.86 line agrees well with the
predicted ratio in the intermediate coupling (Table\,\ref{relative:oii_v10}).
The actual measurement uncertainty of this line could be even larger due to
the weakness.
Another M10 line $\lambda$4085.11 blends with O~{\sc ii} M48b 4f~G[4]$^{\rm
o}_{7/2}$ -- 3d~$^4$F$_{5/2}$ $\lambda$4083.90 and N~{\sc ii} M38b
4f~G[7/2]$_{3}$ -- 3d~$^{3}$F$^{\rm o}_{3}$ $\lambda$4082.89. The O~{\sc ii}
$\lambda$4083.90 contributes about 47 per cent to the total intensity, and
the N~{\sc ii} line is probably negligible. After the correction for the
blend, the intensity of $\lambda$4085.11 agrees well with the newly predicted
value (Table\,\ref{relative:oii_v10}). Measurements of the remaining M10 lines
are difficult: $\lambda$4092.93 (3d~$^{4}$F$_{7/2}$ -- 3p~$^{4}$D$^{\rm
o}_{7/2}$) is partially blended with N~{\sc iii} M1 3p~$^2$P$^{\rm o}_{3/2}$
-- 3s~$^2$S$_{1/2}$ $\lambda$4097.33, which is excited by the Bowen
fluorescence mechanism; $\lambda$4094.14 (3d~$^{4}$F$_{3/2}$ --
3p~$^{4}$D$^{\rm o}_{5/2}$) blends in the N~{\sc iii} $\lambda$4097.33 line,
and $\lambda$4106.02 (3d~$^{4}$F$_{5/2}$ -- 3p~$^{4}$D$^{\rm o}_{7/2}$) is
embedded in the wing of H~{\sc i} $\lambda$4101.

\begin{table}
\centering
\caption{Same as Table,\ref{relative:oii_v1} but for a comparison of the
observed and predicted relative intensities of O~{\sc ii} M10 lines detected
in the spectrum of NGC\,7009. Only the components with the most reliable
measurements are presented.}
\label{relative:oii_v10}
\begin{tabular}{lcccccc}
\hline
Line & $J_2-J_1$ & $I_{\rm LS}$ & $I_{\rm IC}$ & $I_{\rm obs}$ &
$\frac{I_{\rm obs}}{I_{\rm LS}}$ & $\frac{I_{\rm obs}}{I_{\rm IC}}$\\
\hline
$\lambda$4069.89$^a$ & 5/2--3/2 & 0.730 & 0.956 & 0.923 & 1.264 & 0.965\\
$\lambda$4072.16$^b$ & 7/2--5/2 & 0.686 & 0.807 & 0.798 & 1.163 & 0.989\\
$\lambda$4075.86     & 9/2--7/2 & 1.000 & 1.000 & 1.000 & 1.000 & 1.000\\
$\lambda$4078.84     & 3/2--3/2 & 0.112 & 0.141 & 0.130 & 1.161 & 0.922\\
$\lambda$4085.11$^c$ & 5/2--5/2 & 0.146 & 0.162 & 0.165 & 1.129 & 1.018\\
\hline
\end{tabular}
\begin{description}
\item [$^a$] Including the contribution from the O~{\sc ii} M10
3d\,$^4$F$_{3/2}$\,--\,3p\,$^4$D$^{\rm o}_{1/2}$ $\lambda$4069.62 line.
\item [$^b$] Corrected for the contribution from the O~{\sc ii} M48a
4f\,G[5]$^{\rm o}_{9/2}$\,--\,3d\,$^{4}$F$_{7/2}$ $\lambda$4071.23 line, which
is about 9 per cent. Neglecting the N~{\sc ii} M38b
4f\,G[7/2]$_{3}$\,--\,3d\,$^{3}$F$^{\rm o}_{2}$ $\lambda$4073.05 line (about 2
per cent).
\item [$^c$] 
Neglecting N~{\sc ii} M38b 4f\,G[7/2]$_{3}$\,--\,3d\,$^{3}$F$^{\rm o}_{3}$
$\lambda$4082.89 (less than 2 per cent).
\end{description}
\end{table}

\subsubsection{\label{oii_orls:4f-3d}
4f -- 3d transitions}

\begin{figure}
\begin{center}
\includegraphics[width=7.5cm,angle=-90]{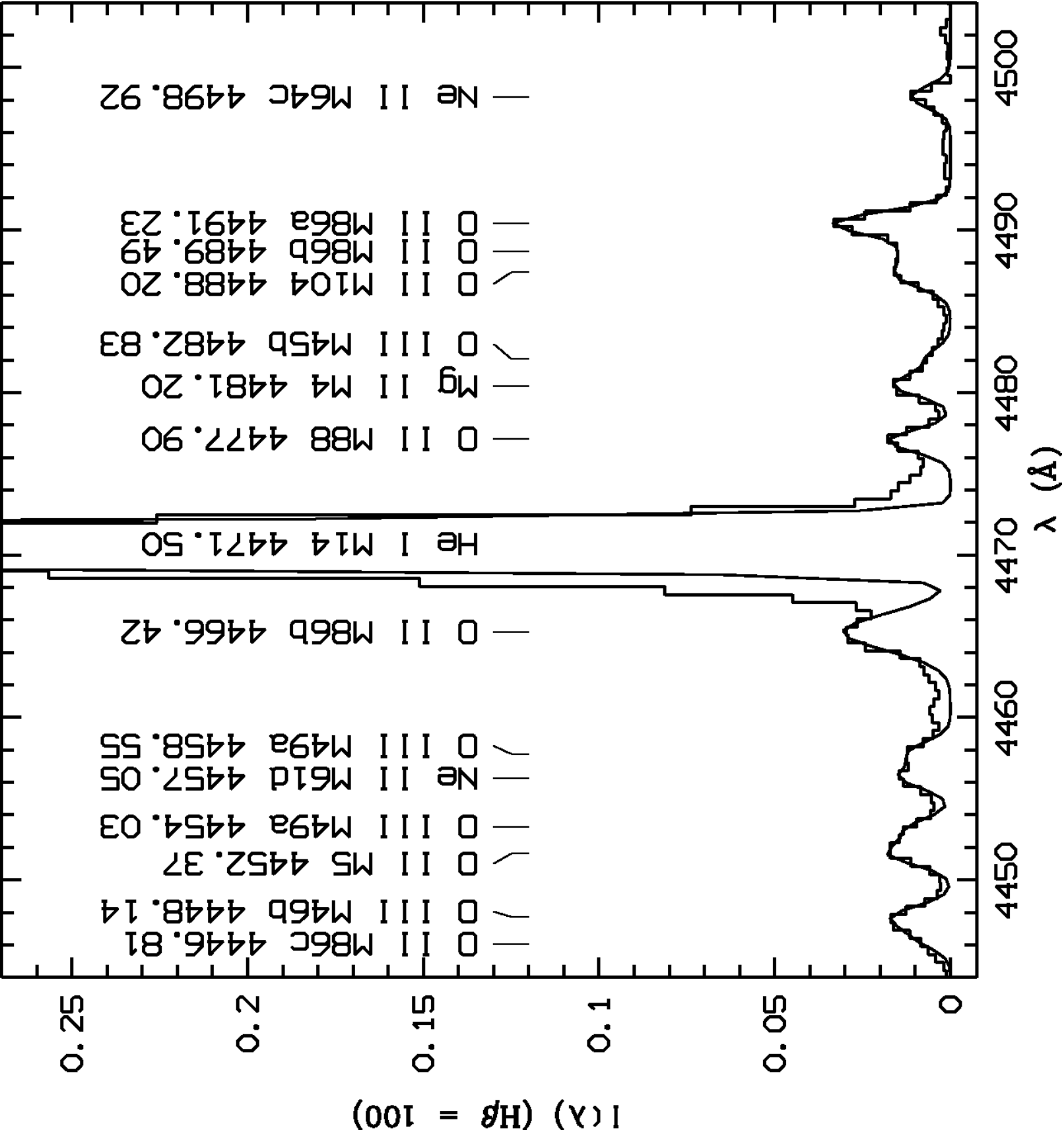}
\caption{Spectrum of NGC\,7009 from 4444 to 4504\,{\AA} showing the O~{\sc ii}
and O~{\sc iii} ORLs. The only Mg~{\sc ii} line detected in NGC\,7009, Mg~{\sc
ii} M4 4f~$^{2}$F$^{\rm o}$ -- 3d~$^{2}$D $\lambda$4481.20, is present. The
continuous curve is the sum of Gaussian profile fits. Continuum has been
subtracted and the spectrum has been normalized such that H$\beta$ has an
integrated flux of 100. Extinction has not been corrected for.}
\label{4444-4504}
\end{center}
\end{figure}

Several dozen transitions of this group were identified and presented in the
emission line list of NGC\,7009 (Paper~I). Table\,\ref{relative:oii_4f-3d}
gives the observed and predicted relative intensities of the 4f\,--\,3d lines
of O~{\sc ii} with the most reliable measurements. For most cases, the
measured intensities agree with both calculations. Here we only present
spectral fits and discussion of the M48a 4f\,G[5]$^{\rm o}$\,--\,3d\,$^{4}$F
multiplet of O~{\sc ii}. Discussion of other multiplets of the O~{\sc ii}
4f\,--\,3d configuration are given in Appendix\,\ref{appendix:oii:4f-3d}.
Figs.\,\ref{4260-4310} and \ref{4444-4504} show some of the detected O~{\sc
ii} ORLs from the 4f\,--\,3d configuration. The strongest 4f\,--\,3d line of
O~{\sc ii} observed in NGC\,7009, $\lambda$4089.29 (M48a 4f\,G[5]$^{\rm
o}_{11/2}$\,--\,3d\,$^{4}$F$_{9/2}$), is shown in Fig.\,\ref{4060-4115}. A
few O~{\sc ii} lines of the 4f\,--\,3d configuration, which are blended with
the N~{\sc ii} lines of the same configuration, are shown in
Fig.\,\ref{4005-4058}.

The $\lambda$4089.29 (M48a 4f\,G[5]$^{\rm o}_{11/2}$\,--\,3d\,$^4$F$_{9/2}$)
line is blended with the $\lambda$4088.27 (M48a 4f\,G[5]$^{\rm
o}_{9/2}$\,--\,3d\,$^4$F$_{9/2}$) and the Si~{\sc iv} M1 4p\,$^2$P$^{\rm
o}_{3/2}$\,--\,4s\,$^2$S$_{1/2}$ $\lambda$4088.86 lines
(Fig.\,\ref{4005-4058}). The $\lambda$4088.27 line contributes less than 2
per cent to the total flux of the blend at $\lambda$4089. The contribution
of the Si~{\sc iv} $\lambda$4088.86 line was estimated from the observed
Si~{\sc iv} M1 4p\,$^2$P$^{\rm o}_{1/2}$\,--\,4s\,$^2$S$_{1/2}$
$\lambda$4116.10, assuming that the relative intensities of the two S~{\sc iv}
M1 lines are as in the pure {\it LS}\,coupling, i.e. 2\,:\,1. The resultant
intensity of the $\lambda$4089.29 line is 0.265$\pm$0.013. The intensity ratio
of $\lambda$4089.29 and the O~{\sc ii} M1 $\lambda$4649.13 line is 0.398,
which agrees with the new theoretical prediction (0.387) within measurement
errors. The other M48a line $\lambda$4071.23 (4f\,G[5]$^{\rm
o}_{9/2}$\,--\,3d\,$^4$F$_{7/2}$) is blended with the O~{\sc ii} M10
3d\,$^{4}$F$_{7/2}$\,--\,3p\,$^{4}$D$^{\rm o}_{5/2}$ $\lambda$4072.16 line
(see Section\,\ref{oii_orls:v10}), which is 10 times stronger.

\begin{table*}
\begin{minipage}{110mm}
\caption{Comparison of the observed and predicted relative intensities of the
O~{\sc ii} 4f\,--\,3d lines detected in the spectrum of NGC\,7009. {\it LSBC}
is the predicted intensities based on the effective recombination coefficients
of Liu et al. \citet{liu1995}, and {\it PJS} is based on the unpublished
effective recombination coefficients of P.~J. Storey. An electron temperature
of 1000~K is assumed for the theoretical predictions.}
\label{relative:oii_4f-3d}
\centering
\begin{tabular}{lcllllr}
\hline
Line & $J_2-J_1$ & $I_{\rm pred}$ & $I_{\rm pred}$ & $I_{\rm obs}$ &
$\frac{I_{\rm obs}}{I_{\rm pred}}$ & $\frac{I_{\rm obs}}{I_{\rm pred}}$\\
 & & {\it LSBC} & {\it PJS} & & {\it LSBC} & {\it PJS}\\
\hline
M48a 4f G[5]$^{\rm o}$ -- 3d $^4$F & & & & &\\
$\lambda$4089.29    & 11/2--9/2  & 1.000 & 1.000 & 1.000 & 1.000 & 1.000\\
M48b 4f G[4]$^{\rm o}$ -- 3d $^4$F & & & & &\\
$\lambda$4083.90     & 7/2--5/2  & 0.285 & 0.316 & 0.326 & 1.141 & 1.032\\
M48c 4f G[3]$^{\rm o}$ -- 3d $^4$F & & & & &\\
$\lambda$4087.15     & 5/2--3/2  & 0.271 & 0.347 & 0.347 & 1.280 & 1.000\\
M50a 4f F[4]$^{\rm o}$ -- 3d $^4$F & & & & &\\
$\lambda$4062.94     & 9/2--9/2  & 0.125 & 0.126 & 0.137 & 1.096 & 1.087\\
M50b 4f F[3]$^{\rm o}$ -- 3d $^4$F & & & & &\\
$\lambda$4048.21     & 7/2--7/2  & 0.063 & 0.068 & 0.076 & 1.206 & 1.120\\
M53a 4f D[3]$^{\rm o}$ -- 3d $^4$P & & & & &\\
$\lambda$4303.83$^a$ & 7/2--5/2  & 0.413 & 0.522 & 0.534 & 1.293 & 1.022\\
M53b 4f D[2]$^{\rm o}$ -- 3d $^4$P & & & & &\\
$\lambda$4294.78$^b$ & 5/2--3/2  & 0.232 & 0.326 & 0.253 & 1.091 & 0.776\\
$\lambda$4307.23     & 3/2--1/2  & 0.105 & 0.118 & 0.108 & 1.031 & 0.919\\
M53c 4f D[1]$^{\rm o}$ -- 3d $^4$P & & & & &\\
$\lambda$4288.82$^c$ & 3/2--1/2  & 0.151 & 0.123 & 0.145 & 0.958 & 1.176\\
M55  4f G[3]$^{\rm o}$ -- 3d $^4$P & & & & &\\
$\lambda$4291.25$^d$ & 7/2--5/2  & 0.156 & 0.188 & 0.221 & 1.414 & 1.176\\
M63a 4f D[3]$^{\rm o}$ -- 3d $^4$D & & & & &\\
$\lambda$4357.25$^e$ & 7/2--5/2  & 0.057 & 0.088 & 0.094 & 1.651 & 1.067\\
M67c 4f F[2]$^{\rm o}$ -- 3d $^4$D & & & & &\\
$\lambda$4282.96$^f$ & 5/2--3/2  & 0.154 & 0.168 & 0.185 & 1.200 & 1.101\\
M76b 4f G[4]$^{\rm o}$ -- 3d $^2$F & & & & &\\
$\lambda$4371.62$^g$ & 9/2--7/2  & 0.097 & 0.109 & 0.127 & 1.303 & 1.159\\
M78a 4f F[4]$^{\rm o}$ -- 3d $^2$F & & & & &\\
$\lambda$4313.44$^h$ & 9/2--7/2  & 0.121 & 0.133 & 0.139 & 1.150 & 1.045\\
M78b 4f F[3]$^{\rm o}$ -- 3d $^2$F & & & & &\\
$\lambda$4285.69     & 7/2--5/2  & 0.189 & 0.264 & 0.229 & 1.208 & 0.869\\
M86a 4f D[3]$^{\rm o}$ -- 3d $^2$P & & & & &\\
$\lambda$4491.23     & 5/2--3/2  & 0.137 & 0.198 & 0.215 & 1.569 & 1.086\\
M86b 4f D[2]$^{\rm o}$ -- 3d $^2$P & & & & &\\
$\lambda$4489.49     & 3/2--1/2  & 0.065 & 0.082 & 0.083 & 1.271 & 1.004\\
M88  4f G[3]$^{\rm o}$ -- 3d $^2$P & & & & &\\
$\lambda$4477.90$^i$ & 5/2--3/2  & 0.086 & 0.109 & 0.113 & 1.316 & 1.108\\
M92a 4f F[4]$^{\rm o}$ -- 3d $^2$D & & & & &\\
$\lambda$4609.44     & 7/2--5/2  & 0.428 & 0.444 & 0.483 & 1.126 & 1.088\\
M92b 4f F[3]$^{\rm o}$ -- 3d $^2$D & & & & &\\
$\lambda$4602.13$^j$ & 5/2--3/2  & 0.171 & 0.194 & 0.195 & 1.139 & 1.003\\
\hline
\end{tabular}
\begin{description}
\item [$^a$] Corrected for the contribution from O~{\sc ii} M65a 4f\,G[5]$^{\rm
o}_{9/2}$\,--\,3d\,$^4$D$_{7/2}$ $\lambda$4303.61 (about 12 per cent).
Neglecting O~{\sc ii} M53a 4f\,D[3]$^{\rm o}_{5/2}$\,--\,3d\,$^4$P$_{5/2}$
$\lambda$4304.08 ($\sim$3 per cent).
\item [$^b$] Including O~{\sc ii} M53b 4f\,D[2]$^{\rm
o}_{3/2}$\,--\,3d\,$^4$P$_{3/2}$ $\lambda$4294.92 ($\sim$12 per cent).
\item [$^c$] Including O~{\sc ii} M53c 4f\,D[1]$^{\rm
o}_{1/2}$\,--\,3d\,$^4$P$_{1/2}$ $\lambda$4288.82.
\item [$^d$] Including O~{\sc ii} M78c 4f\,F[2]$^{\rm
o}_{5/2}$\,--\,3d\,$^{2}$F$_{5/2}$ $\lambda$4292.21. Neglecting O~{\sc ii}
M55 4f\,G[3]$^{\rm o}_{5/2}$\,--\,3d\,$^4$P$_{5/2}$ $\lambda$4291.86 and
O~{\sc ii} M78c 4f\,F[2]$^{\rm o}_{3/2}$\,--\,3d\,$^{2}$F$_{5/2}$
$\lambda$4292.95.
\item [$^e$] Including O~{\sc ii} M63a 4f\,D[3]$^{\rm
o}_{5/2}$\,--\,3d\,$^4$D$_{3/2}$ $\lambda$4357.25. Neglecting O~{\sc ii}
M63a 4f\,D[3]$^{\rm o}_{5/2}$\,--\,3d\,$^4$D$_{5/2}$ $\lambda$4357.52.
\item [$^f$] Including O~{\sc ii} M67c 4f\,F[2]$^{\rm
o}_{5/2}$\,--\,3d\,$^4$D$_{5/2}$ $\lambda$4283.25. Neglecting O~{\sc ii}
M78a 4f\,F[4]$^{\rm o}_{7/2}$\,--\,3d\,$^2$F$_{5/2}$ $\lambda$4282.02.
\item [$^g$] Neglecting O~{\sc ii} M76b 4f\,G[4]$^{\rm
o}_{7/2}$\,--\,3d\,$^{2}$F$_{7/2}$ $\lambda$4371.24 (less than 2 per cent).
\item [$^h$] Corrected for the contribution from the O~{\sc ii} M78a
4f\,F[4]$^{\rm o}_{7/2}$\,--\,3d\,$^{2}$F$_{7/2}$ $\lambda$4312.11 line
($\sim$33 per cent).
\item [$^i$] Neglecting O~{\sc iii} M45a 5g\,H[11/2]$^{\rm
o}_{5,\,6}$\,--\,4f\,G[9/2]$_{5}$ $\lambda$4477.91 (less than 2 per cent).
\item [$^j$] Corrected for the contribution from N\,{\sc ii} M5
3p~$^{3}$P$_{2}$\,--\,3s\,$^{3}$P$^{\rm o}_{1}$ $\lambda$4601.48 (26 per
cent). Neglecting Ne~{\sc ii} M64d 4f\,2[2]$^{\rm
o}_{5/2}$\,--\,3d\,$^{4}$P$_{5/2}$ $\lambda$4600.16.
\end{description}
\end{minipage}
\end{table*}

\begin{figure}
\begin{center}
\includegraphics[width=7.5cm,angle=-90]{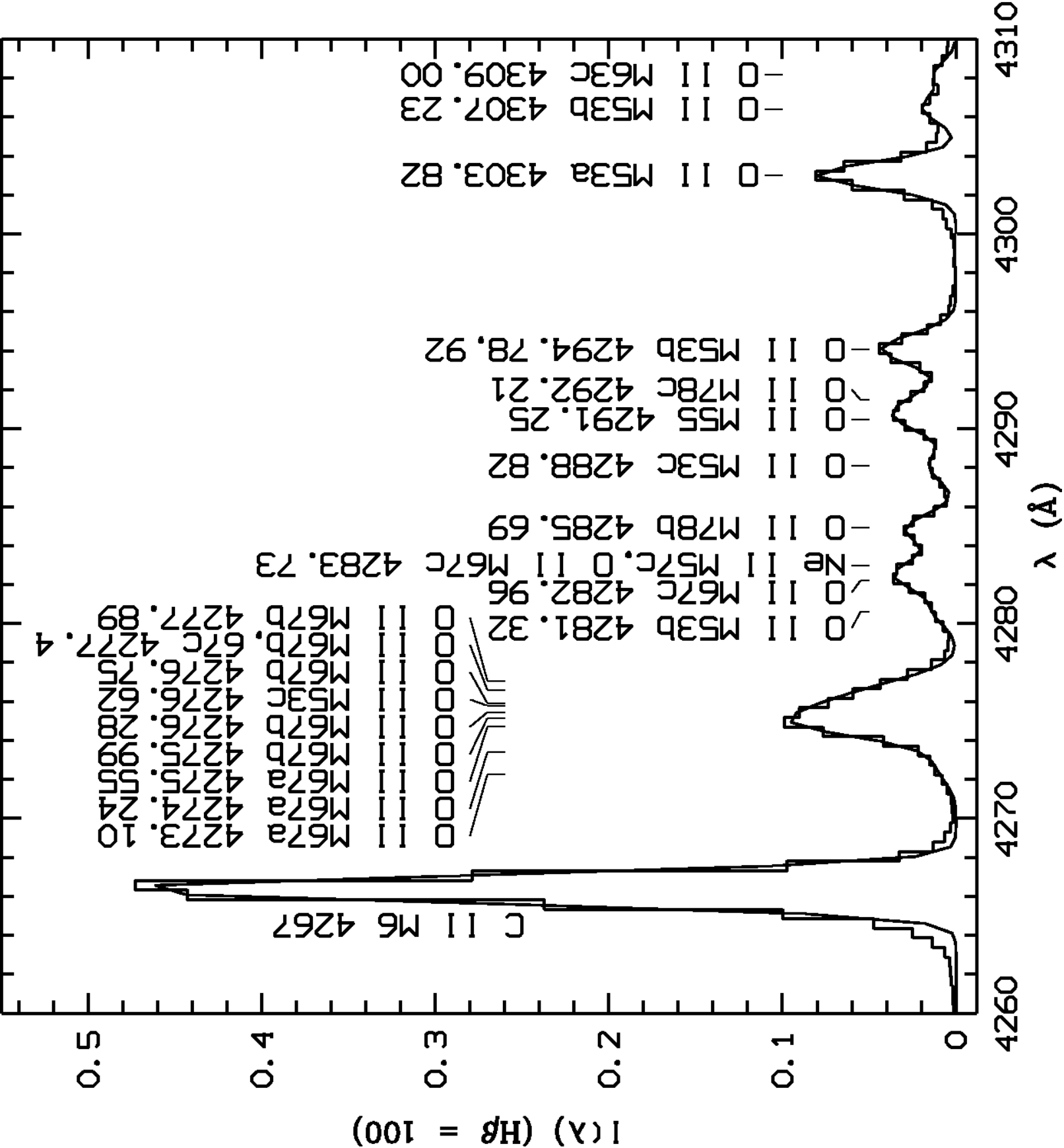}
\caption{Spectrum of NGC\,7009 from 4260 to 4310\,{\AA} showing the O~{\sc ii}
ORLs from the M53a 4f~D[3]$^{\rm o}$ -- 3d~$^4$P, M53b 4f~D[2]$^{\rm o}$ --
3d~$^4$P, M53c 4f~D[1]$^{\rm o}$ -- 3d~$^4$P, M67a 4f~F[4]$^{\rm o}$ --
3d~$^4$D, M67b 4f~F[3]$^{\rm o}$ -- 3d~$^4$D, and M67c 4f~F[2]$^{\rm o}$ --
3d~$^4$D multiplets. The very broad emission feature at $\lambda$4275 is
formed by more than 10 O~{\sc ii} ORLs from the 4f -- 3d configuration blended
together, with positions and wavelengths of the components labeled. The
continuous curve is the sum of Gaussian profile fits. Continuum has been
subtracted and the spectrum has been normalized such that H$\beta$ has an
integrated flux of 100. Extinction has not been corrected for.}
\label{4260-4310}
\end{center}
\end{figure}

\subsubsection{\label{oii_orls:other_parentage}
Multiplets with parentage other than $^3$P}

The O~{\sc ii} ORLs from multiplets with parentage other than $^3$P are
detected in NGC\,7009, and they include M15, M36, M101 and M105, which were
also observed by LSBC. However, the effective recombination coefficients are
only available for two of those multiplets (c.f. discussion in
Section\,\ref{diagnose:niioii}). Table\,\ref{oii_parent:1d} presents the line
intensities. The intensities observed by LSBC are also listed. The O~{\sc ii}
M15 3p$^{\prime}$\,$^{2}$F$^{\rm o}$\,--\,3s$^{\prime}$\,$^{2}$D lines
$\lambda\lambda$4590.97 and 4596.18 are shown in Fig.\,\ref{4555-4625}. The
O~{\sc ii} M36 3d$^{\prime}$\,$^2$G\,--\,3p$^{\prime}$\,$~2$F$^{\rm o}$ lines
$\lambda\lambda$4185.45 and 4189.79 are shown in Fig.\,\ref{4176-4260}.

\begin{table*}
\begin{minipage}{110mm}
\caption{The O~{\sc ii} optical recombination lines in NGC\,7009 with
parentage other than $^{3}$P$_{J}$. Line intensities measured by LSBC are also
presented. All intensities are normalized such that $I$(H$\beta$) = 100.}
\label{oii_parent:1d}
\centering
\begin{tabular}{lcclll}
\hline
Line & Mult.& Transition  &   Current   &   \multicolumn{2}{c}{LSBC}\\
(\AA)      &      &             &             & PA=0$^{\rm o}$ & PA=45$^{\rm o}$\\
\hline
$\lambda$4590.97     &  M15  & ($^{1}$D)3p~$^{2}$F$^{\rm o}_{7/2}$ -- ($^{1}$D)3s~$^{2}$D$_{5/2}$ & 0.087 & 0.0907  & 0.0752\\
$\lambda$4596.18$^a$ &  M15  & ($^{1}$D)3p~$^{2}$F$^{\rm o}_{5/2}$ -- ($^{1}$D)3s~$^{2}$D$_{3/2}$ & 0.062 & 0.0424  & 0.0476\\
$\lambda$4185.45     &  M36  & ($^{1}$D)3d~$^{2}$G$_{7/2}$ -- ($^{1}$D)3p~$^{2}$F$^{\rm o}_{5/2}$ & 0.070 & 0.0572  & 0.0580\\
$\lambda$4189.79$^b$ &  M36  & ($^{1}$D)3d~$^{2}$G$_{9/2}$ -- ($^{1}$D)3p~$^{2}$F$^{\rm o}_{7/2}$ & 0.083 & 0.0703  & 0.0747\\
$\lambda$4253.89$^c$ &  M101 & ($^{1}$D)4f~H[5]$^{\rm o}_{11/2}$ -- ($^{1}$D)3d~$^{1}$G$_{9/2}$   & 0.058 & 0.0318: & 0.0249:\\
$\lambda$4843.37$^d$ &  M105 & ($^{1}$D)4f~P[1]$^{\rm o}_{3/2}$ -- ($^{1}$D)3d~$^{2}$S$_{1/2}$    & 0.021 &         &       \\
\hline
\end{tabular}
\begin{description}
\item [$^a$] Including O~{\sc ii} M15 ($^{1}$D)3p~$^{2}$F$^{\rm o}_{5/2}$ -- ($^{1}$D)3s~$^{2}$D$_{5/2}$ $\lambda$4595.96.
\item [$^b$] Including O~{\sc ii} M36 ($^{1}$D)3d~$^{2}$G$_{7/2}$ -- ($^{1}$D)3p~$^{2}$F$^{\rm o}_{7/2}$ $\lambda$4189.59.
\item [$^c$] Including O~{\sc ii} M101 ($^{1}$D)4f~H[5]$^{\rm o}_{9/2}$ -- ($^{1}$D)3d~$^{1}$G$_{9/2}$ $\lambda$4253.91 and O~{\sc ii} M101 ($^{1}$D)4f~H[5]$^{\rm o}_{9/2}$ -- ($^{1}$D)3d~$^{1}$G$_{7/2}$ $\lambda$4254.12.
\item [$^d$] Including O~{\sc ii} $\lambda$4843.37 M105 ($^{1}$D)4f~P[1]$^{\rm o}_{1/2}$ -- ($^{1}$D)3d~$^{2}$S$_{1/2}$.
\end{description}
\end{minipage}
\end{table*}

\subsubsection{\label{oii_orls;summary}
Comments on the O~{\sc ii} recombination spectrum}

The unpublished effective recombination coefficients of P.~J. Storey used in
the current analysis of the O~{\sc ii} recombination spectrum are accurate at
low temperatures ($T_\mathrm{e}\,<$\,10\,000~K). Appropriate assumptions have
been made in the calculation, as described in Section\,\ref{diagnose:data}.
The O~{\sc ii} recombination spectrum is the richest amongst all the heavy
element ions observed in NGC\,7009. The best observed multiplets of the
3\,--\,3 transitions of O~{\sc ii} are M1
(3p\,$^{4}$D$^{\rm o}$\,--\,3s\,$^{4}$P), M2
(3p\,$^{4}$P$^{\rm o}$\,--\,3s\,$^{4}$P), M10
(3d\,$^{4}$F\,--\,3p\,$^{4}$D$^{\rm o}$), M19
(3d\,$^{4}$P\,--\,3p\,$^{4}$P$^{\rm o}$), M25
(3d\,$^{2}$F\,--\,3p\,$^{2}$D$^{\rm o}$) and M28
(3d\,$^{4}$P\,--\,3p\,$^{4}$S$^{\rm o}$). Although some fine-structure
components of those multiplets are blended with other lines, multi-Gaussian
profile fitting gives reliable intensities for most of them. The 4f\,--\,3d
lines with the most reliable measurements are presented in
Table\,\ref{relative:oii_4f-3d}. The O~{\sc ii} multiplets are not affected
by any other excitation mechanisms (e.g. fluorescence, charge transfer), and
the strongest lines have been used for plasma diagnostics
(Section\,\ref{diagnose:niioii}). Several O~{\sc ii} lines with the parentage
$^{1}$D are detected, as presented in Table\,\ref{oii_parent:1d}. Only the
dielectronic recombination coefficients for the M15 and M36 multiplets are
available from Nussbaumer \& Storey \cite{ns1984}. The calculation of PJS does
not reach such high energy.

\subsection{\label{neii_orls}
The Ne~{\sc ii} optical recombination spectrum}

Several dozen emission features in the spectrum of NGC\,7009 were identified
as the Ne~{\sc ii} permitted lines. In this section, we present spectral fits
and discussion of the results only for the M2 3p\,$^4$D$^{\rm
o}$\,--\,3s\,$^4$P, M13 3d\,$^{4}$F\,--\,3p\,$^{4}$D$^{\rm o}$, M9
3p$^{\prime}$\,$^2$F$^{\rm o}$\,--\,3s$^{\prime}$\,$^2$D, and the M55e
4f\,2[5]$^{\rm o}$\,--\,3d\,$^4$F multiplets. Discussion of other multiplets
of Ne~{\sc ii} are given in Appendix\,\ref{appendix:d}. Line intensities
measured by LLB01 are used for comparison if available. Predicted intensities
that are based on the {\it LS}\,coupling calculations of Kisielius et al.
\cite{kisielius1998} are also used for the analysis.

\subsubsection{\label{neii_orls:v2}
Multiplet 2, 3p $^4$D$^{\rm o}$ -- 3s $^4$P}

This is the strongest multiplet of Ne~{\sc ii}. Fig.\,\ref{3290-3348} shows
that $\lambda$3334.84 (3p~$^4$D$^{\rm o}_{7/2}$ -- 3s~$^4$P$_{5/2}$) is
affected by the O~{\sc iii} fluorescence line M3 3p~$^3$S$_{1}$ --
3s~$^3$P$^{\rm o}_{2}$ $\lambda$3340.76, which is more than 10 times stronger.
The fitted intensity of $\lambda$3334.84 is 0.428, which is higher than LLB01
(0.345). Given that our observational data are the
same as LLB01, such difference is probably due to the different reddening laws
used. Besides, measurements of $\lambda$3334.84 could be overestimated due to
the O~{\sc iii} line.

Another line $\lambda$3355.02 (3p~$^4$D$^{\rm o}_{5/2}$ -- 3s~$^4$P$_{3/2}$)
blends with He~{\sc i} M8 7p~$^1$P$^{\rm o}_{1}$ -- 2s~$^1$S$_{0}$
$\lambda$3354.55 (Fig.\,\ref{3345-3410}), which contributes 43 per cent to the
total intensity, as estimated from the calculations of Benjamin, Skillman \&
Smits \cite{bss99}. The resultant intensity of $\lambda$3355.02 is
0.223$\pm$0.045. The intensity
ratio $\lambda$3355.02/$\lambda$3334.84 agrees well with the predicted value.
The intensity of another line $\lambda$3360.60 (3p~$^4$D$^{\rm o}_{3/2}$ --
3s~$^4$P$_{1/2}$) is 0.040, which is unreliable (Fig.\,\ref{3345-3410}). The
other M2 lines are not observed.

\subsubsection{\label{neii_orls:v13}
Multiplet 13, 3d $^4$F -- 3p $^4$D$^{\rm o}$}

\begin{figure}
\begin{center}
\includegraphics[width=7.5cm,angle=-90]{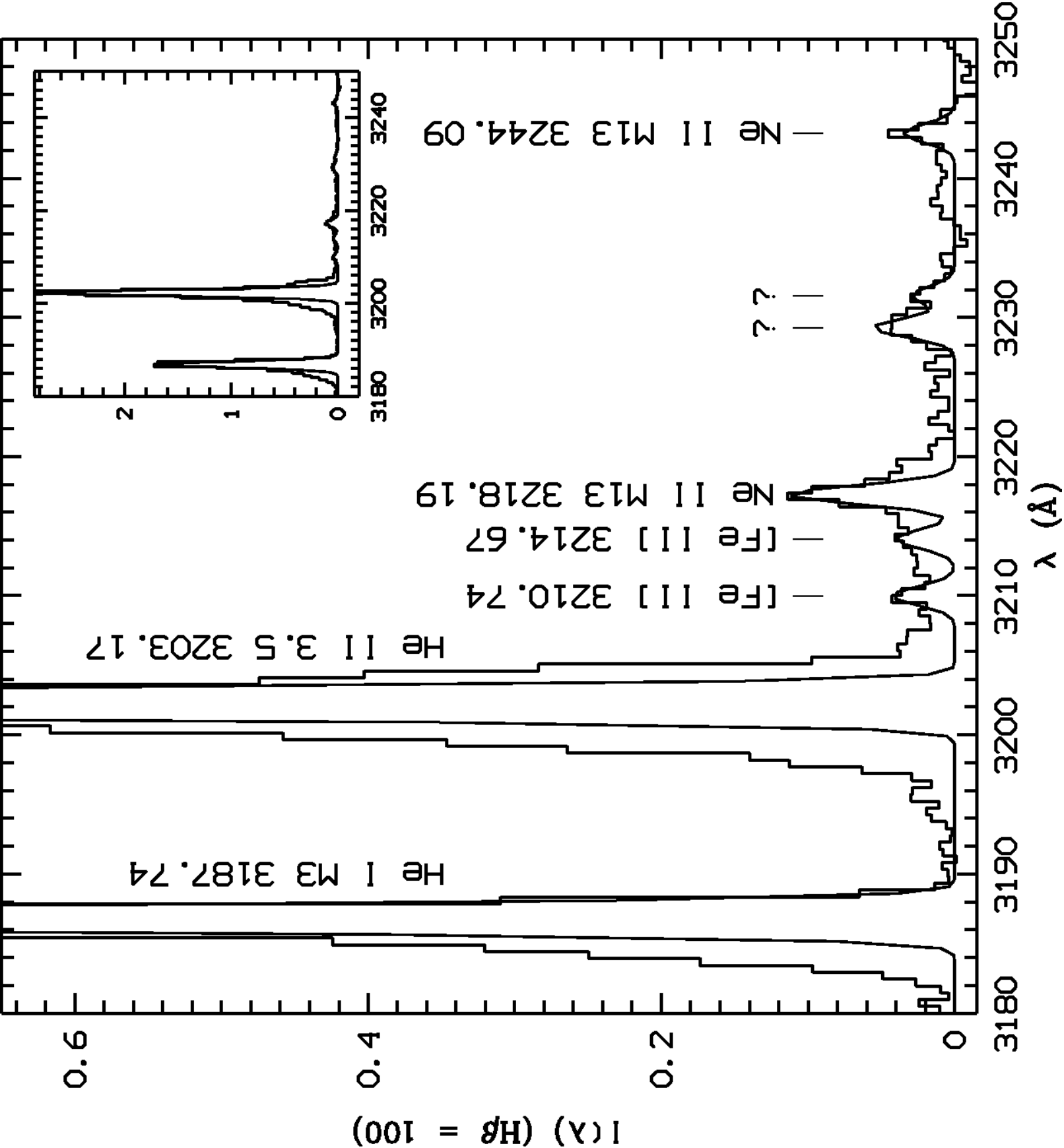}
\caption{Spectrum of NGC\,7009 from 3180 to 3250\,{\AA} showing the Ne~{\sc
ii} M13 lines and other emission features. The inset shows the profiles of
the two strong lines He~{\sc i} M3 4p\,$^3$P$^{\rm o}$\,--\,2s\,$^3$S
$\lambda$3187.74 and He~{\sc ii} 5f\,$^2$F$^{\rm
o}_{7/2}$\,--\,3d\,$^2$D$_{5/2}$ $\lambda$3203.17. The continuous curve is
the sum of Gaussian profile fits. Continuum has been subtracted and the
spectrum has been normalized such that H$\beta$ has an integrated flux of
100. Extinction has not been corrected for.}
\label{3180-3250}
\end{center}
\end{figure}

Ne~{\sc ii} M13 lines locate in the far-blue and they are difficult to observe
due to weakness as well as the relatively poor S/N. Only $\lambda$3218.19
(3d~$^4$F$_{9/2}$ -- 3p~$^4$D$^{\rm o}_{7/2}$) and $\lambda$3244.09
(3d~$^4$F$_{9/2}$ -- 3p~$^4$D$^{\rm o}_{7/2}$) are observed, and they are shown
in Fig.\,\ref{3180-3250}. The fitted intensity of $\lambda$3218.19 is 0.234,
which agrees well with LLB01 (0.229). Here the blended
Ne~{\sc ii} M16 3d~$^4$P$_{1/2}$ -- 3p~$^4$D$^{\rm o}_{3/2}$ $\lambda$3217.30
was assumed to be negligible ($\sim$3 per cent). The intensity ratio of
$\lambda$3218.19 to the Ne~{\sc ii} M2 $\lambda$3334.84 line is 0.546, which
agrees with the predicted ratio 0.594. The intensity of the $\lambda$3244.09
line is 0.074, higher than the measurement of LLB01 (0.0576).

\subsubsection{\label{neii_orls:v9}
Multiplet 9, 3p$^{\prime}$ $^2$F$^{\rm o}$ -- 3s$^{\prime}$ $^2$D}

\begin{figure}
\begin{center}
\includegraphics[width=7.5cm,angle=-90]{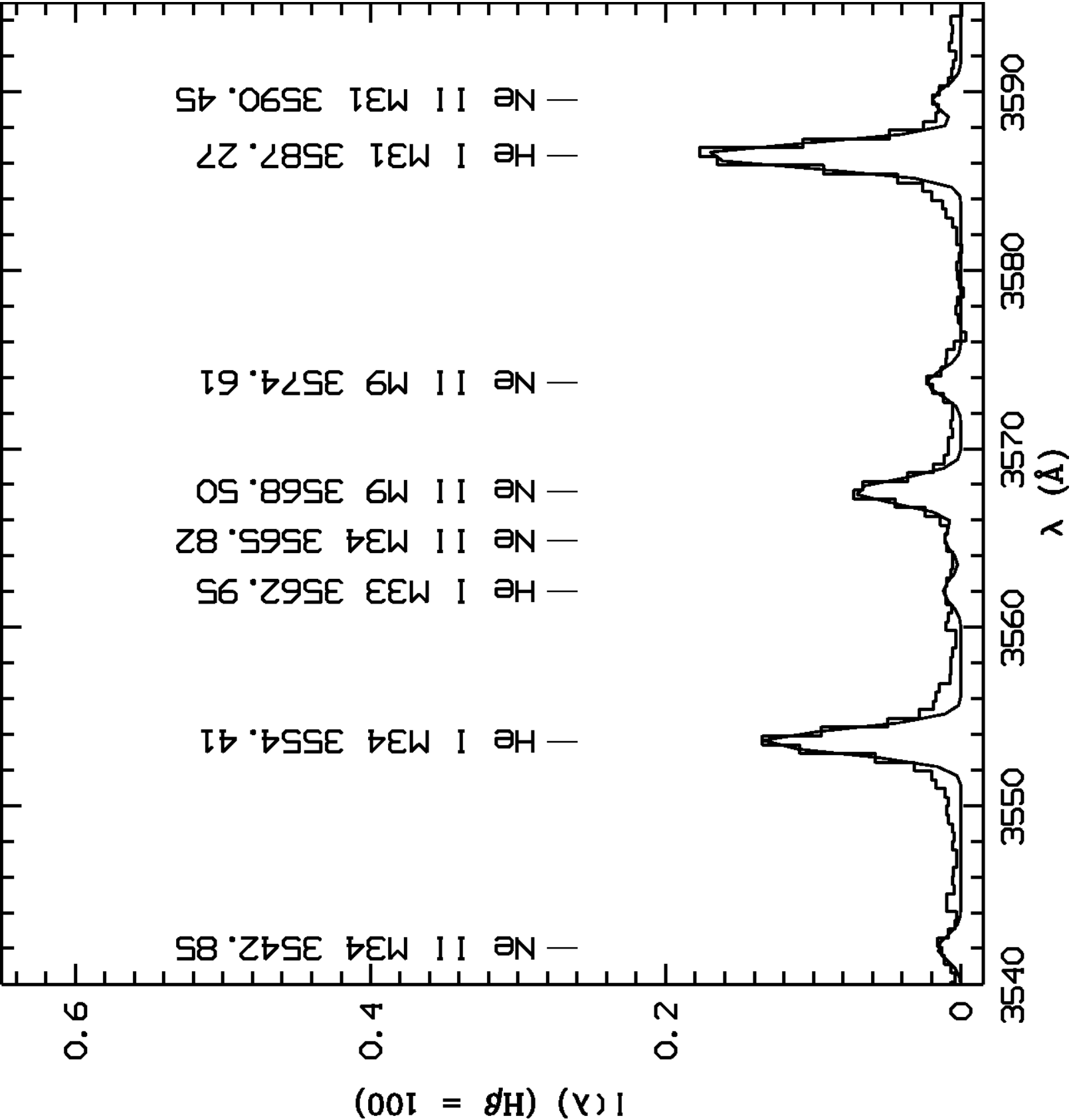}
\caption{Spectrum of NGC\,7009 from 3540 to 3595\,{\AA} showing the Ne~{\sc
ii} M9 and M34 lines as well as other emission features. The continuous curve
is the sum of Gaussian profile fits. Continuum has been subtracted and the
spectrum has been normalized such that H$\beta$ has an integrated flux of 100.
Extinction has not been corrected for.}
\label{3540-3595}
\end{center}
\end{figure}

This is the only Ne~{\sc ii} multiplet of an excited-state parentage
(2s$^{2}$2p$^{4}$\,$^{1}$D) detected in the spectrum of NGC\,7009, and shown
in Fig.\,\ref{3540-3595}. The fitted intensity of $\lambda$3568.50
(3p$^{\prime}$\,$^2$F$^{\rm o}_{7/2}$\,--\,3s$^{\prime}$\,$^2$D$_{5/2}$) is
0.168$\pm$0.016. The measured total intensity of the other two lines
$\lambda\lambda$3574.18,\,61 is 0.053, with an uncertainty of about 10 per
cent. Thus the intensity ratio
($\lambda\lambda$3574.18\,+\,3574.61)/$\lambda$3568.50 is 0.32, which differs
from the {\it LS}\,coupling value 0.75. No measurements of this multiplet are
given in LLB01.

\subsubsection{\label{neii_orls:4f-3d}
4f -- 3d transitions}

Several dozen Ne~{\sc ii} ORLs of the 4f -- 3d configuration are detected or
estimated by deblending techniques (i.e. Gaussian profile fitting). The
measured intensities are 10$^{-4}$ of H$\beta$ or even lower.
Table\,\ref{relative:neii_4f-3d} presents the measured and predicted relative
intensities of the 4f\,--\,3d transitions with the most reliable measurements.
The predicted intensities are based on the preliminary effective recombination
coefficients calculated by P.~J. Storey (private communication) for a few
selected Ne~{\sc ii} 4f\,--\,3d quartet lines. In this section, we only
present spectral fits and results for the M55e 4f\,2[5]$^{\rm
o}$\,--\,3d\,$^4$F multiplet of Ne~{\sc ii}. Discussion of other multiplets
of the 4f\,--\,3d configuration are given in
Appendix\,\ref{appendix:neii:4f-3d}. Some of the Ne~{\sc ii} lines belonging
to the M52a 4f\,2[4]$^{\rm o}$\,--\,3d\,$^4$D and M52b 4f\,2[3]$^{\rm
o}$\,--\,3d\,$^4$D multiplets are shown in Fig\,\ref{4176-4260}.
Fig\,\ref{4380-4445} also shows many Ne~{\sc ii} lines detected in the
spectrum of NGC\,7009.

The $\lambda$4391.99 (M55e 4f\,2[5]$^{\rm o}_{11/2}$\,--\,3d\,$^4$F$_{9/2}$)
line observed in NGC\,7009 is shown in Fig.\,\ref{4380-4445}. Its intensity
is 0.077$\pm$0.008, which agrees with LLB01: 0.0728 (ESO 1.52~m), 0.0779 (WHT
1996) and 0.0713 (WHT 1997). The contribution from the $\lambda$4392.00
(4f\,2[5]$^{\rm o}_{9/2}$\,--\,3d\,$^4$F$_{9/2}$) line of the same multiplet
is negligible. The other M55e line $\lambda$4409.30 (4f\,2[5]$^{\rm
o}_{9/2}$\,--\,3d\,$^4$F$_{7/2}$) is blended with the Ne~{\sc ii}
$\lambda$4409.78 (M55b 4f\,2[3]$^{\rm o}_{7/2}$\,--\,3d\,$^4$F$_{9/2}$) line,
whose intensity is negligible, and the O~{\sc iii} $\lambda$4408.29 (M46a
5g\,H[5]$^{\rm o}_{4}$\,--\,4f\,G[4]$_{3}$) line, which contributes about 10
per cent to the total intensity, as estimated from the effective recombination
coefficients of Kisielius \& Storey \cite{ks1999}. The resultant intensity of
the $\lambda$4409.30 line is 0.060, which agrees with LLB01: 0.0631 (ESO
1.52~m), 0.0615 (WHT 1996) and 0.0624 (WHT 1997). The intensity ratio
$\lambda$4409.30/$\lambda$4391.99 observed in NGC\,7009 is 20 per cent higher
than the predicted value (Table\,\ref{relative:neii_4f-3d}).

\begin{table}
\centering
\caption{Comparison of the observed and predicted relative intensities of
the Ne~{\sc ii} 4f\,--\,3d lines detected in the spectrum of NGC\,7009. The
predicted intensities $I_{\rm pred}$ are based on the preliminary effective
recombination coefficients for some of the strongest Ne~{\sc ii} lines from
the 4f\,--\,3d configuration (P.~J. Storey, private communication).}
\label{relative:neii_4f-3d}
\begin{tabular}{lcllr}
\hline
Line & $J_2-J_1$ & $I_{\rm pred}$ & $I_{\rm obs}$ & $\frac{I_{\rm obs}}{I_{\rm
pred}}$\\
\hline
M55e 4f 2[5]$^{\rm o}$ -- 3d $^4$F & & & &\\
$\lambda$4391.99$^a$ & 11/2--9/2 & 1.000 & 1.000 & 1.000\\
$\lambda$4409.30$^b$ &  9/2--7/2 & 0.665 & 0.812 & 1.221\\
M52a 4f 2[4]$^{\rm o}$ -- 3d $^4$D & & & &\\
$\lambda$4219.75$^c$ &  9/2--7/2 & 0.555 & 0.831 & 1.497\\
$\lambda$4233.85     &  7/2--5/2 & 0.139 & 0.227 & 1.631\\
M52b 4f 2[3]$^{\rm o}$ -- 3d $^4$D & & & &\\
$\lambda$4231.64$^d$ &  7/2--5/2 & 0.131 & 0.370 & 2.811\\
$\lambda$4250.65     &  5/2--3/2 & 0.088 & 0.216 & 2.461\\
M57b 4f 1[4]$^{\rm o}$ -- 3d $^4$F & & & &\\
$\lambda$4397.99     &  7/2--5/2 & 0.346 & 0.325 & 0.941\\
M60c 4f 1[3]$^{\rm o}$ -- 3d $^2$F & & & &\\
$\lambda$4428.64$^e$ &  7/2--5/2 & 0.437 & 0.591 & 1.353\\
M61a 4f 2[4]$^{\rm o}$ -- 3d $^2$D & & & &\\
$\lambda$4430.94$^f$ &  7/2--5/2 & 0.283 & 0.452 & 1.598\\
M61d 4f 2[2]$^{\rm o}$ -- 3d $^2$D & & & &\\
$\lambda$4457.05$^g$ &  5/2--3/2 & 0.098 & 0.361 & 3.662\\
M65  4f 0[3]$^{\rm o}$ -- 3d $^4$P & & & &\\
$\lambda$4413.22$^h$ &  7/2--5/2 & 0.239 & 0.376 & 1.574\\
M66c 4f 1[3]$^{\rm o}$ -- 3d $^4$P & & & &\\
$\lambda$4421.39     &  5/2--3/2 & 0.089 & 0.113 & 1.263\\
\hline
\end{tabular}
\begin{description}
\item [$^a$] Neglecting Ne~{\sc ii} M55e 4f\,2[5]$^{\rm
o}_{9/2}$\,--\,3d\,$^4$F$_{9/2}$ $\lambda$4392.00.
\item [$^b$] Corrected for the contribution from the O~{\sc iii} M46a
5g\,H[5]$^{\rm o}_{4}$\,--\,4f\,G[4]$_{3}$ $\lambda$4408.29 line ($\sim$10
per cent). Neglecting Ne~{\sc ii} M55b 4f\,2[3]$^{\rm
o}_{7/2}$\,--\,3d\,$^4$F$_{9/2}$ $\lambda$4409.78.
\item [$^c$] Including Ne~{\sc ii} M52d 4f\,2[2]$^{\rm
o}_{5/2}$\,--\,3d\,$^4$D$_{5/2}$ $\lambda$4220.89. Neglecting Ne~{\sc ii}
M52a 4f\,2[4]$^{\rm o}_{7/2}$\,--\,3d\,$^4$D$_{7/2}$ $\lambda$4219.37.
\item [$^d$] Including Ne~{\sc ii} M52b 4f\,2[3]$^{\rm
o}_{5/2}$\,--\,3d\,$^4$D$_{5/2}$ $\lambda$4231.53.
\item [$^e$] Including Ne~{\sc ii} M61b 4f\,2[3]$^{\rm
o}_{7/2}$\,--\,3d\,$^2$D$_{5/2}$ $\lambda$4428.52. Neglecting Ne~{\sc ii} M60c
4f\,1[3]$^{\rm o}_{5/2}$\,--\,3d\,$^2$F$_{5/2}$ $\lambda$4428.52 and Ne~{\sc
ii} M61b 4f\,2[3]$^{\rm o}_{5/2}$\,--\,3d\,$^2$D$_{5/2}$ $\lambda$4428.41.
\item [$^f$] Including Ne~{\sc ii} M57a 4f\,1[2]$^{\rm
o}_{5/2}$\,--\,3d\,$^4$F$_{3/2}$ $\lambda$4430.90 and Ne~{\sc ii} M55a
4f\,2[4]$^{\rm o}_{9/2}$\,--\,3d\,$^4$F$_{7/2}$ $\lambda$4430.06. Neglecting
Ne~{\sc ii} M57a 4f\,1[2]$^{\rm o}_{3/2}$\,--\,3d\,$^4$F$_{3/2}$
$\lambda$4431.11.
\item [$^g$] Neglecting Ne~{\sc ii} M61d 4f\,2[2]$^{\rm
o}_{3/2}$\,--\,3d\,$^2$D$_{3/2}$ $\lambda$4457.24 and Ne~{\sc ii} M66c
4f\,1[3]$^{\rm o}_{5/2}$\,--\,3d\,$^4$P$_{5/2}$ $\lambda$4457.24. Including
Ne~{\sc ii} M66c 4f\,1[3]$^{\rm o}_{7/2}$\,--\,3d\,$^4$P$_{5/2}$
$\lambda$4457.36.
\item [$^h$] Neglecting Ne~{\sc ii} M65 4f\,0[3]$^{\rm
o}_{5/2}$\,--\,3d\,$^4$P$_{5/2}$ $\lambda$4413.11. Including Ne~{\sc ii} M57c
4f\,1[3]$^{\rm o}_{5/2}$\,--\,3d\,$^4$F$_{3/2}$ $\lambda$4413.11.
\end{description}
\end{table}

\subsubsection{\label{neii_orls:summary}
Comments on the Ne~{\sc ii} recombination spectrum}

The Ne~{\sc ii} effective recombination coefficients currently used are mainly
from Kisielius et al. \cite{kisielius1998}. This calculation is carried out
for the transitions of $l\,\leq$\,2. Although some effective recombination
coefficients for the Ne~{\sc ii} 4f\,--\,3d transitions are available (P.~J.
Storey, unpublished), only a few selected Ne~{\sc ii} lines are calculated,
and the results are quite preliminary. The best observed Ne~{\sc ii} multiplets
of the 3\,--\,3 transitions is M2 (3p\,$^{4}$D$^{\rm o}$\,--\,3s\,$^{4}$P).
For the Ne~{\sc ii} multiplets, M12 (3d\,$^4$D\,--\,3p\,$^4$D$^{\rm o}$),
M13 (3d\,$^4$F\,--\,3p\,$^4$D$^{\rm o}$), M20 (3d\,$^2$F\,--\,3p\,$^2$D$^{\rm
o}$), M21 (3d\,$^2$D\,--\,3p\,$^2$D$^{\rm o}$), M28
(3d\,$^2$P\,--\,3p\,$^2$S$^{\rm o}$) and M34 (3d\,$^4$P\,--\,3p\,$^4$S$^{\rm
o}$), only the strongest fine-structure components are observed, and the weaker
components are either blended with other lines or not detected. Measurements
of the M1 (3p\,$^{4}$P$^{\rm o}$\,--\,3s\,$^{4}$P), M5 (3p\,$^2$D$^{\rm
o}$\,--\,3s\,$^2$P) and M6 (3p\,$^2$S$^{\rm o}$\,--\,3s\,$^2$P) lines are of
large uncertainties due to line blending. The M9 (3p$^{\prime}$\,$^2$F$^{\rm
o}$\,--\,3s$^{\prime}$\,$^2$D) multiplet is the only Ne~{\sc ii} transition
observed in out spectrum of parentage other than $^{3}$P. The effective
recombination coefficients for this multiplet are available from Kisielius et
al. \cite{kisielius1998}, which are probably unreliable considering that the
calculation is in the {\it LS}\,coupling assumption. The possibility of using
the Ne~{\sc ii} lines to determine electron temperatures is discussed in
Section\,\ref{diagnose:neii}. In NGC\,7009, the effects of the fluorescence
mechanism on the Ne~{\sc ii} M1 and M2 multiplets are probably not important.
Thus those lines could be safe for plasma diagnostics and abundance
determinations.

\subsection{\label{ciii_orls}
The C~{\sc iii} permitted lines}

More than 20 lines in NGC\,7009 were identified as C~{\sc iii} permitted
transitions (Paper~I); some identifications could be questionable. In this
section, we only introduce three multiplets: M1 3p\,$^3$P$^{\rm
o}$\,--\,3s\,$^3$S, M16 5g\,$^3$G\,--\,4f\,$^3$F$^{\rm o}$ and M18
5g\,$^1$G\,--\,4f\,$^1$F$^{\rm o}$. The three C~{\sc iii} M1 lines,
$\lambda\lambda$4647.42, 4650.25 and 4651.47, are blended with the O~{\sc ii}
M1 (3p\,$^{4}$D$^{\rm o}$\,--\,3s\,$^{4}$P) lines $\lambda\lambda$4649.13 and
4650.84, as shown in Fig.\,\ref{4625-4680}. Techniques used to obtain the
total intensity of the C~{\sc iii} M1 multiplet are illustrated in
Section\,\ref{oii_orls:v1}. The intensity ratio of the three lines was assumed
to be as in the {\it LS}\,coupling, i.e. 1\,:\,3\,:\,5. The total intensity
is 0.303, which is accurate to 20 per cent. This intensity agrees with the
measurements of LSBC: 0.274 (PA = 45$^{\rm o}$) and
0.438 (PA = 0$^{\rm o}$).

The C~{\sc iii} M16 lines $\lambda\lambda$4067.94, 4068.91 and 4070.26 are
blended with the [S~{\sc ii}] $\lambda$4068 (3p$^{3}$\,$^{2}$P$^{\rm
o}_{3/2}$\,--\,$^{4}$S$^{\rm o}_{3/2}$) and two O~{\sc ii} M10
(3d\,$^{4}$F\,--\,3p\,$^{4}$D$^{\rm o}$) lines $\lambda\lambda$4069.62 and
4069.89, as shown in Fig.\,\ref{4060-4115}. Details of deriving the total
intensity of the C~{\sc iii} M16 multiplet are given in
Section\,\ref{oii_orls:v10}. The intensity ratio of the three C~{\sc iii} M16
lines was also assumed to be as in the {\it LS}\,coupling, i.e.
1.00\,:\,1.31\,:\,1.71. The total intensity is 0.288, with a large uncertainty
($\sim$40 per cent). The intensity ratio of the C~{\sc iii} M16 $\lambda$4069
and the C~{\sc iii} M1 $\lambda$4650 multiplets is 0.952. The predicted ratio
of the two C~{\sc iii} multiplets is 0.922, which is calculated from the
radiative and dielectronic recombination coefficients of P\'{e}quignot,
Petitjean \& Boisson \cite{ppb1991} and Nussbaumer \& Storey \cite{ns1984},
respectively.

The C~{\sc iii} M18 $\lambda$4186.90 (5g\,$^1$G$_{4}$\,--\,4f\,$^1$F$^{\rm
o}_{3}$) line is blended with the O~{\sc ii} M36 $\lambda$4185.44
(3d$^{\prime}$\,$^2$G$_{7/2}$\,--\,3p$^{\prime}$\,$^2$F$^{\rm o}_{5/2}$) line,
as shown in Fig.\,\ref{4176-4260}. Multi-Gaussian fits yield an intensity of
0.089$\pm$0.018 for the $\lambda$4186.90 line. This intensity agrees with
those given by LSBC: 0.0533 (PA = 45$^{\rm o}$) and 0.102
(PA = 0$^{\rm o}$). The intensity ratio of the C~{\sc iii} M18 $\lambda$4187
and the C~{\sc iii} M1 $\lambda$4650 multiplets is 0.293, which agrees with
the predicted ratio (0.331) within errors.

\subsection{\label{niii_orls}
The N~{\sc iii} permitted lines}

More than 30 lines were identified as the N~{\sc iii} permitted transitions,
including multiplets M1 and M2, which are mainly excited by the Bowen
fluorescence mechanism. Transitions from the states with excited parentage
(i.e. other than $^1$S) are detected. Most N~{\sc iii} lines suffer from
line blending. In this section, we only present intensity measurements and
discussion for the M3 3p$^{\prime}$\,$^{4}$D\,--\,3s$^{\prime}$\,$^{4}$P$^{\rm
o}$ and M18 5g\,$^{2}$G\,--\,4f\,$^{2}$F$^{\rm o}$ multiplets. Discussion of
other multiplets of N~{\sc iii} is given in Appendix\,\ref{appendix:e}.

\subsubsection{\label{niii_orls:v3}
Multiplet 3, 3p$^{\prime}$ $^4$D -- 3s$^{\prime}$ $^4$P$^{\rm o}$}

\begin{figure}
\begin{center}
\includegraphics[width=7.8cm,angle=-90]{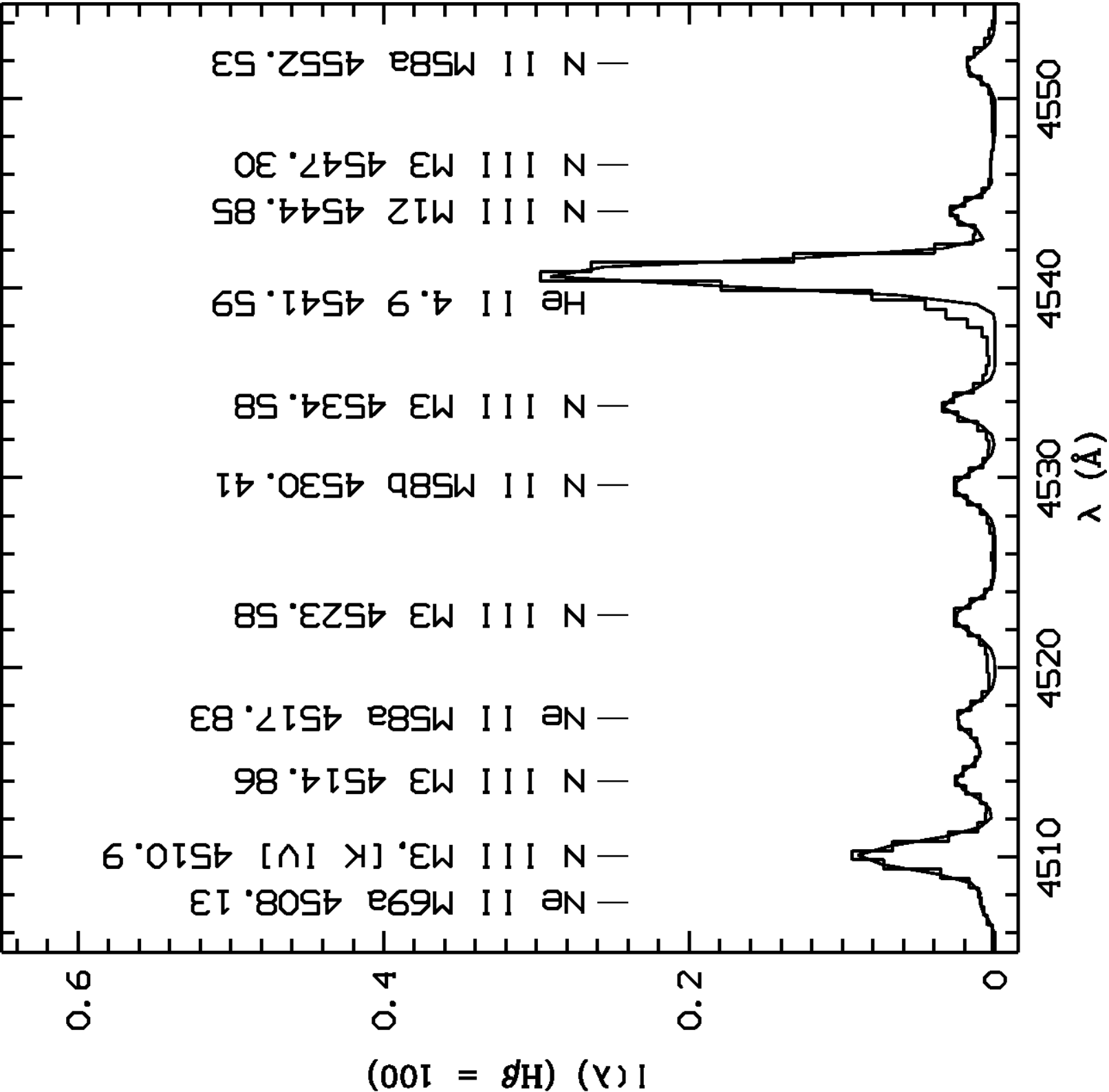}
\caption{Spectrum of NGC\,7009 from 4505 to 4555\,{\AA} showing the N~{\sc iii}
M3 lines and other emission features. The continuous curve is the sum of
Gaussian profile fits. Continuum has been subtracted and the spectrum has been
normalized such that H$\beta$ has an integrated flux of 100. Extinction has
not been corrected for.}
\label{4505-4555}
\end{center}
\end{figure}

$\lambda$4510.91 (3p$^{\prime}$~$^4$D$_{5/2}$ -- 3s$^{\prime}$~$^4$P$^{\rm
o}_{3/2}$ and 3p$^{\prime}$~$^4$D$_{3/2}$ -- 3s$^{\prime}$~$^4$P$^{\rm
o}_{1/2}$) blends with [K~{\sc iv}] 3p$^4$~$^1$S$_{0}$ -- 3p$^4$~$^1$D$_{2}$
$\lambda$4510.92, whose intensity contribution is unknown. Another M3 line
$\lambda$4514.86 (3p$^{\prime}$~$^4$D$_{7/2}$ -- 3s$^{\prime}$~$^4$P$^{\rm
o}_{5/2}$) is partially resolved from Ne~{\sc ii} M58a 4f~2[4]$^{\rm o}_{9/2}$
-- 3d~$^{2}$F$_{7/2}$ $\lambda$4517.83 (Fig.\,\ref{4505-4555}). Several
Ne~{\sc ii} ORLs, M58b $\lambda$4514.88, M64d $\lambda$4516.66 and M58a
$\lambda$4517.83, are also blended in the complex, which makes line
measurements very difficult. Another line $\lambda$4518.15
(3p$^{\prime}$~$^4$D$_{1/2}$ -- 3s$^{\prime}$~$^4$P$^{\rm o}_{1/2}$) blends
with the Ne~{\sc ii} M58a $\lambda$4517.83 line, which probably dominates the
total intensity.

Another two lines $\lambda\lambda$4523.58 (3p$^{\prime}$~$^4$D$_{3/2}$ --
3s$^{\prime}$~$^4$P$^{\rm o}_{3/2}$) and 4547.30 (3p$^{\prime}$~$^4$D$_{3/2}$
-- 3s$^{\prime}$~$^4$P$^{\rm o}_{5/2}$) decay from the same upper level, thus
their intensity ratio depends only on the coupling scheme. The measured line
ratio $\lambda$4547.30/$\lambda$4523.58 is 0.119, which is slightly higher
than the {\it LS}\,coupling value 0.094. Measurements of the $\lambda$4547.30
line could be of large error due to weakness (Fig.\,\ref{4505-4555}).
Another M3 line $\lambda$4534.58 (3p$^{\prime}$~$^4$D$_{5/2}$ --
3s$^{\prime}$~$^4$P$^{\rm o}_{5/2}$) blends with O~{\sc iii} M48 5g~G[4]$^{\rm
o}_{3,\,4}$ -- 4f~D[3]$_{3}$ $\lambda$4534.31 and Ne~{\sc ii} M55b
4f~2[3]$^{\rm o}_{7/2}$ -- 3d~$^4$F$_{5/2}$ $\lambda$4534.64 as well as another
three Ne~{\sc ii} lines M55b $\lambda$4534.52, M55c $\lambda$4535.37, and M55c
$\lambda$4535.57, whose intensity contribution could be negligible. The other
line $\lambda$4530.86 (3p$^{\prime}$~$^4$D$_{1/2}$ -- 3s$^{\prime}$~$^4$P$^{\rm
o}_{3/2}$) blends with N~{\sc ii} M58b 4f~2[5]$_{4}$ -- 3d~$^1$F$^{\rm o}_{3}$
$\lambda$4530.41, which is probably more than 3 times stronger.

\subsubsection{\label{niii_orls:v18}
Multiplet 18, 5g $^2$G -- 4f $^2$F$^{\rm o}$}

$\lambda$4379.11 (5g~$^2$G$_{9/2}$ -- 4f~$^2$F$^{\rm o}_{7/2}$ and
5g~$^2$G$_{7/2}$ -- 4f~$^2$F$^{\rm o}_{7/2}$) is detected in
Fig.\,\ref{4310-4382}. Gaussian profile fitting gives an intensity of 0.367,
with an uncertainty of less than 10 per cent. Measurements
for this line given by LSBC are 0.312 (PA = 45$^{\rm o}$)
and 0.397 (PA = 0$^{\rm o}$). $\lambda$4379.11 blends with Ne~{\sc ii} M60b
4f~1[4]$^{\rm o}_{9/2}$ -- 3d~$^2$F$_{7/2}$ $\lambda$4379.55, which contributes
about 10 per cent to the total intensity, and Ne~{\sc ii} M60b 4f~1[4]$^{\rm
o}_{7/2}$ -- 3d~$^2$F$_{7/2}$ $\lambda$4379.40 and O~{\sc iii} M39c
5g~G[5]$^{\rm o}_{4,5}$ -- 4f~F[4]$_{4}$ $\lambda$4379.58, both of which were
assumed to be negligible. This line is used to derive N$^{3+}$/H$^+$ abundance
ratio.

\subsection{\label{oiii_orls}
The O~{\sc iii} permitted lines}

In Paper~I, about two dozen O~{\sc iii} permitted transitions from the
3\,--\,3 configuration were identified, including multiplets
M2 3p\,$^{3}$D\,--\,3s\,$^{3}$P$^{\rm o}$,
M3 3p\,$^{3}$S\,--\,3s\,$^{3}$P$^{\rm o}$,
M4 3p\,$^{3}$P\,--\,3s\,$^{3}$P$^{\rm o}$,
M12 3d\,$^{3}$P$^{\rm o}$\,--\,3p\,$^{3}$S and
M15 3d\,$^{3}$P$^{\rm o}$\,--\,3p\,$^{3}$P, which are mainly excited by the
Bowen fluorescence or charge-transfer (Liu \& Danziger \citealt{ld1993a}).
All the O~{\sc iii} 5g\,--\,4f lines are blended with other emission features.
In this section, we only present emission line measurements of the
M8 3d\,$^{3}$F$^{\rm o}$\,--\,3p\,$^3$D and M15 3d\,$^{3}$P$^{\rm
o}$\,--\,3p\,$^{3}$P multiplets. Measurement results of other O~{\sc iii}
multiplets are given in Appendix\,\ref{appendix:f}. Discussion of the O~{\sc
iii} fluorescence lines is in Section\,\ref{oiii_orls:fluorescence}.

\subsubsection{\label{oiii_orls:v8}
Multiplet 8, 3d $^3$F$^{\rm o}$ -- 3p $^3$D}

\begin{figure}
\begin{center}
\includegraphics[width=7.8cm,angle=-90]{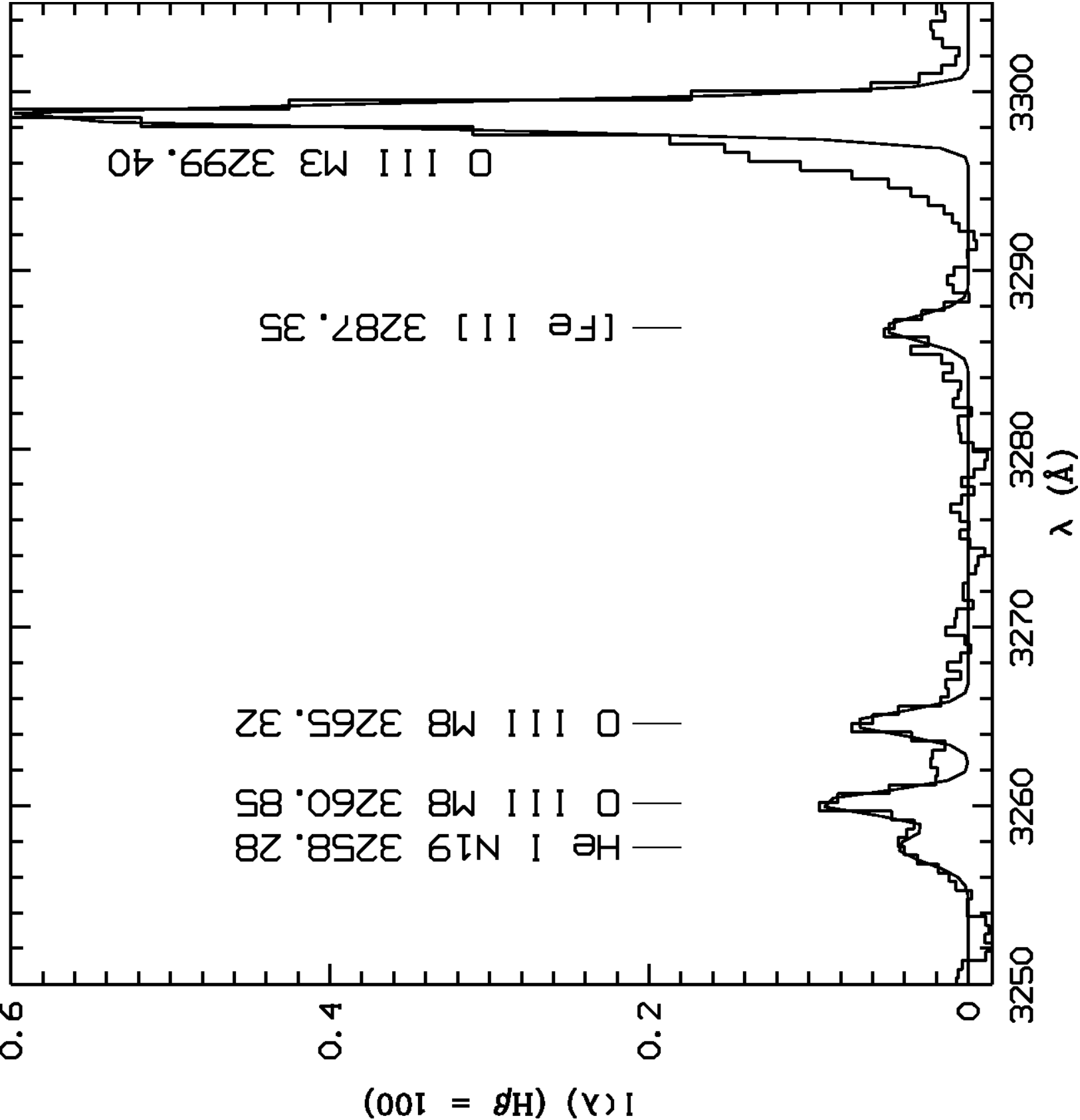}
\caption{Spectrum of NGC\,7009 from 3250 to 3305\,{\AA} showing the O~{\sc
iii} M8 lines $\lambda\lambda$3260.85 and 3265.32. The continuous curve is
the sum of Gaussian profile fits. Continuum has been subtracted and the
spectrum has been normalized such that H$\beta$ has an integrated flux of
100. Extinction has not been corrected for.}
\label{3250-3305}
\end{center}
\end{figure}

Only $\lambda$3260.85 (3d~$^3$F$^{\rm o}_{3}$ -- 3p~$^3$D$_{2}$) and
$\lambda$3265.32 (3d~$^3$F$^{\rm o}_{4}$ -- 3p~$^3$D$_{3}$) are observed
(Fig.\,\ref{3250-3305}). Measurements of the two lines could be of relatively
large error due to poor S/N in the far blue of the spectrum. However, this
multiplet cannot be excited by either the Bowen fluorescence or charge
transfer, and the two lines are still used to determine the O$^{3+}$/H$^{+}$
abundance ratio.

\subsubsection{\label{oiii_orls:v15}
Multiplet 15, 3d $^3$P$^{\rm o}$ -- 3p $^3$P}

\begin{figure}
\begin{center}
\includegraphics[width=7.5cm,angle=-90]{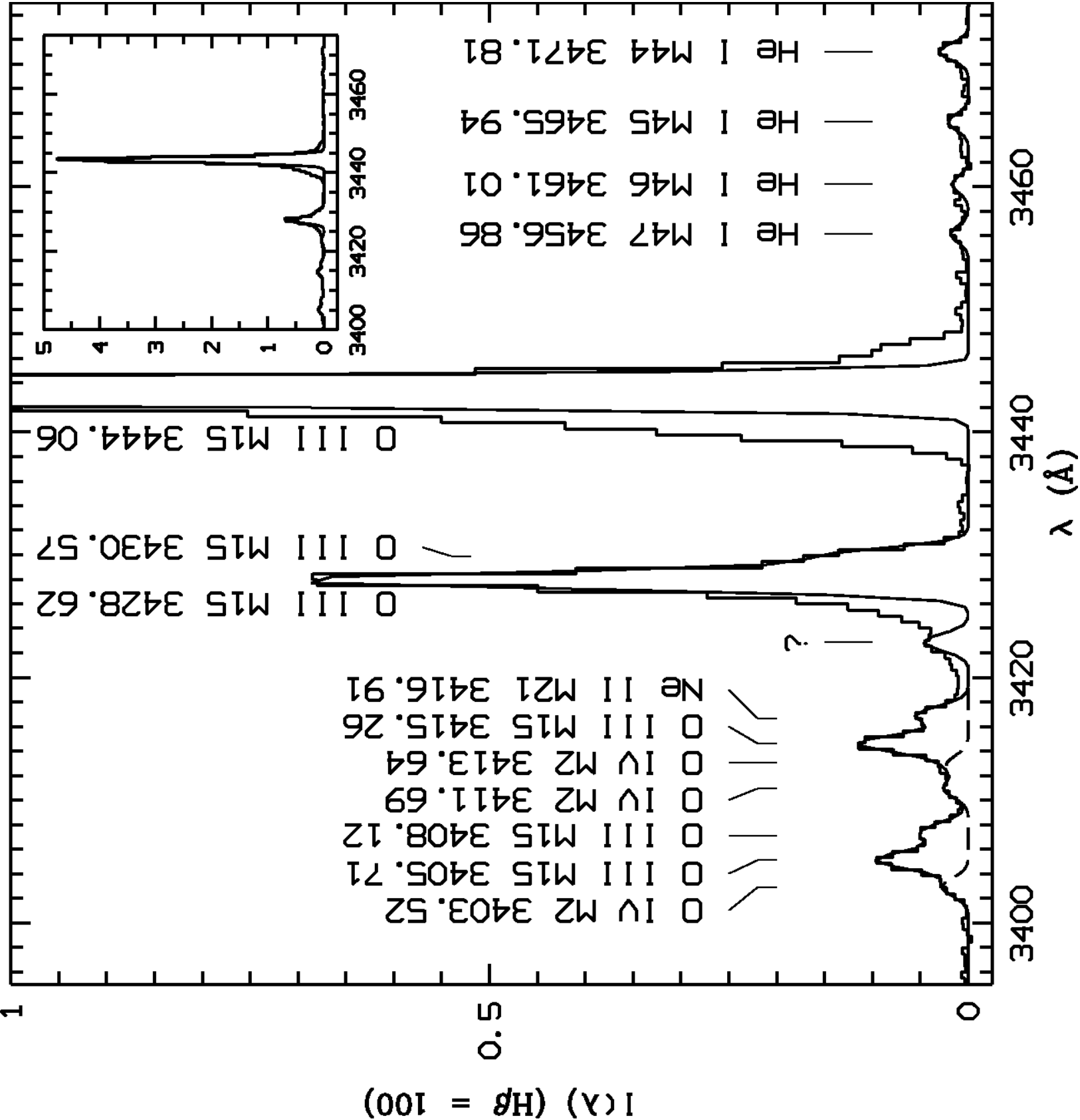}
\caption{Spectrum of NGC\,7009 from 3395 to 3475\,{\AA} showing the O~{\sc
iii} M15 lines. The profile of the O~{\sc iii} M15 $\lambda$3444 line shows
that there is a weak component blended on its right side, which could be the
He~{\sc i} M7 6p\,$^1$P$^{\rm o}_{1}$\,--\,2s\,$^1$S$_{0}$ $\lambda$3447.59
line. Several weak He~{\sc i} lines from the $n$d\,--\,2p transition array
are also detected. The solid continuous curve is the sum of Gaussian profile
fits. The dashed curve is the sum of the Gaussian profiles of the three
O~{\sc iv} M2 3d\,$^{2}$D\,--\,3p\,$^{2}$P$^{\rm o}$ lines
$\lambda\lambda$3403.52, 3411.69 and 3413.64. Continuum has been subtracted
and the spectrum has been normalized such that H$\beta$ has an integrated
flux of 100. Extinction has not been corrected for. The inset shows the
profile of the O~{\sc iii} M15 $\lambda$3444 line.}
\label{3395-3475}
\end{center}
\end{figure}

Fig.\,\ref{3395-3475} shows the O~{\sc iii} M15 lines detected in the spectrum
of NGC\,7009. Single-Gaussian profile fitting yields an intensity of 11.575
for the $\lambda$3444.06 line, with an uncertainty of about 3 per cent. The
intensity contribution from the blended He~{\sc i} M7 $\lambda$3447.59
(6p\,$^1$P$^{\rm o}_{1}$\,--\,2s\,$^1$S$_{0}$) line is only 2 per cent.
Another M15 line $\lambda$3428.62 is blended with the $\lambda$3430.57 line
of the same multiplet, which is marginally resolved in Fig.\,\ref{3395-3475}.
Two Gaussian profiles with the same FWHM were used to fit the feature, and
the resultant intensities of the $\lambda$3428.62 and the $\lambda$3430.57
lines are 1.409 and 0.319, respectively, both with uncertainties of less than
10 per cent.

Another M15 line $\lambda$3415.26 is partially resolved from the Ne~{\sc ii}
M21 $\lambda$3416.91 (3d\,$^2$D$_{5/2}$\,--\,3p\,$^2$D$^{\rm o}_{5/2}$) line;
the other two M15 lines, $\lambda\lambda$3405.71 and 3408.12 are also detected
in the spectrum. The three O~{\sc iv} M2 3d\,$^{2}$D\,--\,3p\,$^{2}$P$^{\rm
o}$ lines, $\lambda\lambda$3403.52, 3411.69 and 3413.64, are blended among the
above three O~{\sc iii} M15 lines, as is shown in Fig.\,\ref{3395-3475}.
Multi-Gaussian profile fitting was carried out for the complex. The intensity
of the $\lambda$3415.26 line is 0.415$\pm$0.021. The intensities of the
$\lambda\lambda$3405.71 and 3408.12 lines are 0.210 and 0.127, respectively,
both with uncertainties of 10 to 15 per cent. Analysis of the measured
intensities of the O~{\sc iv} M2 lines are in Section\,\ref{oiv_orls}.

The observed intensity ratio $\lambda$3428.62/$\lambda$3444.06 is 0.122,
lower than the theoretical prediction (0.336) given by Saraph \& Seaton
\cite{ss1980}, who assumed that the relative intensities within the O~{\sc
iii} M15 multiplet to be as in {\it LS}\,coupling, but agrees with the
intermediate calculations of Kastner et al. \cite{kastner1983}. The observed
intensity ratio of the three O~{\sc iii} M15 lines $\lambda\lambda$3405.71,
3415.26 and 3430.57, which have the common upper level, is
1\,:\,2.102\,:\,1.615. This ratio differs from that in the pure {\it
LS}\,coupling, i.e. 1\,:\,0.75\,:\,1.25. Discussion of the intensity ratios
of the O~{\sc iii} M15 lines is presented in
Section\,\ref{oiii_orls:fluorescence}.

\subsubsection{\label{oiii_orls:5g-4f}
5g -- 4f transitions}

The O~{\sc iii} permitted transitions of 5g\,--\,4f configuration are in the
wavelength range 4300--4600\,{\AA}, where numerous ORLs of the singly-ionized
ions of C, N, O and Ne are located. The typical intensities of the 5g\,--\,4f
lines of O~{\sc iii} are $\sim$10$^{-4}$--10$^{5}$ of H$\beta$, and accurate
measurements of these lines are difficult due to line blending. The strongest
O~{\sc iii} line of the 5g\,--\,4f configuration, $\lambda$4434.60 (M46b
5g\,H[6]$^{\rm o}_{6}$\,--\,4f\,G[5]$_{5}$), is marginally detected in the
spectrum of NGC\,7009, as is shown in Fig.\,\ref{4380-4445}. Multi-Gaussian
profile fitting yields an intensity of 0.024$\pm$0.003, from which we derived
an O$^{3+}$/H$^{+}$ abundance ratio that is higher than that deduced from the
O~{\sc iii} M8 $\lambda$3265 (3d\,$^3$F$^{\rm o}$\,--\,3p\,$^3$D) multiplet
by a factor of 20. Given the relatively more accurate measurements of the
O~{\sc iii} M8 lines (see Section\,\ref{oiii_orls:v8}), this much higher
O$^{3+}$/H$^{+}$ abundance ratio derived from the $\lambda$4434.60 line is
questionable. The effective recombination coefficients calculated by Kisielius
\& Storey \cite{ks1999} for the 5g\,--\,4f transitions of O~{\sc iii} are
utilized to estimate the intensity contributions where necessary.

\subsubsection{\label{oiii_orls:fluorescence}
Fluorescence of the O~{\sc iii} permitted lines}

In Table\,\ref{bowen_ratios} we compare the observed and predicted intensity
ratios for several pairs of O~{\sc iii} Bowen fluorescence lines originating
from the same upper level. The observed intensity ratios (column 2 of
Table\,\ref{bowen_ratios}) are based on the measurements in NGC\,7009 described
in Section\,\ref{oiii_orls}. The errors are estimated from the measurement
errors, including Gaussian profile fit and noise level of the local continuum.
The profiles of the strong O~{\sc iii} Bowen lines deviate from the exact
Gaussian due to charge transfer effect, thus we used Gaussian profiles to fit
these emission lines.
The error from Gaussian profile fit is 5\,--\,10 per cent for strong
O~{\sc iii} Bowen lines. These O~{\sc iii} Bowen lines are close to the blue
end of the spectrum, the uncertainties due to the poor S/N in this wavelength
region are also taken into account. The errors are significant for the two
ratios $\lambda$3791/$\lambda$3755 and $\lambda$3774/$\lambda$3757, because
the flux errors are large for the three relatively weak lines
$\lambda\lambda$3791.28 (about 21 per cent), 3757 (about 20 per cent) and
3774 (about 32 per cent), and thus the resulting uncertainties estimated from
the error propagation formula are comparable to the ratio values. The
observations of Liu \& Danziger \cite{ld1993a} for the same object are also
presented for comparison.

The radiative transition probabilities of O$^{2+}$ have been calculated since
late 1960s (Nussbaumer \citealt{nussbaumer1969}), and later by Saraph \&
Seaton \cite{ss1980} and Luo et al. \cite{luo1989}. In these approaches {\it
LS}\,coupling was assumed. Calculations of the O$^{2+}$ transition
probabilities with the scheme of intermediate coupling are presented by
Kastner et al.
\cite{kastner1983} and Kastner \& Bhatia \cite{kb1990} for some improved
values for cascading from the 2p3d~$^3$P$^{\rm o}_{2}$ level. Fischer
\cite{fischer1994} carried out a non-relativistic configuration interaction
calculation with relativistic correction and showed obvious improvement over
the previous work. Simultaneously, a full relativistic configuration
interaction calculation of Tong et al. \cite{tong1994} gave predicted
O~{\sc iii} Bowen line ratios which were further improvement, especially for
the $\lambda$3133/$\lambda$3444 ratio. Their results are in agreement with
observed ratios within the accuracy of observations. A more recent variational
Breit-Pauli calculation was done by Tachiev \& Fischer \cite{tff2001} for the
carbon-like sequence. The current measurements agree better with the
intermediate coupling values.

\begin{table*}
\begin{minipage}{110mm}
\caption{Comparison of the observed and predicted Bowen fluorescence line
ratios.}
\label{bowen_ratios}
\centering
\begin{tabular}{llllllll}
\hline
Line Ratio & Current obs.       & LD93$^a$        & SS80$^b$ & KBB83$^c$ & FF94$^d$ & TZLL94$^e$ & TFF01$^f$\\
\hline
$\lambda$3133/$\lambda$3444  & 3.261$\pm$0.071 & 3.140$\pm$0.440 & 3.610    & 4.450     & 3.170    & 3.290      & 3.342\\
$\lambda$3299/$\lambda$3341  & 0.252$\pm$0.117 & 0.285$\pm$0.022 & 0.201    & 0.228     & 0.264    & 0.260      & 0.268\\
$\lambda$3312/$\lambda$3341  & 0.698$\pm$0.086 & 0.651$\pm$0.048 & 0.606    & 0.656     & 0.728    & 0.717      & 0.736\\
$\lambda$3428/$\lambda$3444  & 0.122$\pm$0.027 & 0.204$\pm$0.023 & 0.336    & 0.164     & 0.150    & 0.149      & 0.154\\
$\lambda$3791/$\lambda$3755  & 0.271$\pm$0.213 & 0.232$\pm$0.037 & 0.330    & 0.309     & 0.296    & 0.301      & 0.299\\
$\lambda$3774/$\lambda$3757  & 0.539$\pm$0.379 & 0.569$\pm$0.127 & 0.750    & 0.715     & 0.708    & 0.701      & 0.704\\
\hline
\end{tabular}
\begin{description}
\item [$^a$] Liu \& Danziger \cite{ld1993a};
\item [$^b$] Saraph \& Seaton \cite{ss1980};
\item [$^c$] Kastner et al. \cite{kastner1983};
\item [$^d$] Fischer \cite{fischer1994};
\item [$^e$] Tong et al. \cite{tong1994};
\item [$^f$] Tachiev \& Fischer \cite{tff2001}.
\end{description}
\end{minipage}
\end{table*}

\subsection{\label{oiv_orls}
The O~{\sc iv} permitted lines}

Measurements of the O~{\sc iv} lines, $\lambda\lambda$3403.52, 3411.69 and
3413.64 of the M2 3d\,$^2$D\,--\,3p\,$^3$P$^{\rm o}$ multiplet are of large
uncertainty due to weakness and line blending (Fig.\,\ref{3395-3475}). The
intensity of the $\lambda$3411.69 line derived from multi-Gaussian profile
fitting is 0.057$\pm$0.014. The intensity ratio of the other two O~{\sc iv}
M2 lines $\lambda\lambda$3403.52 and 3413.64, which share the same upper
level $^{2}$D$_{3/2}$, is 1.318. This intensity ratio is significantly lower
than the value in the {\it LS}\,coupling assumption, i.e. 5.0. So far the
only effective recombination coefficients available for the O~{\sc iv} lines
are the radiative recombination coefficients given by P\'{e}quignot, Petitjean
\& Boisson \cite{ppb1991} and the dielectronic recombination coefficients
given by Nussbaumer \& Storey \cite{ns1984}. Both calculations were carried
out in the {\it LS}\,coupling scheme. The O~{\sc iv} M2 lines are the only
lines detected for this ion in the spectrum of NGC\,7009, but are not used in
abundance determinations.

\section{\label{abundances}
Ionic and elemental abundances}

In this section, we present the ionic abundances of helium and heavy elements
derived from ORLs. For heavy element ions, the electron temperatures deduced
from the N~{\sc ii} and O~{\sc ii} recombination line ratios are assumed
(Section\,\ref{diagnose:niioii}). For He$^{+}$/H$^{+}$, a value of 5100~K
as deduced from the He~{\sc i} $\lambda$7281/$\lambda$6678 ratio is adopted.
For He$^{++}$/H$^{+}$, an electron temperature of 10\,000~K, roughly the value
deduced from the 5694\,{\AA} discontinuity of the He~{\sc ii} recombination
continuum spectrum, is assumed. The average electron temperature derived from
the various optical CEL ratios is $\sim$10\,000~K (Paper~I), and is used to
determine the forbidden line abundances. Given that the electron densities
derived from the N~{\sc ii} and O~{\sc ii} ORLs are close to those from the
optical CELs, and that emissivities of the heavy element ORLs are only
marginally sensitive to electron density, we have assumed a constant density of
4300~cm$^{-3}$ throughout the abundance determinations.

\subsection{\label{orl_abundances}
Ionic abundances from ORLs}

Ionic abundances of He, C, N, O, Ne, Mg and Si are derived in this Section,
using the extinction-corrected fluxes of ORLs given in Paper~I. A critical
analysis of the C~{\sc ii}, N~{\sc ii}, O~{\sc ii} and Ne~{\sc ii} permitted
lines detected in the spectrum of NGC\,7009 is presented in Section\,3, and
is used to guide the determinations of ORL abundances of heavy element ions.
Again, the new effective recombination coefficients for the N~{\sc ii} and
O~{\sc ii} recombination spectrum are utilized.

\subsubsection{\label{orl_abundances:part1}
He$^+$/H$^+$ and He$^{2+}$/H$^+$}

Numerous calculations have been dedicated to the recombination spectrum of
He~{\sc i} (e.g. Mathis \citealt{mathis57}; Burgess \& Seaton \citealt{bs60a},
\citealt{bs60b}; Pottasch \citealt{pottasch62}; Robbins \citealt{robbins68},
\citealt{robbins70}; Robbins \& Robinson \citealt{rr71}; Brocklehurst
\citealt{b72}; Almog \& Netzer \citealt{an89}; Smits \citealt{smits91},
\citealt{smits96}; Benjamin, Skillman \& Smits \citealt{bss99},
\citealt{bss02}; Porter et al. \citealt{porter05}; Bauman et al.
\citealt{bauman05}). The recombination-cascade spectrum of He~{\sc i}, with
no collision taken into account, was first computed by Brocklehurst \cite{b72}.
A better treatment of emissivities from He~{\sc i} in nebular environment is
that of Smits \cite{smits96}, who used more accurate calculations of radiative
transition rates and photoionization cross-sections, and corrected an error in
Brocklehurst \cite{b72}. Benjamin, Skillman \& Smits \cite{bss99} combined the
detailed He~{\sc i} recombination model of Smits \cite{smits96} with the
collisional transitions of Sawey \& Berrington \cite{sb93} and calculated more
accurate He~{\sc i} recombination line emissivities that include the effects
of collisional excitation from both the 2s\,$^{3}$S and 2s\,$^{1}$S levels.
Benjamin, Skillman \& Smits \cite{bss02} studied the effects of the optical
depth of the 2s\,$^{3}$S metastable level on the He~{s\c i} line intensities.
The availability of these improved atomic data has made it possible to obtain
secure measurements of the ionic and elemental abundances of helium, given
high-quality spectroscopic data. Zhang et al. \cite{zhang05a} developed the
nebular plasma diagnostics based on the He~{\sc i} ORLs, using the theoretical
emissivities of He~{\sc i} lines provided by Benjamin, Skillman \& Smits
\cite{bss99}. In Paper~I, we derived electron temperatures from He~{\sc i}
line ratios using the method of Zhang et al. \cite{zhang05a}. In order to
keep consistency, we derived ionic and elemental abundances of helium in this
Section, using the atomic data of Benjamin, Skillman \& Smits \cite{bss99} and
assuming the temperature to be that yielded by the He~{\sc i} lines
($\sim$5000~K).

Ionic and elemental abundances of helium relative to hydrogen derived from the
He~{\sc i} and He~{\sc ii} ORLs are presented in Table\,\ref{abundances:i}.
The adopted He$^+$/H$^+$ ratio (0.099) is an average of the values derived
from the He~{\sc i} $\lambda\lambda$4471, 5876 and 6678 lines with weights of
1\,:\,3\,:\,1, roughly proportional to their intrinsic intensities (Benjamin,
Skillman \& Smits \citealt{bss99}). Here the effective recombination
coefficients were adopted from Benjamin, Skillman \& Smits \citealt{bss99}.
Case~A recombination was assumed for the triplet lines and Case~B for the
singlet. An electron temperature of 5000~K, as derived from the He~{\sc i}
$\lambda$7281/$\lambda$6678 ratio, and a density of 10\,000~cm$^{-3}$ were
assumed. Under this physical condition, the effective recombination
coefficient for the $\lambda$4471 line given by Benjamin, Skillman \& Smits
\citealt{bss99} is 6 per cent higher than that of Brocklehurst \cite{b72}.
This difference is 7 per cent when the temperature increases to 10\,000~K.
The differences between the calculations of Benjamin, Skillman \& Smits
\citealt{bss99} and Brocklehurst \cite{b72} for the other two He~{\sc i}
lines are less than 2 per cent at 5100~K.
The $\lambda$4471 line suffers most from line blending among the three He~{\sc
i} lines: It is blended with two O~{\sc ii} M86c lines $\lambda\lambda$4469.46
and 4469.48, three O~{\sc iii} lines M49c $\lambda\lambda$4471.02, 4475.17 and
M45b $\lambda$4476.11, and one Ne~{\sc ii} line M61b $\lambda$4468.91. Both
wings of the $\lambda$4471 line are also affected by weak features
(Fig.\,\ref{4444-4504}). All the blending brings an uncertainty of $\sim$10
per cent to the intensity of the $\lambda$4471 line. The $\lambda$4471 line
intensity adopted in the current analysis has been corrected for the
contributions from the blended lines listed above, using the effective
recombination coefficients available for the O~{\sc ii} and O~{\sc iii} lines.
The intensity of the He~{\sc i} $\lambda$6678 line has also been corrected
for the contribution from the blended He~{\sc ii} $\lambda$6683.20
(13h\,$^{2}$H$^{\rm o}$\,--\,5g\,$^{2}$G) line, which brings an uncertainty of
about 3 per cent. The measurement uncertainty of the He~{\sc i} $\lambda$5876
line is less than 2 per cent.

The He$^+$/H$^+$ abundances listed in Table\,\ref{abundances:i} agree with
each other reasonably well, except for those yielded by the triplet line
$\lambda$7065 (3s\,$^{3}$S\,--\,2p\,$^{3}$P$^{\rm o}$) and the singlet line
$\lambda$5016 (3p\,$^{1}$P$^{\rm o}$\,--\,2s\,$^{1}$S). The former one is more
than two times higher than the average abundance, while the latter is nearly
half of the value. The abnormally high abundance value is probably due to the
enhanced $\lambda$7065 line as a result of self-absorption from the metastable
2s\,$^{3}$S level. By comparing the observed intensities of the He~{\sc i}
singlet lines of the $n$s\,$^{1}$S--\,2p\,$^{1}$P$^{\rm o}$,
$n$p\,$^{1}$P$^{\rm o}$\,--\,2s\,$^{1}$S and
$n$d\,$^{1}$D\,--\,2p\,$^{1}$P$^{\rm o}$ series, relative to the $\lambda$4922
(4d\,$^{1}$D$_{2}$\,--\,2p\,$^{1}$P$^{\rm o}_{1}$) line, with the theoretical
predictions (c.f. Section\,$4.5$ in Paper~I), we concluded that the singlet
transitions of He~{\sc i} in NGC\,7009 are close to the Case~B assumption.
Departure from Case~B of the He~{\sc i} singlet lines as a result of the
He~{\sc i} Lyman photons being destroyed by photoionization of neutral
hydrogen and/or absorption by dust grains (Liu et al. \cite{liu2001b}) is
unlikely to be significant. Thus the low ionic abundance yielded by the
$\lambda$5016 line in Table\,\ref{abundances:i} is mainly due to
self-absorption from the metastable 2s\,$^{1}$S level. The $\lambda$5016 line
is blending with the N~{\sc ii} M19 3d\,$^3$F$^{\rm
o}_{2}$\,--\,3p\,$^3$D$_{2}$ $\lambda$5016.39, whose contribution is
negligible ($<$\,1.0 per cent).

The He$^{2+}$/H$^+$ abundance ratios were derived from two He~{\sc ii} lines
$\lambda$3203 (5f\,$^{2}$F$^{\rm o}$\,--\,3d\,$^{2}$D) and $\lambda$4686
(4f\,$^{2}$F$^{\rm o}$\,--\,3d\,$^{2}$D). The effective recombination
coefficients of the two lines were adopted from the hydrogenic calculation
of Storey \& Hummer \cite{sh1995}. Although the electron temperature
($\sim$11\,000~K) derived from the discontinuity at $\lambda$5694\,{\AA}
of the He~{\sc ii} recombination continuum is of large uncertainty due to
weakness of the jump, we assumed an electron temperature of 10\,000~K when
deriving the He$^{2+}$/H$^{+}$ abundance ratio. The total elemental abundance
of helium relative to hydrogen is 0.112, which is calculated from He/H =
He$^+$/H$^+$ + He$^{2+}$/H$^+$. This agrees well with the value of 0.109 given
by LSBC.

Several He~{\sc i} recombination line series have been observed in our
spectrum, and relative intensities of these lines are presented in
Table\,\ref{hei_lines}. Also presented in the Table are the theoretical
predictions given by Benjamin, Skillman \& Smits \cite{bss99}, Brocklehurst
\cite{b72}, and Smits \cite{smits96}. Case~A was assumed for the triplet lines
and Case~B for the singlet. The observed intensities of the
$n$d\,$^{3}$D\,--\,2p\,$^{3}$P$^{\rm o}$ and
$n$d\,$^{1}$D\,--\,2p\,$^{1}$P$^{\rm o}$ series of He~{\sc i}, relative to
the $\lambda$4471 line, agree well with those predicted by recombination
theory. The obvious weakness of the $n$p\,$^{1}$P$^{\rm o}$\,--\,2s\,$^{1}$S
series, compared with theoretical intensities, is caused by self-absorption
from the metastable 2s\,$^{1}$S level. Such self-absorption should at the same
time lead to the enhancement of the $n$s\,$^{1}$S\,--\,2p\,$^{1}$P$^{\rm o}$
series. However, what we observed as shown in Table\,\ref{hei_lines} are
opposite: The $\lambda$7281 line seems weaker than the recent prediction.
The $\lambda$3889 (3p\,$^{3}$P$^{\rm o}$\,--\,2s\,$^{3}$S) line is affected
by self-absorption. Enhancement of the $n$s\,$^{3}$S\,--\,2p\,$^{3}$P$^{\rm
o}$ series, in particular the $\lambda$7065 line, is clearly observed. Similar
patterns in the relative intensities of the He~{\sc i} lines are also observed
in NGC\,6153 (Liu et al. \citealt{liu2000}), M\,1-42 and M\,2-36 (Liu et al.
\citealt{liu2001b}).

\begin{table}
\begin{minipage}{65mm}
\centering
\caption{Recombination line helium abundances.}
\label{abundances:i}
\begin{tabular}{crl}
\hline
He$^{i+}$/H$^{+}$ & Line & Abundance\\
 & (\AA) & \\
\hline
Triplet lines & & \\
He$^{+}$/H$^{+}$ & He~{\sc i} $\lambda$3187.74 & 0.094\\
He$^{+}$/H$^{+}$ & He~{\sc i} $\lambda$3888.64 & 0.089\\
He$^{+}$/H$^{+}$ & He~{\sc i} $\lambda$4026.20 & 0.103\\
He$^{+}$/H$^{+}$ & He~{\sc i} $\lambda$4471.50 & 0.098\\
He$^{+}$/H$^{+}$ & He~{\sc i} $\lambda$5875.60 & 0.103\\
He$^{+}$/H$^{+}$ & He~{\sc i} $\lambda$7065.71 & 0.267\\
\\
Singlet lines & & \\
He$^{+}$/H$^{+}$ & He~{\sc i} $\lambda$4921.93 & 0.099\\
He$^{+}$/H$^{+}$ & He~{\sc i} $\lambda$5015.68 & 0.065\\
He$^{+}$/H$^{+}$ & He~{\sc i} $\lambda$6678.15 & 0.095\\
He$^{+}$/H$^{+}$ & He~{\sc i} $\lambda$7281.35 & 0.090\\
\\
He$^{+}$/H$^{+}$ & Mean & \textbf{0.099}\\
\\
He$^{2+}$/H$^{+}$ & He~{\sc ii} $\lambda$4685.68 & 0.013\\
He$^{2+}$/H$^{+}$ & He~{\sc ii} $\lambda$3203.17 & 0.012\\
\\
He/H & & \textbf{0.112}\\
\hline
\end{tabular}
\end{minipage}
\end{table}

\begin{table}
\begin{minipage}{68mm}
\centering
\caption{The He~{\sc i} lines observed in NGC\,7009. Intensities are
normalized to a scale where $I$(He~{\sc i}~$\lambda$4471) = 1.0. The
theoretical predictions of Benjamin, Skillman \& Smits (\citealt{bss99},
\textbf{BSS99}), Brocklehurst (\citealt{b72}, \textbf{B72}) and Smits
(\citealt{smits96}, \textbf{S96}), at $T_\mathrm{e}$ = 5000~K and
$N_\mathrm{e}$ = 10$^{4}$~cm$^{-3}$, are presented for purpose of comparison.
The Case~A recombination is assumed for the triplets and Case~B for the
singlets.}
\label{hei_lines}
\begin{tabular}{lrlrlr}
\hline
$\lambda_{\rm lab}$ & $n$ & $I_{\rm obs}$ & $I_{\rm pred}$ & $I_{\rm pred}$ & $I_{\rm pred}$\\
(\AA) & & & BSS99 & B72 & S96 \\
\hline
\multicolumn{6}{c}{$n$s\,$^{1}$S\,--\,2p\,$^{1}$P$^{\rm o}$ series}\\
3935.91 & 8 & 0.002 &       & 0.0024 & \\
4023.99 & 7 & 0.005 &       & 0.0037 & \\
4437.55 & 5 & 0.014 & 0.013 & 0.0124 & \\
5047.74 & 4 & 0.033 & 0.031 & 0.030  & \\
7281.35 & 3 & 0.106 & 0.118 & 0.108  & 0.113\\
\\
\multicolumn{6}{c}{$n$p\,$^{1}$P$^{\rm o}$\,--\,2s\,$^{1}$S series}\\
3354.55     & 7 & 0.036 &       & 0.034  & \\
3447.59     & 6 & 0.042 &       & 0.054  & \\
3613.64     & 5 & 0.060 & 0.094 & 0.097  & \\
5015.68$^a$ & 3 & 0.312 & 0.498 & 0.512  & 0.486\\
\\
\multicolumn{6}{c}{$n$d\,$^{1}$D\,--\,2p\,$^{1}$P$^{\rm o}$ series}\\
3926.53     &  8 & 0.029 &       & 0.027  & \\
4009.26     &  7 & 0.049 &       & 0.041  & \\
4143.76     &  6 & 0.059 &       & 0.068  & \\
4387.93     &  5 & 0.125 & 0.120 & 0.124  & 0.120\\
4921.93     &  4 & 0.245 & 0.269 & 0.275  & 0.269\\
6678.15$^b$ &  3 & 0.770 & 0.849 & 0.866  & 0.847\\
\\
\multicolumn{6}{c}{$n$s\,$^{3}$S\,--\,2p\,$^{3}$P$^{\rm o}$ series}\\
3599.32 &  9 & 0.005 &       & 0.0033 & \\
3732.88 &  7 & 0.007 &       & 0.0076 & \\
4120.99 &  5 & 0.054 & 0.028 & 0.0260 & \\
4713.20 &  4 & 0.123 & 0.076 & 0.0649 & 0.075\\
7065.71 &  3 & 0.912 & 0.398 & 0.243  & 0.356\\
\\
\multicolumn{6}{c}{$n$p\,$^{3}$P$^{\rm o}$\,--\,2s\,$^{3}$S series}\\
3187.74     &  4 & 0.742 & 0.748 & 0.747  & 0.748\\
3888.64$^c$ &  3 & 1.714 & 1.891 & 1.895  & 1.865\\
\\
\multicolumn{6}{c}{$n$d\,$^{3}$D\,--\,2p\,$^{3}$P$^{\rm o}$ series}\\
3456.86 & 19 & 0.006 &       & 0.0077 & \\
3461.01 & 18 & 0.008 &       & 0.0089 & \\
3465.94 & 17 & 0.009 &       & 0.0105 & \\
3471.81 & 16 & 0.011 &       & 0.0126 & \\
3478.97 & 15 & 0.018 &       & 0.0153 & \\
3487.72 & 14 & 0.021 &       & 0.0187 & \\
3498.64 & 13 & 0.026 &       & 0.0234 & \\
3512.51 & 12 & 0.034 &       & 0.0297 & \\
3530.49 & 11 & 0.040 &       & 0.0385 & \\
3554.41 & 10 & 0.055 &       & 0.0513 & \\
3587.27 &  9 & 0.073 &       & 0.0707 & \\
3634.23 &  8 & 0.112 &       & 0.101  & \\
3705.00 &  7 & 0.187 &       & 0.154  & \\
3819.60 &  6 & 0.275 &       & 0.251  & \\
4026.20 &  5 & 0.504 & 0.452 & 0.459  & 0.451\\
4471.50 &  4 & 1.000 & 1.000 & 1.000  & 1.000\\
5875.60 &  3 & 3.018 & 2.916 & 3.010  & 2.952\\
\hline
\end{tabular}
\begin{description}
\item [$^a$] Slightly overestimated due to the saturated [O~{\sc iii}]
$\lambda$5007 line.
\item [$^b$] Corrected for the contribution from the He~{\sc ii}
$\lambda$6683.20 (13h\,$^{2}$H$^{\rm o}$\,--\,5g\,$^{2}$G) line.
\item [$^c$] Corrected for the contribution from the H~{\sc i} H8
$\lambda$3889 line.
\end{description}
\end{minipage}
\end{table}

\subsubsection{\label{orl_abundances:part2}
C$^{2+}$/H$^+$ abundances from ORLs}

The C$^{2+}$/H$^+$ abundance ratios derived from the 3\,--\,3 and 4f\,--\,3d
transitions as well as from the $n$g\,--\,4f transition array are presented in
Table\,\ref{abundances:ii}. The effective recombination coefficients of Davey,
Storey \& Kisielius \cite{davey2000} were used. Their calculation was carried
out from 500 to 20\,000~K. As described earlier, an electron temperature of
1000~K deduced from the N~{\sc ii} and O~{\sc ii} ORLs and probably prevalent
in the postulated ``cold" component where the ORLs of heavy elements arise
(Liu et al. \citealt{liu2000}), was assumed in the calculations. Transitions
between doublet states were assumed to be in Case~B given the ground term
C~{\sc ii} $^2$P$^{\rm o}$. For those doublet transitions whose Case~B
effective recombination coefficients are not available in Davey, Storey \&
Kisielius \cite{davey2000}, the more recent calculations of Bastin
\cite{bastin2006} were adopted. Bastin \cite{bastin2006} calculated the
effective recombination coefficients for doublet transitions in both Case~A
and B from 5000 to 50\,000~K. Accurately extrapolating the effective
recombination coefficients of Bastin \cite{bastin2006} to 1000~K is difficult
because the recombination coefficients are not exactly a linear function of
electron temperature. The H~{\sc i} Balmer jump temperature 6500~K and a
density of 4300~cm$^{-3}$ were assumed when we derived the C$^{2+}$/H$^+$
abundance using the coefficients of Bastin \cite{bastin2006}.

The C$^{2+}$/H$^+$ abundance ratio derived from the C~{\sc ii} $\lambda$4267
line is 5.507$\times$10$^{-4}$. In the calculation of Davey, Storey \&
Kisielius \cite{davey2000}, the Case~A effective recombination coefficient
for the C~{\sc ii} $\lambda$4267 line differs from Case~B by only 0.8 per
cent. If a temperature of 10\,000~K is assumed, the C$^{2+}$/H$^+$ ratio
derived will increase to 8.432$\times$10$^{-4}$, which then agrees with the
abundance given by LSBC. At 10\,000~K, the effective
recombination coefficient for the C~{\sc ii} $\lambda$4267 line given by
Davey, Storey \& Kisielius \cite{davey2000} differs from that of Bastin
\cite{bastin2006} by only 1.4 per cent in Case~B, and 1.8 per cent in Case~A.
Thus we expect that the C$^{2+}$/H$^+$ ratio derived from the coefficients
of Bastin \cite{bastin2006} should agree with LSBC.

In Table\,\ref{abundances:ii}, the C$^{2+}$/H$^+$ abundance ratios derived
from the three $n$g\,$^2$G\,--\,4f\,$^2$F$^{\rm o}$ lines (M17.02
$\lambda$9903, M17.04 $\lambda$6462, and M17.06 $\lambda$5342) and the M16.04
6f\,$^2$F$^{\rm o}$\,--\,4d\,$^2$D $\lambda$6151 line are based on the Case~B
effective recombination coefficients of Bastin \cite{bastin2006}. The
abundance ratios all agree with those given by LSBC. The
Case~B effective recombination coefficients for the C~{\sc ii}
$n$g\,$^2$G\,--\,4f\,$^2$F$^{\rm o}$ ($n$\,=\,5, 6 and 7) transitions
calculated by Bastin \cite{bastin2006} are almost identical to the Case~A
values at 6500~K; for the M16.04 6f\,$^2$F$^{\rm o}$\,--\,4d\,$^2$D
$\lambda$6151 transition, the effective recombination coefficients of the two
cases differ by only 2 per cent. The C~{\sc ii} M4
4s\,$^2$S\,--\,3p\,$^2$P$^{\rm o}$ $\lambda$3920 multiplet was not given by
Davey, Storey \& Kisielius \cite{davey2000}, and the Case~B recombination
coefficient of Bastin \cite{bastin2006} was used. The derived C$^{2+}$/H$^+$
abundance ratio agrees with those derived from the three $n$g\,--\,4f
transitions. The adopted C$^{2+}$/H$^+$ abundance ratio in the current paper
is 6.865$\times$10$^{-4}$, which is an average of the abundances derived from
the transitions in Table\,\ref{abundances:ii}.

\begin{table}
\centering
\caption{Recombination line C$^{2+}$/H$^{+}$ abundances. Intensities are
normalized such that $I$(H$\beta$) = 100.}
\label{abundances:ii}
\begin{tabular}{lcll}
\hline
Line & Mult. & $I_{\rm obs}$ & C$^{2+}$/H$^{+}$\\
(\AA)  &  &  & ($\times$10$^{-4}$)\\
\hline
4f $^2$F$^{\rm o}$ -- 3d $^2$D & M6 & & \\
$\lambda$4267 & & 0.875 & 5.507\\
\\
4s $^2$S -- 3p $^2$P$^{\rm o}$ & M4 & & \\
$\lambda$3918.98 & & 0.015 & \\
$\lambda$3920.69 & & 0.033 & \\
sum & & 0.048 & 8.370$^a$\\
\\
3d $^2$D -- 3p $^2$P$^{\rm o}$ & M3 & & \\
$\lambda$7231.32 & & 0.127 & \\
$\lambda$7236.42$^b$ & & 0.262 & \\
sum & & 0.437 & 3.198\\
\\
5g $^2$G -- 4f $^2$F$^{\rm o}$ & M17.02 & & \\
$\lambda$9903.46 & & 0.200 & 8.197$^a$\\
\\
6g $^2$G -- 4f $^2$F$^{\rm o}$ & M17.04 & & \\
$\lambda$6461.95 & & 0.087 & 8.275$^a$\\
\\
7g $^2$G -- 4f $^2$F$^{\rm o}$ & M17.06 & & \\
$\lambda$5342.40 & & 0.037 & 6.771$^a$\\
\\
6f $^2$F$^{\rm o}$ -- 4d $^2$D & M16.04 & & \\
$\lambda$6151.43 & & 0.034 & 7.734$^a$\\
\hline
\end{tabular}
\begin{description}
\item [$^a$] Based on the C~{\sc ii} effective recombination coefficients of
B06. An electron temperature of 6500~K, as derived from the H~{\sc i} Balmer
jump, and a density of 4300~cm$^{-3}$, as derived from the CEL ratios
(Paper~I), are assumed.
\item [$^b$] Including the C~{\sc ii} M3 $\lambda$7237.17 line.
\end{description}
\end{table}

\subsubsection{\label{orl_abundances:part3}
N$^{2+}$/H$^+$ abundances from ORLs}

The N$^{2+}$/H$^+$ abundance ratios derived from the N~{\sc ii} ORLs detected
in the spectrum of NGC\,7009, with the most reliable measurements, are
presented in Table\,\ref{abundances:iii}. The M3 multiplet of the 3p\,--\,3s
configuration is the best observed amongst all the 3\,--\,3 transitions. The
N~{\sc ii} effective recombination coefficients of Kisielius \& Storey
\cite{ks2002} are used when we derive the N$^{2+}$/H$^{+}$ abundance ratios
from the total intensity of an N~{\sc ii} multiplet, which is a sum of line
intensities of all the fine-structure components. If some components are
missing from a multiplet, i.e., not detected due to weakness or line blending,
the total intensity of that multiplet is then calculated by assuming the
relative intensities of the fine-structure components are as in {\it
LS}\,coupling. The most recent N~{\sc ii} effective recombination coefficients
calculated by FSL11 are used when we derive the N$^{2+}$/H$^{+}$ abundances
from the fine-structure components of each N~{\sc ii} multiplet. The Case~B
recombination was assumed for the triplets and Case~A for the singlets. An
electron temperature of 1000~K and a density of 4300~cm$^{-3}$ were assumed
throughout the abundance determinations. Under such physical condition, the
effective recombination coefficient for H$\beta$ adopted is
1.86$\times$10$^{-13}$~cm$^3$\,s$^{-1}$, which is calculated from the
hydrogenic theory of Storey \& Hummer \cite{sh1995}.

The M3 3p\,$^3$D\,--\,3s\,$^3$P$^{\rm o}$ $\lambda$5680 multiplet is the
strongest N~{\sc ii} permitted transition. At $T_\mathrm{e}$ = 1000~K, the
{\it LS}\,coupling effective recombination coefficient of N~{\sc ii} M3
$\lambda$5680 multiplet calculated by Kisielius \& Storey \cite{ks2002} is
case-insensitive, with the Case~B value being only 20 per cent higher than
that for Case~A. In the calculation of FSL11,
the Case~B effective recombination coefficient for the $\lambda$5679.56 line,
which is the strongest component of the N~{\sc ii} M3 multiplet, is 27 per
cent higher than in Case~A. The Case~B effective recombination coefficients
for another two N~{\sc ii} M3 lines, $\lambda\lambda$5666.63 and 5676.02, are
higher than their corresponding Case~A values by 28 and 24 per cent,
respectively. The Case~B coefficients of the other N~{\sc ii} M3 lines are
about 24 to 28 per cent higher than for Case~A. In
Table\,\ref{abundances:iii}, the N$^{2+}$/H$^+$ abundance ratios derived from
the N~{\sc ii} M3 $\lambda$5680 multiplet lines almost agree with the value
derived from the total intensity of this multiplet. The N~{\sc ii} singlet
line 3p\,$^1$D\,--\,3s\,$^1$P$^{\rm o}$ $\lambda$3995 is case-insensitive,
with its Case~B effective recombination coefficient being only 3.6 per cent
higher than for Case~A. The calculation of FSL11 shows that the Case~B
effective recombination coefficient for the $\lambda$3995 line is higher than
the Case~A value by about 5 per cent at 1000~K.

The M5 3p\,$^3$P\,--\,3s\,$^3$P$^{\rm o}$, M20 3d\,$^3$D$^{\rm
o}$\,--\,3p\,$^3$D, M28 3d\,$^3$D$^{\rm o}$\,--\,3p\,$^3$P and M29
3d\,$^3$P$^{\rm o}$\,--\,3p\,$^3$P multiplets of N~{\sc ii} are all
case-sensitive. The Case~B effective recombination coefficient for the M5
3p\,$^3$P\,--\,3s\,$^3$P$^{\rm o}$ $\lambda$4623 multiplet calculated by
Kisielius \& Storey \cite{ks2002} is higher than in Case~A by a factor of 9.
FSL11 shows that the Case~B effective
recombination coefficient for the $\lambda$4630.54 line, which is the
strongest component of the N~{\sc ii} M5 multiplet, is higher than for
Case~A by a factor of 8. The Case~B effective recombination coefficients for
another two N~{\sc ii} M5 lines, $\lambda\lambda$4601.48 and 4621.39, are
8\,--\,9 per cent higher than in Case~A. For the weakest lines in the N~{\sc
ii} M5 multiplet (the $\lambda\lambda$4607.16, 4613.87 and 4643.09 lines),
their Case~B effective recombination coefficients do not differ much from
the Case~A values. In Table\,\ref{abundances:iii}, the N$^{2+}$/H$^+$
abundance ratios derived from the N~{\sc ii} M5 components in the Case~B
assumption agree with those yielded by the case-insensitive N~{\sc ii} M3
$\lambda$5680 lines, which signifies Case~B is a better assumption for the
N~{\sc ii} M5 multiplet.

Kisielius \& Storey \citealt{ks2002} shows that the Case~B effective
recombination coefficients for the N~{\sc ii} M20 3d\,$^3$D$^{\rm
o}$\,--\,3p\,$^3$D $\lambda$4794, M28 3d\,$^3$D$^{\rm o}$\,--\,3p\,$^3$P
$\lambda$5939 and M29 3d\,$^3$P$^{\rm o}$\,--\,3p\,$^3$P $\lambda$5479
multiplets are higher than those for Case~A by a factor of 52, 51 and 25,
respectively. FSL11 reveals that the Case~B effective recombination
coefficients for the strongest fine-structure components in the above three
multiplets are 48, 49 and 16 times higher than those for Case~A, at a given
temperature of 1000~K. Given such large differences between the two cases,
the N$^{2+}$/H$^+$ abundance ratios derived from the three N~{\sc ii}
multiplets (Table\,\ref{abundances:iii}) suggest that Case~B is a better
assumption for the 3\,--\,3 transitions of N~{\sc ii}.

N$^{2+}$/H$^+$ abundances are also derived from the 4f\,--\,3d transitions,
which are case-insensitive. The Case~B effective recombination coefficient for
M39b 4f\,G[9/2]$_{5}$\,--\,3d\,$^3$F$^{\rm o}_{4}$ $\lambda$4041.31, the
strongest 4f\,--\,3d transition of N~{\sc ii}, differs from the Case~A value
by only 0.25 per cent (FSL11). The N$^{2+}$/H$^+$ abundance derived from the
$\lambda$4041.31 line agrees well with that from N~{\sc ii} M39a
4f\,G[7/2]$_{4}$\,--\,3d\,$^3$F$^{\rm o}_{3}$ $\lambda$4043.53. Most of the
other N~{\sc ii} 4f\,--\,3d transitions in Table\,\ref{abundances:iii} yield
abundances close to those from the $\lambda\lambda$4041.31 and 4043.53 lines,
except for M39b 4f\,G[9/2]$_{4}$\,--\,3d\,$^3$F$^{\rm o}_{4}$
$\lambda$4039.35, which gives an abnormally high abundance value. We attribute
this to the large measurement error due to weakness of this line. N~{\sc ii}
M58a 4f\,G[7/2]$_{4}$\,--\,3d\,$^1$F$^{\rm o}_{3}$ $\lambda$4552.53 also
yields a relatively high abundance value, which could be due to the blended
Si~{\sc iii} M2 4p\,$^3$P$^{\rm o}_{2}$\,--\,4s\,$^3$S$_{1}$ $\lambda$4552.62
and Ne~{\sc ii} M55d 4f\,2[2]$^{\rm o}_{5/2}$\,--\,3d\,$^4$F$_{3/2}$
$\lambda$4553.17 lines. The intensity of the N~{\sc ii} M61a
4f\,D[5/2]$_{2}$\,--\,3d\,$^1$P$^{\rm o}_{1}$ $\lambda$4694.64 line is
probably overestimated due to unknown blend, which can be noticed from the
profile of the feature. The three lines $\lambda\lambda$4039.35, 4552.53 and
4694.64 are excluded from calculating the total intensity and the average
abundance ratio. The N$^{2+}$/H$^+$ abundances derived from the
$\lambda\lambda$4041.31,\,4043.53 lines agree with those from the N~{\sc ii}
M3 $\lambda\lambda$5666.63,\,5679.56 lines. The N~{\sc ii} M3 $\lambda$5676.02
line yields a relatively high abundance. Measurements of this line could be
unreliable due to the blended $\lambda$5679.56 line, which is 4 times stronger
(Fig.\,\ref{5650-5760}). The relatively low S/N's at this wavelength region
also affect the measurements.

It has been known for decades that the N~{\sc ii} permitted lines from the
low-lying 3d\,--\,3p and 3p\,--\,3s triplet arrays, whose upper levels are
linked to the ground term 2p$^2$\,$^3$P by resonance lines, can be enhanced by
fluorescence excitation. Grandi \cite{grandi1976} used photoionization models
to study the excitation mechanisms of permitted transitions from common heavy
element ions observed in the spectra of the Orion nebula and PNe NGC\,7027 and
NGC\,7662, and found that while the N~{\sc ii} M28 3d\,$^3$D$^{\rm
o}$\,--\,3p\,$^3$P $\lambda$5942 multiplet is excited by both recombination
and continuum fluorescence of the starlight, emission of the N~{\sc ii} M3
3p\,$^3$D\,--\,3s\,$^3$P$^{\rm o}$ $\lambda$5680, M5
3p\,$^3$P\,--\,3s\,$^3$P$^{\rm o}$ $\lambda$4630 and M30 4s\,$^3$P$^{\rm
o}$\,--\,3p\,$^3$P $\lambda$3838 multiplets are dominated by fluorescence
excitation of the N~{\sc ii} 4s\,$^3$P$^{\rm o}_{1}$ level by the He~{\sc i}
1s8p\,$^1$p$^{\rm o}_{1}$\,--\,1s$^2$\,$^1$S$_{0}$ $\lambda$508.643 resonance
line, which coincides in wavelength with the N~{\sc ii} 2p4s\,$^3$P$^{\rm
o}_{1}$\,--\,2p$^2$\,$^3$P$_{0}$ $\lambda$508.668 line. Fluorescence
excitation, by line or continuum, however, cannot excite the singlet
transitions or transitions from the 3d\,--\,4f configuration.
Escalante \& Morisset \cite{em2005} analyzed the N~{\sc ii} spectrum of the
Orion nebula by using nebular and stellar atmosphere models. Their modeling
shows that the intensity of most of the N~{\sc ii} permitted lines in Orion
could be explained by fluorescence of the starlight continuum. Recombination
of N$^{2+}$ contributes a minor part of the observed intensities of lines
from the 3p and 3d levels connected to the ground state. They confined the
effective temperature of the ionizing star to be lower than 38\,000~K in
order to reproduce the observed line intensities.
Our current analysis shows that the N$^{2+}$/H$^{+}$ abundances derived from
the 3p\,--\,3s transitions agree with those derived from the 4f\,--\,3d
recombination lines (the values in boldface in Table\,\ref{abundances:iii}),
which are unlikely to be affected by the fluorescence mechanisms. That
indicates fluorescence excitation of the N~{\sc ii} 3p\,--\,3s lines, albeit
may still exists, is probably insignificant in NGC\,7009. Fluorescence
enhancement of the N~{\sc ii} lines by starlight could also be negligible,
given the physical condition of NGC\,7009 (C. Morisset, private
communication). However, physical parameters, e.g., UV radiation field of the
central star, optical depth of the resonance transition that connect the
ground state 2p$^2$\,$^3$P and an excited state (e.g. 2p4s\,$^3$P$^{\rm o}$),
column densities of N$^{+}$ and N$^{2+}$ ions, etc., are needed in order to
estimate the enhancement of the N~{\sc ii} lines due to resonance fluorescence
by starlight or by other emission lines. That means consistent modeling of
the central star and nebula is needed.

The average N$^{2+}$/H$^{+}$ abundance from the 3\,--\,3 transitions is
3.70$\times$10$^{-4}$, which agrees with the average value
(3.71$\times$10$^{-4}$) from the 4f\,--\,3d transitions. Here the N~{\sc ii}
transitions (e.g. M29) that yield abnormally high abundances are excluded
from averaging. The N$^{2+}$/H$^{+}$ abundance derived by co-adding the line
intensities of the 4f\,--\,3d transitions is 3.42$\times$10$^{-4}$, which
agrees well with the abundance calculated from the total intensity of the M3
multiplet of N~{\sc ii} (the values in boldface in
Table\,\ref{abundances:iii}). The abundances derived from total intensities
of the 4f\,--\,3d transitions are preferred over the average values of
abundances from individual lines, since strong lines are better detected with
smaller (relative) flux uncertainties. We adopt the mean value
(3.45$\times$10$^{-4}$) derived by averaging the abundances from the M3
multiplet and the total intensity of the 4f\,--\,3d transitions as the
recombination line N$^{2+}$/H$^{+}$ abundance of NGC\,7009. This value is
about 10 per cent higher than 3.10$\times$10$^{-4}$ (slit position angle PA =
45$^{\rm o}$) given by LSBC, who used the N~{\sc ii} M39b $\lambda$4041.31
and M39a $\lambda$4043.53 lines to derive the N$^{2+}$/H$^+$ abundance of
NGC\,7009.

\begin{table}
\centering
\caption{Recombination line N$^{2+}$/H$^{+}$ abundances. Intensities are
normalized such that H$\beta$ = 100. The N~{\sc ii} effective recombination
coefficients of FSL11 and Kisielius \& Storey (\citealt{ks2002}, KS02) are
both used for purpose of comparison.}
\label{abundances:iii}
\begin{tabular}{lclrr}
\hline
Line & Mult. & $I_{\rm obs}$ & \multicolumn{2}{c}{N$^{2+}$/H$^{+}$}\\
 & & & \multicolumn{2}{c}{($\times$10$^{-4}$)}\\
(\AA) & & & FSL11 & KS02\\
\hline
\multicolumn{5}{c}{3 -- 3 transitions}\\
$\lambda$5666.63 & M3 & 0.064 & 3.442 & \\
$\lambda$5676.02 & M3 & 0.036 & 4.075 & \\
$\lambda$5679.56 & M3 & 0.130 & 3.319 & \\
$\lambda$5686.21 & M3 & 0.024 & 4.695 & \\
$\lambda$5710.77 & M3 & 0.020 & 2.886 & \\
\textbf{M3 3p~$^3$D -- 3s~$^3$P$^{\rm o}$} & & 0.280 & \textbf{3.482} & \textbf{3.358}\\
\\
$\lambda$4601.48     & M5 & 0.016 & 3.087 & \\
$\lambda$4621.39$^a$ & M5 & 0.020 & 4.423 & \\
$\lambda$4630.54     & M5 & 0.067 & 3.549 & \\
\textbf{M5 3p~$^3$P -- 3s~$^3$P$^{\rm o}$} & & 0.102 & \textbf{3.598} & \textbf{2.242}\\
\\
$\lambda$3994.99 & M12 & 0.033 & 6.655 & \\
\textbf{M12 3p~$^1$D -- 3s~$^1$P$^{\rm o}$} & & 0.033 & \textbf{6.655} & \textbf{6.776}\\
\\
$\lambda$4803.29 & M20 & 0.032 & 2.723 & \\
\textbf{M20 3d~$^3$D$^{\rm o}$ -- 3p~$^3$D} & & 0.078 & \textbf{4.412} & \textbf{3.464}\\
\\
$\lambda$5941.65$^b$ & M28 & 0.030 & 1.795 & \\
\textbf{M28 3d~$^3$D$^{\rm o}$ -- 3p~$^3$P} & & 0.063 & \textbf{3.038} & \textbf{2.119}\\
\\
$\lambda$5480.06$^c$ & M29 & 0.012 & 9.409 & \\
\textbf{M29 3d~$^3$P$^{\rm o}$ -- 3p~$^3$P} & & 0.061 & \textbf{14.949} & \textbf{7.590}\\
\\
Average & & & \textbf{3.695} & \\
\\
\multicolumn{5}{c}{4f -- 3d transitions}\\
$\lambda$4035.08     & M39a & 0.035 & 2.926 &\\
$\lambda$4041.31$^d$ & M39b & 0.081 & 3.174 &\\
$\lambda$4043.53     & M39a & 0.035 & 3.168 &\\
$\lambda$4171.61     & M43b & 0.023 & 3.033 &\\
$\lambda$4176.16     & M43a & 0.021 & 3.739 &\\
$\lambda$4236.91$^e$ & M48a & 0.036 & 3.727 &\\
$\lambda$4241.78$^f$ & M48a & 0.093 & 3.689 &\\
$\lambda$4179.67     & M50a & 0.010 & 3.508 &\\
$\lambda$4432.74     & M55a & 0.036 & 3.089 &\\
$\lambda$4442.02     & M55a & 0.011 & 3.204 &\\
$\lambda$4552.53$^g$ & M58a & 0.032 & 6.381 &\\
$\lambda$4530.41     & M58b & 0.046 & 3.792 &\\
$\lambda$4678.14     & M61b & 0.012 & 3.049 &\\
$\lambda$4694.64     & M61a & 0.020 & 5.393 &\\
\\
Sum & & 0.466 & \textbf{3.419} &\\
Average & & & \textbf{3.705} &\\
\hline
\end{tabular}
\begin{description}
\item [$^a$] Including O~{\sc ii} M92c lines 4f\,F[2]$^{\rm
o}_{5/2}$\,--\,3d\,$^2$D$_{5/2}$ $\lambda$4621.27. Neglecting O~{\sc ii} M92c
4f\,F[2]$^{\rm o}_{3/2}$\,--\,3d\,$^2$D$_{5/2}$ $\lambda$4622.14.
\item [$^b$] Neglecting the N~{\sc ii} M28 3d\,$^3$D$^{\rm
o}_{1}$\,--\,3p\,$^3$P$_{1}$ $\lambda$5940.24 line, which contributes less
than 1 per cent to the total intensity.
\item [$^c$] Including N~{\sc ii} M29 3d\,$^3$P$^{\rm
o}_{2}$\,--\,3p\,$^3$P$_{1}$ $\lambda$5478.10.
\item [$^d$] Corrected for the contribution from the O~{\sc ii} M50c
4f\,F[2]$^{\rm o}_{5/2}$\,--\,3d\,$^4$F$_{5/2}$ $\lambda$4041.28 line (7 per
cent).
\item [$^e$] Corrected for the contribution from the N~{\sc ii} M48b
4f\,F[7/2]$_{3}$\,--\,3d\,$^3$D$^{\rm o}_{2}$ $\lambda$4237.05 line ($\sim$30
per cent).
\item [$^f$] Including N~{\sc ii} M48b 4f\,F[7/2]$_{4}$\,--\,3d\,$^3$D$^{\rm
o}_{3}$ $\lambda$4241.78.
\item [$^g$] Including the contribution from the Ne~{\sc ii} M55d
4f\,2[2]$^{\rm o}_{5/2}$\,--\,3d\,$^4$F$_{3/2}$ $\lambda$4553.17 line.
Neglecting Ne~{\sc ii} M55d 4f\,2[2]$^{\rm o}_{3/2}$\,--\,3d\,$^4$F$_{3/2}$
$\lambda$4553.40.
\end{description}
\end{table}

\subsubsection{\label{orl_abundances:part4}
O$^{2+}$/H$^+$ abundances from ORLs}

In the spectrum of NGC\,7009, O~{\sc ii} has the most abundant optical
recombination spectrum amongst all the heavy element ions detected. The
most prominent multiplets of the O~{\sc ii} transitions are presented in
Section	s\,\ref{oii_orls} and \ref{appendix:c}. The spectrum analyzed in
the current paper covers very broad wavelength range (3040--11\,100\,{\AA})
and is among the deepest CCD spectra ever taken for an emission line nebula,
and also has higher quality than that published by LSBC. In the wavelength
ranges 3040--4048\,{\AA} and 3990--4980\,{\AA}, where the most prominent
recombination lines of O~{\sc ii} are located, our data quality is as good
as that of Liu et al. \cite{liu2000} for PN NGC\,6153, which was observed
using the same instruments mounted on the ESO 1.5~m telescope.
The O$^{2+}$/H$^+$ abundance ratios derived from the O~{\sc ii} ORLs with
the most reliable measurements are presented in Tables\,\ref{abundances:iv}
(the 3d\,--\,3p and 3p\,--\,3s transitions) and \ref{abundances:v} (the
4f\,--\,3d transitions). The Case~B effective recombination coefficients of
O~{\sc ii} calculated by PJS are adopted for the abundance determinations.
An electron temperature of 1000~K is assumed. For purpose of comparison,
the effective recombination coefficients of Storey \cite{storey1994} for
the 3p\,--\,3s transitions, and LSBC for the 3d\,--\,3p and 4f\,--\,3d
transitions, are also used. Case~B is assumed for the quartet transitions,
and Case~A for the doublets. An electron temperature of 5000~K is assumed
when using the data of Storey \cite{storey1994}, whose calculation is valid
from 5000 to 20\,000~K. Since the calculation of Storey \cite{storey1994} is
only spectral term-resolved, we deduce the effective recombination
coefficients for each fine-structure component of a multiplet with the
assumption that their relative intensities are as in the {\it LS}\,coupling.
As in the case of the N~{\sc ii} lines (Section\,\ref{orl_abundances:part3}),
for each multiplet in Table\,\ref{abundances:iv}, we calculate the abundance
value using the co-added intensities from all fine-structure components
observed; for the O~{\sc ii} 4f\,--\,3d transitions in
Table\,\ref{abundances:v}, we also calculate the abundance after co-adding
the intensities of all detected lines.

The O~{\sc ii} M1 3p\,$^4$D$^{\rm o}$\,--\,3s\,$^4$P $\lambda$4650 multiplet
is case-insensitive. At 5000~K, the Case~B effective recombination coefficient
of the M1 $\lambda$4650 multiplet given by Storey \cite{storey1994} is only
3.4 per cent higher than for Case~A. P\'{e}quignot, Petitjean \& Boisson
\cite{ppb1991} shows that the difference between the effective radiative
recombination coefficients in the two cases for this multiplet is 2.5 per
cent. The most recent calculation of PJS reveals that the difference between
the effective recombination coefficients in the two cases for the strongest
O~{\sc ii} M1 line $\lambda$4649.13 is less than 5 per cent. The coefficients
in the two cases for the other O~{\sc ii} M1 fine-structure lines are of
similar order. The O$^{2+}$/H$^+$ abundances derived from the individual
O~{\sc ii} M1 lines agree with each other, except for $\lambda\lambda$4638.86
and 4673.73, which yield very high abundance values. The average
O$^{2+}$/H$^+$ from O~{\sc ii} M1 is 16.2$\times$10$^{-4}$, which agrees well
with the values derived from the recombination coefficients of P\'{e}quignot,
Petitjean \& Boisson \cite{ppb1991} and Storey \cite{storey1994}. The current
measurements of the multiplet also agrees with LSBC, who gives
15.3$\times$10$^{-4}$ (PA = 0$^{\rm o}$) and 13.5$\times$10$^{-4}$ (PA =
45$^{\rm o}$).

Of the seven observed 3d\,--\,3p multiplets in Table\,\ref{abundances:iv},
the intensities of those from the upper terms 2p$^{2}$3d\,${4}$F and
2p$^{2}$3d\,$^{4}$D are almost independent of the assumption of Case~A or
Case~B. The M2 3p\,$^4$P$^{\rm o}$\,--\,3s\,$^4$P
$\lambda$4341, M6 3p\,$^2$P$^{\rm o}$\,--\,3s\,$^2$P $\lambda$3967,
M10 3d\,$^4$F\,--\,3p\,$^4$D$^{\rm o}$ $\lambda$4075,
M12 3d\,$^4$D\,--\,3p\,$^4$D$^{\rm o}$ $\lambda$3867,
M20 3d\,$^4$D\,--\,3p\,$^4$P$^{\rm o}$ $\lambda$4111 and
M26 3d\,$^2$D\,--\,3p\,$^2$D$^{\rm o}$ $\lambda$4385 multiplets of O~{\sc ii}
are case-insensitive, with their Case~B or Case~C\footnote{Defined with
reference to the O~{\sc ii} recombination spectrum. In Case~B, lines
terminating on the 2p$^3$\,$^4$S$^{\rm o}$ term are assumed to be optically
thick and no radiative decays are permitted to this state. In Case~C,
radiative decays to both the 2p$^3$\,$^4$S$^{\rm o}$ and $^2$D$^{\rm o}$
terms are excluded.} effective recombination coefficients being 2.3--30 per
cent higher than those for Case~A (Storey \citealt{storey1994}). The
O$^{2+}$/H$^{+}$ abundance ratios derived from the M1 and M2 multiplets
agree with those derived from the 3d\,--\,3p and 4f\,--\,3d transitions,
while in LSBC and Liu et al. \cite{liu2000} for NGC\,6153, the abundances
from those two multiplets are lower by about 40 per cent and a factor of 2.
respectively.
For Case~C to apply to the doublets, it requires transitions to the
2p$^{3}$\,$^{2}$D$^{\rm o}$ state of the ground configuration to be optically
thick, which is unlikely in the physical condition of NGC\,7009. Thus the
doublets (e.g. the M5 3p\,$^2$D$^{\rm o}$\,--\,3s\,$^2$P $\lambda$4418,
M6 3p\,$^{2}$P$^{\rm o}$\,--\,3s\,$^{2}$P $\lambda$3967 and M25
3d\,$^2$F\,--\,3p\,$^2$D$^{\rm o}$ $\lambda$4704 multiplets) can be regarded
as case-insensitive. The M11 3d\,$^4$P\,--\,3p\,$^4$D$^{\rm o}$ $\lambda$3903,
M19 3d\,$^4$P\,--\,3p\,$^4$P$^{\rm o}$ $\lambda$4152 and M28
3d\,$^4$P\,--\,3p\,$^4$S$^{\rm o}$ $\lambda$4913 multiplets, which have the
same upper term that can decay to the 2p$^{3}$\,$^{4}$S$^{\rm o}$ ground
state of O~{\sc ii} via resonance transitions, are expected to be very
case-sensitive: their Case~B effective recombination coefficients are more
than 20 times higher than the Case~A values.

Table\,\ref{abundances:iv} shows that in our current analysis, the
O$^{2+}$/H$^{+}$ abundance ratios derived from the quartets M10, M12 and M20,
all case-insensitive, are systematically higher than those derived from the
case-sensitive multiplets M19 and M28, a phenomenon discovered by Liu et al.
\cite{liu2000} for NGC\,6153. As pointed out by Liu et al. \cite{liu2000},
it is possible that there is a small departure from the assumed Case~B towards
Case~A, which would increase the derived abundances for the multiplets that
decay from the 2p$^{2}$3d\,$^{4}$P upper term. The three doublets presented
in Talbe\,\ref{abundances:iv} all yield abundance values that are consistent
with those derived from the case-insensitive quartets (M10, M12 and M20),
except for the M6 doublets, which yield systematically higher abundances.
The O~{\sc ii} M28 lines yield relatively lower abundance ratios, indicating
that Case~B might not be a good approximation for this multiplet. The
O$^{2+}$/H$^{+}$ abundance ratios derived from the $\lambda$4132.80 and
$\lambda$4153.30 line of the M19 multiplet of O~{\sc ii} are obviously lower
than the other O~{\sc ii} multiplets. The observed $\lambda$4156.53 line of
M19 is too strong compared to other components of this multiplet, with the
derived abundance value being higher than those deduced from other multiplets
by nearly a factor of two. That was also observed in NGC\,6153 (Liu et al.
\citealt{liu2000}) and by LSBC for the same object. No convincing candidates
for lines which might be blended with the O~{\sc ii} M19 $\lambda$4156.53
line, and thus cause the discrepancy, were found by LSBC. In NGC\,7009,
the intensity ratio of the $\lambda$4156.53 ($J$ = 3/2\,--\,5/2) and
$\lambda$4132.80 ($J$ = 3/2\,--\,1/2) lines of M19, which decay from the same
upper level, is 1.06, which is about a factor of two lower than the values of
1.7 (slit position PA = 0$^{\rm o}$) and 2.1 (PA = 45$^{\rm o}$) found for
the same object by LSBC, but agrees with the values of
0.6 (for the minor axis) and 1.1 (for the whole nebula) given by Liu et al.
\cite{liu2000} for NGC\,6153. The $\lambda$4132.80 line detected in our
spectrum of NGC\,7009 has an FWHM of 1.85\,{\AA}, which is broader than the
$\lambda$4156.53 line (FWHM$\sim$1.60\,{\AA}). Such a line width of the
$\lambda$4132.80 line might be contributed by line blending, which results
in a line ratio that is lower than the previous measurements. Besides,
different data quality might also contribute to the discrepancy between LSBC
and our current measurements. For the $\lambda$4156.53 line, we still do not
know which feature it is blended with.

In general, the O$^{2+}$/H$^{+}$ abundance ratios deduced using the Case~B
effective recombination coefficients of PJS are lower than those deduced
using the radiative recombination coefficients of LSBC, by 30--50 per cent
for the 3d\,--\,3p transitions, and by 20--30 per cent for most of the
4f\,--\,3d transitions.  The O$^{2+}$/H$^{+}$ abundance ratios calculated,
by co-adding the line intensities of each multiplet of the 3p\,--\,3s
configuration, using the effective recombination coefficients of PJS, do not
differ much from those deduced using the coefficients of Storey
\cite{storey1994}. However, the abundances deduced from each fine-structure
components of the 3p\,--\,3s multiplets, using the effective recombination
coefficients of PJS, are systematically lower than those deduced using the
coefficients of Storey \cite{storey1994}. This difference is more obvious for
the two doublets M5 and M6, with the abundances deduced based on the
coefficients of PJS being lower by about 10 per cent. If we use the Case~A
effective recombination coefficients of Storey \cite{storey1994}, the derived
abundances will be lower than those deduced using the data of PJS by a factor
of 7. The 4f\,--\,3d transitions are essentially case-insensitive, and the
O$^{2+}$/H$^+$ abundance ratios derived from those lines using the data
of PJS agree well. The abundance value calculated by co-adding the 4f\,--\,3d
line intensities in Table\,\ref{abundances:iv} is 1.330$\times$10$^{-3}$,
which agrees with the average value (1.403$\times$10$^{-3}$) of the 3\,--\,3
transitions. The mean O$^{2+}$/H$^{+}$ abundance ratios derived by averaging
the values from all 3\,--\,3 multiplets (excluding the values that are
abnormally high) plus the co-added 4f\,--\,3d transitions is
1.417$\times$10$^{-3}$, which is slightly lower than the recombination line
abundances given by LSBC: 17.0$\pm$1.0$\times$10$^{-4}$ (PA = 45$^{\rm o}$)
and 17.6$\pm$1.7$\times$10$^{-4}$ (PA = 0$^{\rm o}$). This value is adopted
as the recombination line O$^{2+}$/H$^{+}$ abundance in NGC\,7009.

\begin{table}
\centering
\caption{Recombination line O$^{2+}$/H$^{+}$ abundances derived from the
3\,--\,3 transitions. Intensities are normalized such that $I$(H$\beta$) =
100. The effective recombination coefficients of Storey \citet{storey1994}
(for the 3p\,--\,3s transitions) and LSBC (for the 3d\,--\,3p and 4f\,--\,3d
transitions) are also used for purpose of comparison.}
\label{abundances:iv}
\begin{tabular}{lllrr}
\hline
Line & Mult. & $I_{\rm obs}$ & \multicolumn{2}{c}{O$^{2+}$/H$^{+}$}\\
 & & & \multicolumn{2}{c}{($\times$10$^{-4}$)}\\
(\AA) & & & PJS & LSBC\\
\hline
$\lambda$4638.86 & M1 & 0.335 & 25.002 &  33.581\\
$\lambda$4641.81 & M1 & 0.437 & 14.549 &  17.355\\
$\lambda$4649.13 & M1 & 0.666 & 13.677 &  13.886\\
$\lambda$4650.84 & M1 & 0.175 & 12.731 &  17.542\\
$\lambda$4661.63 & M1 & 0.217 & 14.490 &  16.946\\
$\lambda$4673.73 & M1 & 0.052 & 21.607 &  25.815\\
$\lambda$4676.24 & M1 & 0.158 & 15.125 &  14.642\\
$\lambda$4696.35 & M1 & 0.015 & 13.528 &  12.510\\
\textbf{M1 3p~$^4$D$^{\rm o}$ -- 3s~$^4$P} & & 2.056 & \textbf{15.241} & \textbf{17.369}\\
\\
$\lambda$4317.14 & M2 & 0.091 & 12.821 &  12.383\\
$\lambda$4319.63 & M2 & 0.063 & 14.010 &   7.938\\
$\lambda$4325.76 & M2 & 0.029 & 17.471 &  19.731\\
$\lambda$4336.86 & M2 & 0.053 & 18.008 &  22.536\\
$\lambda$4345.56 & M2 & 0.141 & 15.455 &  19.187\\
$\lambda$4349.43 & M2 & 0.195 & 15.876 &  10.530\\
$\lambda$4366.89 & M2 & 0.085 & 11.471 &  10.710\\
\textbf{M2 3p~$^4$P$^{\rm o}$ -- 3s~$^4$P} & & 0.657 & \textbf{14.589} & \textbf{12.577}\\
\\
$\lambda$4414.90 & M5 & 0.100 & 19.665 &  22.885\\
$\lambda$4416.97 & M5 & 0.064 & 17.370 &  26.363\\
$\lambda$4452.37 & M5 & 0.014 & 20.859 &  28.835\\
\textbf{M5 3p~$^2$D$^{\rm o}$ -- 3s~$^2$P} & & 0.178 & \textbf{18.855} & \textbf{24.440}\\
\\
$\lambda$3945.04 & M6 & 0.026 & 56.459 &  61.094\\
$\lambda$3954.36 & M6 & 0.030 & 31.310 &  35.245\\
$\lambda$3973.26 & M6 & 0.065 & 27.719 &  30.546\\
$\lambda$3982.71 & M6 & 0.010 & 20.411 &  23.497\\
\textbf{M6 3p~$^2$P$^{\rm o}$ -- 3s~$^2$P} & & 0.131 & \textbf{30.785} & \textbf{29.451}\\
\\
$\lambda$4069.89$^a$ & M10 & 0.635 & 16.777 & 23.914\\
$\lambda$4072.16$^b$ & M10 & 0.549 & 17.051 & 20.461\\
$\lambda$4075.86     & M10 & 0.687 & 16.687 & 19.241\\
$\lambda$4078.84     & M10 & 0.089 & 15.899 & 11.811\\
$\lambda$4085.11     & M10 & 0.107 & 16.746 & 23.176\\
$\lambda$4092.93$^c$ & M10 & 0.077 & 17.567 & 22.941\\
\textbf{M10 3d~$^4$F -- 3p~$^4$D$^{\rm o}$} & & 2.144 & \textbf{16.732} & \textbf{20.841}\\
\\
$\lambda$3907.45     & M11 & 0.010 & 16.364 & 9.291\\
\textbf{M11 3d~$^4$P -- 3p~$^4$D$^{\rm o}$} & & 0.026$^d$ & \textbf{15.861} &  \textbf{8.411}\\
\\
$\lambda$3842.82     & M12 & 0.015 & 12.406 & 28.465\\
$\lambda$3851.03     & M12 & 0.017 & 12.351 & 19.397\\
$\lambda$3882.19$^e$ & M12 & 0.053 & 14.929 & 14.462\\
\textbf{M12 3d~$^4$D -- 3p~$^4$D$^{\rm o}$} & & 0.099 & \textbf{13.877} & \textbf{10.972}\\
\\
$\lambda$4129.32     & M19 & 0.032 & 21.489 & 45.444\\
$\lambda$4132.80     & M19 & 0.078 &  9.736 & 13.237\\
$\lambda$4140.70$^f$ & M19 & 0.003 & 45.123 & 23.923\\
$\lambda$4153.30     & M19 & 0.112 &  9.596 & 13.302\\
$\lambda$4156.53$^g$ & M19 & 0.079 & 35.714 & 59.101\\
$\lambda$4169.22     & M19 & 0.082 & 17.892 & 28.689\\
\textbf{M19 3d~$^4$P -- 3p~$^4$P$^{\rm o}$} & & 0.386     & \textbf{13.760} & \textbf{17.267}\\
\\
$\lambda$4110.79     & M20 & 0.025 & 14.198 & 9.655\\
\textbf{M20 3d~$^4$D -- 3p~$^4$P$^{\rm o}$} & & 0.259$^d$ & \textbf{14.162} & \textbf{10.973}\\
\hline
\end{tabular}
\end{table}

\addtocounter{table}{-1}
\begin{table}
\centering
\caption{Continued.}
\label{abundances:iv}
\begin{tabular}{lllrr}
\hline
Line & Mult. & $I_{\rm obs}$ & \multicolumn{2}{c}{O$^{2+}$/H$^{+}$}\\
 & & & \multicolumn{2}{c}{($\times$10$^{-4}$)}\\
(\AA) & & & PJS & LSBC\\
\hline
$\lambda$4699.22 & M25 & 0.024 & 25.809 & 33.771\\
$\lambda$4705.35 & M25 & 0.021 & 16.924 & 17.753\\
\textbf{M25 3d~$^2$F -- 3p~$^2$D$^{\rm o}$} & & 0.046$^h$ & \textbf{20.735} & \textbf{24.293}\\
\\
$\lambda$4890.86 & M28 & 0.013 &  6.360 & 10.377\\
$\lambda$4906.83 & M28 & 0.046 & 10.907 & 17.114\\
$\lambda$4924.53 & M28 & 0.074 & 10.609 & 16.190\\
\textbf{M28 3d~$^4$P -- 3p~$^4$S$^{\rm o}$} & & 0.133     & \textbf{10.049} & \textbf{15.311}\\
\\
Average & & & \textbf{14.032} & \textbf{21.896}\\
\hline
\end{tabular}
\begin{description}
\item [$^a$] Including the O~{\sc ii} M10
3d\,$^4$F$_{3/2}$\,--\,3p\,$^4$D$^{\rm o}_{1/2}$ $\lambda$4069.62 line.
\item [$^b$] Corrected for the contribution from the O~{\sc ii} M48a
4f\,G[5]$^{\rm o}_{9/2}$\,--\,3d\,$^{4}$F$_{7/2}$ $\lambda$4071.23 line
($\sim$10 per cent). Neglecting the N~{\sc ii} M38b
4f\,G[7/2]$_{3}$\,--\,3d\,$^{3}$F$^{\rm o}_{2}$ $\lambda$4073.05 line
($\sim$2 per cent).
\item [$^c$] Overestimated due to N~{\sc iii} M1 3p\,$^{2}$P$^{\rm
o}_{3/2}$\,--\,3s\,$^{2}$S$_{1/2}$ $\lambda$4097.33 line which is more than
20 times stronger.
\item [$^d$] Assuming the relative intensities of this multiplet are the
predicted values based on the effective recombination coefficients of PJS.
\item [$^e$] Corrected for the contribution from the O~{\sc ii} M11
3d\,$^4$P$_{3/2}$\,--\,3p\,$^4$D$^{\rm o}_{3/2}$ $\lambda$3882.45 line
($\sim$17 per cent). Neglecting the O~{\sc ii} M12
3d\,$^{4}$D$_{5/2}$\,--\,3p\,$^{4}$D$^{\rm o}_{7/2}$ $\lambda$3883.13 line
($\sim$2 per cent).
\item [$^f$] Overestimated due to the much stronger He~{\sc i} M53
6d\,$^{1}$D$_{2}$\,--\,2p\,$^{1}$P$^{\rm o}_{1}$ $\lambda$4143.76 line.
\item [$^g$] Neglecting the N~{\sc ii} M50b
4f\,D[3/2]$_{2}$\,--\,3d\,$^3$D$^{\rm o}_{1}$ $\lambda$4156.39 and the N~{\sc
ii} M50b 4f\,D[2]$_{1}$\,--\,3d\,$^3$D$^{\rm o}_{1}$ $\lambda$4157.01 lines
($\sim$4 per cent in total).
\item [$^h$] Not including the O~{\sc ii} M25
3d\,$^2$F$_{5/2}$\,--\,3p\,$^2$D$^{\rm o}_{5/2}$ $\lambda$4741.71 line.
\end{description}
\end{table}

\begin{table}
\centering
\caption{Recombination line O$^{2+}$/H$^{+}$ abundances from the 4f\,--\,3d
transitions. Intensities are normalized such that $I$(H$\beta$) = 100. The
O~{\sc ii} effective recombination coefficients of LSBC are also used for
purpose of comparison.}
\label{abundances:v}
\begin{tabular}{lllrr}
\hline
Line & Mult. & $I_{\rm obs}$ & \multicolumn{2}{c}{O$^{2+}$/H$^{+}$}\\
 & & & \multicolumn{2}{c}{($\times$10$^{-4}$)}\\
(\AA) & & & PJS & LSBC\\
\hline
$\lambda$4089.29$^a$ & M48a & 0.264 & 12.901 & 15.308\\
$\lambda$4083.90     & M48b & 0.094 & 13.310 & 19.077\\
$\lambda$4087.15$^b$ & M48c & 0.091 & 13.675 & 19.461\\
$\lambda$4062.94$^c$ & M50a & 0.036 & 13.975 & 16.592\\
$\lambda$4048.21$^d$ & M50b & 0.021 & 14.363 & 19.236\\
$\lambda$4303.83$^e$ & M53a & 0.120 & 13.844 & 16.847\\
$\lambda$4294.78$^f$ & M53b & 0.067 & 13.025 & 16.756\\
$\lambda$4307.23     & M53b & 0.029 & 12.415 & 16.053\\
$\lambda$4288.82$^g$ & M53c & 0.026 & 17.477 & 15.659\\
$\lambda$4282.96$^h$ & M67c & 0.063 & 17.615 & 23.690\\
$\lambda$4466.42$^i$ & M86b & 0.032 & 12.247 & 20.368\\
$\lambda$4489.49     & M86b & 0.022 & 12.958 & 19.542\\
$\lambda$4491.23     & M86a & 0.057 & 14.606 & 24.119\\
$\lambda$4609.44$^j$ & M92a & 0.128 & 13.969 & 17.309\\
$\lambda$4602.13$^k$ & M92b & 0.052 & 12.938 & 17.628\\
\\
sum & & 1.103 & \textbf{13.303} & \textbf{17.556}\\
Average & & & \textbf{13.955} & \textbf{18.510}\\
\hline
\end{tabular}
\begin{description}
\item [$^a$] Corrected for the contribution from Si~{\sc iv} M1
4p~$^{2}$P$^{\rm o}_{3/2}$\,--\,4s~$^{2}$S$_{1/2}$ $\lambda$4088.86, which is
about 15 per cent. Neglecting O~{\sc ii} M48a 4f~G[5]$^{\rm
o}_{9/2}$\,--\,3d~$^{4}$F$_{9/2}$ $\lambda$4088.27 (less than 2 per cent).
\item [$^b$] Neglecting N~{\sc ii} M38a 4f~F[5/2]$_{3}$\,--\,3d~$^3$F$^{\rm
o}_{3}$ $\lambda$4087.30 (about 3 per cent).
\item [$^c$] Including Ne~{\sc ii} M53 4f~0[3]$^{\rm
o}_{7/2}$\,--\,3d~$^{4}$D$_{5/2}$ $\lambda$4062.97.
\item [$^d$] Neglecting the contribution from O~{\sc ii} M50b 4f~F[3]$^{\rm
o}_{5/2}$\,--\,3d~$^{4}$F$_{7/2}$ $\lambda$4047.80, which is probably 7 per
cent.
\item [$^e$] Corrected for the contribution from O~{\sc ii} M65a 4f~G[5]$^{\rm
o}_{9/2}$\,--\,3d~$^4$D$_{7/2}$ $\lambda$4303.61, which is about 20 per cent.
\item [$^f$] The contribution from O~{\sc ii} M53b 4f~D[2]$^{\rm
o}_{3/2}$\,--\,3d~$^4$P$_{3/2}$ $\lambda$4294.92 to the blend at
$\lambda$4295, which is estimated to be 28 per cent, has been subtracted.
\item [$^g$] A blend of O~{\sc ii} M53c 4f~D[1]$^{\rm
o}_{1/2}$\,--\,3d~$^4$P$_{1/2}$ $\lambda$4288.82 and O~{\sc ii} M53c
4f~D[1]$^{\rm o}_{3/2}$\,--\,3d~$^4$P$_{1/2}$ $\lambda$4288.82.
\item [$^h$] Corrected for the contributions from O~{\sc ii} M67c 4f~F[2]$^{\rm
o}_{3/2}$\,--\,3d~$^4$D$_{3/2}$ $\lambda$4283.73 (about 35 per cent) and O~{\sc
ii} M67c 4f~F[2]$^{\rm o}_{5/2}$\,--\,3d~$^4$D$_{5/2}$ $\lambda$4283.25 (about
6 per cent). Neglecting O~{\sc ii} M78a 4f~F[4]$^{\rm
o}_{7/2}$\,--\,3d~$^{2}$F$_{5/2}$ $\lambda$4282.02 (less than 2 per cent).
Including Ne~{\sc ii} M57c 4f~1[3]$^{\rm o}_{7/2}$\,--\,3d~$^{4}$F$_{7/2}$
$\lambda$4283.73, whose contribution to the total intensity is unknown due to
the lack of atomic data.
\item [$^i$] Corrected for the contribution from O~{\sc ii} M86b 4f~D[2]$^{\rm
o}_{3/2}$\,--\,3d~$^{2}$P$_{3/2}$ $\lambda$4466.59, which is about 20 per cent.
\item [$^j$] Corrected for the contribution from O~{\sc ii} M92c 4f~F[2]$^{\rm
o}_{5/2}$\,--\,3d~$^{2}$D$_{3/2}$ $\lambda$4610.20, which is about 20 per cent.
\item [$^k$] Corrected for the contribution from N~{\sc ii} M5
3p~$^{3}$P$_{2}$\,--\,3s~$^{3}$P$^{\rm o}_{1}$ $\lambda$4601.48, which is
about 25 per cent.
\end{description}
\end{table}

\subsubsection{\label{orl_abundances:part5}
Ne$^{2+}$/H$^+$ abundances from ORLs}

Tables\,\ref{abundances:vi} and \ref{abundances:vii} present the recombination
line Ne$^{2+}$/H$^+$ abundances derived from the 3\,--\,3 and 4f\,--\,3d
transitions, respectively. For the 3d\,--\,3p and 3p\,--\,3s transitions, the
effective recombination coefficients calculated in the {\it LS}\,coupling
assumption by Kisielius et al. \cite{kisielius1998} are adopted. Case~A is
assumed for the quartet transitions and Case~B for the doublets. The
calculation of Kisielius et al. \cite{kisielius1998} is valid from 1000 to
20\,000~K, and four density cases, 10$^{2}$, 10$^{4}$, 10$^{5}$ and
10$^{6}$~cm$^{-3}$, were calculated. We assumed an electron density of
10$^4$~cm$^{-3}$ and a temperature of 1000~K in the abundance determinations.
For purpose of comparison, also presented in Tables\,\ref{abundances:vi} and
\ref{abundances:vii} are the recombination line Ne$^{2+}$/H$^{+}$ abundances
derived by LLB01, who used the same CCD spectrum as analyzed in the current
paper but assumed an electron temperature of 7100~K, as derived from the
Balmer discontinuity. In general, the Ne$^{2+}$/H$^+$ abundance ratios derived
from the 3d\,--\,3p and 3p\,--\,3s transitions in the current work are lower
than those given by LLB01. This is mainly due to the different temperatures
adopted. For the 4f\,--\,3d transitions, the abundances derived by us are
systematically lower than those given by LLB01 by about 13 per cent, except
the strongest lines of this transition array, e.g. $\lambda$4391.99 (M55e
4f\,2[5]$^{\rm o}_{11/2}$\,--\,3d\,$^4$F$_{9/2}$) and $\lambda$4409.30 (M55e
4f\,2[5]$^{\rm o}_{9/2}$\,--\,3d\,$^4$F$_{7/2}$), which yield close
Ne$^{2+}$/H$^{+}$ abundance ratios from the two analyses. The difference
between the abundances given by the two analyses are partially contributed by
the different extinctions used: The logarithmic extinction at H$\beta$,
$c$(H$\beta$), derived by LLB01 is 0.07, while ours is 0.174 (Paper~I).

The Ne~{\sc ii} M21 3d\,$^2$D$_{5/2}$\,--\,3p\,$^2$D$^{\rm o}_{5/2}$
$\lambda$3416.91 line yields relatively higher abundance, probably due to a
blend of the Ne~{\sc ii} M19 3d\,$^4$F$_{7/2}$\,--\,3p\,$^2$D$^{\rm o}_{5/2}$
$\lambda$3417.69 line. The nearby O~{\sc iii} Bowen fluorescence line M15
3d\,$^3$P$^{\rm o}_{1}$\,--\,3p\,$^3$P$_{1}$ $\lambda$3415.26 also affects
the measurement of the $\lambda$3416.91 line (Fig.\,\ref{3395-3475}). The
Ne~{\sc ii} lines of the M9 3p$^{\prime}$\,$^{2}$F$^{\rm
o}$\,--\,3s$^{\prime}$\,$^{2}$D multiplet, $\lambda\lambda$3568.50 and 3574.61
are detected in the spectrum of NGC\,7009 (Fig.\,\ref{3540-3595}). The
Ne$^{2+}$/H$^{+}$ abundance derived from this multiplet is much higher than
yielded by other multiplets. Similar results are also observed by LLB01 and
in another two PNe, M\,1-42 and M\,2-36 (Liu et al. \citealt{liu2001b}).
LLB01 pointed out that the high abundance yielded by M9 is possibly due to
the underestimated effective recombination coefficients for this multiplet.
The 3p\,$^{4}$D$^{\rm o}_{5/2}$\,--\,3s\,$^{4}$P$_{3/2}$ $\lambda$3355.02
line of multiplet M2 is blended with the He~{\sc i} M8 7p\,$^1$P$^{\rm
o}_{1}$\,--\,2s\,$^1$S$_{0}$ $\lambda$3354.55 line, and is also partially
blended with the [Cl~{\sc iii}] 3p$^{3}$\,$^{2}$P$^{\rm
o}_{1/2}$\,--\,3p$^{3}$\,$^{4}$S$^{\rm o}_{3/2}$ $\lambda$3353.17 line, as
shown in Fig.\,\ref{3345-3410}. The intensity of the [Cl~{\sc iii}] line was
obtained from line fitting with two Gaussian profiles. The intensity
contribution from the He~{\sc i} line was corrected for, as LLB01 did, using
the observed intensity of the He~{\sc i} M6 5p\,$^1$P$^{\rm
o}_{1}$\,--\,2s\,$^1$S$_{0}$ $\lambda$3613.64 line, assuming the line ratio
$I$($\lambda$3354.55)/$I$($\lambda$3613.64) = 0.35 (in Case~B assumption), as
predicted by Brocklehurst \cite{b72}. Here an electron temperature of 5000~K,
as derived from the He~{\sc i} line ratios (Paper~I), and a density of
10\,000~cm$^{-3}$ were assumed. The correction for the He~{\sc i} line amounts
to 26 per cent, close to the result of LLB01 (30 per cent). The reason that
we use the He~{\sc i} $\lambda$3613.64 line, instead of the He~{\sc i} M48
4d\,$^{1}$D$_{2}$\,--\,2p\,$^{1}$P$^{\rm o}_{1}$ $\lambda$4921.93 line, to
correct for the intensity contribution from the He~{\sc i} $\lambda$3354.55
line is that given the small wavelength span between the $\lambda$3354.55,
3613.64 lines, measurements of their intensity ratio are much less sensitive
to any uncertainties in reddening corrections and flux calibration.

The Ne~{\sc ii} 4f\,--\,3d recombination lines in Table\,\ref{abundances:vii}
are those whose preliminary effective recombination coefficients are
available (P.~J. Storey, private communication). The Ne$^{2+}$/H$^{+}$
abundances derived from the 4f\,--\,3d transitions are systematically higher
than those derived from the 3\,--\,3 transitions by about 50 per cent. The
difference between the abundances derived from the 3\,--\,3 and 4f\,--\,3d
transitions is mainly due to the inadequacy of the Ne~{\sc ii} effective
recombination coefficients. The average Ne$^{2+}$/H$^+$ abundance from the
4f\,--\,3d transitions is 8.5$\times$10$^{-4}$, about 0.15 dex higher than
the average value deduced from the individual lines of the 3\,--\,3
transition array. Here the two abnormally high abundances yielded by the
Ne~{\sc ii} M52b $\lambda$4250.65 (4f\,2[3]$^{\rm
o}_{5/2}$\,--\,3d\,$^{4}$D$_{3/2}$) and M61d $\lambda$4457.05 (4f\,2[2]$^{\rm
o}_{5/2}$\,--\,3d\,$^{2}$D$_{3/2}$) lines are excluded from averaging. We
adopt the abundance value 8.42$\times$10$^{-4}$, which is calculated by
co-adding the intensities of the 4f\,--\,3d transitions, as the recombination
line Ne$^{2+}$/H$^{+}$ abundance of NGC\,7009. The two Ne~{\sc ii} lines
$\lambda\lambda$4250.65 and 4457.05 that yield abnormally high abundances
are excluded from the abundance calculation.

\begin{table}
\centering
\caption{Recombination line Ne$^{2+}$/H$^+$ abundances derived from the
3\,--\,3 transitions. Intensities are normalized such that $I$(H$\beta$)
= 100. The abundances of LLB01 are presented for purpose of comparison.}
\label{abundances:vi}
\begin{tabular}{lclrr}
\hline
Line & Mult. & $I_{\rm obs}$ & \multicolumn{2}{c}{Ne$^{2+}$/H$^{+}$}\\
 & & & \multicolumn{2}{c}{($\times$10$^{-4}$)}\\
(\AA) & & & Current & LLB01\\
\hline
$\lambda$3694.21 & M1 & 0.254 & 7.932 & 8.97\\
$\lambda$3709.62 & M1 & 0.105 & 8.290 & 8.49\\
$\lambda$3777.14 & M1 & 0.048 & 3.859 & 3.54\\
\textbf{M1 3p~$^4$P$^{\rm o}$ -- 3s~$^4$P} & & 0.686 & \textbf{7.544} & \textbf{7.45}\\
\\
$\lambda$3334.84     & M2 & 0.414 & 5.847 & 5.14\\
$\lambda$3344.40     & M2 & 0.129 & 8.805 & \\
$\lambda$3355.02$^a$ & M2 & 0.195 & 5.278 & 5.89\\
\textbf{M2 3p~$^4$D$^{\rm o}$ -- 3s~$^4$P} & & 1.035 & \textbf{5.850} & \textbf{5.39}\\
\\
$\lambda$3713.08 & M5 & 0.297 & 6.141 & \\
\textbf{M5 3p~$^2$D$^{\rm o}$ -- 3s~$^2$P} & & 0.495 & \textbf{6.141} & \\
\\
$\lambda$3481.93 & M6 & 0.043 & 4.795 & 5.52\\
\textbf{M6 3p~$^2$S$^{\rm o}$ -- 3s~$^2$P} & & 0.066 & \textbf{4.795} & \textbf{5.52}\\
\\
$\lambda$3047.56 & M8 & 0.120 & 7.720 & \\
\textbf{M8 3d~$^4$D -- 3p~$^4$P$^{\rm o}$} & & 0.571 & \textbf{7.692} & \\
\\
$\lambda$3329.16 & M12 & 0.051 & & 7.67\\
$\lambda$3357.82 & M12 & 0.020 & & \\
$\lambda$3362.94 & M12 & 0.026 & & \\
$\lambda$3374.06 & M12 & 0.014 & & \\
$\lambda$3390.55 & M12 & 0.016 & & \\
\textbf{M12 3d~$^4$D -- 3p~$^4$D$^{\rm o}$} & & 0.149 & \textbf{8.051} & \textbf{7.67}\\
\\
$\lambda$3218.19 & M13 & 0.227 & 4.836 & 5.45\\
$\lambda$3244.09 & M13 & 0.072 & 2.253 & 2.02\\
\textbf{M13 3d~$^4$F -- 3p~$^4$D$^{\rm o}$} & & 0.636 & \textbf{4.840} & \textbf{3.68}\\
\\
$\lambda$3367.22 & M20 & 0.102 & 2.820 & 3.38\\
\textbf{M20 3d~$^2$F -- 3p~$^2$D$^{\rm o}$} & & 0.179 & \textbf{2.832} & \textbf{3.38}\\
\\
$\lambda$3416.91$^b$ & M21 & 0.075 & 11.818 & 13.00\\
$\lambda$3453.07     & M21 & 0.018 &  4.448 &  4.25\\
\textbf{M21 3d~$^2$D -- 3p~$^2$D$^{\rm o}$} & & 0.050 & \textbf{3.529} & \textbf{4.25}\\
\\
$\lambda$3542.85 & M34 & 0.030 & & 2.89\\
$\lambda$3565.82 & M34 & 0.024 & & \\
$\lambda$3594.16 & M34 & 0.011 & & \\
\textbf{M34 3d~$^4$P -- 3p~$^4$S$^{\rm o}$} & & 0.065 & \textbf{3.006} & \textbf{3.05}\\
\\
$\lambda$3568.50     & M9 & 0.168 & & \\
$\lambda$3574.61$^c$ & M9 & 0.053 & & \\
\textbf{M9 3p$^{\prime}$~$^{2}$F$^{\rm o}$ -- 3s$^{\prime}$~$^{2}$D} & & 0.221 & \textbf{36.125} & \\
\\
Average & & & \textbf{6.040}\\
\hline
\end{tabular}
\begin{description}
\item [$^a$] Corrected for the contribution from the He~{\sc i} M8
7p\,$^1$P$^{\rm o}_{1}$\,--\,2s\,$^1$S$_{0}$ $\lambda$3354.55 line.
\item [$^b$] Probably overestimated due to Ne~{\sc ii} M19
3d\,$^4$F$_{7/2}$\,--\,3p\,$^2$D$^{\rm o}_{5/2}$ $\lambda$3417.69. Excluded
from calculating the average Ne$^{2+}$/H$^+$ abundance ratio.
\item [$^c$] Including the Ne~{\sc ii} M9 3p$^{\prime}$\,$^{2}$F$^{\rm
o}_{5/2}$\,--\,3s$^{\prime}$\,$^{2}$D$_{5/2}$ $\lambda$3574.18 line.
\end{description}
\end{table}

\begin{table}
\centering
\caption{Recombination line Ne$^{2+}$/H$^+$ abundances from the 4f\,--\,3d
transitions. Intensities are normalized such that $I$(H$\beta$) = 100. The
abundances of LLB01 are presented for purpose of comparison.}
\label{abundances:vii}
\begin{tabular}{lclrr}
\hline
Line & Mult. & $I_{\rm obs}$ & \multicolumn{2}{c}{Ne$^{2+}$/H$^{+}$}\\
 & & & \multicolumn{2}{c}{($\times$10$^{-4}$)}\\
(\AA) & & & Current & LLB01\\
\hline
$\lambda$4391.99$^a$ & M55e & 0.072  &  7.451 &  7.53\\
$\lambda$4409.30     & M55e & 0.058  &  9.039 &  9.80\\
$\lambda$4219.75$^b$ & M52a & 0.049  &  9.135 & 12.60\\
$\lambda$4233.85     & M52a & 0.011  &  8.230 & 11.40\\
$\lambda$4231.64     & M52b & 0.009  &  7.084 & 19.20\\
$\lambda$4250.65$^c$ & M52b & 0.016  & 18.853 & 19.90\\
$\lambda$4397.99     & M57b & 0.024  &  7.174 &  6.56\\
$\lambda$4379.55$^d$ & M60b & 0.050  &  8.367 & \\
$\lambda$4428.64$^e$ & M60c & 0.037  &  8.775 & 11.50\\
$\lambda$4430.94$^f$ & M61a & 0.025  &  9.158 & 12.10\\
$\lambda$4457.05$^{c,g}$ & M61d & 0.026  & 27.305 & 26.30\\
$\lambda$4413.22$^h$ & M65  & 0.021  &  9.099 & 12.90\\
$\lambda$4421.39     & M66c & 0.008  &  9.723 & \\
\\
Sum & & 0.364$^i$ & \textbf{8.425} & \textbf{9.93}\\
Average & & & \textbf{8.476} & \textbf{9.77}\\
\hline
\end{tabular}
\begin{description}
\item [$^a$] Neglecting the contribution from Ne~{\sc ii} M55e
4f\,2[5]$^{\rm o}_{9/2}$\,--\,3d\,$^4$F$_{9/2}$ $\lambda$4392.00.
\item [$^b$] Neglecting the contribution from Ne~{\sc ii} M52a
4f\,2[4]$^{\rm o}_{7/2}$\,--\,3d\,$^4$D$_{7/2}$ $\lambda$4219.37.
\item [$^c$] Not included in calculating the total intensity or the average
abundance value.
\item [$^d$] Neglecting the contribution from Ne~{\sc ii} M60b
4f\,1[4]$^{\rm o}_{7/2}$\,--\,3d\,$^2$F$_{7/2}$ $\lambda$4379.40.
\item [$^e$] Neglecting the contribution from Ne~{\sc ii} M60c
4f\,1[3]$^{\rm o}_{5/2}$\,--\,3d\,$^2$F$_{5/2}$ $\lambda$4428.52.
\item [$^f$] The contribution from Ne~{\sc ii} M57a 4f\,1[2]$^{\rm
o}_{5/2}$\,--\,3d\,$^4$F$_{3/2}$ $\lambda$4430.90, which is about 30 per
cent, has been subtracted. The contribution from Ne~{\sc ii} M57a
4f\,1[2]$^{\rm o}_{3/2}$\,--\,3d\,$^4$F$_{3/2}$ $\lambda$4431.11 is
negligible.
\item [$^g$] Including the contribution from the three Ne~{\sc ii} lines,
M66c 4f\,1[3]$^{\rm o}_{5/2}$\,--\,3d\,$^4$P$_{5/2}$ $\lambda$4457.24,
M61d 4f\,2[2]$^{\rm o}_{3/2}$\,--\,3d\,$^2$D$_{3/2}$ $\lambda$4457.24 and
M66c 4f\,1[3]$^{\rm o}_{7/2}$\,--\,3d\,$^4$P$_{5/2}$ $\lambda$4457.36.
\item [$^h$] Including the contribution from Ne~{\sc ii} M57c
4f\,1[3]$^{\rm o}_{5/2}$\,--\,3d\,$^4$F$_{3/2}$ $\lambda$4413.11. Neglecting
Ne~{\sc ii} M65 4f\,0[3]$^{\rm o}_{5/2}$\,--\,3d\,$^4$P$_{5/2}$
$\lambda$4413.11.
\item [$^i$] Excluding Ne~{\sc ii} M52b 4f\,2[3]$^{\rm
o}_{5/2}$\,--\,3d\,$^{4}$D$_{3/2}$ $\lambda$4250.65 and Ne~{\sc ii} M61d
4f\,2[2]$^{\rm o}_{5/2}$\,--\,3d\,$^{2}$D$_{3/2}$ $\lambda$4457.05.
\end{description}
\end{table}

\subsubsection{\label{orl_abundances:part7}
ORL abundances of other ions}

Table\,\ref{abundances:viii} presents the recombination line C$^{3+}$/H$^+$,
C$^{4+}$/H$^+$, N$^{3+}$/H$^+$, O$^{3+}$/H$^+$, and O$^{4+}$/H$^+$ abundances.
The C$^{3+}$/H$^+$ abundance ratios are derived from the M1 $\lambda$4650, M16
$\lambda$4069 and M18 $\lambda$4187 multiplets. The abundances from the three
multiplets agree with each other, and the adopted C$^{3+}$/H$^+$ ratio in
NGC\,7009 is an average from them. The C$^{4+}$/H$^+$ abundance ratio
2.20$\times$10$^{-5}$ is derived from the C~{\sc iv} M8 $\lambda$4658 line,
which is slightly contaminated by [Fe~{\sc iii}] $\lambda$4658. This abundance
agrees with those given by LSBC: 0.239$\times$10$^{-4}$
(PA = 45$^{\rm o}$) and 0.182$\times$10$^{-4}$ (PA = 0$^{\rm o}$).

The N$^{3+}$/H$^+$ abundance ratio is derived from N~{\sc iii} M18
$\lambda$4379. The N~{\sc iii} M17 lines, $\lambda\lambda$3998.63 and
4003.58,72 are observed, but only the effective dielectronic recombination
coefficients for this multiplet are available (Nussbaumer \& Storey
\citealt{ns1984}). If we adopt the dielectronic data, the derived
N$^{3+}$/H$^+$ abundance from the M17 multiplet will be more than one order
of magnitude higher than that from the N~{\sc iii} M18 $\lambda$4379 line
(Table\,\ref{abundances:viii}). This indicates that the excitation of the
N~{\sc ii} M17 lines is probably dominated by radiative recombination. We
adopt the N$^{3+}$/H$^+$ ratio (1.31$\times$10$^{-4}$) derived from the N~{\sc
ii} M18 $\lambda$4379 line as the abundance in NGC\,7009. This N$^{3+}$/H$^+$
abundance agrees with those given by LSBC who also used
the $\lambda$4379 line: 1.34$\times$10$^{-4}$ (PA = 45$^{\rm o}$) and
1.71$\times$10$^{-4}$ (PA = 0$^{\rm o}$). We observed the N~{\sc iii} M1 and
M2 lines, which are excited by the secondary Bowen fluorescence mechanism.
Also detected in the spectra of NGC\,7009 are the N~{\sc iii} transitions with
the $^3$P$^{\rm o}$ parentage: M3 3p$^{\prime}$~$^4$D --
3s$^{\prime}$~$^4$P$^{\rm o}$, M6 3p$^{\prime}$~$^2$D --
3s$^{\prime}$~$^2$P$^{\rm o}$ and M9 3d$^{\prime}$~$^4$F$^{\rm o}$ --
3p$^{\prime}$~$^4$D. Analyses of these lines are in Section\,\ref{niii_orls}.

Some strong O~{\sc iii} lines from the 3\,--\,3 arrays are excited by the
fluorescence mechanism or radiative charge-transfer reaction of O$^{3+}$
and H$^0$ (Liu \& Danziger \citealt{ld1993a}; Liu, Danziger \& Murdin
\citealt{ldm93}). The O~{\sc iii} M14 3d~$^3$D$^{\rm o}$ -- 3p~$^3$P
$\lambda$3713 lines cannot be excited by fluorescence or charge-transfer
reaction, but is likely to be excited only by recombination and therefore a
useful abundance indicator for O$^{3+}$/H$^+$. Unfortunately, no recombination
coefficients are available for this multiplet. The O~{\sc iii} M8
3d~$^3$F$^{\rm o}$\,--\,3p~$^3$D $\lambda$3265 lines originate from radiative
and dielectronic recombination (Liu \& Danziger \cite{ld1993a}), and the
effective dielectronic and radiative recombination coefficients for this
multiplet are available from Nussbaumer \& Storey \cite{ns1984} and
P\'{e}quignot, Petitjean \& Boisson \cite{ppb1991}, respectively. Thus the
O$^{3+}$/H$^+$ abundances are derived from two of the O~{\sc iii} M8 lines
$\lambda\lambda$3260.85 and 3265.32, which are observed in NGC\,7009
(Fig.\,\ref{3250-3305}). Another M8 line $\lambda$3267.20 is blended with
the $\lambda$3265.32 line, which is expected to be the strongest in M8, but
its contribution is probably negligible. The O$^{3+}$/H$^+$ abundance ratio
derived from O~{\sc iii} M8 is about 6.6$\times$10$^{-5}$
(Table\,\ref{abundances:viii}).

A very faint O~{\sc iv} line M2 3d~$^2$D$_{5/2}$ -- 3p~$^2$P$^{\rm o}_{3/2}$
$\lambda$3411.69 is also observed to partially blended with O~{\sc iii} M15
$\lambda$3415.26 (Fig.\,\ref{3395-3475}). Another M2 line $\lambda$3413.64
is blended in between. The other line $\lambda$3403.52 is blended with
O~{\sc iii} M15 $\lambda$3405.71 (Fig.\,\ref{3395-3475}), and its accurate
measurements are difficult. Multi-Gaussian profile fits give an intensity of
0.0514 for the $\lambda$3411.69 line, and 0.0567 for the $\lambda$3413.64
line, both have an uncertainty of more than 20 per cent (Table\,$7$
in Paper~I). The intensity ratio of the two lines differ from the pure
{\it LS}\,coupling ratio, i.e. 9\,:\,1. We use $\lambda$3411.69 to derive the
O$^{4+}$/H$^+$ abundance ratio because its measurement is probably more
reliable. Here the effective radiative and dielectronic recombination
coefficients from P\'{e}quignot, Petitjean \& Boisson \cite{ppb1991} and
Nussbaumer \& Storey \cite{ns1984}, respectively, are used.

\begin{table}
\centering
\caption{Recombination line C$^{3+}$/H$^+$, C$^{4+}$/H$^+$, N$^{3+}$/H$^+$,
O$^{3+}$/H$^+$, and O$^{4+}$/H$^+$ abundances. Intensities are normalized
such that $I$(H$\beta$) = 100.}
\label{abundances:viii}
\begin{tabular}{lclr}
\hline
Line & Mult. & $I_{\rm obs}$ & C$^{3+}$/H$^+$\\
(\AA) & & & ($\times$10$^{-4}$)\\
\hline
$\lambda$4647.42 & M1 & 0.170 & \\
$\lambda$4650.25 & M1 & 0.100 & \\
$\lambda$4651.47 & M1 & 0.034 & \\
\textbf{M1 3p~$^3$P$^{\rm o}$ -- 3s~$^3$S} & & 0.304 & \textbf{1.659}\\
$\lambda$4067.94 & M16 & 0.071 & \\
$\lambda$4068.91 & M16 & 0.093 & \\
$\lambda$4070.26 & M16 & 0.120 & \\
\textbf{M16 5g~$^3$G -- 4f~$^3$F$^{\rm o}$} & & 0.284 & \textbf{1.472}\\
\textbf{M18 5g~$^1$G -- 4f~$^1$F$^{\rm o}$} & $\lambda$4186.90 & 0.088 & \textbf{1.309}\\
\\
\hline
Wavelength & Mult. & $I_{\rm obs}$ & C$^{4+}$/H$^+$\\
(\AA) & & & ($\times$10$^{-4}$)\\
\hline
\textbf{M8 6h~$^2$H$^{\rm o}$ -- 5g~$^2$G} & $\lambda$4658.30$^a$ & 0.147 & \textbf{0.220}\\
\\
\hline
Wavelength & Mult. & $I_{\rm obs}$ & N$^{3+}$/H$^+$\\
(\AA) & & & ($\times$10$^{-4}$)\\
\hline
$\lambda$3998.63     & M17 & 0.025 & \\
$\lambda$4003.58$^b$ & M17 & 0.040 & \\
\textbf{M17 5f~$^2$F$^{\rm o}$ -- 4d~$^2$D} & & 0.065 & \textbf{12.390}\\
\textbf{M18 5g~$^2$G -- 4f~$^2$F$^{\rm o}$} & $\lambda$4379.11 & 0.310 & \textbf{1.313}\\
\\
\hline
Wavelength & Mult. & $I_{\rm obs}$ & O$^{3+}$/H$^+$\\
(\AA) & & & ($\times$10$^{-4}$)\\
\hline
$\lambda$3260.85 & M8 & 0.179 & \\
$\lambda$3265.32 & M8 & 0.139 & \\
\textbf{M8 3d~$^3$F$^{\rm o}$ -- 3p~$^3$D} & & 0.324 & \textbf{0.659}\\
\\
\hline
Wavelength & Mult. & $I_{\rm obs}$ & O$^{4+}$/H$^+$\\
(\AA) & & & ($\times$10$^{-4}$)\\
\hline
$\lambda$3411.69$^c$ & M2 & 0.057 & \\
\textbf{M2 3d~$^2$D -- 3p~$^2$P$^{\rm o}$} & & 0.114 & \textbf{0.158}\\
\hline
\end{tabular}
\begin{description}
\item [$^a$] Including [Fe~{\sc iii}] $\lambda$4658.05.
\item [$^b$] Including N~{\sc iii} M17 $\lambda$4003.72 (5f\,$^2$F$^{\rm
o}_{5/2}$\,--\,4d\,$^2$D$_{5/2}$).
\item [$^c$] Could be of large uncertainty due to line blending.
\end{description}
\end{table}

The adopted C, N, O and Ne ionic abundances from optical recombination lines
are summarized in Table\,\ref{orlabundances:adopted}. They are mostly averaged
from the abundance ratios that are calculated by co-adding the line
intensities of individual multiplets (or transition arrays). Atomic data
references used for the ORL analysis are listed in Table\,\ref{references:orl}.

\begin{table}
\centering
\caption{Adopted recombination line abundances for the C, N, O, and Ne ions.}
\label{orlabundances:adopted}
\begin{tabular}{lrr}
\hline
Ion & \multicolumn{2}{c}{Abundances}\\
 & ($\times$10$^{-4}$) & $\log$[X$^{i+}$/H$^+$]+12\\
\hline
C$^{2+}$/H$^+$  &  6.865 & 8.796\\
C$^{3+}$/H$^+$  &  1.480 & 8.170\\
C$^{4+}$/H$^+$  &  0.220 & 7.342\\
N$^{2+}$/H$^+$  &  3.450 & 8.538\\
N$^{3+}$/H$^+$  &  1.313 & 8.118\\
O$^{2+}$/H$^+$  & 14.176 & 9.152\\
O$^{3+}$/H$^+$  &  0.659 & 7.819\\
O$^{4+}$/H$^+$  &  0.158 & 7.198\\
Ne$^{2+}$/H$^+$ &  8.425 & 8.926\\
\hline
\end{tabular}
\end{table}

Several permitted lines emitted by silicon ions were observed or deblended
(Paper~I). As the first and the second ionization potentials of atomic silicon
are 8.15 and 16.35~eV, respectively, we expect that the main ionization stages
of silicon are doubly and triply ionized, while the amount of Si$^+$ is
assumed to be negligible and Si$^{4+}$ should exist but is of much lower
abundance compared to Si$^{2+}$ and Si$^{3+}$.

The Si$^{3+}$/H$^+$ abundance ratios derived from Si~{\sc iii} M2 and M5
lines are presented in Table\,\ref{abundances:ix}. The Si~{\sc iv} M1
$\lambda$4116.10 line is also observed (Fig.\,\ref{4110-4175}), which means
that $\lambda$4088.86 of the same multiplet should also exist. Since only the
effective dielectronic recombination coefficients for a few selected
Si~{\sc iii} transitions are available from Nussbaumer \& Storey \cite{ns1986},
we only present the Si$^{3+}$/H$^+$ abundance ratios. The averaged
Si$^{3+}$/H$^+$ ratio is 6.04$\times$10$^{-6}$.

The Mg~{\sc ii} M4 $\lambda$4481 line is observed (Fig.\,\ref{4444-4504}),
and the Mg$^{2+}$/H$^+$ abundance derived is presented in
Table\,\ref{abundances:ix}. Since the ionization potentials of neutral
Mg$^0$ and Mg$^{2+}$ are 7.65 and 80.14~eV, respectively, we assume that
magnesium in NGC\,7009 is mainly doubly ionized. Unfortunately no effective
recombination coefficients for Mg~{\sc ii} lines are available. Given the
similarity between the atomic structure of Mg~{\sc ii} and C~{\sc ii}, we
assumed that the effective recombination coefficient of the Mg~{\sc ii} M4
4f~$^2$F$^{\rm o}$ -- 3d~$^2$D $\lambda$4481 line is equal to, or at least is
close to, that of the C~{\sc ii} M6 4f~$^2$F$^{\rm o}$ -- 3d~$^2$D
$\lambda$4267 transition. The effective recombination coefficient (in Case~B)
for the C~{\sc ii} $\lambda$4267 line is adopted from Bastin \cite{bastin2006},
with the assumption of $T_\mathrm{e}$ = 10\,000~K and $N_\mathrm{e}$ =
10\,000~cm$^{-3}$. The calculation of Davey, Storey \& Kisielius
\cite{davey2000} differs
from that of Bastin \cite{bastin2006} by 1.5 per cent for the  C~{\sc ii}
$\lambda$4267 line. The Mg$^{2+}$/H$^+$ abundance derived from the
$\lambda$4481 line is 3.18$\times$10$^{-5}$.

%
%
\begin{table}
\centering
\caption{Recombination line Si$^{3+}$/H$^+$ and Mg$^{2+}$/H$^+$ abundances.
Intensities are normalized such that $I$(H$\beta$) = 100.}
\label{abundances:ix}
\begin{tabular}{lcll}
\hline
Line & Mult. & $I_{\rm obs}$ & Si$^{3+}$/H$^+$\\
(\AA) & & & ($\times$10$^{-5}$)\\
\hline
$\lambda$4552.62 & M2 & 0.0175 & \\
$\lambda$4567.82 & M2 & 0.0114 & \\
$\lambda$4574.76 & M2 & 0.0041 & \\
\textbf{M2 4p~$^3$P$^{\rm o}$ -- 4s~$^3$S} & & 0.0325 & \textbf{0.770}\\
$\lambda$3806.54 & M5 & 0.022 & \\
\textbf{M5 4d~$^3$D -- 4p~$^3$P$^{\rm o}$} & & 0.0396 & \textbf{0.438}\\
\\
\hline
Wavelength & Mult. & $I_{\rm obs}$ & Mg$^{2+}$/H$^+$\\
(\AA) & & & ($\times$10$^{-5}$)\\
\hline
$\lambda$4481.20$^a$ & M4 & 0.0303 & \\
\textbf{M4 4f~$^2$F$^{\rm o}$ -- 3d~$^2$D} & & 0.0309 & \textbf{3.179}\\
\hline
\end{tabular}
\begin{description}
\item [$^a$] We assume that the Mg~{\sc ii} 4f\,--\,3d $\lambda$4481 line has
an effective recombination coefficient equal to that of the C~{\sc ii}
4f\,--\,3d $\lambda$4267 line, given the similarity between the atomic
structure of Mg~{\sc ii} and C~{\sc ii}.
\end{description}
\end{table}

\begin{table*}
\begin{minipage}{90mm}
\caption{References for the ORL atomic data.}
\label{references:orl}
\centering
\begin{tabular}{lll}
\hline
ion & \multicolumn{2}{c}{ORLs}\\
 & Effec.recomb. coefficients & Comments\\
\hline
H~{\sc i}    &   Storey \& Hummer \cite{sh1995}        &   Case B\\
He~{\sc i}   &   Benjamin, Skillman \& Smits \cite{bss99}   &   Case B; singlets\\
             &   Brocklehurst \cite{b72}  &   Case A; triplets\\
He~{\sc ii}  &   Storey \& Hummer \cite{sh1995}      &   Case B\\
C~{\sc i}    &   Escalante \& Victor \cite{ev1990}   &   Case A; singlets\\
             &   Escalante \& Victor \cite{ev1990}   &   Case B; triplets\\
C~{\sc ii}   &   Davey, Storey \& Kisielius \cite{davey2000}       &   Case B\\
C~{\sc iii}  &   P\'{e}quignot, Petitjean \& Boisson \cite{ppb1991}  &   Case A\\
             &   Nussbaumer \& Storey \cite{ns1984}  &   Dielectronic recombination\\
C~{\sc iv}   &   P\'{e}quignot, Petitjean \& Boisson \cite{ppb1991}  &   Case A\\
N~{\sc i}    &   P\'{e}quignot, Petitjean \& Boisson \cite{ppb1991}  &   Case A; doublets\\
             &   P\'{e}quignot, Petitjean \& Boisson \cite{ppb1991}  &   Case B; quartets\\
N~{\sc ii}   &   FSL11   &   Case B\\
N~{\sc iii}  &   P\'{e}quignot, Petitjean \& Boisson \cite{ppb1991}  &   Case A\\
             &   Nussbaumer \& Storey \cite{ns1984}   &   Dielectronic recombination\\
O~{\sc i}    &   P\'{e}quignot, Petitjean \& Boisson \cite{ppb1991}  &   Case A\\
O~{\sc ii}   &   P.~J. Storey (PJS, private communication)   &   Case B\\
O~{\sc iii}  &   P\'{e}quignot, Petitjean \& Boisson \cite{ppb1991}  &   Case A\\
O~{\sc iv}   &   P\'{e}quignot, Petitjean \& Boisson \cite{ppb1991}  &   Case A\\
             &   Nussbaumer \& Storey \cite{ns1984}   &   Dielectronic recombination\\
Ne~{\sc ii}  &   Kisielius et al. \cite{kisielius1998}   &   Case B; doublets\\
             &   Storey (unpublished)   &   Case A; quartets\\
             &   Nussbaumer \& Storey \cite{ns1987}   &   Dielectronic recombination\\
Mg~{\sc ii}  &   Davey, Storey \& Kisielius \cite{davey2000}$^a$   &   Case B\\
Si~{\sc ii}  &   Nussbaumer \& Storey \cite{ns1986}   &   Dielectronic recombination\\
Si~{\sc iii} &   Nussbaumer \& Storey \cite{ns1986}   &   Dielectronic recombination\\
\hline
\end{tabular}
\begin{description}
\item [$^a$] Given the similarity between the atomic structure of Mg~{\sc ii}
and C~{\sc ii}, we have assumed that the Mg~{\sc ii} M4 4f~$^2$F$^{\rm o}$ --
3d~$^2$D $\lambda$4481 line has an effective recombination coefficient equal
to that of the C~{\sc ii} M6 4f~$^2$F$^{\rm o}$ -- 3d~$^2$D $\lambda$4267 line
(Zhang et al. \citealt{zhang05b}).
\end{description}
\end{minipage}
\end{table*}

\subsection{\label{cel_abundances}
Ionic abundances from CELs}

\subsubsection{\label{cel_abundances:part1}
Ionic abundances from the optical CELs}

The ionic abundances derived from optical CELs detected in the spectrum of
NGC\,7009 are presented in Table\,\ref{abundances:cels}. An electron
temperature of 10\,000~K, which is an average from different CEL diagnostic
ratios (Paper~I), and a constant density of 4300~cm$^{-3}$, an average
derived from a variety of optical CEL ratios, are assumed throughout the
abundance determinations. In addition to the ionic abundances of N, O and Ne,
abundances are also derived for the ions of F, Mg, Si, S, Cl, Ar from the
CELs detected in the spectrum of NGC\,7009. The atomic data references used
for CEL analysis are listed in Table\,\ref{references:cel}.

\begin{table*}
\begin{minipage}{115mm}
\caption{Ionic abundances derived from optical CELs. Line intensities are
normalized such that $I$(H$\beta$) = 100.}
\label{abundances:cels}
\centering
\begin{tabular}{llrccc}
\hline
\multicolumn{2}{c}{Lines} & $I_{\rm obs}$ & X$^{i+}$/H$^+$ & \multicolumn{2}{c}{Abundances}\\
Ions & Lines (\AA) & & & X$^{i+}$/H$^+$ & $\log$[X$^{i+}$/H$^+$]+12\\
\hline
$[$C~{\sc i}$]$    & $\lambda$$\lambda$9824.13,9850.26 &    0.035 & C$^{0}$/H$^+$   & 8.483$\times$10$^{-9}$  & 3.929\\
$[$N~{\sc i}$]$    & $\lambda$$\lambda$5197.90,5200.26 &    0.091 & N$^{0}$/H$^+$   & 8.435$\times$10$^{-8}$  & 4.926\\
$[$O~{\sc i}$]$    & $\lambda$$\lambda$6300.30,6363.78 &    0.740 & O$^{0}$/H$^+$   & 8.395$\times$10$^{-7}$  & 5.924\\
$[$N~{\sc ii}$]$   & $\lambda$5754.64                   &    0.390 & N$^+$/H$^+$     & 2.876$\times$10$^{-6}$  & 6.459\\
$[$N~{\sc ii}$]$   & $\lambda$$\lambda$6548.04,6583.46 &   20.600 & N$^+$/H$^+$     & 2.725$\times$10$^{-6}$  & 6.435\\
$[$O~{\sc ii}$]$   & $\lambda$$\lambda$3726.03,3728.81 &   19.971 & O$^+$/H$^+$     & 8.779$\times$10$^{-6}$  & 6.943\\
$[$O~{\sc ii}$]$   & $\lambda$$\lambda$7319.99,7330.73 &    2.300 & O$^+$/H$^+$     & 1.996$\times$10$^{-5}$  & 7.300\\
$[$O~{\sc iii}$]$  & $\lambda$4363.21                   &    7.300 & O$^{2+}$/H$^+$  & 2.435$\times$10$^{-4}$  & 8.387\\
$[$O~{\sc iii}$]$  & $\lambda$4931.23                   &    0.120 & O$^{2+}$/H$^+$  & 2.455$\times$10$^{-4}$  & 8.390\\
$[$O~{\sc iii}$]$  & $\lambda$$\lambda$4958.91,5006.84 & 1550.753 & O$^{2+}$/H$^+$  & 3.231$\times$10$^{-4}$  & 8.509\\
$[$F~{\sc ii}$]$   & $\lambda$4789.45                   &    0.017 & F$^+$/F$^+$     & 2.579$\times$10$^{-8}$  & 4.411\\
$[$F~{\sc iv}$]$   & $\lambda$4059.90                   &    0.012 & F$^{3+}$/F$^+$  & 4.387$\times$10$^{-9}$  & 3.642\\
$[$Ne~{\sc iii}$]$ & $\lambda$3342.50                   &    0.755 & Ne$^{2+}$/H$^+$ & 3.079$\times$10$^{-4}$  & 8.488\\
$[$Ne~{\sc iii}$]$ & $\lambda$$\lambda$3868.76         &  118.837 & Ne$^{2+}$/H$^+$ & 1.296$\times$10$^{-4}$  & 7.928\\
$[$Ne~{\sc iii}$]$ & $\lambda$4012.01                   &    0.014 & Ne$^{2+}$/H$^+$ & 2.161$\times$10$^{-4}$  & 8.335\\
$[$Ne~{\sc iv}$]$  & $\lambda$$\lambda$4724.17,4725.67 &    0.042 & Ne$^{3+}$/H$^+$ & 1.501$\times$10$^{-5}$  & 7.176\\
$[$Ne~{\sc iv}$]$  & $\lambda$$\lambda$4714.17,4715.66 &    0.067 & Ne$^{3+}$/H$^+$ & 3.845$\times$10$^{-5}$  & 7.585\\
$[$S~{\sc ii}$]$   & $\lambda$$\lambda$4068.60,4076.35 &    0.960 & S$^+$/H$^+$     & 1.085$\times$10$^{-7}$  & 5.035\\
$[$S~{\sc ii}$]$   & $\lambda$$\lambda$6716.44,6730.82 &    3.700 & S$^+$/H$^+$     & 1.192$\times$10$^{-7}$  & 5.076\\
$[$S~{\sc iii}$]$  & $\lambda$3721.69                   &    1.045 & S$^{2+}$/H$^+$  & 2.781$\times$10$^{-6}$  & 6.444\\
$[$S~{\sc iii}$]$  & $\lambda$6312.10                   &    1.400 & S$^{2+}$/H$^+$  & 2.265$\times$10$^{-6}$  & 6.355\\
$[$S~{\sc iii}$]$  & $\lambda$$\lambda$9068.60,9530.60 &   64.000 & S$^{2+}$/H$^+$  & 1.965$\times$10$^{-6}$  & 6.293\\
$[$Cl~{\sc ii}$]$  & $\lambda$6161.84                   &    0.006 & Cl$^+$/H$^+$    & 3.434$\times$10$^{-8}$  & 4.536\\
$[$Cl~{\sc ii}$]$  & $\lambda$8578.69,9123.60           &    0.075 & Cl$^+$/H$^+$    & 4.994$\times$10$^{-9}$  & 3.698\\
$[$Cl~{\sc iii}$]$ & $\lambda$3353.17                   &    0.076 & Cl$^{2+}$/H$^+$ & 1.388$\times$10$^{-7}$  & 5.142\\
$[$Cl~{\sc iii}$]$ & $\lambda$$\lambda$5517.72,5537.89 &    1.000 & Cl$^{2+}$/H$^+$ & 5.513$\times$10$^{-8}$  & 4.741\\
$[$Cl~{\sc iii}$]$ & $\lambda$8480.85                   &    0.020 & Cl$^{2+}$/H$^+$ & 8.319$\times$10$^{-8}$  & 4.920\\
$[$Cl~{\sc iv}$]$  & $\lambda$5323.28                   &    0.012 & Cl$^{3+}$/H$^+$ & 2.525$\times$10$^{-8}$  & 4.402\\
$[$Cl~{\sc iv}$]$  & $\lambda$$\lambda$7530.80,8045.63 &    0.990 & Cl$^{3+}$/H$^+$ & 5.475$\times$10$^{-8}$  & 4.738\\
$[$Ar~{\sc iii}$]$ & $\lambda$3109.17                   &    0.175 & Ar$^{2+}$/H$^+$ & 5.956$\times$10$^{-7}$  & 5.775\\
$[$Ar~{\sc iii}$]$ & $\lambda$5191.82                   &    0.100 & Ar$^{2+}$/H$^+$ & 8.584$\times$10$^{-7}$  & 5.934\\
$[$Ar~{\sc iii}$]$ & $\lambda$$\lambda$7135.80,7751.10 &   18.300 & Ar$^{2+}$/H$^+$ & 1.027$\times$10$^{-6}$  & 6.012\\
$[$Ar~{\sc iv}$]$  & $\lambda$$\lambda$4711.37,4740.17 &    7.600 & Ar$^{3+}$/H$^+$ & 5.592$\times$10$^{-7}$  & 5.748\\
$[$Ar~{\sc iv}$]$  & $\lambda$$\lambda$7237.40,7262.76 &    0.372 & Ar$^{3+}$/H$^+$ & 2.713$\times$10$^{-6}$  & 6.433\\
$[$Ar~{\sc v}$]$   & $\lambda$$\lambda$6435.10,7005.67 &    0.065 & Ar$^{4+}$/H$^+$ & 6.654$\times$10$^{-9}$  & 3.823\\
$[$K~{\sc iv}$]$   & $\lambda$$\lambda$6101.83,6795.10 &    0.196 & K$^{3+}$/H$^+$  & 1.146$\times$10$^{-8}$  & 4.059\\
\hline
\end{tabular}
\end{minipage}
\end{table*}

\begin{table*}
\begin{minipage}{90mm}
\caption{References for the CEL atomic data.}
\label{references:cel}
\centering
\begin{tabular}{lll}
\hline
ion & \multicolumn{2}{c}{CELs}\\
 & Transition probabilities & Collision strengths\\
\hline
C~{\sc ii}   &  Nussbaumer \& Storey \cite{ns1981a} & Blum \& Pradhan \cite{bp1992}\\
C~{\sc iii}  &  Keenan et al. \cite{keenan1992} & Keenan et al. \cite{keenan1992}\\
             &  Fleming et al. \cite{fleming1996} & \\
C~{\sc iv}   &  Wiese et al. \cite{wiese1966} & Gau \& Henry \cite{gh1977}\\
N~{\sc i}    &  Zeippen \cite{zeippen1982} & Berrington \& Burke \cite{bb1981}\\
N~{\sc ii}   &  Nussbaumer \& Rusca \cite{nr1979} & Stafford et al. \cite{stafford1994}\\
O~{\sc i}    &  Baluja \& Zeippen \cite{bz1988} & Berrington \cite{berrington1988}\\
             & & Berrington \& Burke \cite{bb1981}\\
O~{\sc ii}   &  Zeippen \cite{zeippen1982} & Pradhan \cite{pradhan1976}\\
O~{\sc iii}  &  Nussbaumer \& Storey \cite{ns1981b} & Aggarwal \cite{aggarwal1983}\\
F~{\sc ii}   &  Baluja \& Zeippen \cite{bz1988} & Butler \& Zeippen \cite{bz1994}\\
F~{\sc iv}   &  Fischer \& Saha \cite{fs1985} & Lennon \& Burke \cite{lb1994}\\
Ne~{\sc ii}  &  Mendoza \cite{mendoza1983} & Bayes et al. \cite{bayes1985}\\
Ne~{\sc iii} &  Mendoza \cite{mendoza1983} & Butler \& Zeippen \cite{bz1994}\\
Ne~{\sc iv}  &  Zeippen \cite{zeippen1982} & Giles \cite{giles1981}\\
Ne~{\sc v}   &  Fischer \& Saha \cite{fs1985} & Lennon \& Burke \cite{lb1994}\\
S~{\sc ii}   &  Mendoza \& Zeippen \cite{mz1982b} & Keenan et al. \cite{keenan1996}\\
             &  Keenan et al. \cite{keenan1993} & \\
S~{\sc iii}  &  Mendoza \& Zeippen \cite{mz1982a} & Mendoza \cite{mendoza1983}\\
S~{\sc iv}   &  Storey (unpublished) & Saraph \& Storey \cite{ss1999}\\
Cl~{\sc ii}  &  Mendoza \cite{mendoza1983} & Mendoza \cite{mendoza1983}\\
Cl~{\sc iii} &  Mendoza \cite{mendoza1983} & Mendoza \cite{mendoza1983}\\
Cl~{\sc iv}  &  Mendoza \& Zeippen \cite{mz1982b} & Butler \& Zeippen \cite{bz1989}\\
Ar~{\sc ii}  &  Pelan \& Berrington \cite{pb1995} & Vujnovic \& Wiese \cite{vw1992}\\
Ar~{\sc iii} &  Mendoza \& Zeippen \cite{mz1983} & Johnson \& Kingston \cite{jk1990}\\
Ar~{\sc iv}  &  Mendoza \& Zeippen \cite{mz1982b} & Zeippen et al. \cite{zeippen1987}\\
Ar~{\sc v}   &  Mendoza \& Zeippen \cite{mz1982a} & Mendoza \cite{mendoza1983}\\
Fe~{\sc iii} &  Nahar \& Pradhan \cite{np1996} & Zhang \cite{zhang1996}\\
Fe~{\sc iv}  &  Garstang \cite{garstang1958} & Zhang \& Pradhan \cite{zp1997}\\
             &  Fischer \& Rubin \cite{fr2004} & \\
Fe~{\sc v}   &   & Berrington \cite{berrington1995}\\
Fe~{\sc vi}  &  Nussbaumer \& Storey \cite{ns1978} & Nussbaumer \& Storey \cite{ns1978}\\
Fe~{\sc vii} &  Nussbaumer \& Storey \cite{ns1982} & Keenan \& Norrington \cite{kn1987}\\
             &   & Berrington et al. \cite{berrington2000}\\
\hline
\end{tabular}
\end{minipage}
\end{table*}

\subsubsection{\label{cel_abundances:part2}
Ionic abundances from the IR and UV CELs}

NGC\,7009 has been observed in wavelength range other than optical: the {\it
IUE}\,Short Wavelength Prime (SWP) and Long Wavelength Redundant (LWR)
observations by Perinotto \& Benvenuti \cite{pb1981}, the {\it IRAS} Low
Resolution Spectrometer (LRS) observations by Pottasch et al.
\cite{pottasch1986}, the {\it ISO} Short Wavelength Spectrometer (SWS) and
Long Wavelength Spectrometer (LWS) observations by Liu et al. \cite{liu2001a},
and the Kuiper Airborne Observatory ({\it KAO}) observations by Rubin et al.
\cite{rubin1997}. Ionic abundances derived from ten near- to far-infrared
lines and seven ultraviolet lines are presented in
Table\,\ref{abundances_ir_uv}.

The dereddened and normalized intensities of three infrared lines, the
[Ne~{\sc ii}] 12.8$\mu$m and the [Ne~{\sc iii}] 15.5 and 36.0$\mu$m, are
adopted from LLB01. The Ne$^+$/H$^+$ abundance ratio is derived from the
[Ne~{\sc ii}] 12.8$\mu$m line, assuming a temperature of 10\,020~K and a
density of 4300~cm$^{-3}$. The derived Ne$^+$/H$^+$ ratio is
1.32$\times$10$^{-5}$, which agrees with 1.38$\times$10$^{-5}$ given by
LLB01, as is expected.
The critical densities of the [Ne~{\sc iii}] $^3$P$_{1}$ and $^3$P$_{0}$
levels are 2.1$\times$10$^5$ and 3.1$\times$10$^4$~cm$^{-3}$ (at
$T_\mathrm{e}$ = 10\,000~K; Osterbrock \& Ferland \citealt{of2006}),
respectively, much larger than the average electron
density. The Ne$^{2+}$/H$^+$ abundance ratios deduced from the [Ne~{\sc iii}]
15.5 and 36$\mu$m lines are 1.67$\times$10$^{-4}$ and 1.48$\times$10$^{-4}$,
respectively. The two ratio values agree with each other within errors. We
adopt a value of 1.65$\times$10$^{-4}$, which is derived from the sum of
the intensities of the two [Ne~{\sc iii}] infrared (IR) fine-structure lines,
as the Ne$^{2+}$/H$^+$ ratio in NGC\,7009. It agrees with the value of
1.63$\times$10$^{-4}$ given by LLB01. An electron temperature of $9980$~K and
a density of 3930~cm$^{-3}$ were assumed in LLB01.

The observed line fluxes, in units of erg\,cm$^{-2}$\,s$^{-1}$, of the
[N~{\sc iii}] 57$\mu$m and the [O~{\sc iii}] 52 and 88$\mu$m fine-structure
lines are adopted from Liu et al. \cite{liu2001a}. These fluxes were
normalized using the observed total H$\beta$ flux,
10$^{-9.63}$~erg\,cm$^{-2}$\,s$^{-1}$. The extinction of the three IR lines,
as pointed out by Liu et al. \cite{liu2001a}, should be negligible. Since the
critical density of the [N~{\sc iii}] $^2$P$^{\rm o}_{3/2}$ level is
1.5$\times$10$^3$~cm$^{-3}$, comparable to the density of NGC\,7009, we have
assumed a density of $1260$~cm$^{-3}$, deduced from the [O~{\sc iii}]
52$\mu$m/88$\mu$m line ratio, in deriving the N$^{2+}$/H$^+$ abundance ratio
from the [N~{\sc iii}] 57$\mu$m line. Here an electron temperature of
10\,020~K is again assumed. The derived N$^{2+}$/H$^+$ ratio is
4.97$\times$10$^{-5}$, in close agreement with 4.91$\times$10$^{-5}$ given
by Liu et al. \cite{liu2001a}.

The O$^{2+}$/H$^+$ abundance ratios derived from the [O~{\sc iii}] 52 and
88$\mu$m lines are 2.79$\times$10$^{-4}$ and 2.76$\times$10$^{-4}$,
respectively. Here an electron temperature of 9800~K derived from the [O~{\sc
iii}] $\lambda$4959/$\lambda$4363 ratio, and a density of 1260~cm$^{-3}$
derived from the [O~{\sc iii}] 52$\mu$m/88$\mu$m ratio, were assumed. Given
that the critical densities of the [O~{\sc iii}] $^3$P$_{1}$ and $^3$P$_{2}$
fine-structure levels are 5.1$\times$10$^2$ and 3.6$\times$10$^3$~cm$^{-3}$,
respectively (Osterbrock \& Ferland \citealt{of2006}), the density value of
1260~cm$^{-3}$ for the O~{\sc iii} IR-line abundances is appropriate.
We adopt a value of 2.99$\times$10$^{-4}$, which is derived from the sum of
the two [O~{\sc iii}] IR lines, as the O$^{2+}$/H$^+$ abundance ratio in
NGC\,7009. This abundance ratio agrees well with the value of
2.96$\times$10$^{-4}$ given by Liu et al. \cite{liu2001a}.

The flux of the [O~{\sc iv}] 25.9$\mu$m line were estimated from the
$F$([O~{\sc iv}]~25.9$\mu$m)/$F$([S~{\sc iii}]~18.7$\mu$m) flux ratio given
by Rubin et al. \cite{rubin1997}, who obtained far-IR observations of the PNe
NGC\,7009, NGC\,7027 and NGC\,6210 with the Kuiper Airborne Observatory
({\it KAO}). The {\it ISO} SWS observations by X.-W. Liu (unpublished)
during the {\it ISO} Orbit\,$\#344$ in 1996 gives the $F$([O~{\sc
iv}]~25.9$\mu$m)/$F$([S~{\sc iii}]~18.7$\mu$m) ratio that differs from that
of Rubin et al. \cite{rubin1997} by more than 30 per cent. The O$^{4+}$/H$^+$
abundance ratio derived from the above two observations are given in
Table\,\ref{abundances_ir_uv}. Here the flux of the [S~{\sc iii}] 18.7$\mu$m
line was adopted from Pottasch et al. \cite{pottasch1986}.

The {\it IRAS} fluxes (in units of erg\,cm$^{-2}$\,s$^{-1}$) of the [Ne~{\sc
v}] 14.3$\mu$m, the [S~{\sc iii}] 18.7$\mu$m and the [S~{\sc iv}] 10.52$\mu$m
lines, as well as the total flux of H$\beta$, are adopted from Pottasch et
al. \cite{pottasch1986}. The Ne$^{4+}$/H$^+$ abundance ratio derived from the
[Ne~{\sc v}] 14.3$\mu$m line is 6.11$\times$10$^{-7}$. Here an electron
temperature of 10\,020~K is assumed. The critical densities for the [Ne~{\sc
v}] $^3$P$_{1}$ and $^3$P$_{2}$ fine-structure levels are 6.2$\times$10$^3$
and 3.5$\times$10$^4$~cm$^-3$ (Osterbrock \& Ferland \citealt{of2006}),
respectively. Thus the electron density of 1260~cm$^{-3}$ from
the [O~{\sc iii}] 52$\mu$m/88$\mu$m ratio is again assumed. If we adopt a
density value of 4300~cm$^{-3}$, the derived Ne$^{4+}$/H$^+$ abundance ratio
slightly increases to 6.87$\times$10$^{-7}$.

The S$^{2+}$/H$^+$ abundance ratio derived from the [S~{\sc iii}] 18.7$\mu$m
line is 8.36$\times$10$^{-7}$. Here an electron temperature of 10\,020~K and
a density of 1260~cm$^{-3}$ are assumed. Since the critical densities of the
[S~{\sc iii}] $^3$P$_{1}$ and $^3$P$_{2}$ fine-structure levels are
1.98$\times$10$^3$ and 1.54$\times$10$^4$~cm$^{-3}$, respectively, a density
of 1260~cm$^{-3}$ is reasonable. If we increase the density value to
4300~cm$^{-3}$, the S$^{2+}$/H$^+$ ratio derived then increases to
9.93$\times$10${-7}$. The S$^{3+}$/H$^+$ abundance ratio derived from the
[S~{\sc iv}] 10.52$\mu$m line is 7.34$\times$10$^{-6}$. The same temperature
is assumed.

The observed fluxes (in unit of erg\,cm$^{-2}$\,s$^{-1}$) for the seven
ultraviolet (UV) lines in Table\,\ref{abundances_ir_uv} are adopted from
Perinotto \& Benvenuti
\cite{pb1981}. The fluxes are normalized using their H$\beta$ flux, which
should be multiplied by a factor 0.48, the fraction of the H$\beta$ flux
entering into the {\it IUE} slot in position of SWP and LWR images. Using the
logarithmic reddening constant $c({\rm H}\beta)$ = 0.174 derived in Paper~I
and the extinction curve of Howarth \cite{howarth1983}, we derived the
dereddened intensities of those UV lines. The ionic abundances are presented
in Table\,\ref{abundances_ir_uv}. Here an electron temperature of 10\,020~K
and a density of 4300~cm$^{-3}$ are assumed in the abundance calculations.

\begin{table*}
\begin{minipage}{112mm}
\caption{Ionic abundances derived from the UV and far-IR fine-structure
CELs. Intensities are normalized such that $I$(H$\beta$) = 100.}
\label{abundances_ir_uv}
\centering
\begin{tabular}{lrrllcl}
\hline
Ions & Lines & $I_{\rm obs}$ & & X$^{i+}$/H$^+$ & $\log$(X$^{i+}$/H$^+$)+12 & Ref.\\
\hline
\multicolumn{2}{c}{IR Lines} & & & & & \\
\hline
$[$N~{\sc iii}$]$  & 57~$\mu$m   &   24.315 & N$^{2+}$/H$^{+}$  & 4.974$\times$10$^{-5}$  & 8.065 & (1)\\
$[$O~{\sc iii}$]$  & 52~$\mu$m   &  165.940 & O$^{2+}$/H$^{+}$  & 2.792$\times$10$^{-4}$  & 8.446 & (1)\\
$[$O~{\sc iii}$]$  & 88~$\mu$m   &   55.029 & O$^{2+}$/H$^{+}$  & 2.762$\times$10$^{-4}$  & 8.441 & (1)\\
$[$O~{\sc iv}$]$   & 25.9~$\mu$m & 20.29$\pm$1.10 & O$^{3+}$/H$^{+}$  & 1.275$\times$10$^{-5}$  & 7.105 & (2)$^a$\\
$[$O~{\sc iv}$]$   & 25.9~$\mu$m & 14.16$\pm$0.76 & O$^{3+}$/H$^{+}$  & 8.893$\times$10$^{-6}$  & 6.950 & (2)$^b$\\
$[$Ne~{\sc ii}$]$  & 12.8~$\mu$m & 9.9$\pm$3.0    & Ne$^+$/H$^{+}$ & 1.322$\times$10$^{-5}$  & 7.121 & (3)\\
$[$Ne~{\sc iii}$]$ & 15.5~$\mu$m & 250$\pm$12     & Ne$^+$/H$^{+}$ & 1.667$\times$10$^{-4}$  & 8.222 & (3)\\
$[$Ne~{\sc iii}$]$ & 36.0~$\mu$m & 18.0$\pm$1.8   & Ne$^{2+}$/H$^{+}$ & 1.477$\times$10$^{-4}$  & 8.169 & (3)\\
$[$Ne~{\sc v}$]$   & 14.3~$\mu$m  & 8.696         & Ne$^{4+}$/H$^{+}$ & 6.107$\times$10$^{-7}$  & 5.786 & (4)\\
$[$S~{\sc iii}$]$  & 18.7~$\mu$m  & 7.826         & S$^{2+}$/H$^{+}$  & 8.356$\times$10$^{-7}$  & 5.922 & (4)\\
$[$S~{\sc iv}$]$   & 10.52~$\mu$m & 230.434       & S$^{3+}$/H$^{+}$  & 7.342$\times$10$^{-6}$  & 6.866 & (4)\\
\hline
\multicolumn{2}{c}{UV Lines (\AA)} & & & & & \\
\hline
C~{\sc ii}$]$  & $\lambda$2326 &  3.821 & C$^{+}$/H$^{+}$  & 4.555$\times$10$^{-6}$ & 5.629 & (5)\\
C~{\sc iii}$]$ & $\lambda$1908 & 44.832 & C$^{2+}$/H$^{+}$ & 1.308$\times$10$^{-4}$ & 8.117 & (5)\\
N~{\sc iii}$]$ & $\lambda$1751 &  6.959 & N$^{2+}$/H$^{+}$ & 5.535$\times$10$^{-5}$ & 7.743 & (5)\\
N~{\sc iv}$]$  & $\lambda$1486 &  6.197 & N$^{3+}$/H$^{+}$ & 6.151$\times$10$^{-5}$ & 7.789 & (5)\\
O~{\sc iii}$]$ & $\lambda$1663 &  4.313 & O$^{2+}$/H$^{+}$ & 2.387$\times$10$^{-4}$ & 8.378 & (5)\\
O~{\sc iv}$]$  & $\lambda$1403 &  0.705 & O$^{3+}$/H$^{+}$ & 3.977$\times$10$^{-5}$ & 7.600 & (5)\\
$[$Ne~{\sc iv}$]$ & $\lambda$2424 & 7.789 & Ne$^{3+}$/H$^{+}$ & 1.895$\times$10$^{-5}$ & 7.278 & (5)\\
\hline
\end{tabular}
\begin{description}
\item [(1)] The observed flux is from Liu et al. \cite{liu2001a}.
\item [(2)$^a$] The observed flux is estimated from the flux ratio
$F$([O~{\sc iv}]~25.9$\mu$m)/$F$([S~{\sc iii}]~18.7$\mu$m) adopted from
the ISO/LWS observations during ISO $\#$344 Orbit in 1996 (Liu et al.,
unpublished), using the [S~{\sc iii}] 18.7$\mu$m flux adopted from
Pottasch et al. \cite{pottasch1986}.
\item [(2)$^b$] The observed flux is estimated from the flux ratio
$F$([O~{\sc iv}]~25.9$\mu$m)/$F$([S~{\sc iii}]~18.7$\mu$m) adopted from
the KAO observations of Rubin et al. \cite{rubin1997}, using the
[S~{\sc iii}] 18.7$\mu$m flux adopted from Pottasch et al.
\cite{pottasch1986}.
\item [(3)] The observed flux is from LLB01.
\item [(4)] The observed flux is from Pottasch et al. \cite{pottasch1986}.
\item [(5)] The observed flux is from Perinotto \& Benvenuti \cite{pb1981}.
\end{description}
\end{minipage}
\end{table*}

\subsection{\label{compare}
Comparison of the ORL and CEL abundances}

\subsubsection{\label{compare:part1}
Ionic abundances}

\begin{figure}
\begin{center}
\includegraphics[width=8.0cm,angle=-90]{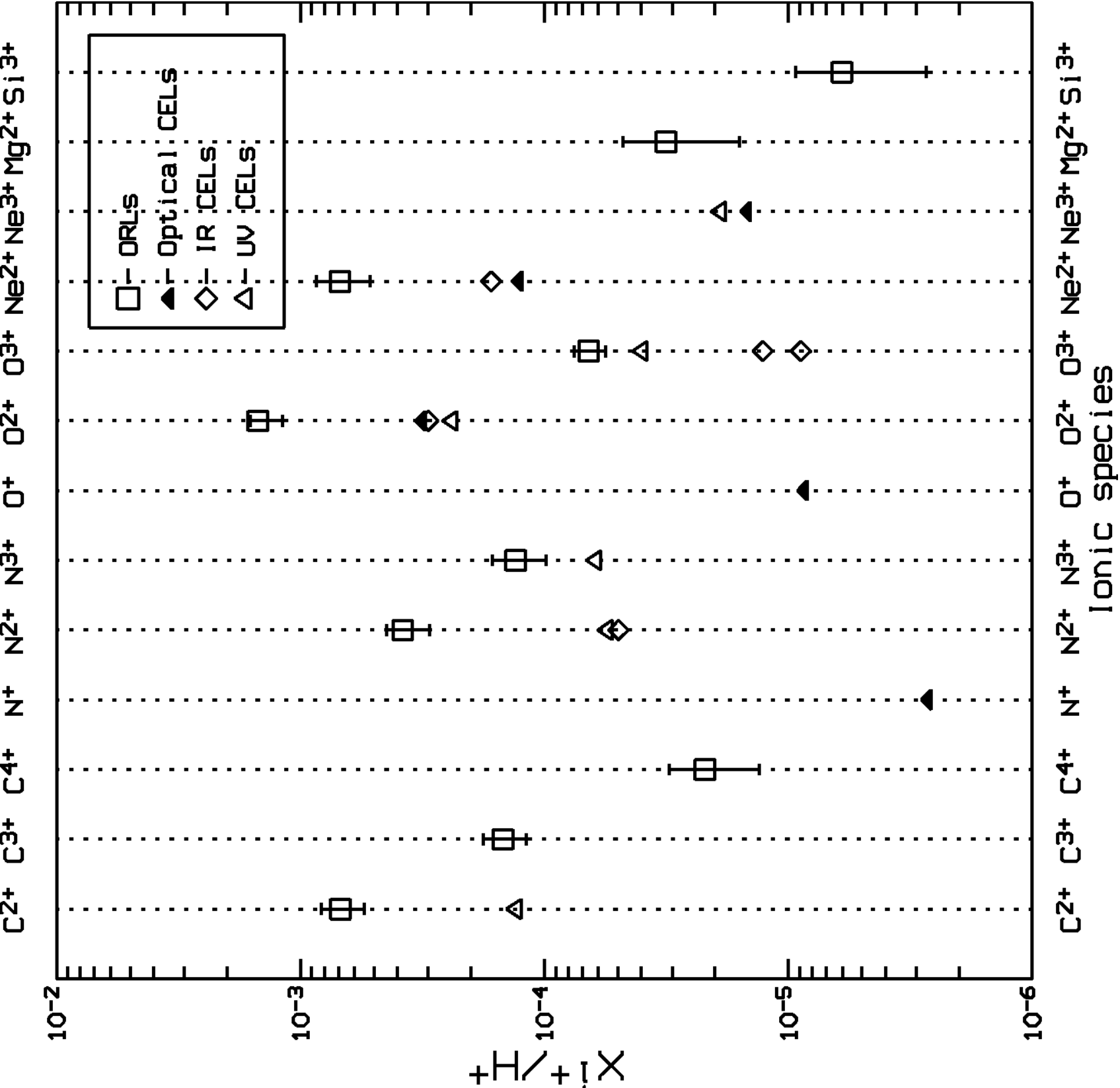}
\caption{Comparison of the ionic abundances derived from ORLs and the optical,
UV and IR CELs. Error bars on the ORL abundances are from the propagation
calculations of the ORL measurement errors.}
\label{compare:orlcel}
\end{center}
\end{figure}

In Fig.\,\ref{compare:orlcel}, the ionic abundances of C, N, O and Ne derived
from ORLs are compared with the corresponding values derived from the optical,
UV and far-IR CELs. Here the ionic abundances of C, N, O and Ne derived from
ORLs are from Table\,\ref{orlabundances:adopted}, and the ionic abundances
derived from CELs are from Tables\,\ref{abundances:cels} (optical) and
\ref{abundances_ir_uv} (UV and IR). The IR fine-structure line fluxes are
adopted from the recent {\it ISO}\,observations (Liu et al.
\citealt{liu2001a}). We also make use of the IR line fluxes from observations
of {\it IRAS} in the literature. The UV line fluxes from the
{\it IUE}\,observations are dereddened using the extinction derived in
Paper~I, and they are used to derive ionic abundances for highly ionized heavy
elemental ions. The recombination line C$^{2+}$/H$^{+}$, N$^{2+}$/H$^{+}$,
O$^{2+}$/H$^{+}$ and Ne$^{2+}$/H$^{+}$ abundances are all higher than the
abundance ratios derived from CELs by nearly a factor of 5 (i.e. ADF$\sim$5),
in agreement with what was observed by LSBC (for C, N and O) and LLB01 (for
Ne). However, the ADF differs from 5 when the abundances derived from UV CELs
are used: The N$^{3+}$/H$^{+}$ abundance derived from ORL is higher
than the value derived from the N~{\sc iv}$]$ $\lambda$1486 UV line by a
factor of 2. The recombination line O$^{3+}$/H$^{+}$ abundance is higher than
the abundance derived from the O~{\sc iv}$]$ $\lambda$1403 UV line by only 65
per cent. This is probably mainly due to systematic difference in flux
calibrations, given that the UV data are adopted from the early {\it IUE}
observations of Perinotto \& Benvenuti \cite{pb1981}. The ADF value of
O$^{3+}$ is close to 5 when the abundance derived from the [O~{\sc iv}]
25.9$\mu$m IR line is used.

\subsubsection{\label{compare:part2}
Total elemental abundances}

Elemental abundances derived from ORLs and CELs are compared in
Table\,\ref{abundances:total}. Abundance errors (numbers in brackets) are
estimated from measurement uncertainties, which is calculated by
quadratically adding line flux errors from Gaussian profile fitting and the
systematic uncertainties in line measurements e.g. subtraction of the
continuum. Uncertainties introduced by ionization correction method, i.e. the
ionization correction factors ($icf$'s), are not taken into account in error
estimate. Also given in this table are the solar
abundances from Asplund et al. \cite{asplund09} and average abundances
of the Galactic disc and bulge PNe from literature. The C/H, N/H, O/H and
Ne/H elemental abundances derived from ORLs are higher than the corresponding
values derived from CELs by a factor of 5.4, 6.9, 4.7 and 5.3, respectively.
This result is similar to LSBC, who derived ADF values of 6.1, 4.1 and 4.7
for C, N and O respectively in NGC\,7009. The Ne abundance discrepancy
derived by LLB01 is about 4. In the analyses of LSBC, the elemental abundances
of C and N derived from CELs were adopted from the earlier observations of
Barker \cite{barker1983}, who first discussed the large discrepancy between
the C$^{2+}$/H$^{+}$ abundances derived from the C~{\sc ii} $\lambda$4267
recombination line and from the UV CEL C~{\sc iii}$]$
$\lambda\lambda$1907,\,1909.

Whenever available, the ionization correction factors ($icf$'s) given by
Kingsburg \& Barlow \cite{kb1994} were used. In
Section\,\ref{cel_abundances:part2}, we have derived 16 ionic
abundances from UV and IR data available from the literature
(Table\,\ref{abundances_ir_uv}). These ionic abundances can be used
as an aid to derive total elemental abundances. When they are not necessary,
we can use them to check the elemental abundance values derived from $icf$'s,
by adding these UV and IR abundances with the optical abundance ratios in
Table\,\ref{abundances:cels} and compare with the ionization corrected total
abundances.

The forbidden line O/H abundance ratio was calculated from the O$^+$/H$^+$
ratio derived from the [O~{\sc ii}] $\lambda\lambda$3726 and 3729
lines\footnote{Recombination excitation of the [O~{\sc ii}]
$\lambda$$\lambda$3726, 3729 doublet is neglected. Given the small ionic
concentration of O$^+$, compared to O$^{2+}$, the errors introduced to the
total O/H elemental abundances deduced below, both from CELs and from ORLs,
should be negligible.} and the O$^{2+}$/H$^+$ ratio derived from the
[O~{\sc iii}] $\lambda\lambda$4959 and 5007 lines, correcting for the unseen
O$^{3+}$ in optical using

\begin{equation}
  \label{icf1}
\frac{\rm O}{\rm H} =
icf(\rm O)\times(\frac{{\rm O}^{+}}{{\rm H}^{+}} + \frac{{\rm O}^{2+}}{{\rm H}^{+}})\\
= [\frac{{\rm He}^{+} + {\rm He}^{2+}}{{\rm He}^{+}}]^{2/3}
\times(\frac{{\rm O}^{+}}{{\rm H}^{+}} + \frac{{\rm O}^{2+}}{{\rm H}^{+}}).
\end{equation}
From the He$^+$ and He$^{2+}$ abundances given in Table\,\ref{abundances:i},
we have $icf(\rm O)$ = 1.086. Thus derived O/H ratio is 3.603$\times$10$^{-4}$,
close to the value 3.716$\times$10$^{-4}$, which is derived by combining the
O ions from UV, IR and optical lines, O/H = O$^+$/H$^+$ + O$^{2+}$/H$^+$ +
O$^{3+}$/H$^+$. Here the O$^{3+}$/H$^+$ is derived from the O~{\sc iv}$]$
$\lambda$1403 UV line (Section\,\ref{cel_abundances:part2}).

The recombination line abundance O$^+$/H$^+$ is not available, thus in order
to make use of the above equation, we assume that the recombination line
O$^+$/O$^{2+}$ ratio is the same as that derived from the CELs. Given the
small ionic concentration of O$^+$ (less than 10 per cent in NGC\,7009), the
errors introduced should be negligible. The O$^{2+}$/H$^+$, O$^{3+}$/H$^+$ and
O$^{4+}$/H$^+$ abundance ratios derived from ORLs are available from
Table\,\ref{orlabundances:adopted}. The recombination line O/H abundance ratio
thus derived is 1.667$\times$10$^{-3}$.

Both C$^{2+}$/H$^+$ and C$^{3+}$/H$^+$ abundance ratios have been derived
from ORLs and are presented in Table\,\ref{orlabundances:adopted}. The
C$^{4+}$/H$^+$ abundance ratio has also been derived from the C~{\sc iv} M8
$\lambda$4658.30 line, but probably is not acceptable because it blends with
[Fe~{\sc iii}] $\lambda$4658.05. The unseen C$^+$/H$^+$ are corrected for
using the equation of LSBC,

\begin{equation}
  \label{icf2}
\frac{\rm C}{\rm H} =
\frac{{\rm C}^{2} + {\rm C}^{3}}{{\rm H}^{+}} \times icf(\rm C)
\end{equation}
and

\begin{equation}
  \label{icf3}
icf(\rm C) = [\frac{{\rm He}^{+} + {\rm He}^{2+}}{{\rm He}^+}]^{1/3}
   \times [\frac{{\rm O}^{+} + {\rm O}^{2+}}{{\rm O}^{2+}}].
\end{equation}

The ionic abundances of O$^+$ and O$^{2+}$ in the above equation are assumed
to be those derived from the forbidden line measurements. In NGC\,7009, most
of the carbon should exist in the form of C$^{2+}$ and C$^{3+}$, and the
correction required for unobserved C$^+$ and C$^{4+}$ is quite small. The
Equation~\ref{icf3} gives $icf(\rm C)$ = 1.07. Thus derived recombination line
C/H abundance ratio is 8.932$\times$10$^{-4}$, which is only 3 per cent lower
than the value given by LSBC.

For the collisionally excited lines, C$^+$/H$^+$ and C$^{2+}$/H$^+$ are
derived from UV lines (Table\,\ref{abundances_ir_uv}). We assume
C$^{3+}$/C$^{2+}$ = 0.216, as given by the ORL abundance ratios. Thus the
CEL ratio of carbon is C/H = C$^+$/H$^+$ + C$^{2+}$/H$^+$ + C$^{3+}$/H$^+$,
which is 1.636$\times$10$^{-4}$.

Recombination line abundances are available for the N$^{2+}$/H$^+$ and
N$^{3+}$/H$^+$ ratios (Table\,\ref{orlabundances:adopted}) but not for
N$^+$/H$^+$. The latter is available from the collisionally excited
[N~{\sc ii}] $\lambda\lambda$6548 and 6584 lines. The N$^{2+}$/H$^+$ ratio
derived from the UV collisionally excited N~{\sc iii}$]$ $\lambda$1751 line
is about 10 per cent higher than that deduced from the [N~{\sc iii}] 57~$\mu$m
far-IR fine-structure line (Table\,\ref{abundances_ir_uv}). Given the weakness
of the $\lambda$1751 and the relative new observations of the 75\,$\mu$m line,
we adopt the N$^{2+}$/H$^+$ abundance ratio derived from the far-IR line.
N$^{2+}$/H$^+$ from the 57~$\mu$m and N$^+$/H$^+$ from the
$\lambda\lambda$6548 and 6584 lines yield N$^+$/N$^{2+}$ = 0.0548. We assume
that this is also valid for corresponding abundances derived from ORLs. The
N~{\sc iv} optical recombination lines are not clearly detected in the
spectrum of NGC\,7009, but the we detected the O~{\sc iv} M2 3d~$^2$D --
3p~$^2$P$^{\rm o}$ $\lambda$3412 line, and the O$^{4+}$/H$^+$ abundance ratio
is derived from it (Table\,\ref{abundances:viii}). Given the ionization
potential of N$^{3+}$ (77.472~eV) is the same to that of O$^{3+}$ (77.4~eV),
we assume N$^{4+}$/N = O$^{4+}$/O. Thus the total recombination line N/H
abundance is given by

\begin{equation}
  \label{icf4}
\frac{\rm N}{\rm H} =
(1.0548\times\frac{{\rm N}^{2+}}{{\rm H}^{+}} + \frac{{\rm N}^{3+}}{{\rm H}^{+}})
/ (1 - \frac{{\rm O}^{4+}}{\rm O}).
\end{equation}
To obtain the forbidden-line N/H abundance, we correct for the unseen
N$^{3+}$/H$^+$ assuming N$^{3+}$/N$^{2+}$ = 0.380, as given by ORLs. so that

\begin{equation}
  \label{icf5}
\frac{\rm N}{\rm H} =
\frac{{\rm N}^{+}}{{\rm H}^{+}} + 1.380\times\frac{{\rm N}^{2+}}{{\rm H}^{+}}.
\end{equation}
The total CEL N/H abundance thus derived is 7.01$\times$10$^{-5}$. Here the
N$^{4+}$/H$^+$ is neglected, given its very low abundance, if we assume that
the N$^{4+}$/N$^{2+}$ ratio from CELs is the same as that given by the ORL
values. If we take into account the N$^{3+}$/H$^+$ abundance ratio derived
from the UV N~{\sc iv}$]$ $\lambda$1486 line (Table\,\ref{abundances_ir_uv}),
the total CEL abundance ratio is then N/H = N$^+$/H$^+$ + N$^{2+}$/H$^+$ +
N$^{3+}$/H$^+$ = 1.140$\times$10$^{-4}$, which is 64 per cent higher than the
value derived by the ionization correction Equation\,\ref{icf5}. Since the
recent IR data should be more reliable, we adopt the ratio derived from
Equation\,\ref{icf5}.

Ne$^+$/H$^+$, Ne$^{2+}$/H$^+$ and Ne$^{3+}$/H$^+$ ionic abundances are
available from IR (Table\,\ref{abundances_ir_uv}) and optical CELs
(Table\,\ref{abundances:cels}). For the Ne$^{2+}$/H$^+$ ratio, we adopt the
abundance derived from the [Ne~{\sc iii}] $\lambda$3868\footnote{Another
[Ne~{\sc iii}] nebular line $\lambda$3967 is saturated in the higher
resolution spectrum, and is not saturated but blended with the H~{\sc i}
$\lambda$3970 line in the low-resolution spectrum.} optical line, which is
only about 21 per cent lower than that derived from the ISO SWS observation
of the [Ne~{\sc iii}] 15.5+36~$\mu$m IR fine-structure lines. For the
Ne$^{3+}$/H$^+$ ratio, we adopt the value derived from the [Ne~{\sc iv}]
$\lambda\lambda$4724.17 and 4725.67 optical lines. From the IUE observation
by Perinotto \& Benvenuti \cite{pb1981}, we also derived the Ne$^{3+}$/H$^+$
abundance ratio from their observed Ne~{\sc iv}$]$ $\lambda\lambda$2422 and
2424 lines (Table\,\ref{abundances_ir_uv}), which is 26 per cent higher than
that from optical. Thus the CEL Ne/H ratio obtained from the equation

\begin{equation}
  \label{icf6}
\frac{\rm Ne}{\rm H} =
\frac{{\rm Ne}^{+}}{{\rm H}^{+}} + \frac{{\rm Ne}^{2+}}{{\rm H}^{+}} +
\frac{{\rm Ne}^{3+}}{{\rm H}^{+}}
\end{equation}
is 1.578$\times$10$^{-4}$ (Table\,\ref{abundances:total}). This is only 9
per cent lower than the Ne/H ratio derived by LLB01.

Only the Ne$^{2+}$/H$^+$ ratio is available from ORLs. The Ne$^+$/Ne$^{2+}$
ratio from CELs is 0.102, with and Ne$^{3+}$/Ne$^{2+}$ from optical CELs is
0.116. Assuming these ionization ratios from CELs are the same as given by
ORLs, we obtain a recombination line Ne/H abundance ratio value of
8.398$\times$10$^{-4}$ by using

\begin{equation}
  \label{icf7}
\frac{\rm Ne}{\rm H} = (0.102 + 1.0 + 0.116)\times
\frac{{\rm Ne}^{2+}}{{\rm H}^{+}}.
\end{equation}
This ratio value is 21 per cent higher than that given by LLB01.

CELs emitted by [F~{\sc ii}] and [F~{\sc iv}] ions are detected, with the
F$^+$/H$^+$ and F$^{3+}$/H$^+$ abundance ratios derived from the [F~{\sc ii}]
$\lambda$4789 and [F~{\sc iv}] $\lambda$4060 lines respectively
(Table\,\ref{abundances:cels}). O$^+$ has ionization potential comparable to
F$^+$ and Ne$^{3+}$ has that comparable to F$^{3+}$. Zhang \& Liu
\cite{zl2005} suggested that F/O = F$^{2+}$/O$^{2+}$ for low-excitation PNe,
and F/O = (F$^{3+}$/Ne$^{3+}$)(Ne/O) for high-excitation PNe. Here we adopt
the assumption of Zhang \& Liu \cite{zl2005}. Given that NGC\,7009 is a
medium excitation PN, we derive the Ne$^{2+}$/H$^+$ abundance ratio from the
equation

\begin{equation}
  \label{icf8}
\frac{\rm F}{\rm H} =
(\frac{{\rm F}^{+}}{{\rm H}^{+}} + \frac{{\rm F}^{3+}}{{\rm H}^{+}}) /
(1 - \frac{{\rm O}^{2+}}{\rm O}).
\end{equation}

The derived F/H abundance ratio is 2.923$\times$10$^{-7}$. The F$^{4+}$/H$^+$
should be negligible because the ionization potential of F$^{3+}$ is 87~eV,
which is too high. No recombination line from the fluorine ions are detected.

For the elements heavier than Ne, we detected CELs emitted by S, Cl, Ar and
K ions (Table\,\ref{abundances:cels}), and ORLs by Si and Mg ions
(Table\,\ref{abundances:ix}). Considering that the first ionization
potential of silicon atom is only 8.2~eV, and the ionization potential of
Si$^{4+}$ is 166.8~eV, which is a huge jump from that of Si$^{3+}$, 45~eV,
we assume that the S$^+$/H$^+$ abundance is negligible, and the main
ionization stages of Si in NGC\,7009 are Si$^{2+}$, Si$^{3+}$ and Si$^{4+}$.
Several Si~{\sc iii} multiplets and the Si~{\sc iv} multiplet V1
are detected from NGC\,7009. Only effective dielectronic recombination
coefficients for Si~{\sc iii} multiplets are available from Nussbaumer \&
Storey \cite{ns1984}, thus only the S$^{3+}$/H$^+$ abundance ratio from
S~{\sc iii} ORLs is derived. The S$^{3+}$/H$^+$ ratio is derived from the
Si~{\sc iii} multiplets M2 4p~$^3$P$^{\rm o}$ -- 4s~$^3$S and M5 4d~$^3$D --
4p~$^3$P$^{\rm o}$ (Table\,\ref{abundances:ix}). Since the ionization
potential of S$^+$ (16.3~eV) is close to that of F (17.4~eV), and the
ionization potential of Si$^{3+}$ (45~eV) is comparable to that of N$^{2+}$
(47.4~eV), we assume the relation: Si$^{2+}$/Si = F$^+$/F = O$^+$/O, and
Si$^{4+}$/Si = N$^{3+}$/N. Thus the total Si/H abundance ratio from ORLs can
be derived from the equation

\begin{equation}
\label{icf9}
\frac{\rm Si}{\rm H} =
\frac{{\rm Si}^{3+}}{{\rm H}^{+}} /
(1 - \frac{{\rm O}^{+}}{\rm O} - \frac{{\rm N}^{3+}}{\rm N}) =
\frac{{\rm Si}^{3+}}{{\rm H}^{+}} /
(1 - \frac{{\rm F}^{+}}{\rm F} - \frac{{\rm N}^{3+}}{\rm N}).
\end{equation}
The Si/H abundance ratio thus derived is 8.278$\times$10$^{-6}$. In
Equation\,\ref{icf9}, the O$^+$/O ratio is from optical CELs, and the
N$^{3+}$/N ratio is from ORLs.

The Mg$^{2+}$/H$^+$ abundance ratio is derived from the Mg~{\sc ii} M4
4f~$^2$F$^{\rm o}$ -- 3d~$^2$D $\lambda$4481 line. The ionization potentials
of Mg$^0$ and Mg$^{2+}$ are 7.65~eV and 80.1~eV, respectively, thus we assume
that magnesium in NGC\,7009 are mainly doubly ionized, and Mg$^+$ and
Mg$^{3+}$ are negligible. No effective recombination coefficients for Mg~{\sc
ii} lines are available. Given the similarity between the atomic structure of
Mg~{\sc ii} and C~{\sc ii}, we assume that the Mg~{\sc ii} $\lambda$4481 line
has an effective recombination coefficient equal to that of C~{\sc ii} M6
$\lambda$4267. Thus we have Mg/H = Mg$^{2+}$/H$^+$.

For S, we have S$^+$/H$^+$ derived from the [S~{\sc ii}] $\lambda\lambda$6716
and 6731 optical lines, S$^{2+}$/H$^+$ derived from the [S~{\sc iii}]
$\lambda$6312 optical line. The S$^{2+}$/H$^+$ is also available from the
[S~{\sc iii}] 18.7~$\mu$m IR line. S$^{3+}$/H$^+$ is derived from the [S~{\sc
iv}] 10.5$\mu$m far-IR fine-structure line, which was adopted from the IRAS
observations by Pottasch et al. \cite{pottasch1986}. We use the S$^{2+}$/H$^+$
ratio derived from the $\lambda$6312 line. S$^{4+}$ is not observed. Since
S$^{3+}$ has an ionization potential of 47.3~eV, very close to the value of
47.4~eV for N$^{2+}$, we assume that S$^{4+}$/S = N$^{3+}$/N. The N$^{3+}$/N
ratio is from ORLs. Thus the S/H ratio is obtained from

\begin{equation}
\label{icf10}
\frac{\rm S}{\rm H} =
(\frac{{\rm S}^{+}}{{\rm H}^{+}} + \frac{{\rm S}^{2+}}{{\rm H}^{+}} +
\frac{{\rm S}^{3+}}{{\rm H}^{+}}) / (1 - \frac{{\rm N}^{3+}}{\rm N}).
\end{equation}
The derived S/H ratio is 1.299$\times$10$^{-5}$. If we adopt the
S$^{2+}$/H$^+$ ratio derived from the [S~{\sc iii}] 18.7~$\mu$m IR line,
then the S/H ratio is 1.10$\times$10$^{-5}$, about 15 per cent lower than
the case when we adopt the optical value.

The Cl$^+$/H$^+$, Cl$^{2+}$/H$^+$, and Cl$^{3+}$/H$^+$ abundances are derived
from optical CELs (Table\,\ref{abundances:cels}). Given that the ionization
potential of Cl$^{3+}$ (53.5~eV)is similar to that of He$^+$ (54.4~eV), and
that the ionization potential of Cl$^{4+}$ (67.7~eV) is comparable to that of
Ne$^{2+}$ (63.5~eV), we assumed Cl$^{4+}$/Cl = He$^{2+}$/He = 0.013/0.112 =
0.116 and Cl$^{5+}$/Cl = Ne$^{3+}$/Ne = 0.095. Here the Ne$^{3+}$/Ne ratio is
from the CEL abundances. Thus the Cl/H abundance ratio can be obtained from
the equation

\begin{equation}
  \label{icf11}
\frac{\rm Cl}{\rm H} = (\frac{{\rm Cl}^{+}}{{\rm H}^{+}} +
\frac{{\rm Cl}^{2+}}{{\rm H}^{+}} + \frac{{\rm Cl}^{3+}}{{\rm H}^{+}}) /
(1 - \frac{{\rm He}^{2+}}{\rm He} - \frac{{\rm Ne}^{3+}}{\rm Ne}).
\end{equation}
The derived total Cl/H abundance is 1.927$\times$10$^{-7}$.

Ar$^{2+}$/H$^+$ is derived from the [Ar~{\sc iii}] $\lambda\lambda$7136 and
7751 lines, Ar$^{3+}$/H$^+$ is from [Ar~{\sc iv}] $\lambda\lambda$4711 and
4740, and Ar$^{4+}$/H$^+$ is from the [Ar~{\sc v}] $\lambda\lambda$6435 and
7006 lines (Table\,\ref{abundances:cels}). The unseen Ar$^+$ is corrected for
assuming Ar$^+$/Ar = N$^+$/N, where the N$^+$/N is derived from optical CELs.
The total Ar/H ratio can be obtained using

\begin{equation}
  \label{icf12}
\frac{\rm Ar}{\rm H} =
(\frac{{\rm Ar}^{2+}}{{\rm H}^{+}} + \frac{{\rm Ar}^{3+}}{{\rm H}^{+}} +
\frac{{\rm Ar}^{4+}}{{\rm H}^{+}}) / (1 - \frac{{\rm N}^{+}}{\rm N}).
\end{equation}
The finally derived Ar/H abundance ratio is 2.570$\times$10$^{-6}$.

We have derived the K$^{3+}$/H$^+$ abundance ratio from the [K~{\sc iv}]
$\lambda\lambda$6102 and 6795 lines. A very faint feature, which might be the
[K~{\sc v}] $\lambda$4163 line is detected in the spectrum of NGC\,7009, but
we could not confirm that. K$^+$ is probably negligible in NGC\,7009, judging
from the very low ionization potential of K. Since the ionization potential of
K$^+$ (31.6~eV) is comparable to that of N$^+$ (29.6~eV), and ionization
potential of K$^{3+}$ (60.9~eV) is comparable to that of F$^{2+}$ (62.7~eV), we
assume K$^{2+}$/K = N$^{2+}$/N and K$^{4+}$/K = F$^{3+}$/F. Here the N$^{2+}$/N
ratio is from the ORLs, and the F$^{3+}$/F ratio is from CELs. We use the
following equation to correct for the unseen K$^{2+}$ and K$^{4+}$ ions,

\begin{equation}
  \label{icf13}
\frac{\rm K}{\rm H} =
\frac{{\rm K}^{3+}}{{\rm H}^{+}} / (1 - \frac{{\rm N}^{2+}}{\rm N} -
\frac{{\rm F}^{3+}}{\rm F}).
\end{equation}
The derived the total K/H abundance ratio is 4.242$\times$10$^{-8}$.

\begin{table*}
\begin{minipage}{150mm}
\caption{Total elemental abundances derived from ORLs and CELs, in units such
that $\log{N({\rm H})}$ = 12.}
\label{abundances:total}
\centering
\begin{tabular}{lccrrrrrrrr}
\hline
Element & \multicolumn{2}{c}{X/H} & \multicolumn{2}{c}{$\log$[X/H]+12} &
\multicolumn{2}{c}{TLW$^a$} & \multicolumn{2}{c}{TLW$^b$} & KB94$^c$ & Solar$^d$\\
 & ORLs & CELs & ORLs & CELs & ORLs & CELs & ORLs & CELs & & \\
\hline
He     & 0.112                  &                        & 11.049 &       & 11.02 &      & 11.06 &      & 11.06 & 10.93\\
C      & 8.93($\pm$0.45)$\times$10$^{-4}$ & 1.64($\pm$0.33)$\times$10$^{-4}$ &  8.95 & 8.21 &  9.09 & 8.52 &  9.03 & 8.56 &  8.74 &  8.43\\
N      & 4.92($\pm$0.20)$\times$10$^{-4}$ & 7.14($\pm$0.35)$\times$10$^{-5}$ &  8.69 & 7.85 &  8.94 & 8.17 &  9.14 & 8.34 &  8.38 &  7.83\\
O      & 1.67($\pm$0.06)$\times$10$^{-3}$ & 3.60($\pm$0.07)$\times$10$^{-4}$ &  9.22 & 8.56 &  9.22 & 8.60 &  9.32 & 8.70 &  8.66 &  8.69\\
F      &                                  & 2.92($\pm$0.58)$\times$10$^{-7}$ &       & 5.47 &       &      &       &      &       &  4.56\\
Ne     & 8.40($\pm$0.42)$\times$10$^{-4}$ & 1.58($\pm$0.08)$\times$10$^{-4}$ &  8.92 & 8.20 &  9.06 & 7.99 &  9.07 & 8.13 &  8.06 &  7.93\\
Mg$^e$ & 3.18($\pm$0.16)$\times$10$^{-5}$ &                                  &  7.50 &      &  7.56 &      &  7.71 &      &       &  7.60\\
Si$^f$ & 8.28($\pm$0.80)$\times$10$^{-6}$ &                                  &  6.92 &      &       &      &       &      &       &  7.51\\
S      &                                  & 1.30($\pm$0.11)$\times$10$^{-5}$ &       & 7.11 &       & 6.84 &       & 7.05 &  6.99 &  7.12\\
Cl     &                                  & 1.93($\pm$0.21)$\times$10$^{-7}$ &       & 5.28 &       & 5.35 &       & 5.29 &       &  5.50\\
Ar     &                                  & 2.57($\pm$0.38)$\times$10$^{-6}$ &       & 6.41 &       & 6.20 &       & 6.34 &  6.51 &  6.40\\
K$^g$  &                                  & 4.24($\pm$0.84)$\times$10$^{-8}$ &       & 4.63 &       &      &       &      &       &  5.03\\
\hline
\end{tabular}
\begin{description}
\item [$^a$] Average abundances for 23 Galactic bulge PNe of Wang \& Liu
\cite{wl2007} plus Galactic bulge PNe M\,1-42 and M\,2-36 analyzed by Liu
et al. \cite{liu2001b}.
\item [$^b$] Average abundances given by Wang \& Liu \cite{wl2007} for 58
Galactic disc PNe which were selected from Tsamis et al.
(\citealt{tsamis2003}, \citealt{tsamis2004}), Liu et al.
(\citealt{liu2004a},\,b) and Wesson, Liu \& Barlow \cite{wlb2005}.
\item [$^c$] Average abundances of Galactic disc and bulge PNe (Kingsburg \&
Barlow \citealt{kb1994}; Exter, Barlow \& Walton \citealt{exter04}), all
based on CEL analyses except for helium for which ORLs were used.
\item [$^d$] Solar values from Asplund et al. \cite{asplund09}.
\item [$^e$] Mg$^+$/H$^+$ and Mg$^{3+}$/H$^+$ are neglected in calculating
the total ORL abundance.
\item [$^f$] Si$^+$/H$^+$ is neglected in calculating the total ORL abundance.
\item [$^g$] K$^+$/H$^+$ and K$^{4+}$/H$^+$ are neglected in calculating the
total CEL abundance.
\end{description}
\end{minipage}
\end{table*}

\section{\label{discussion}
Discussion}

Average elemental abundances for the Galactic disc and bulge PNe taken from
Kingsburg \& Barlow \cite{kb1994} and Exter, Barlow \& Walton \cite{exter04}
are presented in Table\,\ref{abundances:total} for purpose of comparison.
Also presented in this table are the average abundances for 23 Galactic bulge
PNe of Wang \& Liu \cite{wl2007} plus two bulge PNe M\,1-42 and M\,2-36
studied by Liu et al. \cite{liu2001b}, and 58 Galactic disc PNe selected from
Tsamis et al. (\citealt{tsamis2003}, \citealt{tsamis2004}), Liu et al.
(\citealt{liu2004a},\,b) and Wesson et al. \cite{wlb2005}. The helium
abundance derived from the current analyses agrees well with the average
value of the 58 disc sample, but 0.12 dex higher than the most recent solar
value (Asplund et al. \citealt{asplund09}). The recombination line C/H
abundance 8.95 derived for NGC\,7009 also agrees with the average value
(9.03) of the disc sample, but 0.14 dex lower than the 23 bulge sample of
Wang \& Liu \cite{wl2007}. The forbidden-line C/H abundance for NGC\,7009
is lower than all the values quoted from literature. The elemental N/H
abundance derived from CELs for NGC\,7009 agrees well with the the Sun,
but lower than all other average values. Our recombination line N/H abundance
is lower than the average values of the bulge and dsic samples compiled by
Wang \& Liu \cite{wl2007}, but is higher than the average value of the sample
from Kingsburg \& Barlow \cite{kb1994} plus Exter, Barlow \& Walton
\cite{exter04}. The forbidden-line O/H abundance for NGC\,7009 is 0.13 dex
low than the solar value, and lower than the average value of the disc sample
of Wang \& Liu \cite{wl2007} by the same amount. Our C/O abundance ratio
derived from CELs is 0.96, slightly lower than all other ratios from
literature, indicating that NGC\,7009 might be enriched in oxygen, which is
consistent with the fact that the spectrum of NGC\,7009 is obviously rich in
oxygen emission lines. The forbidden-line neon abundance of NGC\,7009 is
0.27 dex ($\sim$) higher than the solar value, and agrees with the average
value (8.13) of the disc sample of Wang \& Liu \cite{wl2007}. The
forbidden-line Ne/O ratio observed in NGC\,7009 is higher than the solar
ratio by a factor of 2.5. However, Wang \& Liu \cite{wl2008} suggests that
the solar Ne/O ratio of Asplund, Grevesse \& Sauval \cite{asplund05} be
revised upwards by 0.22 dex, i.e. increased to 0.37. Similarly high
forbidden-line Ne/O ratio (0.34--0.36) is also found in NGC\,6153 by Liu et
al. \cite{liu2000}. The sulfur and argon abundances of NGC\,7009 both agree
with the solar values. The Mg/H abundance of NGC\,7009 agrees with the
average abundance of the bulge PNe sample of Wang \& Liu \cite{wl2007},
and is 0.1 dex lower than the solar value.

Elemental abundances derived from ORLs and CELs observed in NGC\,7009 are
presented in Table\,\ref{abundances:total}. By using the IR and UV observed
line fluxes from the literature and correcting for extinction, we are able
to derive elemental abundances of C, N, O, and Ne relative to hydrogen from
both ORLs and CELs. The resultant ORL abundances of the above four elements
are all higher than the abundance ratios derived from CELs, by a factor of
5--7. After deep spectroscopic observations of the ORLs detected in the
spectrum of NGC\,7009 by LSBC (for C, N and O) and LLB01 (for Ne), those
heavy-element abundance discrepancy problems are again studied
quantitatively, using the deepest CCD spectrum ever taken for a gaseous
nebula and the new effective recombination coefficients calculated for the
N~{\sc ii} and O~{\sc ii} recombination spectra under the nebular conditions
in the intermediate coupling scheme. The analyses of the spectrum are
carried out in a very consistent manner, i.e. an electron temperature of
1000~K as derived from the N~{\sc ii} and O~{\sc ii} ORL ratios is assumed
throughout analyses of the heavy-element ORLs, while a temperature of about
10\,000~K as yielded by CELs is assumed in calculating the CEL abundances.
In the previous deep spectroscopy of NGC\,7009 by LSBC, the electron
temperature ($\sim$10\,000~K) derived from the [O~{\sc iii}]
nebular-to-auroral line ratio was adopted in calculating the ORL abundances.
In the analyses of the neon abundance in NGC\,7009 by LLB01, the electron
temperature (7100~K) derived from the H~{\sc i} Balmer jump was used to
calculate the recombination line neon abundance. Liu et al. \cite{liu2000}
assumed a temperature of 9100~K, as derived from the [O~{\sc iii}] CEL ratio,
in calculating the recombination line abundances for the C, N, O and Ne ions
in NGC\,6153. In the analyses of Galactic bulge PNe M\,1-42 and M\,2-36, Liu
et al. \cite{liu2001b} used the H~{\sc i} Balmer jump temperature (3560~K
for M\,1-42 and 5900~K for M\,2-36) to derived the recombination line
abundances. However, the recombination line abundances
derived in the studies mentioned above are questionable, given that
the temperatures assumed are not derived from the heavy-element ORLs.
The effective recombination coefficients of the permitted transitions of
heavy element ions, which are mainly excited by radiative recombination,
usually decrease as the electron temperature increases, except for the
dielectronic transitions (with high-lying parent), whose effective
recombination coefficients increases as the temperature increases. When we
use the recombination lines that are mainly excited by radiative
recombination, such as those used by LSBC and LLB01, to calculate the ionic
abundances of C, N, O and Ne, using the forbidden line temperature e.g.
$T_\mathrm{e}$([O~{\sc iii}]) or the H~{\sc i} Balmer jump temperature will
result in overestimated recombination line abundances, because those two
temperatures are both much higher than the temperatures yielded by the heavy
element ORLs as revealed in the current analyses as well as many previous
studies (e.g. Liu \citealt{liu2011} for a recent review). Here we take the
O~{\sc ii} recombination lines as examples. When the electron temperature
increases from 1000 to 10\,000~K, the effective recombination coefficient
for the O~{\sc ii} M1 $\lambda$4649.13 line decreases by a factor of 7.5
(PJS), and the effective recombination coefficient for H$\beta$ decreases by
a factor of 6 (Storey \& Hummer \citealt{sh1995}), as a consequence the
O$^{2+}$/H$^{+}$ ionic abundance, which is calculated from the ratio of the
effective recombination coefficients of the two lines
$\alpha_{\rm eff}$(H$\beta$)/$\alpha_{\rm eff}$($\lambda$4649), will
increase by 22 per cent.

The ADFs observed in NGC\,7009, although moderate compared with those found
in PNe NGC\,6153 (ADF$\sim$10, Liu et al. \citealt{liu2000}), M\,1-42
(ADF$\sim$20, Liu et al. \citealt{liu2001b}), NGC\,1501 (ADF$\sim$30,
Ercolano et al. \citealt{ercolano04}) and Hf\,2-2 (ADF$\sim$70, Liu et al.
\citealt{lbzbs06}), is high compared with most of the other Milky Way PNe
so far spectroscopically studied (ADFs$\sim$1.6--3.2). Ever since the
observation of an ADF value of about $5$ from the deep spectroscopy of
NGC\,7009 by LSBC, it has been wondered that possible systematic effects
in the classic plasma analysis procedure based on CEL measurements might
cause the observed discrepancies, if inaccuracies in the effective
recombination coefficients or contamination of the recombination lines by
processes such as stellar continuum resonance fluorescence as the origins
of the discrepancies can be ruled out. After observations of the [O~{\sc
iii}] $\lambda$4931/$\lambda$4959 line ratio in seven Milky Way PNe and
the Orion Nebula by Mathis \& Liu \cite{ml1999}, the measurement errors in
CELs as the culprit has also been ruled out. Clearly the temperature and
abundance discrepancies are real, and the two categories of emission lines
probably arise from regions with different physical conditions. A detailed
study of NGC\,6153 by Liu et al. \cite{liu2000} yielded an ADF of about 10
for that object and, based on the empirical composite models, Liu et al.
\cite{liu2000} proposed a bi-abundance nebular model as a possible
explanation to such high abundance discrepancy.

Ever since the detection of a Balmer jump temperature as low as 3560~K for
M\,1-42, 5660~K lower than the [O~{\sc iii}] forbidden line temperature
for the same nebula (Liu et al. \citealt{liu2001b}), it has become
increasingly clear that PNe, at least those exhibiting large ADFs, must
contain another component of previously unknown ionized gas. This component
of gas mainly emits ORLs yet is essentially invisible in strong CELs
because the electron temperature prevailing in the component is too low to
excite any UV or optical CELs. The low temperature condition in that
component is probably due to the much enhanced cooling by the IR
fine-structure lines of the heavy element ions, which is a consequence of
of a very high metallicity (i.e. H-deficient). This physical idea is
supported by detailed photoionization modeling of P\'{e}quignot et al.
\cite{pequignot03} and Tylenda \cite{tylenda03} and also by direct
measurements of the average electron temperatures under which various
types of emission lines are emitted
(c.f. Liu \citealt{liu2003}, \citealt{liu2006a},\,b and \citealt{liu2011}
for reviews; Zhang et al. \citealt{zhang2004} for the plasma diagnostics
based on the H~{\sc i} recombination spectrum and Zhang et al.
\citealt{zhang05a} for the plasma diagnostics based on the He~{\sc i}
recombination spectrum). The discovery of a dramatically high ADF value
of $\sim$70 and the remarkably low Balmer jump temperature ($\sim$900~K)
in PN Hf\,2-2 by Liu et al. \cite{liu2006a} strengthens the validity of the
bi-abundance nebular model. In order to reproduce the multi-waveband
spectroscopic and imaging observations of NGC\,6153 and investigate the
nature and origin of the H-deficient inclusions, Yuan et al. \cite{yuan2011}
constructed three-dimensional photoionization models, using the Monte Carlo
photoionization code MOCASSIN developed by Ercolano et al. \cite{ercolano03}.
Modeling of NGC\,6153 showed that chemically homogeneous models yielded small
electron temperature fluctuations and failed to reproduce the strengths of
the ORLs of heavy element ions. In contrast, bi-abundance models
incorporating a small amount of metal-rich inclusions ($\sim$1.3 per cent of
the total nebular mass) are able to match all the observations within
measurement uncertainties. The metal-rich inclusions, cooled down in a very
low temperature ($\sim$800~K) by ionic IR fine-structure lines, dominate the
emission of heavy-element ORLs, but contribute almost nil to the emission of
most CELs. The current analyses of the optical recombination spectrum of
NGC\,7009 are carried out under the context of the bi-abundance nebular
model, and the results of plasma diagnostics based on various types of
emission lines and abundance determinations are consistent with that context:
the temperature sequence $T_\mathrm{e}$([O~{\sc iii}]) $\gtrsim$
$T_\mathrm{e}$(H~{\sc i}~BJ) $\gtrsim$ $T_\mathrm{e}$(He~{\sc i}) $\gtrsim$
$T_\mathrm{e}$(N~{\sc ii}~\&~O~{\sc ii}~ORLs) is consistent with predictions
from the bi-abundance model; the C$^{2+}$/H$^{+}$, N$^{2+}$/H$^{+}$,
O$^{2+}$/H$^{+}$ and Ne$^{2+}$/H$^{+}$ ionic abundances derived from ORLs,
using the new effective recombination coefficients and the electron
temperature yielded by the N~{\sc ii} and O~{\sc ii} ORLs, are systematically
higher, by about a factor of 5, than the corresponding abundances derived
from CELs. It has been shown from optical observations that the ADF varies
with position in several high-ADF PNe and is highest close to the central
star. The ``cold" inclusions should be cooled via the IR fine-structure lines
of heavy element ions. Thus it is interesting to see if the IR fine-structure
line fluxes relative to optical/UV CELs peak where the ADF peaks in PNe with
large ADFs. Recently, Herschel and Hubble observations of NGC\,7009 have been
carried out and the results show that within the first $\sim$5~arcsec from the
central star of NGC\,7009, the [O~{\sc iii}] 88$\mu$m/$\lambda$5007 flux
ratio seems to increase towards center (R.~H. Rubin, private communication).
With the very deep spectrum and the high-quality atomic data now available,
we can derive more precise physical properties e.g. total mass, spatial
distribution of the ``cold", metal-rich inclusions in NGC\,7009 through
three-dimensional photoionization modeling.

Given that the ADFs found in PNe are all larger than unity, the metal-rich
(H-deficient) inclusions are probably a real feature of PNe. However, the
the presence of those ``cold" inclusions is not predicted by the current
theories of stellar evolution. Iben, Kaler \& Truran \cite{iben83} proposed
that an evolved star undergoing a very late helium flash (the so-called
'born-again' PNe) may harbour H-deficient material, such as the H-deficient
knots detected in the two 'born-again' PNe Abell\,30 and Abell\,58. Wesson,
Liu \& Barlow \cite{wlb2003} and Wesson et al. \cite{wesson08} found that the
H-deficient knots in Abell\,30 and Abell\,58 are O-rich, in contradiction
with the expectation of the 'born-again' scenario. The ADFs are found to be
high in the PNe with Wolf-Rayet central stars, and that can be explained by
the scenario of a single post-asymptotic giant branch (post-AGB) star
experiencing a list helium shell flash, but not all PNe with large ADFs have
an H-deficient central star, such as the case of NGC\,7009 studied in the
current paper. Garc\'{i}a-Rojas, Pe\~{n}a \& Peimbert \cite{garcia09}
observed the faint ORLs in Galactic PNe with [WC] nucleus and found the
results that argue against the presence of H-deficient knots coming from a
late thermal pulse event. De~Marco \cite{demarco08} suggested a binary
scenario to explain the observations that are in contradiction with the
theory of the single post-AGB evolution. In this respect, it is interesting
to note that Abell\,58 have experienced a nova-like outburst (Clayton \&
De~Marco \citealt{cd97}). Lutz et al. \cite{lutz98} has also found the
central star of Hf\,2-2, a PN with the largest ADF value ($\sim$70, Liu et
al. \citealt{liu2006a}) ever found for an emission line nebulae, to be a
close binary system. An alternative scenario of the origin of the
H-deficient inclusions is that they evolve from metal-rich planetary
material, such as icy planetesimals left over from the debris of planetary
system of the progenitor star of the PN (Liu \citealt{liu2003},
\citealt{liu2006a}). Both high spectral- and spatial-resolution observations
in the future, in combination with detailed three-dimensional photoionization
modeling as has been carried out for NGC\,6153 (Yuan et al.
\citealt{yuan2011}), will help to reveal the possible astrophysical origins
of the H-deficient inclusions in NGC\,7009.

\section{\label{summary}
Summary and conclusions}

Nearly two decades after the first analysis of the O~{\sc ii} optical
recombination spectrum of the bright PN NGC\,7009 (LSBC), once again we focus
on the rich ORLs of heavy element ions detected in very deep CCD spectrum of
the same object. Thanks to much advance in observational techniques, which
enables accurate detection of the weak ORLs of heavy element ions, and the
steady improvements in atomic data, especially the recombination theories of
heavy element ions in the physical conditions of photoionized nebulae, we are
now clear that the long-standing dichotomy between nebular plasma diagnostics
and abundance determinations using CELs on the one hand and ORLs on the other,
are real rather than cause by, e.g., observational uncertainties or errors in
atomic data. Unremitting efforts in nebular research of the past 40 years
gradually lead to new understanding of the problems in nebular astrophysics.
Various mechanisms (e.g. temperature fluctuations and/or density
inhomogeneities, abundance inhomogeneities, non-Maxwell-Boltzmann equilibrium
electrons e.g. the $\kappa$-distribution of electron energies) have been
proposed to explain the discrepancies in plasma diagnostics and abundance
determinations, and debate over these mechanisms are still going on.

In the context of bi-abundance nebular model postulated by Liu et al.
\cite{liu2000}, we present a comprehensive and critical analysis of the rich
optical recombination spectrum of NGC\,7009. Transitions from individual
multiplets of heavy element ions, e.g. C~{\sc ii}, N~{\sc ii}, O~{\sc ii} and
Ne~{\sc ii}, are checked carefully for line blending, and accurate dereddened
line fluxes of the most prominent transitions of those ions are obtained
through multi-Gaussian profile fitting. In addition to the accurate
observations of ORLs, we have finished new calculations of the effective
recombination coefficients for the N~{\sc ii} recombination spectrum. The
new effective recombination coefficients for the nebular O~{\sc ii} lines
calculated by P.~J. Storey (unpublished) help to enlarge our current atomic
dataset for nebular recombination line study. Both calculations were carried
out in the intermediate coupling scheme, and have taken into account the
density-dependence of the relative populations of the ground fine-structure
levels of recombining ions (i.e. N$^{2+}$ $^{2}$P$^{\rm o}_{1/2}$ and
$^2$P$^{\rm o}_{3/2}$ for N~{\sc ii}, and O$^{2+}$ $^{3}$P$_{0}$,
$^{3}$P$_{1}$ and $^{3}$P$_{2}$ for O~{\sc ii}). The new effective
recombination coefficients of N~{\sc ii} and O~{\sc ii} are of high quality
and make nebular density diagnostics using the ORLs of heavy element ions
possible for the first time.

The observed relative intensities of ORLs are compared with theoretical
predictions that are based on the new effective recombination coefficients.
At a given electron temperature ($T_\mathrm{e}$ = 1000~K) as yielded by the
ORL ratios of N~{\sc ii} and O~{\sc ii}, the predicted relative intensities
of ORLs agree with the observed values. Plasma diagnostics based on the
best observed N~{\sc ii} and O~{\sc ii} ORLs (i.e. the
$I$(M3~$\lambda$5679)/$I$(M39b~$\lambda$4041) ratio of N~{\sc ii} and the
$I$(M1~$\lambda$4649)/$I$(M48a~$\lambda$4089) ratio of O~{\sc ii}) both yield
electron temperatures close to 1000~K, which is lower than those derived from
the CEL ratios by nearly one order of magnitude. The low temperatures yielded
by the N~{\sc ii} and O~{\sc ii} ORLs indicate that the recombination lines
of heavy element ions originate from very cold regions. The electron
temperatures derived from the intensity ratios of the O~{\sc ii}
high-excitation recombination lines M15~$\lambda$4591 and M36~$\lambda$4189,
which are formed from recombination of excited-state parent (i.e. O$^{2+}$
2p$^{2}$\,$^{1}$D), relative to the O~{\sc ii} M1~$\lambda$4649 line agree
with eath other ($\sim$3600~K), and is consistent with the fact that very
cold ($\leq$1000~K) inclusions probably exist in the nebula. The electron
temperature ($\sim$3000~K) yielded by the C~{\sc ii}
$I$(M28.01~$\lambda$8794)/$I$(M6~$\lambda$4267) dielectronic-to-radiative
recombination line ratio also agrees with the conjecture of very cold
inclusions. The C$^{2+}$/H$^+$, N$^{2+}$/H$^+$, O$^{2+}$/H$^+$ and
Ne$^{2+}$/H$^+$ ionic abundance ratios derived from ORLs, using the new
effective recombination coefficients of N~{\sc ii} and O~{\sc ii}, are
consistently higher than the corresponding values derived from CELs, by about
a factor of 5. An electron temperature of 1000~K, which is yielded by the
best observed N~{\sc ii} and O~{\sc ii} recombination line ratios and as a
consequence presumably represents the physical condition prevailing in the
regions where the heavy element ORLs arise, has been assumed throughout the
recombination-line abundance determinations. The results of plasma diagnostics
and abundance determinations for NGC\,7009 points to the existence of
``cold", metal-rich inclusions in NGC\,7009, and is thus consistent with the
context of the current spectral analyses, i.e. the bi-abundance nebular model.

Recombination line analysis for NGC\,7009 helps to assess the new atomic data.
The agreement between the observed and predicted relative intensities of the
N~{\sc ii} and O~{\sc ii} ORLs indicates that the current calculations of the
recombination spectra of those two ionic species well represent the physical
processes, i.e. radiative and dielectronic recombination, under nebular
conditions. Our nebular analysis also shows that the recombination lines of
different multiplets, or different $J$-resolved fine-structure components of
a multiplet yield consistent ionic abundances (e.g. N$^{2+}$/H$^{+}$ and
O$^{2+}$/H$^{+}$). This is another evidence that the new effective
recombination coefficients are reliable. However, the Ne$^{2+}$/H$^{+}$
abundance ratio derived from the total intensity of the 4f\,--\,3d transitions
is higher, by nearly 0.2 dex, than the average value derived from the
multiplets of the 3\,--\,3 configuration. That indicates new calculations of
the effective recombination coefficients for the Ne~{\sc ii} lines are needed.


\section*{Acknowledgements}
We thank Dr. P.~J. Storey for making the effective recombination coefficients
for the O~{\sc ii} lines and the Ne~{\sc ii} 4f\,--\,3d transitions available
prior to publication. We thank Dr. R.~H. Rubin for fruitful discussions. We
would also like to thank Dr. O. De~Marco for her valuable comments and
suggestions which have greatly improved the quality of this paper. This work
is supported by the Natural Science Foundation of China (No. 10933001).


\clearpage
\appendix

\section{The C~{\sc ii} optical recombination spectrum}\label{appendix:a}

\subsection{Multiplet 6,
4f\,$^2$F$^{\rm o}$ -- 3d\,$^2$D}\label{appendix:cii:v6}

See Section\,\ref{orls:cii:v6}.

\subsection{Multiplet 3,
3d\,$^2$D -- 3p\,$^2$P$^{\rm o}$}\label{appendix:cii:v3}

Only the $\lambda$7231.32 (3d~$^2$D$_{3/2}$ -- 3p~$^2$P$^{\rm o}_{1/2}$) line
is detected; the other two components $\lambda$7236.42 (3d~$^2$D$_{5/2}$ --
3p~$^2$P$^{\rm o}_{3/2}$) and $\lambda$7237.17 (3d~$^2$D$_{3/2}$ --
3p~$^2$P$^{\rm o}_{3/2}$) are blended with the [Ar~{\sc iv}] 3p3~$^2$P$^{\rm
o}_{3/2}$ -- 3p3~$^2$D$^{\rm o}_{5/2}$ $\lambda$7237.40 line. The intensity
of $\lambda$7231.32 is 0.130, with an uncertainty of less than 5 per cent.
Assuming that the intensity ratios of the C~{\sc ii} M3 lines are as in the
{\it LS}\,coupling, i.e. 1\,:\,5\,:\,9, we obtain an total intensity
ratio of M3 $\lambda$7235 to C~{\sc ii} M6 $\lambda$4267 of 0.442, which is
lower than the predicted ratio $1.204$ in Case~B but much higher than the
Case~A value ($0.017$). Those theoretical ratios are based on the C~{\sc ii}
effective recombination coefficients of B06. Since the calculation of B06 for
the C~{\sc ii} recombination spectrum is valid from 5000~K, here an electron
temperature of 10\,000~K and a density of 10\,000~cm$^{-3}$ were assumed.
In this section, an electron temperature of 5000 or 10\,000~K is assumed when
we use the effective recombination coefficients of B06.

The calculation of Davey, Storey \& Kisielius \cite{davey2000} is valid from
500 to 20\,000~K. Using their C~{\sc ii} effective recombination coefficients,
we derived a theoretical C~{\sc ii}
$I$(M3~$\lambda$7235)/$I$(M6~$\lambda$4267) ratio of $0.013$ in Case~A and
$0.907$ in Case~B. An electron temperature of 1000~K and a density of
10\,000~cm$^{-3}$ are assumed when we use the atomic data of Davey, Storey
\& Kisielius \cite{davey2000} in the current paper. Given that the effective
recombination coefficients are insensitive to electron density at low
temperature, we expect that the intensity ratio of those two C~{\sc ii}
multiplets at 10\,000~cm$^{-3}$ does not differ much from that at
2000\,--\,4000~cm$^{-3}$, a density range yielded by the N~{\sc ii} and O~{\sc
ii} recombination line diagnostics (see Figs.\,\ref{nii_te:v3v39} and
\ref{oii_te:v1v48}).
Thus the C~{\sc ii} $I$(M3~$\lambda$7235)/$I$(M6~$\lambda$4267) ratio observed
in NGC\,7009 (0.442) lies between the two cases.

\subsection{Multiplet 4,
4s\,$^2$S -- 3p\,$^2$P$^{\rm o}$}\label{appendix:cii:v4}

The $\lambda$3918.98 (4s~$^2$S$_{1/2}$ -- 3p~$^2$P$^{\rm o}_{1/2}$) component
is blending with the O~{\sc ii} M17 3p$^{\prime}$~$^2$P$^{\rm o}_{1/2}$ --
3s$^{\prime}$~$^2$D$_{3/2}$ $\lambda$3919.29 and N~{\sc ii} M17 3d~$^1$P$^{\rm
o}_{1}$ -- 3p~$^1$P$_{1}$ $\lambda$3919.00 lines, and is partially blended
with the other M4 line $\lambda$3920.69 (4s~$^2$S$_{1/2}$ -- 3p~$^2$P$^{\rm
o}_{3/2}$). Two Gaussian profiles were used to fit the complex. Using the
O~{\sc ii} effective recombination coefficients of PJS, we estimated that the
O~{\sc ii} line contributes 42 per cent to the total intensity of the blend at
$\lambda$3919. The contribution of the N~{\sc ii} line is negligible according
to the calculation of FSL11. The intensities of
the $\lambda$3918.98 and $\lambda$3920.69 lines have 0.016$\pm$0.003 and
0.034$\pm$0.004, respectively. Thus the intensity ratio of C~{\sc ii} M4
$\lambda$3920 to the C~{\sc ii} M6 $\lambda$4267 multiplet is 0.057, which
agrees with the predicted ratio in Case~B derived from the coefficients of
B06, but higher than the Case~A value 0.018. An electron temperature of
10\,000~K was assumed. At 5000~K, the lowest temperature of the calculation
by B06, this C~{\sc ii} ratio is $0.011$ in Case~A and $0.035$ in Case~B. The
C~{\sc ii} M4 multiplet is not presented in the calculation of Davey, Storey
\& Kisielius \cite{davey2000}.

\subsection{Multiplet 16.04,
6f\,$^2$F$^{\rm o}$ -- 4d\,$^2$D}\label{appendix:cii:v16.04}

All fine-structure components of this multiplet have an identical laboratory
wavelength 6151.43\,{\AA}. Single Gaussian profile fitting to this feature
gives an intensity of 0.028$\pm$0.004. The contribution from the blended
N~{\sc ii} M36 4p~$^3$D$_{2}$\,--\,3d~$^3$F$^{\rm o}_{2}$ $\lambda$6150.75
line is negligible. Thus the intensity ratio of this multiplet to
C~{\sc ii} M6 $\lambda$4267 is 0.031, agrees with the predicted ratio 0.027
deduced from the coefficients of Davey, Storey \& Kisielius \cite{davey2000}.
The predicted ratio of the two C~{\sc ii} multiples, based on the coefficients
of B06, is 0.037. Both Davey, Storey \& Kisielius \cite{davey2000} and B06
show that the Case~A effective recombination coefficient for the $\lambda$6151
multiplet differs from its Case~B value by only 2 per cent. For the C~{\sc ii}
transitions of large $l$ ($l\,>\,4$), only Case~A coefficients are presented by
Davey, Storey \& Kisielius \cite{davey2000}. Thus in this section, the
theoretical intensity ratios of those C~{\sc ii} transitions relative to the
M6 $\lambda$4267 multiplet are all presented in Case~A.

\subsection{Multiplet 17.02,
5g\,$^2$G -- 4f\,$^2$F$^{\rm o}$}\label{appendix:cii:v17.02}

This multiplet ($\lambda$9903) is free of blending, and has an intensity
0.213$\pm$0.020. The intensity ratio of M17.02 $\lambda$9903 to the C~{\sc ii}
M6 $\lambda$4267 multiplet is 0.234, which is slightly lower than the
predicted ratio 0.302  deduced from the coefficients of Davey, Storey \&
Kisielius \cite{davey2000}. At 10\,000~K, the effective recombination
coefficient for this multiplet given by Davey, Storey \& Kisielius
\cite{davey2000} differs from that of B06 by only 3 per cent.

\subsection{Multiplet 17.04,
6g\,$^2$G -- 4f\,$^2$F$^{\rm o}$}\label{appendix:cii:v17.04}

This multiplet ($\lambda$6462) has an intensity of 0.078$\pm$0.004. The
intensity ratio of $\lambda$6462 to the C~{\sc ii} M6 $\lambda$4267 multiplet
is 0.088, which agrees with the predicted value 0.076 (deduced from the atomic
data of Davey, Storey \& Kisielius \cite{davey2000}) within errors.

\subsection{Multiplet 17.06,
7g\,$^2$G -- 4f\,$^2$F$^{\rm o}$}\label{appendix:cii:v17.06}

The C~{\sc ii} M17.06 $\lambda$5342 multiplet is partially blending with a
very weak feature, which was identified as [Kr~{\sc iv}] $\lambda$5346.02
(Paper~I). Two Gaussian profiles were used to fit the complex, and that gives
an intensity of 0.037$\pm$0.002 for $\lambda$5342. The intensity ratio of
$\lambda$5342 to the C~{\sc ii} M6 $\lambda$4267 multiplet is 0.043, which
agrees well with the predicted ratio 0.045 deduced from the data of Davey,
Storey \& Kisielius \cite{davey2000}).

\subsection{Multiplet 28.01, 3d$^{\prime}$\,$^2$F$^{\rm o}$ --
3p$^{\prime}$\,$^2$D}\label{appendix:cii:v28.01}

See Section\,\ref{orls:cii:v28.01}.

\section{The N~{\sc ii} optical recombination spectrum}\label{appendix:b}

\subsection{Multiplet 3,
3p\,$^3$D -- 3s\,$^3$P$^{\rm o}$}\label{appendix:nii:v3}

See Section\,\ref{nii_orls:v3}.

\subsection{Multiplet 5,
3p\,$^3$P -- 3s\,$^3$P$^{\rm o}$}\label{appendix:nii:v5}

The strongest component of the N~{\sc ii} M5 multiplet, $\lambda$4630.54
(3p\,$^3$P$_{2}$\,--\,3s\,$^3$P$^{\rm o}_{2}$) is partially blended with the
N~{\sc iii} M2 3d~$^2$D$_{3/2}$ -- 3p~$^2$P$^{\rm o}_{1/2}$ $\lambda$4634.14
fluorescence line (Fig.\,\ref{4625-4680}). Gaussian profile fitting yields an
intensity of 0.067$\pm$0.013 for $\lambda$4630.54. The intensity ratio of
the $\lambda$4630.54 line and the N~{\sc ii} M3 $\lambda$5679.56 line is
0.512, which agrees well with the predicted ratio $0.491$. Another M5 line
$\lambda$4601.48 (3p~$^3$P$_{2}$ -- 3s~$^3$P$^{\rm o}_{1}$) is blended with
the O~{\sc ii} M92b 4f~F[3]$^{\rm o}_{5/2}$ -- 3d~$^2$D$_{3/2}$
$\lambda$4602.13 line, which contributes about 70 per cent to the total
intensity, as estimated from the O~{\sc ii} effective recombination
coefficients of PJS.

\begin{figure}
\begin{center}
\includegraphics[width=7.5cm,angle=-90]{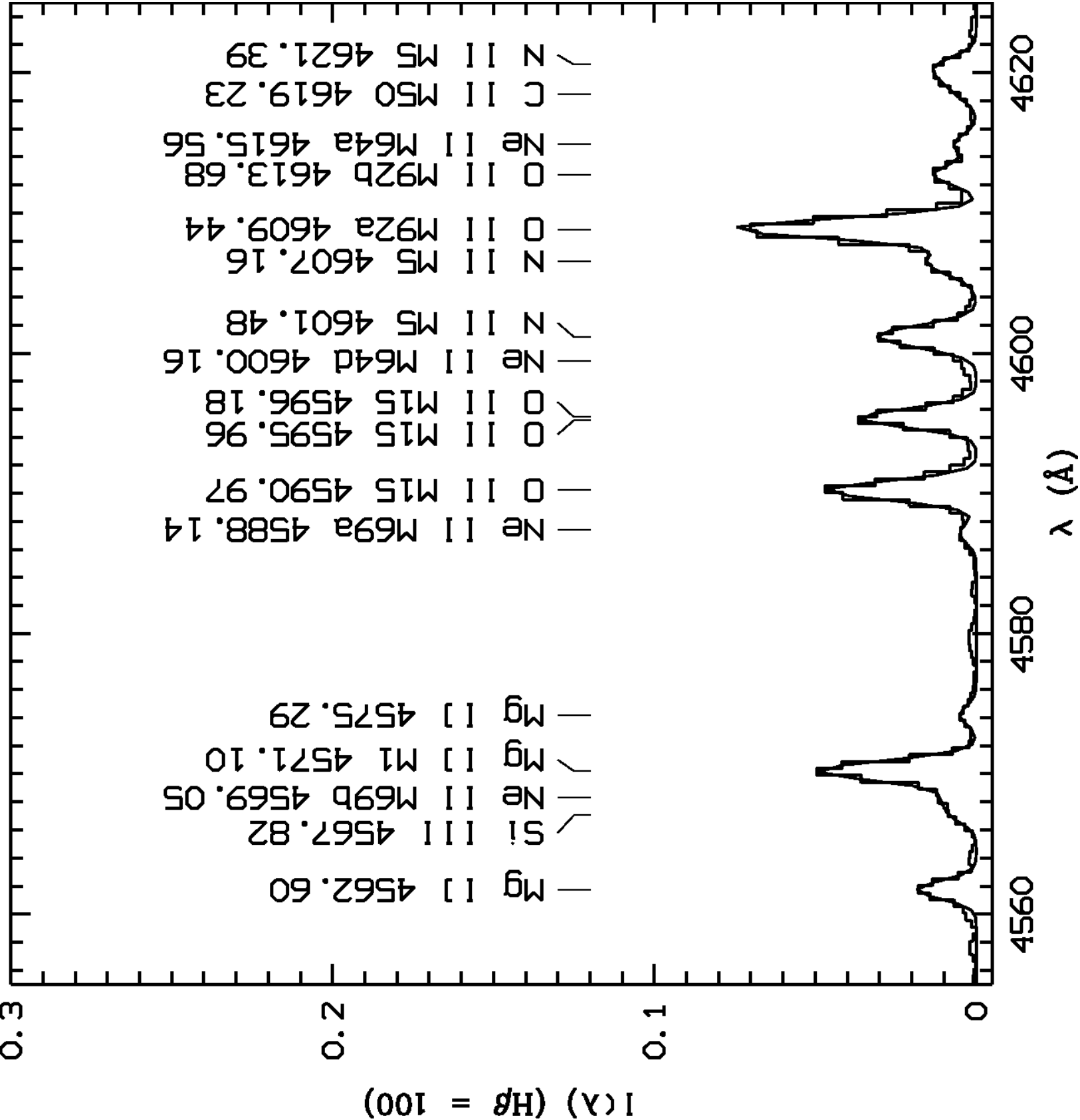}
\caption{Spectrum of NGC\,7009 from 4555 to 4625\,{\AA} showing the N~{\sc ii}
M5 lines and other emission features. Note the detection of the three Mg~{\sc
i}$]$ lines of the 3s3p\,$^{3}$P$^{\rm o}$\,--\,3s$^{2}$\,$^{1}$S multiplet.
The continuous curve is the sum of Gaussian profile fits. Continuum has been
subtracted and the spectrum has been normalized such that H$\beta$ has an
integrated flux of 100. Extinction has not been corrected for.}
\label{4555-4625}
\end{center}
\end{figure}

The other four M5 lines all suffer from serious line blending: (1) The
$\lambda$4607.16 (3p\,$^3$P$_{1}$\,--\,3s\,$^3$P$^{\rm o}_{0}$) line is
blended with the [Fe~{\sc iii}] $\lambda$4607.03 line, whose contribution
to the total intensity is uncertain. The blend at $\lambda$4607 is again
partially blended with a much stronger feature centered at $\lambda$4609,
which is a blend of O~{\sc ii} M92a 4f\,F[4]$^{\rm
o}_{7/2}$\,--\,3d\,$^2$D$_{5/2}$ $\lambda$4609.44 and O~{\sc ii} M92c
4f\,F[2]$^{\rm o}_{5/2}$\,--\,3d\,$^2$D$_{3/2}$ $\lambda$4610.20
(Fig.\,\ref{4555-4625}). (2) The $\lambda$4613.87
(3p\,$^3$P$_{1}$\,--\,3s\,$^3$P$^{\rm o}_{1}$) line is blended with at least
three lines, O~{\sc ii} M92b 4f\,F[3]$^{\rm o}_{5/2}$\,--\,3d\,$^2$D$_{5/2}$
$\lambda$4613.14, O~{\sc ii} M92b 4f\,F[3]$^{\rm
o}_{7/2}$\,--\,3d\,$^2$D$_{5/2}$ $\lambda$4613.68, and Ne~{\sc ii} M64b
4f\,2[3]$^{\rm o}_{7/2}$\,--\,3d\,$^4$P$_{5/2}$) $\lambda$4612.93. (3) The
$\lambda$4621.39 (3p\,$^3$P$_{0}$\,--\,3s\,$^3$P$^{\rm o}_{1}$) line is
blended with at least two lines, O~{\sc ii} M92c 4f\,F[2]$^{\rm
o}_{5/2}$\,--\,3d\,$^2$D$_{5/2}$ $\lambda$4621.27 and O~{\sc ii} M92c
4f\,F[2]$^{\rm o}_{3/2}$\,--\,3d\,$^2$D$_{5/2}$ $\lambda$4622.14. (4) The
$\lambda$4643.09 (3p\,$^3$P$_{1}$\,--\,3s\,$^3$P$^{\rm o}_{2}$) line is
blended with the N~{\sc iii} M2 $\lambda\lambda$4640.64 and 4641.85
fluorescence lines (Fig.\,\ref{4625-4680}). Accurate measurements of these
four N~{\sc ii} M5 lines are difficult.


\subsection{Multiplet 12,
3p\,$^1$D -- 3s\,$^1$P$^{\rm o}$}\label{appendix:nii:v12}

This singlet ($\lambda$3994.99) is shown in Fig.\,\ref{3940-4006}. Gaussian
profile fitting yields an intensity of 0.033$\pm$0.004. The intensity ratio
of the $\lambda$3994.99 line to the N~{\sc ii} M3 $\lambda$5679.56 line is
0.257, which is higher than the theoretical prediction $0.123$.

\subsection{Multiplet 19,
3d\,$^3$F$^{\rm o}$ -- 3p\,$^3$D}\label{appendix:nii:v19}

The M19 $\lambda$5004 lines are blended with or seriously affected by the
saturated [O~{\sc iii}] $\lambda$5007 line.

\subsection{Multiplet 20,
3d\,$^3$D$^{\rm o}$ -- 3p\,$^3$D}\label{appendix:nii:v20}

The M20 $\lambda$4803.29 (3d\,$^3$D$^{\rm o}_{3}$\,--\,3p\,$^3$D$_{3}$) line
is blended with the C~{\sc ii} M17.08 8g\,$^2$G\,--\,4f\,$^2$F$^{\rm o}$
$\lambda$4802.70 line, which contributes $\sim$50 per cent to the total
intensity of the blend at $\lambda$4803, as estimated from the C~{\sc ii}
effective recombination coefficients of Davey, Storey \& Kisielius
\cite{davey2000}. The intensity of the $\lambda$4803.29 line is
0.032$\pm$0.016. The intensity ratio of $\lambda$4803.29 and the N~{\sc ii} M3
$\lambda$5679.56 line is 0.248, which agrees with the predicted value 0.282
within errors. The other M20 lines are not detected.

\subsection{Multiplet 28,
3d\,$^3$D$^{\rm o}$ -- 3p\,$^3$P}\label{appendix:nii:v28}

See Section\,\ref{nii_orls:v28}.

\subsection{Multiplet 29,
3d\,$^3$P$^{\rm o}$ -- 3p\,$^3$P}\label{appendix:nii:v29}

Accurate measurements of this multiplet are difficult. Only the
$\lambda$5454.22 (3d\,$^3$P$^{\rm o}_{0}$\,--\,3p\,$^3$P$_{1}$) and
$\lambda$5480.06 (3d\,$^3$P$^{\rm o}_{1}$\,--\,3p\,$^3$P$_{2}$) lines are
detected. The former line is blended with the S~{\sc ii} M6 4p\,$^4$D$^{\rm
o}_{7/2}$\,--\,4s\,$^4$P$_{5/2}$ $\lambda$5453.83 line and the latter one is
blended with the $\lambda$5478.10 (3d\,$^3$P$^{\rm
o}_{2}$\,--\,3p~$^3$P$_{1}$) line of the same multiplet.

\subsection{4f -- 3d transitions}\label{appendix:nii:4f-3d}

\subsubsection{Multiplet 39a,
4f\,G[7/2] -- 3d\,$^3$F$^{\rm o}$:}\label{appendix:nii:v39a}

See Section\,\ref{nii:4f-3d:v39a}.

\subsubsection{Multiplet 39b,
4f\,G[9/2] -- 3d\,$^3$F$^{\rm o}$:}\label{appendix:nii:v39b}

See Section\,\ref{nii:4f-3d:v39b}.

\subsubsection{Multiplet 38a,
4f\,F[5/2] -- 3d\,$^3$F$^{\rm o}$:}\label{appendix:nii:v38a}

Accurate measurements of this multiplet are difficult. The $\lambda$4076.91
(4f\,F[5/2]$_{2}$\,--\,3d\,$^3$F$^{\rm o}_{2}$) and $\lambda$4077.40
(4f\,F[5/2]$_{3}$\,--\,3d\,$^3$F$^{\rm o}_{2}$) lines are blended with the
[S~{\sc ii}] $\lambda$4076.35 (3p$^3$\,$^2$P$^{\rm
o}_{1/2}$\,--\,3p$^3$\,$^4$S$^{\rm o}_{3/2}$) line. The $\lambda$4086.83
(4f\,F[5/2]$_{2}$\,--\,3d\,$^3$F$^{\rm o}_{3}$) and $\lambda$4087.30
(4f\,F[5/2]$_{3}$\,--\,3d\,$^3$F$^{\rm o}_{3}$) lines are blended with the
O~{\sc ii} M48c 4f\,G[3]$^{\rm o}_{5/2}$\,--\,3d\,$^4$F$_{3/2}$
$\lambda$4087.15 line. The other M38a line $\lambda$4100.97
(4f\,F[5/2]$_{3}$\,--\,3d\,$^3$F$^{\rm o}_{4}$) line is blended with the
H~{\sc i} $\lambda$4101 line.

\subsubsection{Multiplet 38b,
4f\,F[7/2] -- 3d\,$^3$F$^{\rm o}$:}\label{appendix:nii:v38b}

Accurate measurements of the M38b lines are difficult. The $\lambda$4073.04
(4f\,F[7/2]$_{3}$\,--\,3d\,$^3$F$^{\rm o}_{2}$) line is blended with the much
stronger O~{\sc ii} M10 $\lambda$4072.16 line. The $\lambda$4095.90
(4f\,F[7/2]$_{4}$\,--\,3d\,$^3$F$^{\rm o}_{4}$) and $\lambda$4096.58
(4f\,F[7/2]$_{3}$\,--\,3d\,$^3$F$^{\rm o}_{4}$) lines are blended with the
N~{\sc iii} M1 3p\,$^2$P$^{\rm o}_{3/2}$\,--\,3s\,$^2$S$_{1/2}$
$\lambda$4097.33 fluorescence line. The other lines are not observed.

\subsubsection{Multiplet 43a,
4f\,F[5/2] -- 3d\,$^1$D$^{\rm o}$:}\label{appendix:nii:v43a}

The intensity of the $\lambda$4176.16 (4f\,F[5/2]$_{3}$\,--\,3d\,$^1$D$^{\rm
o}_{2}$) line is 0.028$\pm$0.004, which includes the contribution from the
N~{\sc ii} M43a 4f\,F[5/2]$_{2}$\,--\,3d\,$^1$D$^{\rm o}_{2}$ $\lambda$4175.66
line (about 20 per cent).

\subsubsection{Multiplet 43b,
4f\,F[7/2] -- 3d\,$^1$D$^{\rm o}$:}\label{appendix:nii:v43b}

The $\lambda$4171.61 (4f\,F[7/2]$_{3}$\,--\,3d\,$^1$D$^{\rm o}_{2}$) line is
partially blended with the O~{\sc ii} M19 $\lambda$4169.22 line. The
$\lambda$4171.61 line has an intensity of 0.023$\pm$0.003, which agrees well
with the predicted value (Table\,\ref{relative:nii_4f-3d}).

\subsubsection{Multiplet 48a,
4f\,F[5/2] -- 3d\,$^3$D$^{\rm o}$:}\label{appendix:nii:v48a}

The $\lambda$4241.78 (4f\,F[5/2]$_{3}$\,--\,3d\,$^3$D$^{\rm o}_{2}$) line
coincides in wavelength with the N~{\sc ii} M48b
4f\,F[7/2]$_{4}$\,--\,3d\,$^3$D$^{\rm o}_{3}$ $\lambda$4241.78 line.
The $\lambda$4236.91 (4f\,F[5/2]$_{2}$\,--\,3d\,$^3$D$^{\rm o}_{1}$) line is
blended with the N~{\sc ii} M48b 4f\,F[7/2]$_{3}$\,--\,3d\,$^3$D$^{\rm o}_{2}$
$\lambda$4237.05 line (Fig.\,\ref{4176-4260}). Gaussian profile fitting yields
an intensity of 0.055$\pm$0.011 for the $\lambda$4236.91 line, which includes
the contribution from the N~{\sc ii} M48b $\lambda$4237.05 line ($\sim$30 per
cent). The $\lambda$4236.91 line intensity is 24 per cent higher than what is
predicted (Table\,\ref{relative:nii_4f-3d}). Measurements of the other two
M48a lines, $\lambda$4246.71 (4f\,F[5/2]$_{2}$\,--\,3d\,$^3$D$^{\rm o}_{3}$)
and $\lambda$4247.20 (4f\,F[5/2]$_{3}$\,--\,3d\,$^3$D$^{\rm o}_{3}$) are
difficult.

\subsubsection{Multiplet 48b,
4f\,F[7/2] -- 3d\,$^3$D$^{\rm o}$:}\label{appendix:nii:v48b}

The $\lambda$4241.78 (4f\,F[7/2]$_{4}$\,--\,3d\,$^3$D$^{\rm o}_{3}$) line
coincides in wavelength with the N~{\sc ii} M48a
4f\,F[5/2]$_{3}$\,--\,3d\,$^3$D$^{\rm o}_{2}$ $\lambda$4241.78 line. The total
intensity of the $\lambda$4242 line agrees well with the predicted value
(Table\,\ref{relative:nii_4f-3d}). The contribution from the blended Ne~{\sc
ii} M52c 4f\,2[1]$^{\rm o}_{3/2}$\,--\,3d\,$^4$D$_{1/2}$ $\lambda$4242.04 line
($\sim$10--15 per cent) is included. The contributions from the N~{\sc ii}
M48a 4f\,F[5/2]$_{2}$\,--\,3d\,3D$^{\rm o}_{2}$ $\lambda$4241.24 and M48b
$\lambda$4242.49 (4f\,F[7/2]$_{3}$\,--\,3d\,$^3$D$^{\rm o}){3}$) lines are
negligible. The other M48b line $\lambda$4237.05
(4f\,F[7/2]$_{3}$\,--\,3d\,$^3$D$^{\rm o}_{2}$) is blended with the N~{\sc ii}
M48a 4f\,F[5/2]$_{2}$\,--\,3d\,$^3$D$^{\rm o}_{1}$ $\lambda$4236.91 line,
which contributes $\sim$70 per cent to the total intensity.

\subsubsection{Multiplet 49a,
4f\,G[7/2] -- 3d\,$^3$D$^{\rm o}$:}\label{appendix:nii:v49a}

The $\lambda$4195.97 (4f\,G[7/2]$_{3}$\,--\,3d\,$^3$D$^{\rm o}_{2}$) line is
blended with the N~{\sc iii} M6
3p$^{\prime}$\,$^2$D$_{3/2}$\,--\,3s$^{\prime}$\,$^2$P$^{\rm o}_{1/2}$
$\lambda$4195.76 line (Fig.\,\ref{4176-4260}), and contributes 8 per cent to
the total intensity. The other two M49a lines, $\lambda$4199.98
(4f\,G[7/2]$_{4}$\,--\,3d\,$^3$D$^{\rm o}_{3}$) and $\lambda$4201.35
(4f\,G[7/2]$_{3}$\,--\,3d\,$^3$D$^{\rm o}_{3}$) are blended with the much
stronger He~{\sc ii} 11g\,$^2$G\,--\,4f\,$^2$F$^{\rm o}$ $\lambda$4199.83
line.

\subsubsection{Multiplet 49b,
4f\,G[9/2] -- 3d\,$^3$D$^{\rm o}$:}\label{appendix:nii:v49b}

The $\lambda$4181.10 (4f\,G[9/2]$_{4}$\,--\,3d\,$^3$D$^{\rm o}_{3}$) line is
blended with the N~{\sc ii} M50a $\lambda$4178.86
(4f\,D[5/2]$_{2}$\,--\,3d\,$^3$D$^{\rm o}_{3}$) and M50a $\lambda$4179.67
(4f\,D[5/2]$_{3}$\,--\,3d\,$^3$D$^{\rm o}_{3}$) lines, whose intensity
contribution is less than 40 per cent. [Fe~{\sc v}] $\lambda$4180.60 is
probably also blended (Fig.\,\ref{4176-4260}), but its intensity is uncertain.

\subsubsection{Multiplet 50a,
4f\,D[5/2] -- 3d\,$^3$D$^{\rm o}$:}\label{appendix:nii:v50a}

The $\lambda$4169.38 (4f\,D[5/2]$_{2}$\,--\,3d\,$^3$D$^{\rm o}_{1}$) line is
blended with the O~{\sc ii} M19 $\lambda$4169.22 line, which contributes
$\sim$70 per cent to the total intensity. The $\lambda$4178.86
(4f\,D[5/2]$_{2}$\,--\,3d\,$^3$D$^{\rm o}_{3}$) and $\lambda$4179.67
(4f\,D[5/2]$_{3}$\,--\,3d\,$^3$D$^{\rm o}_{3}$) lines are blended with the
N~{\sc ii} M49b 4f\,2[5]$_{4}$\,--\,3d\,$^3$D$^{\rm o}_{3}$ $\lambda$4181.10
line, which contributes more than 60 per cent to total intensity. The other
two M50a lines $\lambda$4173.57 (4f\,D[5/2]$_{2}$\,--\,3d\,$^3$D$^{\rm
o}_{2}$) and $\lambda$4174.38 (4f\,D[5/2]$_{3}$\,--\,3d\,$^3$D$^{\rm o}_{2}$)
are not observed.

\subsubsection{Multiplet 50b,
4f\,D[3/2] -- 3d\,$^3$D$^{\rm o}$:}\label{appendix:nii:v50b}

The $\lambda$4156.39 (4f\,D[3/2]$_{2}$\,--\,3d\,$^3$D$^{\rm o}_{1}$) and
$\lambda$4157.01 (4f\,D[3/2]$_{1}$\,--\,3d\,$^3$D$^{\rm o}_{1}$) lines are
blended with the O~{\sc ii} M19 $\lambda$4156.53
(3d\,$^4$P$_{3/2}$\,--\,3p\,$^4$P$^{\rm o}_{5/2}$) line, which contributes
$\sim$75 per cent to the total intensity of the blend at $\lambda$4157. The
other M50b lines are not observed.

\subsubsection{Multiplet 55a,
4f\,D[5/2] -- 3d\,$^3$P$^{\rm o}$:}\label{appendix:nii:v55a}

The $\lambda$4442.02 (4f\,D[5/2]$_{2}$\,--\,3d\,$^3$P$^{\rm o}_{1}$) line
is blended with the Ne~{\sc ii} M60b
4f\,1[4]$^{\rm o}_{7/2}$\,--\,3d\,$^2$F$_{5/2}$ $\lambda$4442.69 and O~{\sc
iii} M49b 5g\,F[3]$^{\rm o}_{2,\,3}$\,--\,4f\,D[3]$_{2}$ $\lambda$4442.02
lines. The Ne~{\sc ii} line contributes $\sim$40\,--\,50 per cent to the total
intensity of the blend at $\lambda$4442, while the contribution of the O~{\sc
iii} line is negligible. The corrected intensity of the $\lambda$4442.02 line
is higher than the theoretical prediction (Table\,\ref{relative:nii_4f-3d}).
The other two M55a lines, $\lambda$4431.82
(4f\,D[5/2]$_{2}$\,--\,3d\,$^3$P$^{\rm o}_{2}$) and $\lambda$4432.74
(4f\,D[5/2]$_{3}$\,--\,3d\,$^3$P$^{\rm o}_{2}$) are blended with several weak
Ne~{\sc ii} line of the 4f\,--\,3d transition array, whose effective
recombination coefficients are unavailable.

\subsubsection{Multiplet 58a,
4f\,G[7/2] -- 3d\,$^1$F$^{\rm o}$:}\label{appendix:nii:v58a}

The $\lambda$4552.53 (4f\,G[7/2]$_{4}$\,--\,3d\,$^1$F$^{\rm o}_{3}$) line is
blended with the Ne~{\sc ii} M55d 4f\,2[2]$^{\rm
o}_{5/2}$\,--\,3d\,$^4$F$_{3/2}$ $\lambda$4553.17 and Si~{\sc iii} M2
4p\,$^3$P$^{\rm o}_{2}$\,--\,4s\,$^3$S$_{1}$ $\lambda$4552.62 lines, whose
intensities are unknown. The other M58a line $\lambda$4554.10
(4f\,G[7/2]$_{3}$\,--\,3d\,$^1$F$^{\rm o}_{3}$) is not observed.

\subsubsection{Multiplet 58b,
4f\,G[9/2] -- 3d\,$^1$F$^{\rm o}$:}\label{appendix:nii:v58b}

The $\lambda$4530.41 (4f\,G[9/2]$_{4}$\,--\,3d\,$^1$F$^{\rm o}_{3}$) line is
blended with the N~{\sc iii} M3
3p$^{\prime}$\,$^4$D$_{1/2}$\,--\,3s$^{\prime}$\,$^4$P$^{\rm
o}_{3/2}$ $\lambda$4530.86 (Fig.\,\ref{4505-4555}), which contributes $\sim$12
per cent to the total intensity. The intensity of the $\lambda$4530.41 line
is 0.049$\pm$0.007, which agrees with the theoretical prediction
(Table\,\ref{relative:nii_4f-3d}).

\subsubsection{Multiplet 61a,
4f\,D[5/2] -- 3d\,$^1$P$^{\rm o}$:}\label{appendix:nii:v61a}

The $\lambda$4694.64 (4f\,D[5/2]$_{2}$\,--\,3d\,$^1$P$^{\rm o}_{1}$) line is
partially blended with the weak O~{\sc ii} M1 $\lambda$4696.35 (3p\,$^4$D$^{\rm
o}_{3/2}$\,--\,3s\,$^4$P$_{5/2}$) line. The $\lambda$4694.64 line has an
intensity that is 50 per cent higher than what is predicted
(Table\,\ref{relative:nii_4f-3d}). The intensity of this line is unreliable
due to the weakness.

\section{The O~{\sc ii} optical recombination spectrum}\label{appendix:c}

\subsection{Multiplet 1,
3p\,$^4$D$^{\rm o}$ -- 3s\,$^4$P}\label{appendix:oii:v1}

See Section\,\ref{oii_orls:v1}.

\subsection{Multiplet 2,
3p\,$^4$P$^{\rm o}$ -- 3s\,$^4$P}\label{appendix:oii:v2}

\begin{figure}
\begin{center}
\includegraphics[width=7.5cm,angle=-90]{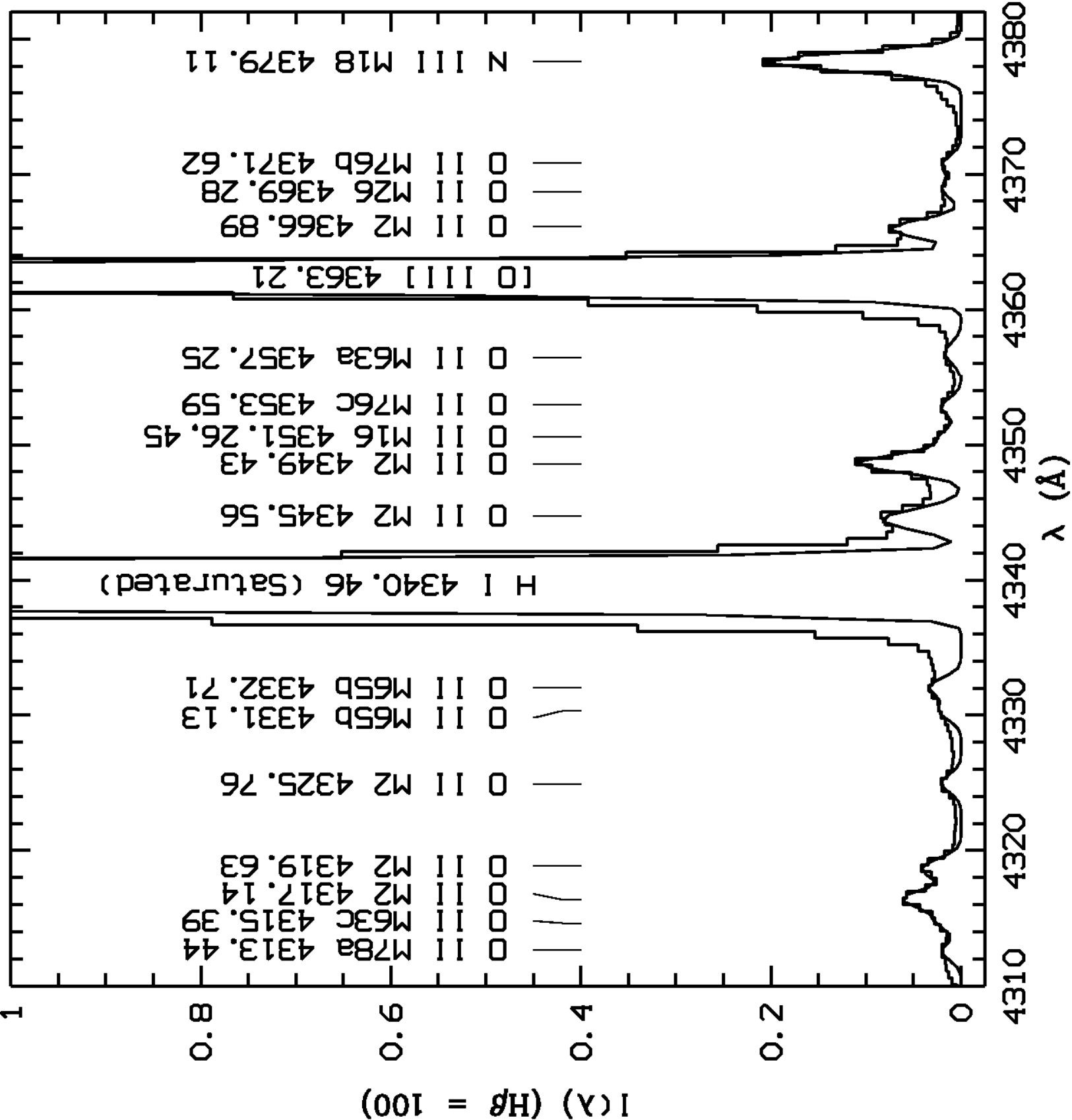}
\caption{Spectrum of NGC\,7009 from 4310 to 4382\,{\AA} showing the O~{\sc ii}
M2 lines and other O~{\sc ii} ORLs. The continuous curve, which is the sum
of Gaussian profiles, only shows the positions and possible profiles of some
weak O~{\sc ii} lines whose accurate measurements are difficult due to the
strong [O~{\sc iii}] $\lambda$4363 and the saturated H~{\sc i} $\lambda$4340
lines, both of which are assumed to be Gaussian. Global continuum has been
subtracted and the spectrum has been normalized such that H$\beta$ has an
integrated flux of 100. Extinction has not been corrected for.}
\label{4310-4382}
\end{center}
\end{figure}

The observed and predicted relative intensities of O~{\sc ii} M2 lines with
most reliable measurements are presented in Table\,\ref{relative:oii_v2}. The
features of these lines as well as results of Gaussian profile fitting are
shown in Fig.\,\ref{4310-4382}. $\lambda$4349.43 is clearly observed. A
single Gaussian profile fit to the feature of $\lambda$4349.43 gives an
intensity of 0.195$\pm$0.029. The intensity ratio of the $\lambda$4349.43 line
to the O~{\sc ii} M1 $\lambda$4649.13 line is 0.292, slightly higher than the
predicted value 0.274. The intensity of the $\lambda$4349.43 line could be
overestimated due to the saturated H~{\sc i} $\lambda$4340 line
(Fig.\,\ref{4310-4382}). The O~{\sc ii} M16 3p$^{\prime}$\,$^2$D$^{\rm
o}_{5/2}$ -- 3s$^{\prime}$\,$^2$D$_{5/2}$ $\lambda$4351.26 and
3p$^{\prime}$\,$^2$D$^{\rm o}_{5/2}$ -- 3s$^{\prime}$\,$^2$D$_{3/2}$
$\lambda$4351.45 lines are also blended with the $\lambda$4349.43 line, but
their contributions are probably insignificant.

Another three M2 lines, $\lambda\lambda$4317.14, 4319.63 and 4325.76, are less
affected by the saturated H~{\sc i} $\lambda$4340 line. $\lambda$4317.14 blends
with O~{\sc ii} M53a 4f~D[3]$^{\rm o}_{5/2}$ -- 3d~$^4$P$_{3/2}$
$\lambda$4317.70, and this blended feature is partially resolved from
$\lambda$4319.63 (Fig.\,\ref{4310-4382}). Three O~{\sc ii} M63c and two O~{\sc
ii} M78b lines are blended with the left wing of $\lambda$4317.14, and we
treated these five O~{\sc ii} lines as a single component since they are all
weak and have close wavelengths. Three Gaussian profiles were used to fit the
complex. The resultant intensities of $\lambda$4317.14 and $\lambda$4319.63
both agree better with the newly predicted values
(Table\,\ref{relative:oii_v2}). The intensity contribution of the blended
O~{\sc ii} M53a $\lambda$4317.70 line was estimated from the most recent
effective recombination coefficients and was subtracted. $\lambda$4325.76 is
free of blend, and its fitted intensity (with an uncertainty of 10--20 per
cent) also agrees better with the intermediate coupling prediction. The main
source of uncertainties in the $\lambda$4325.76 intensity is due to the
saturated H~{\sc i} $\lambda$4340 line, while those in $\lambda\lambda$4317.14
and 4319.63 are mainly due to blending.

Measurements of the remaining three M2 lines, $\lambda\lambda$4336.86, 4345.56
and 4366.89, are difficult: the former two are blended with the saturated
H~{\sc i} $\lambda$4340 line, and the latter one is affected by [O~{\sc iii}]
$\lambda$4363 (Fig.\,\ref{4310-4382}). The measurement of $\lambda$4366.89
seems to agree with the predicted intensity (Table\,\ref{relative:oii_v2}).
The fitting error of the $\lambda$4363 intensity is 12 per cent, but the actual
uncertainty could be even larger. The contributions of O~{\sc iii} M39a
5g~G[4]$^{\rm o}_{3,4}$ -- 4f~F[3]$_{3}$ $\lambda$4366.99 and O~{\sc ii} M75a
4f~D[3]$^{\rm o}_{7/2}$ -- 3d~2F$_{5/2}$ $\lambda$4366.53 were assumed to be
negligible.

\begin{table}
\centering
\caption{Comparison of the observed and predicted relative intensities of
the O~{\sc ii} M2 lines detected in the spectrum of NGC\,7009. Only lines
with the most reliable measurements are presented. Here an electron
temperature of 1000~K and a density of 4300~cm$^{-3}$ are assumed for the
theoretical intensities $I_{\rm IC}$.}
\label{relative:oii_v2}
\begin{tabular}{lcccccc}
\hline
Line & $J_2-J_1$ & $I_{\rm LS}$ & $I_{\rm IC}$ & $I_{\rm obs}$ &
$\frac{I_{\rm obs}}{I_{\rm LS}}$ & $\frac{I_{\rm obs}}{I_{\rm IC}}$\\
\hline
$\lambda$4317.14$^a$ & 3/2--1/2 & 0.397 & 0.553 & 0.551 & 1.388 & 0.996\\
$\lambda$4319.63     & 5/2--3/2 & 0.429 & 0.365 & 0.322 & 0.750 & 0.882\\
$\lambda$4325.76     & 1/2--1/2 & 0.079 & 0.119 & 0.147 & 1.861 & 1.235\\
$\lambda$4349.43     & 5/2--5/2 & 1.000 & 1.000 & 1.000 & 1.000 & 1.000\\
$\lambda$4366.89     & 3/2--5/2 & 0.429 & 0.578 & 0.640 & 1.492 & 1.107\\
\hline
\end{tabular}
\begin{description}
\item [$^a$] Corrected for the contribution from O~{\sc ii} M53a 4f~D[3]$^{\rm
o}_{5/2}$ -- 3d~$^{4}$P$_{3/2}$ $\lambda$4317.70, which is about 15 per cent.
\end{description}
\end{table}

\subsection{Multiplet 5,
3p\,$^2$D$^{\rm o}$ -- 3s\,$^2$P}\label{appendix:oii:v5}

\begin{figure}
\begin{center}
\includegraphics[width=7.5cm,angle=-90]{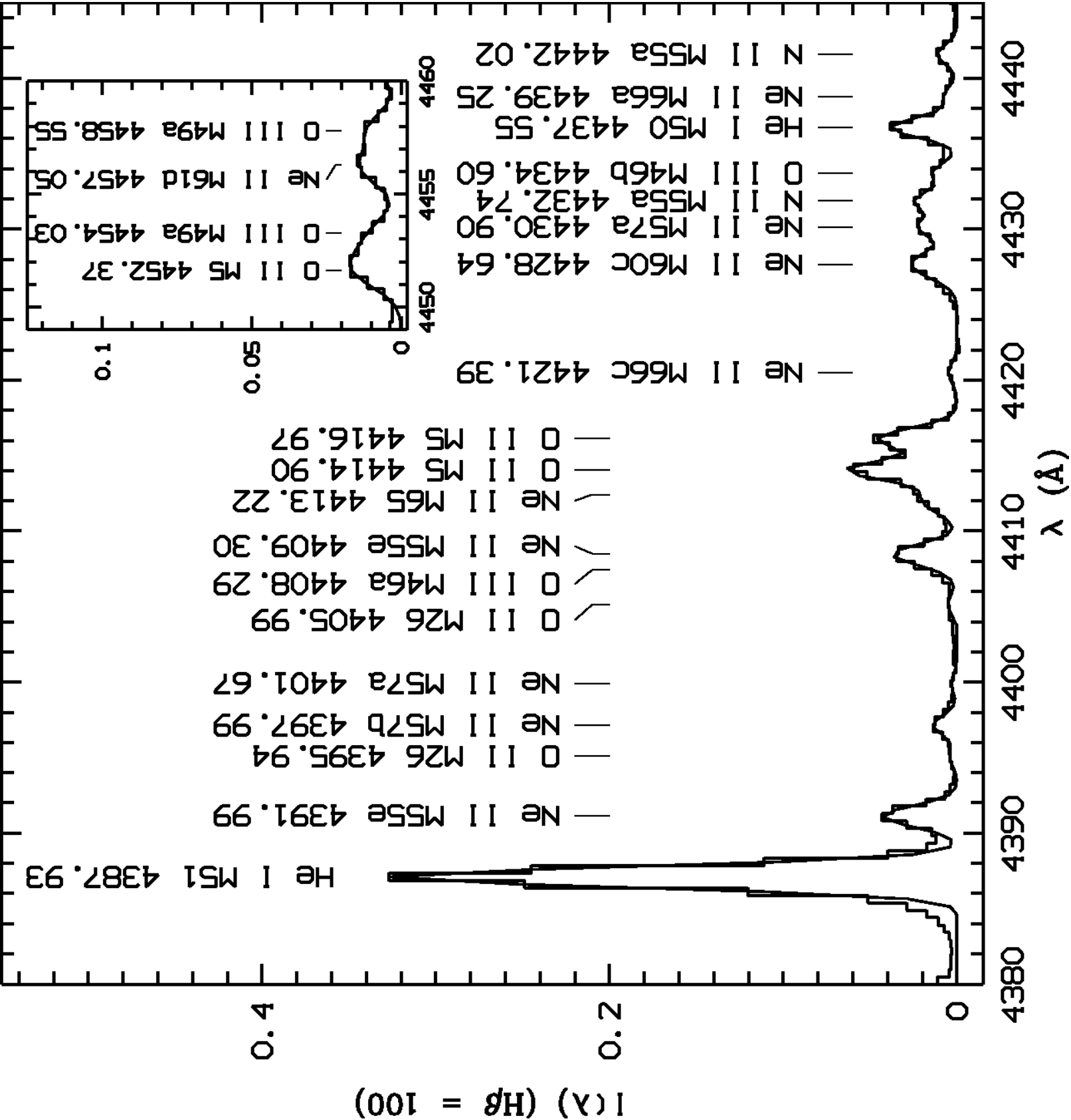}
\caption{Spectrum of NGC\,7009 from 4380 to 4445\,{\AA} showing the O~{\sc ii}
M5 lines and other emission features. Many Ne~{\sc ii} lines of the 4f\,--\,3d
configuration are also detected. The inset shows the spectral region
4449--4460\,{\AA}, where the O~{\sc ii} M5 $\lambda$4452.37 line is located.
The continuous curve is the sum of Gaussian profile fits. Continuum has been
subtracted and the spectrum has been normalized such that H$\beta$ has an
integrated flux of 100. Extinction has not been corrected for.}
\label{4380-4445}
\end{center}
\end{figure}

Accurate measurements of the three M5 lines are difficult. $\lambda$4414.90
(3p\,$^2$D$^{\rm o}_{5/2}$\,--\,3s\,$^2$P$_{3/2}$) is partially resolved
from another M5 line $\lambda$4416.97 (3p\,$^2$D$^{\rm
o}_{3/2}$\,--\,3s\,$^2$P$_{1/2}$), as is shown in Fig.\,\ref{4380-4445}.
$\lambda$4414.90 coincides in wavelength O~{\sc iii} M46a 5g\,H[5]$^{\rm
o}_{5}$\,--\,4f\,G[4]$_{4}$ $\lambda$4414.85, whose contribution to the
total flux of the blend at $\lambda$4415 was estimated from the effective
recombination coefficients of Kisielius \& Storey \cite{ks1999}.
$\lambda$4414.90 is also affected by at
least three weak Ne~{\sc ii} lines from the 4f\,--\,3d configuration, and for
most these Ne~{\sc ii} lines, the effective recombination coefficients are
not available. The $\lambda$4416.97 line is blended with N~{\sc ii} M55b
4f\,D[3/2]$_{2}$\,--\,3d\,$^3$P$^{\rm o}_{2}$ $\lambda$4417.07, N~{\sc ii}
M55b 4f\,D[3/2]$_{1}$\,--\,3d\,$^3$P$^{\rm o}_{2}$ $\lambda$4417.82, Ne~{\sc
ii} M61d 4f\,2[2]$^{\rm o}_{5/2}$\,--\,3d\,$^2$D$_{5/2}$ $\lambda$4416.76
and Ne~{\sc ii} M61d 4f\,2[2]$^{\rm o}_{3/2}$\,--\,3d\,$^2$D$_{5/2}$
$\lambda$4416.77. The contributions of the two N~{\sc ii} lines to the total
flux of the blend at $\lambda$4417 were estimated from the new N~{\sc ii}
effective recombination coefficients, while those of the two Ne~{\sc ii}
lines could be negligible. The other line $\lambda$4452.37 (3p\,$^2$D$^{\rm
o}_{3/2}$\,--\,3s\,$^2$P$_{3/2}$) is blended with O~{\sc iii} M49a
5g\,F[4]$^{\rm o}_{3,4}$\,--\,4f\,D[3]$_{3}$ $\lambda$4454.03 and [Fe~{\sc
ii}] $\lambda$4452.10 (Fig.\,\ref{4380-4445}). The contribution from the
O~{\sc iii} $\lambda$4454.03 line was estimated from the effective
recombination coefficients of Kisielius \& Storey \cite{ks1999}, but that
of the [Fe~{\sc ii}] line is uncertain.

\subsection{Multiplet 6,
3p\,$^2$P$^{\rm o}$ -- 3s\,$^2$P}\label{appendix:oii:v6}

\begin{figure}
\begin{center}
\includegraphics[width=7.5cm,angle=-90]{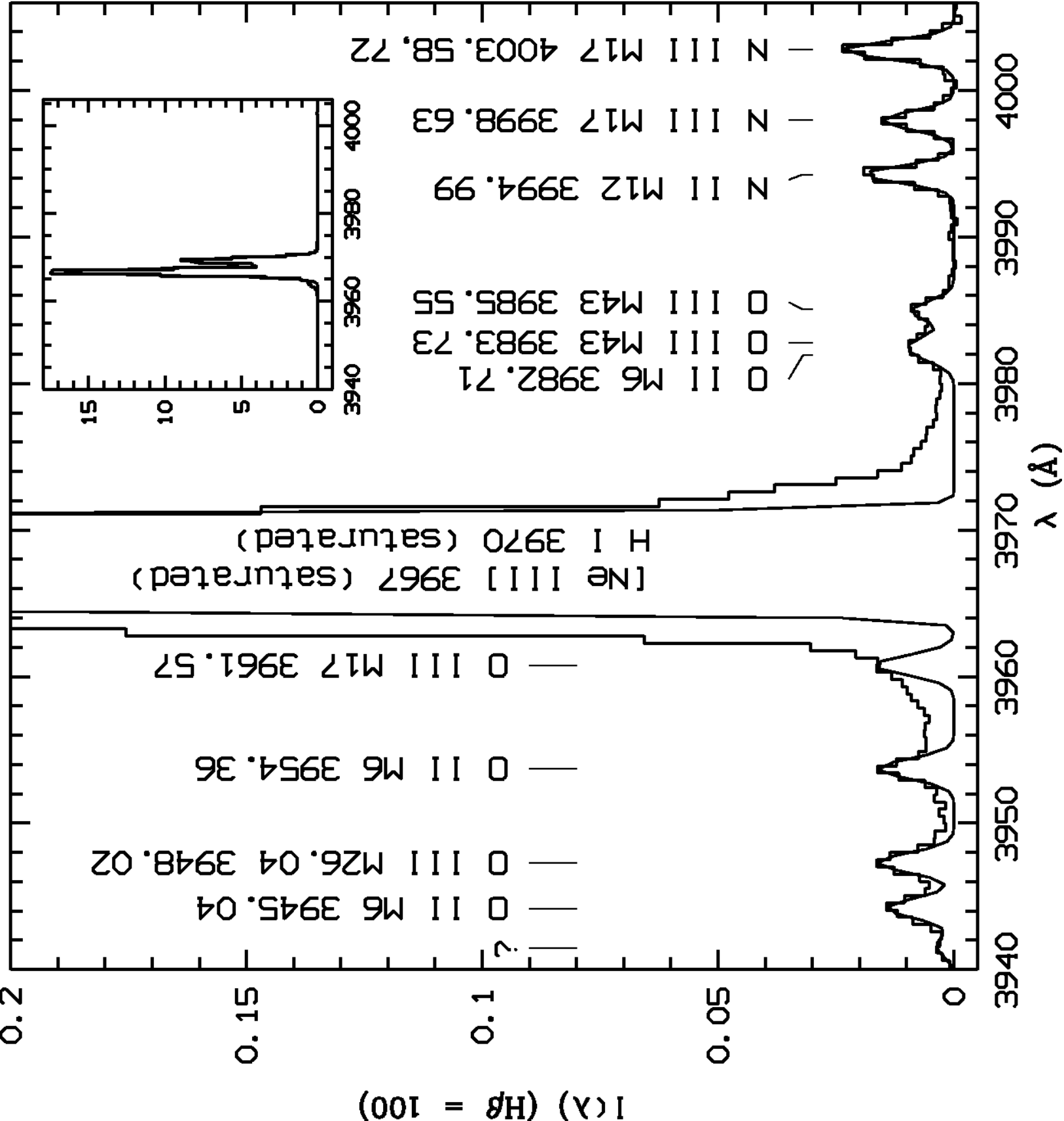}
\caption{Spectrum of NGC\,7009 from 3940 to 4006\,{\AA} showing the O~{\sc ii}
M6 lines and other emission features. The continuous curve is the sum of
Gaussian profile fits. The profile of the O~{\sc ii} M17 $\lambda$3961.57
line, whose accurate measurement is impossible due to the saturated [Ne~{\sc
iii}] $\lambda$3967 line, is only an estimate. The inset shows the profiles
of the [Ne~{\sc iii}] $\lambda$3967 and H~{\sc i} $\lambda$3970 lines, both
of which are saturated but assumed to be Gaussian. Continuum has been
subtracted and the spectrum has been normalized such that H$\beta$ has an
integrated flux of 100. Extinction has not been corrected for.}
\label{3940-4006}
\end{center}
\end{figure}

The strongest M6 line $\lambda$3973.26 (3p~$^2$P$^{\rm o}_{3/2}$ --
3s~$^2$P$_{3/2}$) is blended with the saturated H~{\sc i} $\lambda$3970 line
(Fig.\,\ref{3940-4006}). $\lambda\lambda$3945.04 (3p~$^2$P$^{\rm o}_{3/2}$ --
3s~$^2$P$_{1/2}$) and 3954.36 (3p~$^2$P$^{\rm o}_{1/2}$ -- 3s~$^2$P$_{1/2}$)
are free of blend, but are affected by the saturated [Ne~{\sc iii}]
$\lambda$3967 line (Fig.\,\ref{3940-4006}). The measured intensities of
$\lambda$3945.04 and $\lambda$3954.36 are higher than the predicted values (in
IC) by 45 and 83 per cent, respectively. Their measurements are of large
uncertainties ($\gtrsim$\,30 per cent).

Accurate measurements of the other M6 line $\lambda$3982.71 (3p~$^2$P$^{\rm
o}_{1/2}$ -- 3s~$^2$P$_{3/2}$) are difficult due to the saturated H~{\sc i}
$\lambda$3970 line. It is also partially resolved from a weak feature that was
identified as O~{\sc iii} M43 3d$^{\prime}$~$^3$D$_{1}$ -- 4d~$^3$P$^{\rm
o}_{1}$ $\lambda$3985.55. If this identification is correct, $\lambda$3982.71
should blend with another O~{\sc iii} M43 line $\lambda$3983.73. Unfortunately
no effective recombination coefficients for O~{\sc iii} M43 are available for
us to confirm the conjecture.

\subsection{Multiplet 10,
3d\,$^4$F -- 3p\,$^4$D$^{\rm o}$}\label{appendix:oii:v10}

See Section\,\ref{oii_orls:v10}.

\subsection{Multiplet 12,
3d\,$^4$D -- 3p\,$^4$D$^{\rm o}$}\label{appendix:oii:v12}

\begin{figure}
\begin{center}
\includegraphics[width=7.5cm,angle=-90]{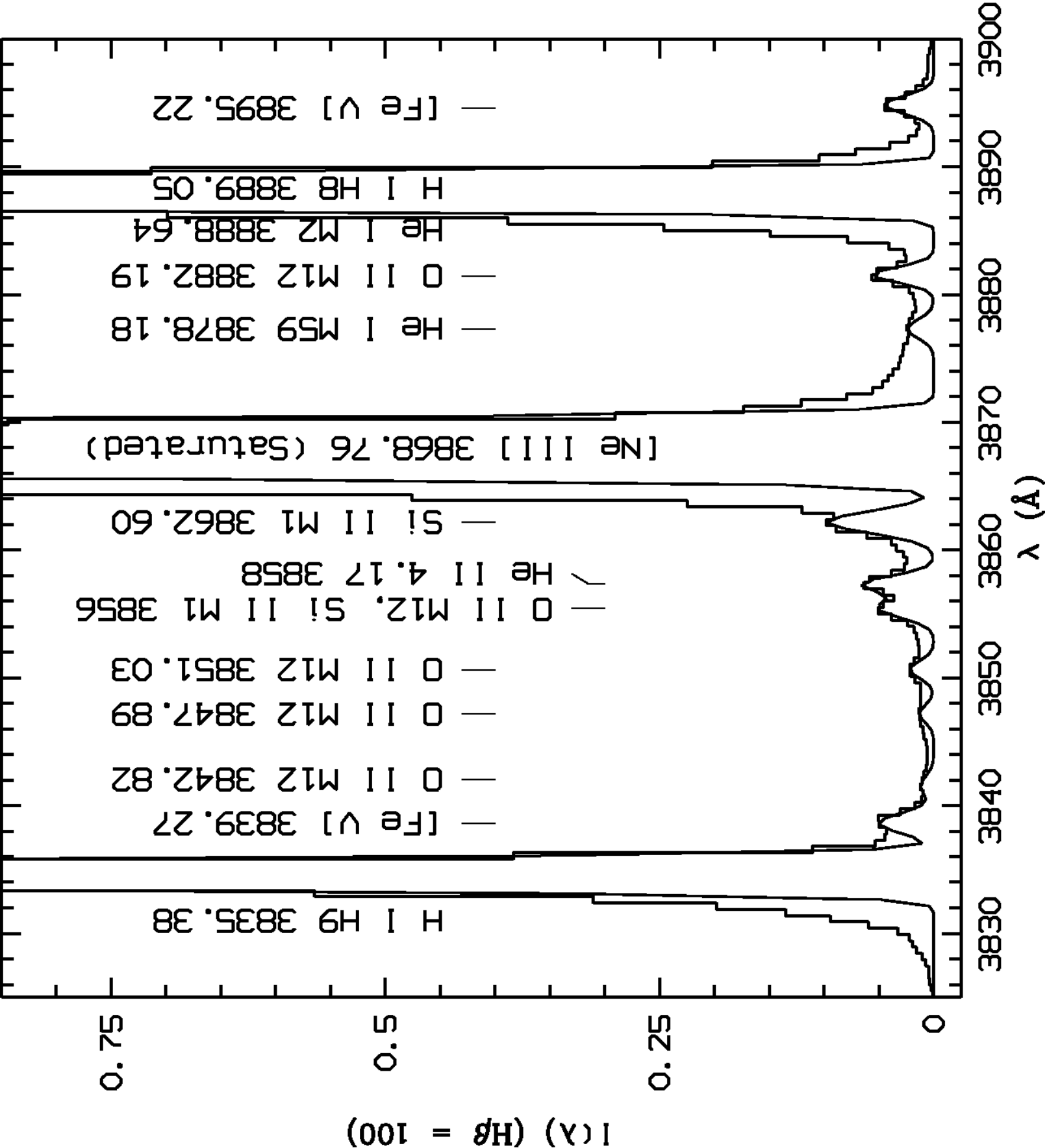}
\caption{Spectrum of NGC\,7009 from 3825 to 3900\,{\AA} showing the O~{\sc ii}
M12 features. The continuous curve, which is the sum of Gaussian profiles, only
shows the positions and possible profiles of the weak lines whose accurate
measurements are difficult due to the strong H~{\sc i} $\lambda$3889 and the
saturated [Ne~{\sc iii}] $\lambda$3869 lines. Global continuum has been
subtracted and the spectrum has been normalized such that H$\beta$ has an
integrated flux of 100. Extinction has not been corrected for.}
\label{3825-3900}
\end{center}
\end{figure}

The features of O~{\sc ii} M12 lines are shown in Fig.\,\ref{3825-3900}.
$\lambda$3882.19 (3d~$^{4}$D$_{7/2}$ -- 3p~$^{4}$D$^{\rm o}_{7/2}$) is close
to H~{\sc i} $\lambda$3889, which may affect its measurement. A single-Gaussian
profile fit yields an intensity value of 0.084$\pm$0.005 for the
$\lambda$3882.19 line. The intensity of the
$\lambda$3882 line, which is more than 50 per cent higher than the predicted
value, is overestimated due to the saturated [Ne~{\sc iii}] $\lambda$3868. The
other M12 lines are either much affected by the saturated [Ne~{\sc iii}] line
or are not observed due to weakness.

\subsection{Multiplet 19,
3d\,$^4$P -- 3p\,$^4$P$^{\rm o}$}\label{appendix:oii:v19}

\begin{figure}
\begin{center}
\includegraphics[width=7.5cm,angle=-90]{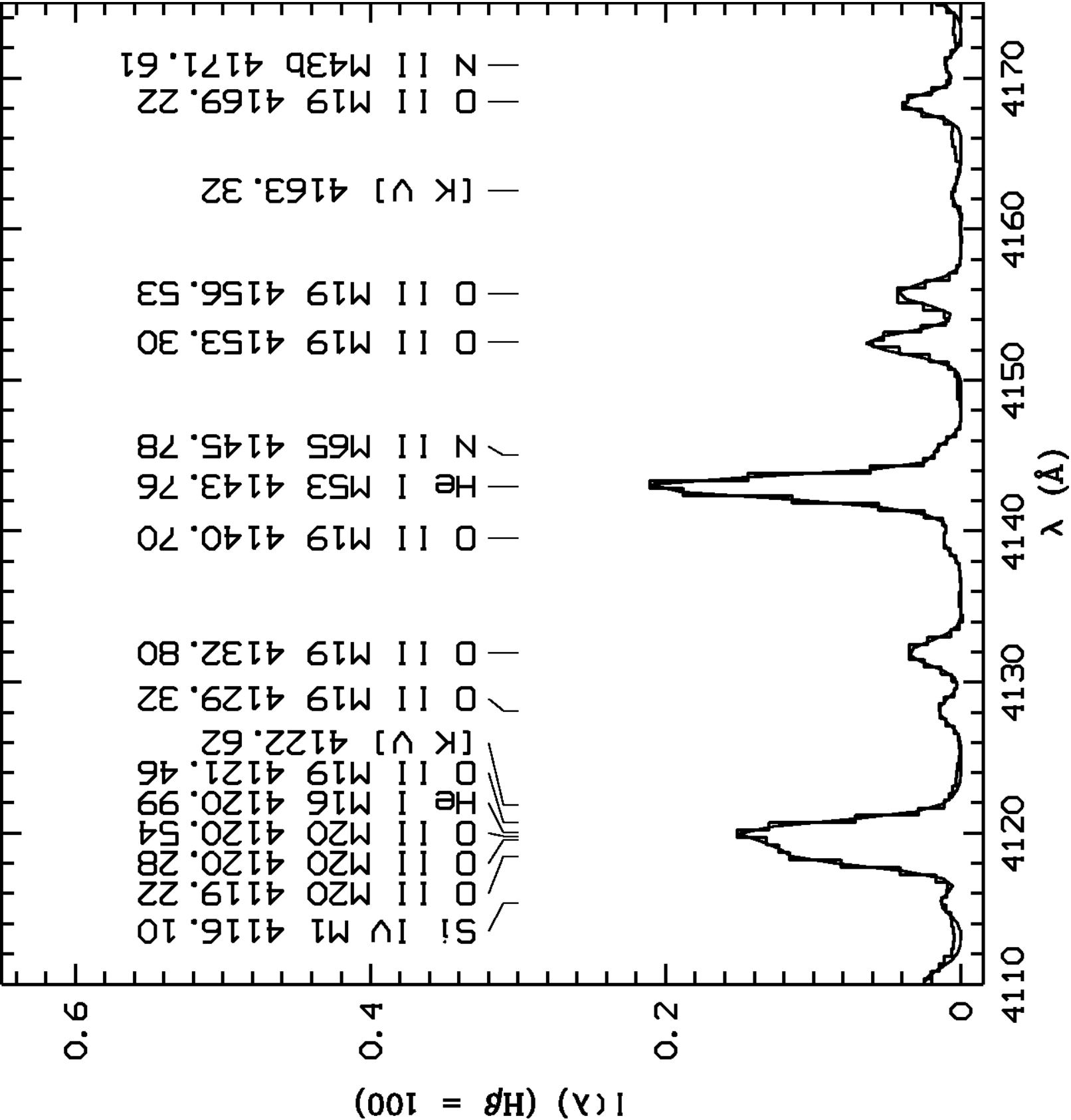}
\caption{Spectrum of NGC\,7009 from 4110 to 4175\,{\AA} showing the O~{\sc ii}
M19 lines and other emission features. The continuous curve is the sum of
Gaussian profile fits. Continuum has been subtracted and the spectrum has been
normalized such that H$\beta$ has an integrated flux of 100. Extinction has
not been corrected for.}
\label{4110-4175}
\end{center}
\end{figure}

The O~{\sc ii} M19 lines are observed in Fig.\,\ref{4110-4175}, and the
observed and predicted relative intensities are presented in
Table\,\ref{relative:oii_v19}. $\lambda$4169.22 blends with He~{\sc i} M52
6s~$^1$S$_{0}$ -- 2p~$^1$P$^{\rm o}_{1}$ $\lambda$4168.97 and N~{\sc ii} M50a
4f~2[3]$_{2}$ -- 3d~$^3$D$^{\rm o}_{1}$ $\lambda$4169.38. The He~{\sc i}
$\lambda$4168.97 line contributes 30 per cent to the total intensity, which
was estimated from the theoretical He~{\sc i} line intensities of Benjamin,
Skillman \& Smits \citealt{bss99}, while the N~{\sc ii} line was assumed to be
negligible. The resultant intensity of the $\lambda$4169.22 line is
0.082$\pm$0.008. The predicted
intensity of this line is 0.090, which is derived from the most recent O~{\sc
ii} effective recombination coefficients (the measured intensity of O~{\sc ii}
M1 $\lambda$4649.13 was used). The measured intensity ratio of $\lambda$4153.30
to $\lambda$4169.22 is much higher than the predicted value. No explanation can
be given for this large discrepancy except there is unknown blend.

Although the measured intensity of
$\lambda$4156.53 is also higher than theory, its actual intensity should agree
well with the predicted value, given the fact that it is blended with N~{\sc
ii} M50b 4f~D[3/2]$_{2}$ -- 3d~$^3$D$^{\rm o}_{1}$ $\lambda$4156.39 and N~{\sc
ii} M50b 4f~D[3/2]$_{1}$ -- 3d~$^3$D$^{\rm o}_{1}$ $\lambda$4157.01, which
contribute about 25 per cent to the total flux of the blend at $\lambda$4156.
The measurements of another two M19 lines $\lambda\lambda$4129.32 and 4132.80
both agree with the newly predicted intensities. These two lines blend with
three Si~{\sc ii} M3 4f~$^2$F$^{\rm o}$ -- 3d~$^2$D lines
$\lambda\lambda$4128.07, 4130.87 and 4130.89, whose contributions to the total
flux are negligible. $\lambda$4140.70 is overestimated due to the strength of
He~{\sc i} M53 6d~$^1$D$_{2}$ -- 2p~$^1$P$^{\rm o}_{1}$ $\lambda$4143.76, which
is more than 20 times stronger. Accurate measurements of the other line
$\lambda$4121.46 is impossible because it blends with He~{\sc i} M16 5s~$^3$S
-- 2p~$^3$P$^{\rm o}$ $\lambda$4120.99 and three O~{\sc ii} M20 lines
(Fig.\,\ref{4110-4175}).

\begin{table}
\centering
\caption{Same as Table\,\ref{relative:oii_v2} but for the O~{\sc ii} M1 lines
detected in the spectrum of NGC\,7009.}
\label{relative:oii_v19}
\begin{tabular}{lccclcc}
\hline
Line & $J_2-J_1$ & $I_{\rm LS}$ & $I_{\rm IC}$ & $I_{\rm obs}$ &
$\frac{I_{\rm obs}}{I_{\rm LS}}$ & $\frac{I_{\rm obs}}{I_{\rm IC}}$\\
\hline
$\lambda$4129.32     & 1/2--3/2 & 0.397 & 0.335 & 0.389 & 0.981 & 1.162\\
$\lambda$4132.80$^a$ & 3/2--1/2 & 0.397 & 0.768 & 0.918 & 2.311 & 1.195\\
$\lambda$4140.70$^b$ & 3/2--3/2 & 0.127 & 0.028 & 0.040 & 0.316 & 1.415\\
$\lambda$4153.30     & 5/2--3/2 & 0.429 & 0.909 & 1.363 & 3.180 & 1.500\\
$\lambda$4156.53$^c$ & 3/2--5/2 & 0.429 & 0.474 & 0.688 & 2.229 & 2.017\\
$\lambda$4169.22     & 5/2--5/2 & 1.000 & 1.000 & 1.000 & 1.000 & 1.000\\
\hline
\end{tabular}
\begin{description}
\item [$^a$] Probably including unknown blend.
\item [$^b$] Overestimated due to He~{\sc i} M53 He~{\sc i} M53 6d~$^1$D$_{2}$
-- 2p~$^1$P$^{\rm o}_{1}$ $\lambda$4143.76, which is stronger by early two
orders of magnitude.
\item [$^c$] Neglecting N~{\sc ii} M50b 4f~D[3/2]$_{2}$ -- 3d~$^3$D$^{\rm
o}_{1}$ $\lambda$4156.39 and N~{\sc ii} M50b 4f~D[2]$_{1}$ -- 3d~$^3$D$^{\rm
o}_{1}$ $\lambda$4157.01, which contribute about 4 per cent to the total
intensity.
\end{description}
\end{table}

\subsection{Multiplet 20,
3d\,$^4$D -- 3p\,$^4$P$^{\rm o}$}\label{appendix:oii:v20}

Three O~{\sc ii} M20 lines $\lambda\lambda$4119.22, 4120.28 and 4120.54 blend
with He~{\sc i} M16 5s~$^3$S -- 2p~$^3$P$^{\rm o}$ $\lambda$4120.99 and O~{\sc
ii} M19 $\lambda$4121.46. Accurate measurements of the lines are difficult.
Three Gaussian profiles (the He~{\sc i} $\lambda$4120.99 and O~{\sc ii} M19
$\lambda$4121.46 lines are treated as a single component, and the two O~{\sc
ii} $\lambda$4120 lines are treated as another one) were used to fit the
feature. The intensity ratio of the three O~{\sc ii} M20 lines
($\lambda$4120.28~+~$\lambda$4120.54)/$\lambda$4119.22 were assumed to be the
predicted value (0.278), and the intensity of the He~{\sc i} $\lambda$4120.99
line was estimated from the atomic data of Benjamin, Skillman \& Smits
\cite{bss99}. The fitting results agree well with the observed spectrum, as
is shown in Fig.\,\ref{4110-4175}. Accurate measurements of $\lambda$4110.79
are also difficult due to the strong H~{\sc i} $\lambda$4101
(Fig.\,\ref{4060-4115}). The other M20 lines are not observed.

\subsection{Multiplet 25,
3d\,$^2$F -- 3p\,$^2$D$^{\rm o}$}\label{appendix:oii:v25}

\begin{figure}
\begin{center}
\includegraphics[width=7.5cm,angle=-90]{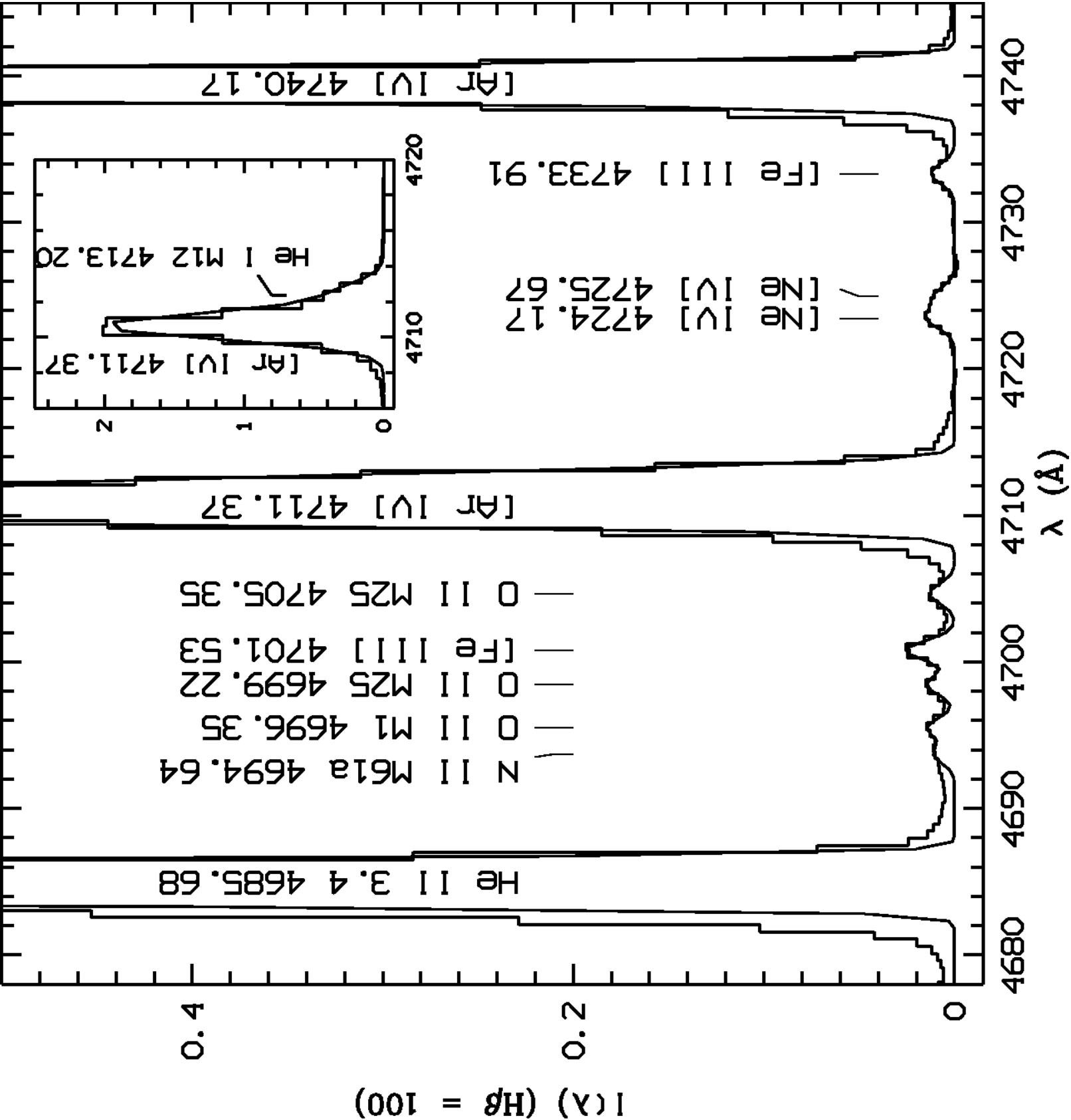}
\caption{Spectrum of NGC\,7009 from 4678 to 4745\,{\AA} showing the O~{\sc ii}
M25 lines and other emission features. The inset shows the [Ar~{\sc iv}]
$\lambda$4711 line blends with He~{\sc i} M12 $\lambda$4713.20, which is 6
times fainter. The continuous curve is the sum of Gaussian profile fits.
Continuum has been subtracted and the spectrum has been normalized such that
H$\beta$ has an integrated flux of 100. Extinction has not been corrected
for.}
\label{4678-4745}
\end{center}
\end{figure}

Fig.\,\ref{4678-4745} shows two O~{\sc ii} M25 lines $\lambda\lambda$4699.22
(5/2 -- 3/2) and 4705.35 (7/2 -- 5/2). $\lambda$4699.22 is partially resolved
from [Fe~{\sc iii}] $\lambda$4701.53, and it also blends with O~{\sc ii} M40
3d$^{\prime}$~$^2$F$_{7/2}$ -- 3p$^{\prime}$~$^2$D$^{\rm o}_{5/2}$
$\lambda$4699.00 which contributes about 30 per cent to the total flux of the
blend at $\lambda$4699. $\lambda$4705.35 is affected by the much stronger
[Ar~{\sc iv}] $\lambda$4711. The measured intensity ratio
$\lambda$4699.22\,:\,$\lambda$4705.35 is 0.779\,:\,1.00, which agrees with the
newly predicted ratio 0.765\,:\,1, and differs from the value in pure {\it
LS}\,coupling, i.e. 0.7\,:\,1.0. Here the contribution of the blended O~{\sc
ii} M40 $\lambda$4699.00 line has been subtracted.

%

\subsection{Multiplet 28,
3d\,$^4$P -- 3p\,$^4$S$^{\rm o}$}\label{appendix:oii:v28}

\begin{figure}
\begin{center}
\includegraphics[width=7.5cm,angle=-90]{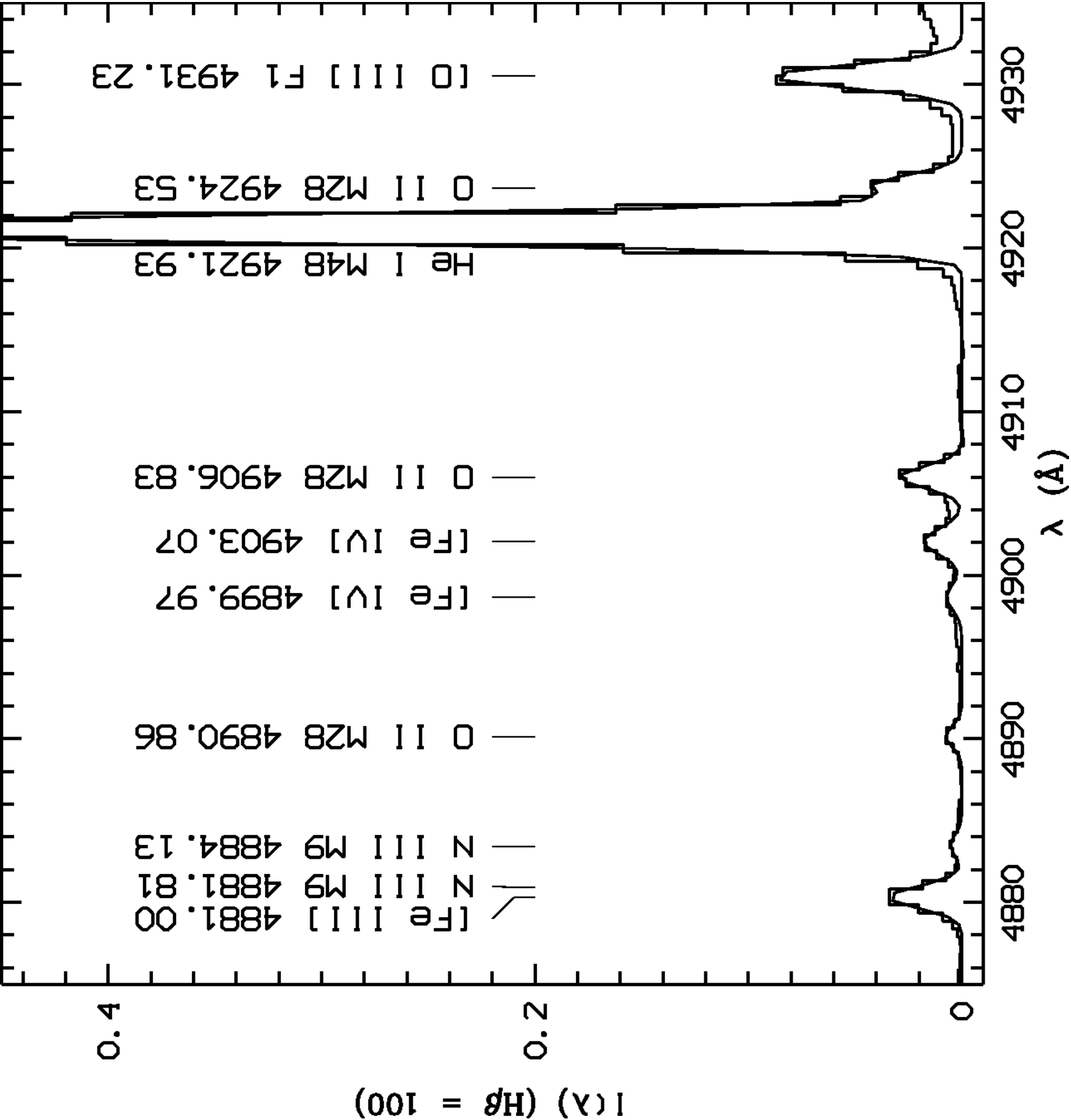}
\caption{Spectrum of NGC\,7009 from 4875 to 4935\,{\AA} showing the O~{\sc
ii} M28 lines and other emission features. The continuous curve is the sum
of Gaussian profile fits. Continuum has been subtracted and the spectrum has
been normalized such that H$\beta$ has an integrated flux of 100. Extinction
has not been corrected for.}
\label{4875-4935}
\end{center}
\end{figure}

The three M28 lines are shown in Fig.\,\ref{4875-4935}. $\lambda$4924.53
(5/2 -- 3/2) is partially resolved from He~{\sc i} M48 4d~$^1$D$_{2}$ --
2p~$^1$P$^{\rm o}_{1}$ $\lambda$4921.93, and the other two $\lambda$4890.86
(1/2 -- 3/2) and $\lambda$4906.83 (3/2 -- 3/2) are free of blend. The observed
relative intensity of the three lines is 1.0\,:\,3.52\,:\,5.73, which deviates
from the pure {\it LS}\,coupling ratio 1\,:\,2\,:\,3, and also differs from
the newly predicted ratio in the intermediate coupling scheme is
3.49\,:\,2.09\,:\,1.0.

\subsection{4f -- 3d transitions}\label{appendix:oii:4f-3d}

\subsubsection{Multiplet 48a,
4f\,G[5]$^{\rm o}$ -- 3d\,$^4$F}\label{appendix:oii:v48a}

See Section\,\ref{oii_orls:4f-3d}.

\subsubsection{Multiplet 48b,
4f\,G[4]$^{\rm o}$ -- 3d\,$^4$F}\label{appendix:oii:v48b}

$\lambda$4083.90 (4f~G[4]$^{\rm o}_{7/2}$ -- 3d~$^4$F$_{5/2}$) blends with
O~{\sc ii} M10 3d~$^4$F$_{5/2}$ -- 3p~$^4$D$^{\rm o}_{5/2}$ $\lambda$4085.11,
which contributes more than 50 per cent to the total intensity. The other lines
are not observed.

\subsubsection{Multiplet 48c,
4f\,G[3]$^{\rm o}$ -- 3d\,$^4$F}\label{appendix:oii:v48c}

$\lambda$4087.15 (4f~G[3]$^{\rm o}_{5/1}$ -- 3d~$^4$F$_{3/2}$) lies between
the features of O~{\sc ii} M48a $\lambda$4089.29 and O~{\sc ii} M10
$\lambda$4085.11. Its intensity given by the fit agrees with the predicted
value (Table\,\ref{relative:oii_4f-3d}), while the measurement uncertainty is
large due to weakness. Here the contributions of the blended N~{\sc ii} M38a
4f~F[5/2]$_{2}$ -- 3d~$^{3}$F$^{\rm o}_{3}$ $\lambda$4086.83 and N~{\sc ii}
M38a 4f~F[5/2]$_{3}$ -- 3d~$^{3}$F$^{\rm o}_{3}$ $\lambda$4087.30 lines were
assumed to be negligible. The other M48c lines are not observed.

\subsubsection{Multiplet 50a,
4f\,F[4]$^{\rm o}$ -- 3d\,$^4$F}\label{appendix:oii:v50a}

Only $\lambda$4062.94 (4f~F[4]$^{\rm o}_{9/2}$ -- 3d~$^4$F$_{9/2}$) is
observed, but its measurements are quite uncertain due to weakness.

\subsubsection{Multiplet 50b,
4f\,F[3]$^{\rm o}$ -- 3d\,$^4$F}\label{appendix:oii:v50b}

Only $\lambda$4048.21 (4f~F[3]$^{\rm o}_{7/2}$ -- 3d~$^4$F$_{7/2}$) is
observed. $\lambda$4048.21 blends with $\lambda$4047.80 (4f~F[3]$^{\rm
o}_{5/2}$ -- 3d~$^4$F$_{7/2}$), whose intensity was assumed to be negligible.
The measured intensity of $\lambda$4048.21 is slightly higher than the current
prediction, but its uncertainty could be large due to weakness.
$\lambda$4035.07 (4f~F[3]$^{\rm o}_{5/2}$ -- 3d~$^4$F$_{5/2}$) and
$\lambda$4035.46 (4f~F[3]$^{\rm o}_{7/2}$ -- 3d~$^4$F$_{5/2}$) blend with
N~{\sc ii} M39a 4f~G[7/2]$_{3}$ -- 3d~$^3$F$^{\rm o}_{2}$ $\lambda$4035.08
(Section\,\ref{nii:4f-3d:v39a}), which contributes more than 80 per cent to the
total intensity of the blend at $\lambda$4035.

\subsubsection{Multiplet 50c,
4f\,F[2]$^{\rm o}$ -- 3d\,$^4$F}\label{appendix:oii:v50c}

$\lambda$4032.48 (4f~F[2]$^{\rm o}_{5/2}$ -- 3d~$^4$F$_{3/2}$) and
$\lambda$4033.16 (4f~F[2]$^{\rm o}_{3/2}$ -- 3d~$^4$F$_{3/2}$) blend with
O~{\sc ii} M50a 4f~F[4]$^{\rm o}_{7/2}$ -- 3d~$^4$F$_{5/2}$ $\lambda$4032.25,
and measurements are difficult. $\lambda$4041.28 (4f~F[2]$^{\rm o}_{5/2}$ --
3d~$^4$F$_{5/2}$) and $\lambda$4041.95 (4f~F[2]$^{\rm o}_{3/2}$ --
3d~$^4$F$_{5/2}$) blended with N~{\sc ii} M39b 4f~G[9/2]$_{5}$ --
3d~$^3$F$^{\rm o}_{4}$ $\lambda$4041.31, which contributes more than 90 per
cent to the total intensity of the blend at $\lambda$4041. The other line
$\lambda$4054.08 (4f~F[2]$^{\rm o}_{5/2}$ -- 3d~$^4$F$_{7/2}$) is not observed.

\subsubsection{Multiplet 53a,
4f\,D[3]$^{\rm o}$ -- 3d\,$^4$P}\label{appendix:oii:v53a}

Only $\lambda$4303.82 (4f~D[3]$^{\rm o}_{7/2}$ -- 3d~$^4$P$_{5/2}$) is
observed, which is shown in Fig.\,\ref{4260-4310}. It blends with another M53a
line $\lambda$4304.08 (4f~D[3]$^{\rm o}_{5/2}$ -- 3d~$^4$P$_{5/2}$) and O~{\sc
ii} M65a 4f~G[5]$^{\rm o}_{9/2}$ -- 3d~$^4$D$_{7/2}$ $\lambda$4303.61, whose
total intensity contribution to the blend amounts to about 15 per cent.
The resultant measured intensity of $\lambda$4303.83 agrees well with the
most recent predicted value (Table\,\ref{relative:oii_4f-3d}). The other line
$\lambda$4317.70 (4f~D[3]$^{\rm o}_{5/2}$ -- 3d~$^4$P$_{3/2}$) is blended with
O~{\sc ii} M2 3p~$^{4}$P$^{\rm o}_{3/2}$ -- 3s~$^{4}$P$_{1/2}$ $\lambda$4317.14
which is 5\,--\,6 times stronger.

\subsubsection{Multiplet 53b,
4f\,D[2]$^{\rm o}$ -- 3d\,$^4$P}\label{appendix:oii:v53b}

$\lambda$4294.78 (4f~D[2]$^{\rm o}_{5/2}$ -- 3d~$^4$P$_{3/2}$) coincides in
wavelength with $\lambda$4294.92 (4f~D[2]$^{\rm o}_{3/2}$ -- 3d~$^4$P$_{3/2}$)
which contributes about 12 per cent to the total intensity of the feature at
$\lambda$4295. The total intensity of the two $\lambda$4295 lines agrees with
the new prediction. Another line $\lambda$4307.23 (4f~D[2]$^{\rm o}_{3/2}$ --
3d~$^4$P$_{1/2}$) is too weak to perform accurate measurements although it is
observed (Fig.\,\ref{4260-4310}). The other M53b lines are not observed.

\subsubsection{Multiplet 53c,
4f\,D[1]$^{\rm o}$ -- 3d\,$^4$P}\label{appendix:oii:v53c}

$\lambda$4288.82 is a blend of two M53c lines (4f~D[1]$^{\rm o}_{3/2}$ --
3d~$^4$P$_{1/2}$ and 4f~D[1]$^{\rm o}_{1/2}$ -- 3d~$^4$P$_{1/2}$). The measured
intensity of $\lambda$4288.82 agrees with the newly predicted value, but is
much lower than the one predicted based on the previous calculations
(Table\,\ref{relative:oii_4f-3d}). The other lines are not observed.

\subsubsection{Multiplet 55,
4f\,G[3]$^{\rm o}$ -- 3d\,$^4$P}\label{appendix:oii:v55}

$\lambda$4291.25 (4f~G[3]$^{\rm o}_{7/2}$ -- 3d~$^4$P$_{5/2}$) blends with
O~{\sc ii} M78c 4f~F[2]$^{\rm o}_{5/2}$ -- 3d~$^2$F$_{5/2}$ $\lambda$4292.21,
which contributes about 38 per cent to the total intensity. The contribution
from O~{\sc ii} $\lambda$4291.86 (4f~G[3]$^{\rm o}_{5/2}$ -- 3d~$^4$P$_{5/2}$)
was assumed to be negligible. The other line $\lambda$4305.39 is not observed.

\subsubsection{Multiplet 63a,
4f\,D[3]$^{\rm o}$ -- 3d\,$^4$D}\label{appendix:oii:v63a}

Measurements of $\lambda$4357.25 (4f~D[3]$^{\rm o}_{7/2}$ -- 3d~$^4$D$_{5/2}$
and 4f~D[3]$^{\rm o}_{5/2}$ -- 3d~$^4$D$_{3/2}$) are affected by [O~{\sc iii}]
$\lambda$4363, as well as the saturated H~{\sc i} $\lambda$4340 line
(Fig.\,\ref{4310-4382}). The other M63a lines are not observed.

\subsubsection{Multiplet 65a,
4f\,G[5]$^{\rm o}$ -- 3d\,$^4$D}\label{appendix:oii:v65a}

$\lambda$4303.61 (4f~G[5]$^{\rm o}_{9/2}$ -- 3d~$^4$D$_{7/2}$) blends with
O~{\sc ii} M53a 4f~D[3]$^{\rm o}_{7/2}$ -- 3d~$^{4}$P$_{5/2}$ $\lambda$4303.82,
which is about 7\,--\,8 times stronger, and O~{\sc ii} M53a 4f~D[3]$^{\rm
o}_{5/2}$ -- 3d~$^{4}$P$_{5/2}$ $\lambda$4304.08, which contributes about 3
per cent to the total intensity of the blend at $\lambda$4304.

\subsubsection{Multiplet 65b,
4f\,G[4]$^{\rm o}$ -- 3d\,$^4$D}\label{appendix:oii:v65b}

Measurements of this multiplet are difficult due to the saturated H~{\sc i}
$\lambda$4340 line.

\subsubsection{Multiplet 65c,
4f\,G[3]$^{\rm o}$ -- 3d\,$^4$D}\label{appendix:oii:v65c}

Measurements of this multiplet are difficult due to the saturated H~{\sc i}
$\lambda$4340 line.

\subsubsection{Multiplet 67a,
4f\,F[4]$^{\rm o}$ -- 3d\,$^4$D}\label{appendix:oii:v67a}

In Fig.\,\ref{4260-4310}, the three lines of this multiplet blend with several
O~{\sc ii} ORLs of the 4f -- 3d transition array. Using the effective
recombination coefficients of PJS, we estimated that O~{\sc ii} M67a
$\lambda$4275.55 contributes about 35 per cent to the broad feature at
$\lambda$4275, which is formed by more than 10 O~{\sc ii} ORLs from the 4f --
3d configuration. The other two M67a lines contribute very little.

\subsubsection{Multiplet 67b,
4f\,F[3]$^{\rm o}$ -- 3d\,$^4$D}\label{appendix:oii:v67b}

Measurements of this multiplet are difficult due to the reason given in
Section\,\ref{appendix:oii:v67a}. Using the new O~{\sc ii} effective
recombination coefficients, we estimated that the $\lambda$4276.75
(4f\,F[3]$^{\rm o}_{7/2}$\,--\,3d\,$^4$D$_{5/2}$) line, which is the strongest
in O~{\sc ii} M67b, contributes about 17 per cent to the total intensity of
the feature at $\lambda$4275.

\subsubsection{Multiplet 67c,
4f\,F[2]$^{\rm o}$ -- 3d\,$^4$D}\label{appendix:oii:v67c}

Measurements of this multiplet are difficult due to blending.

\subsubsection{Multiplet 76b,
4f\,G[4]$^{\rm o}$ -- 3d\,$^2$F}\label{appendix:oii:v76b}

$\lambda$4371.62 (4f~G[4]$^{\rm o}_{9/2}$ -- 3d~$^2$F$_{7/2}$) blends with
$\lambda$4371.24 (4f~G[4]$^{\rm o}_{7/2}$ -- 3d~$^2$F$_{7/2}$) whose intensity
contribution was assumed to be negligible. The measurement of $\lambda$4371.62
agrees with the predicted intensity, but this intensity could be of large
uncertainty due to weakness as well as the strength of [O~{\sc iii}]
$\lambda$4363.

\subsubsection{Multiplet 76c,
4f\,G[3]$^{\rm o}$ -- 3d\,$^2$F}\label{appendix:oii:v76c}

Only $\lambda$4353.59 (4f~G[3]$^{\rm o}_{7/2}$ -- 3d~$^2$F$_{5/2}$) is
observed, which is shown in Fig.\,\ref{4310-4382}. Its fitted intensity is
higher than the newly predicted value. This measurement is overestimated due
to the saturated H$\gamma$. The intensity contribution from $\lambda$4354.18
(4f~G[3]$^{\rm o}_{5/2}$ -- 3d~$^2$F$_{5/2}$) was assumed to be negligible
(only about 6 per cent). The other two M76c lines $\lambda\lambda$4384.70 and
4385.32 are blended with He~{\sc i} M51 5d~$^1$D$_{2}$ -- 2p~$^1$P$^{\rm
o}_{1}$ $\lambda$4387.93 (Fig.\,\ref{4380-4445}).

\subsubsection{Multiplet 78a,
4f\,F[4]$^{\rm o}$ -- 3d\,$^2$F}\label{appendix:oii:v78a}

The measured intensity of $\lambda$4313.44 (4f~F[4]$^{\rm o}_{9/2}$ --
3d~$^2$F$_{7/2}$) agrees with the most recent prediction, but is probably of
large uncertainty due to weakness. Here the contribution from $\lambda$4312.11
(4f~F[4]$^{\rm o}_{7/2}$ -- 3d~$^2$F$_{7/2}$), which is about 32 per cent, has
been corrected for. The other line $\lambda$4282.01 (4f~F[4]$^{\rm o}_{7/2}$
-- 3d~$^2$F$_{5/2}$) is not observed.

\subsubsection{Multiplet 78b,
4f\,F[3]$^{\rm o}$ -- 3d\,$^2$F}\label{appendix:oii:v78b}

$\lambda$4285.69 (4f~F[3]$^{\rm o}_{7/2}$ -- 3d~$^2$F$_{5/2}$) blends with
$\lambda$4285.21 (4f~F[3]$^{\rm o}_{5/2}$ -- 3d~$^2$F$_{5/2}$) which
contributes only about 1 per cent to the total intensity. It is also partially
resolved from a feature at $\lambda$4283, which is a blend of O~{\sc ii} M67c
4f~F[2]$^{\rm o}_{5/2}$ -- 3d~$^{4}$D$_{3/2}$ $\lambda$4282.96 and Ne~{\sc ii}
M57c 4f~1[3]$^{\rm o}_{7/2}$ -- 3d~$^{4}$F$_{7/2}$ $\lambda$4283.73. The fitted
intensity of $\lambda$4285.69 agrees with predicted value. The other lines are
not observed.

\subsubsection{Multiplet 86a,
4f\,D[3]$^{\rm o}$ -- 3d\,$^2$P}\label{appendix:oii:v86a}

In Fig.\,\ref{4444-4504}, $\lambda$4491.23 (4f~D[3]$^{\rm o}_{5/2}$ --
3d~$^2$P$_{3/2}$) is partially resolved from a feature at $\lambda$4488 which
is a blend of the O~{\sc ii} M104 4f$^{\prime}$~D[2]$^{\rm o}_{3/2}$ --
3d$^{\prime}$~$^{2}$P$_{1/2}$ lines ($\lambda$4487.72 3/2 -- 1/2, and
$\lambda$4488.20 5/2 -- 3/2 and 3/2 -- 3/2) and O~{\sc ii} M86b 4f~D[2]$^{\rm
o}_{3/2}$ -- 3d~$^2$P$_{1/2}$ $\lambda$4489.49. Four Gaussian profiles were
used to fit the complex. Since the effective recombination coefficients for
the O~{\sc ii} M104 lines are not available, we assumed that their relative
intensities are as in pure {\it LS}\,coupling, i.e. 5\,:\,9\,:\,1. The
resultant intensity of $\lambda$4491.23 agrees well with the most recent
theoretical value.

\subsubsection{Multiplet 86b,
4f\,D[2]$^{\rm o}$ -- 3d\,$^2$P}\label{appendix:oii:v86b}

$\lambda$4489.49 (4f~D[2]$^{\rm o}_{3/2}$ -- 3d~$^2$P$_{1/2}$) blends with
O~{\sc ii} M104 $\lambda$4488.20 and O~{\sc ii} M86a $\lambda$4491.23. The
fitted intensity of $\lambda$4489.49 agrees with the newly predicted value
quite well. Details of line fits are given in Section\,\ref{appendix:oii:v86a}.
Accurate measurements of $\lambda$4466.41,\,59 are difficult due to He~{\sc i}
$\lambda$4471, which is stronger by more than two orders of magnitude.

\subsubsection{Multiplet 88,
4f\,G[3]$^{\rm o}$ -- 3d\,$^2$P}\label{appendix:oii:v88}

$\lambda$4477.90 (4f~G[3]$^{\rm o}_{5/2}$ -- 3d~$^2$P$_{3/2}$) is observed
in Fig.\,\ref{4444-4504}. It is blended with O~{\sc iii} M45a
5g\,H[11/2]$^{\rm o}_{5,\,6}$\,--\,4f\,G[9/2]$_{5}$ $\lambda$4477.91, which
contributes about 1--2 per cent to the total intensity and thus is negligible
(an estimate based on the effective recombination coefficients for the O~{\sc
iii} 5g -- 4f recombination spectrum given by Kisielius \& Storey
\citealt{ks1999}). The measured intensity of $\lambda$4477.90 agrees well
with the newly predicted value (Table\,\ref{relative:oii_4f-3d}).

\subsubsection{Multiplet 92a,
4f\,F[4]$^{\rm o}$ -- 3d\,$^2$D}\label{appendix:oii:v92a}

$\lambda$4609.44 (4f~F[4]$^{\rm o}_{7/2}$ -- 3d~$^2$D$_{5/2}$) is observed in
Fig.\,\ref{4555-4625}. It blends with O~{\sc ii} M92c 4f~F[2]$^{\rm o}_{5/2}$
-- 3d~$^{2}$D$_{3/2}$ $\lambda$4610.20 and O~{\sc ii} M92c 4f~F[2]$^{\rm
o}_{3/2}$ -- 3d~$^{2}$D$_{3/2}$ $\lambda$4611.07, which contribute about 20
per cent to the total intensity. After subtracting the blend, the resultant
intensity of $\lambda$4609.44 agrees well with the new prediction
(Table\,\ref{relative:oii_4f-3d}).

\subsubsection{Multiplet 92b,
4f\,F[3]$^{\rm o}$ -- 3d\,$^2$D}\label{appendix:oii:v92b}

$\lambda$4602.13 (4f~F[3]$^{\rm o}_{5/2}$ -- 3d~$^2$D$_{3/2}$) is observed
in Fig.\,\ref{4555-4625}. It blends with N~{\sc ii} M5 3p~$^{3}$P$_{2}$ --
3s~$^{3}$P$^{\rm o}_{1}$ $\lambda$4601.48, which contributes about 26 per
cent to the total intensity. Subtraction of the blend gives an intensity of
$\lambda$4602.13 that agrees well with the most recent prediction. The
contribution from Ne~{\sc ii} M64d 4f~2[2]$^{\rm o}_{5/2}$ --
3d~$^{4}$P$_{5/2}$ $\lambda$4600.16 was assumed to be negligible.
Fig.\,\ref{4555-4625} also shows the weak feature of $\lambda$4613.14
(4f~F[3]$^{\rm o}_{5/2}$ -- 3d~$^2$D$_{5/2}$) which blends with
$\lambda$4613.68 (4f~F[3]$^{\rm o}_{7/2}$ -- 3d~$^2$D$_{5/2}$). The total
measured intensity of the two lines agrees with both of the predicted values
(Table\,\ref{relative:oii_4f-3d}). Correction has been made for the
contribution from N~{\sc ii} M5 3p~$^{1}$P$_{1}$ -- 3s~$^{3}$P$^{\rm o}_{1}$
$\lambda$4613.87, which is about 7 per cent, and the contribution from Ne~{\sc
ii} M64b 4f~2[3]$^{\rm o}_{7/2}$ -- 3d~$^{4}$P$_{5/2}$ $\lambda$4612.93 was
assumed to be negligible.

\section{The Ne~{\sc ii} optical recombination spectrum}\label{appendix:d}

\subsection{Multiplet 1,
3p\,$^4$P$^{\rm o}$ -- 3s\,$^4$P}\label{appendix:neii:v1}

Fig.\,\ref{3680-3745} shows that $\lambda$3694.21 (3p~$^4$P$^{\rm o}_{5/2}$ --
3s~$^4$P$_{5/2}$) lies between the two H~{\sc i} lines H18 $\lambda$3691.55
and H17 $\lambda$3697.15. The fitted intensity of $\lambda$3694.21 is
0.254$\pm$0.026. This measurement
agrees with the values of LLB01: 0.294 (ESO 1.52~m) and 0.224 (WHT). Another
M1 line $\lambda$3777.14 (3p~$^4$P$^{\rm o}_{3/2}$ -- 3s~$^4$P$_{1/2}$) is
observed in Fig.\,\ref{3740-3825}. Its measured intensity is 0.050, with an
uncertainty of about 20 to 25 per cent, and also agrees with LLB01: 0.045 (ESO
1.52~m) and 0.059 (WHT). The intensity ratio $\lambda$3777.14/$\lambda$3694.21
is 0.195, lower than the {\it LS}\,coupling ratio 0.397. The other lines are
not observed.

\subsection{Multiplet 2,
3p\,$^4$D$^{\rm o}$ -- 3s\,$^4$P}\label{appendix:neii:v2}

See Section\,\ref{neii_orls:v2}.

\subsection{Multiplet 5,
3p\,$^2$D$^{\rm o}$ -- 3s\,$^2$P}\label{appendix:neii:v5}

Fig.\,\ref{3680-3745} shows that $\lambda$3713.08 (3p~$^2$D$^{\rm o}_{5/2}$ --
3s~$^2$P$_{3/2}$) is blended with H~{\sc i} H15 $\lambda$3711.97, which is more
than 5 times stronger. Also blended here is O~{\sc iii} M3 3d~$^{3}$D$^{\rm
o}_{3}$ -- 3p~$^{3}$P$_{2}$ $\lambda$3715.08. Three Gaussian profiles were used
to fit the complex, and the fitted intensity of $\lambda$3713.08 is 0.306,
which could be of large uncertainty (over 50 per cent).
The intensity ratio of $\lambda$3713.08 to Ne~{\sc ii} M2 $\lambda$3334.84
agrees with the {\it LS}\,coupling ratio. The other M2 lines are not observed.

\subsection{Multiplet 6,
3p\,$^2$S$^{\rm o}$ -- 3s\,$^2$P}\label{appendix:neii:v6}

$\lambda$3481.93 (3p~$^2$S$^{\rm o}_{1/2}$ -- 3s~$^2$P$_{3/2}$) is partially
resolved from He~{\sc i} M43 15d~$^3$D -- 2p~$^3$P$^{\rm o}$ $\lambda$3478.97,
which is about 2 times stronger. The fitted intensity of $\lambda$3481.93 is
0.043, which is slightly lower than LLB01 (0.051). The other
line $\lambda$3557.81 is not observed.

\subsection{Multiplet 8,
3d\,$^4$D -- 3p\,$^4$P$^{\rm o}$}\label{appendix:neii:v8}

The wavelengths of the Ne~{\sc ii} M8 lines are in the range
3017\,--\,3054\,{\AA}. The blue cutoff of our spectrum is around 3040\,{\AA},
and only three lines have the wavelengths longer than this cutoff. Measurements
of these lines are difficult: The $\lambda\lambda$3045.56 and 3047.56 lines
are blended with the much stronger O~{\sc iii} M4 $\lambda$3047.12
(3p\,$^{3}$P$_{2}$\,--\,3s\,$^{3}$P$^{\rm o}_{2}$) line, which is affected by
the Bowen fluorescence mechanism; the $\lambda$3054.67 line is not observed.

\subsection{Multiplet 12,
3d\,$^4$D -- 3p\,$^4$D$^{\rm o}$}\label{appendix:neii:v12}

\begin{figure}
\begin{center}
\includegraphics[width=7.5cm,angle=-90]{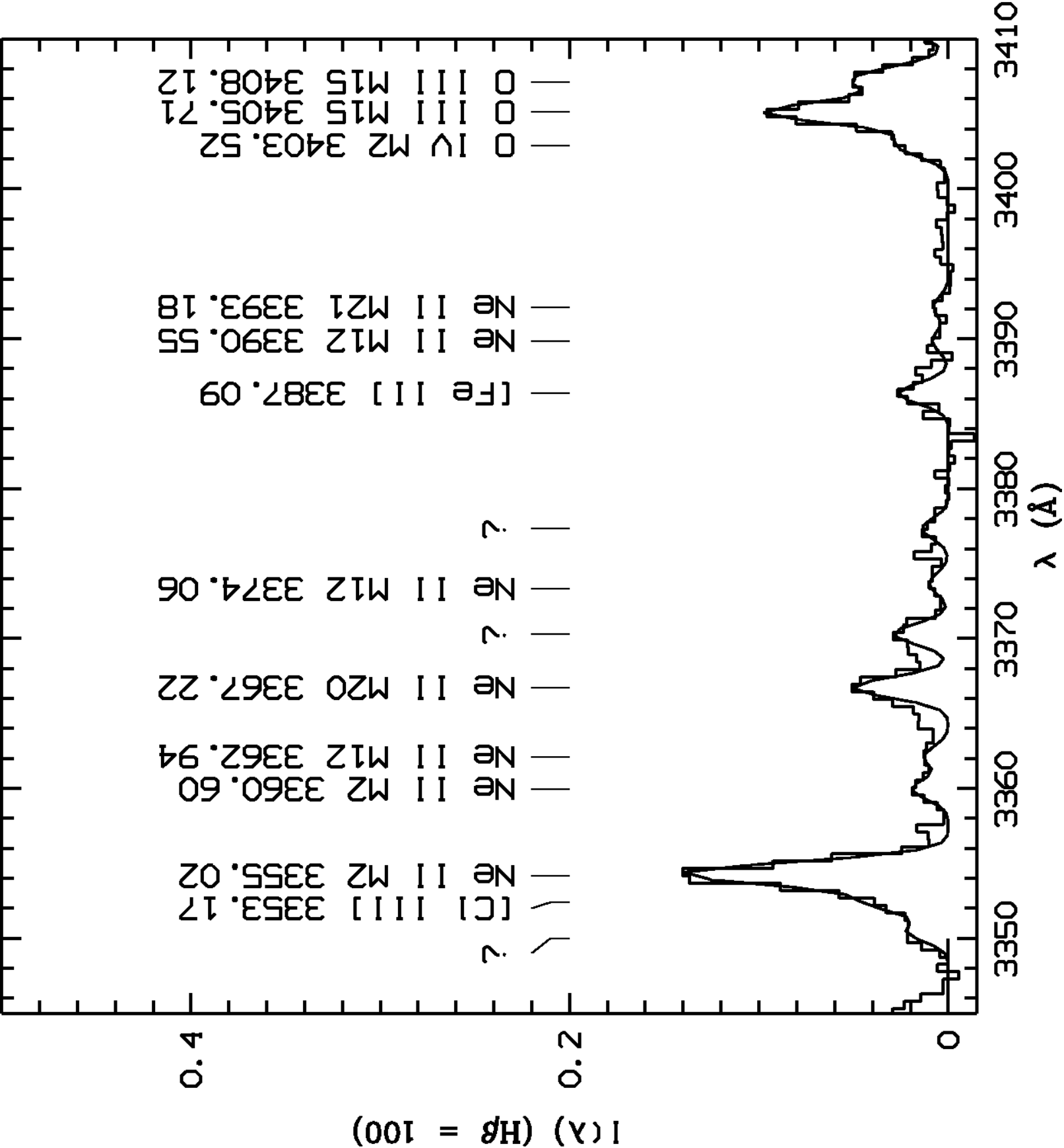}
\caption{Spectrum of NGC\,7009 from 3345 to 3410\,{\AA} showing the Ne~{\sc
ii} M12 3d\,$^{4}$D\,--\,3p\,$^{4}$D$^{\rm o}$ lines. Also detected in this
region are the Ne~{\sc ii} lines of the M2 3p\,$^{4}$D$^{\rm
o}$\,--\,3s\,$^{4}$P, M20 3d\,$^{2}$F\,--\,3p\,$^{2}$D$^{\rm o}$ and M21
3d\,$^{2}$D\,--\,3p\,$^{2}$D$^{\rm o}$ multiplets. The continuous curve is
the sum of Gaussian profile fits. Continuum has been subtracted and the
spectrum has been normalized such that H$\beta$ has an integrated flux of
100. Extinction has not been corrected for.}
\label{3345-3410}
\end{center}
\end{figure}

The intensity of $\lambda$3329.16 is 0.053, which is
slightly higher than that of LLB01 (0.0464). This measurement is of large
uncertainty due to weakness of the line, as is shown in Fig.\,\ref{3290-3348}.
Another M12 line $\lambda$3366.98 blends with Ne~{\sc ii} M20 3d~$^2$F$_{7/2}$
-- 3p~$^2$D$^{\rm o}_{5/2}$ $\lambda$3367.22 (Fig.\,\ref{3345-3410}), which
contributes more than 90 per cent to the total intensity. The other M12 lines
are not observed.

\subsection{Multiplet 13,
3d\,$^4$F -- 3p\,$^4$D$^{\rm o}$}\label{appendix:neii:v13}

See Section\,\ref{neii_orls:v13}.

\subsection{Multiplet 20,
3d\,$^2$F -- 3p\,$^2$D$^{\rm o}$}\label{appendix:neii:v20}

$\lambda$3367.22 (3d~$^2$F$_{7/2}$ -- 3p~$^2$D$^{\rm o}_{5/2}$) blends with
Ne~{\sc ii} M12 3d~$^4$D$_{7/2}$ -- 3p~$^4$D$^{\rm o}_{5/2}$ $\lambda$3366.98,
whose intensity contribution is negligible (Section\,\ref{appendix:neii:v12}).
The fitted intensity of $\lambda$3367.22 is 0.106$\pm$0.016, which agrees with
that of LLB01 (0.110). The other M20 lines are not observed.

\subsection{Multiplet 21,
3d\,$^2$D -- 3p\,$^2$D$^{\rm o}$}\label{appendix:neii:v21}

Fig.\,\ref{3395-3475} shows that $\lambda$3416.91 (3d~$^2$D$_{5/2}$ --
3p~$^2$D$^{\rm o}_{5/2}$) is partially resolved from O~{\sc iii} M15
3d~$^3$P$^{\rm o}_{1}$ -- 3p~$^3$P$_{1}$ $\lambda$3415.26, which is excited by
the Bowen fluorescence mechanism. It also blends with Ne~{\sc ii} M19
3d~$^4$F$_{7/2}$ -- 3p~$^2$D$^{\rm o}_{5/2}$ $\lambda$3417.69, whose intensity
contribution is unknown due to the lack of atomic data. The total intensity
of the feature at $\lambda$3417 given by fit is 0.114, higher than the
measurement of LLB01 (0.0825). The other M21 lines are not
observed.

\subsection{Multiplet 28,
3d\,$^2$P -- 3p\,$^2$S$^{\rm o}$}\label{appendix:neii:v28}

The fitted intensity of $\lambda$3456.61 (3d~$^2$P$_{3/2}$ -- 3p~$^2$S$^{\rm
o}_{1/2}$) is 0.031, which is slightly lower than the intensity value 0.0334
given by LLB01. Here the contribution from He~{\sc i}
19d~$^3$D -- 2p~$^3$P$^{\rm o}$ $\lambda$3456.86 was assumed to be negligible.
The other line $\lambda$3503.58 is not observed.

\subsection{Multiplet 34,
3d\,$^4$P -- 3p\,$^4$S$^{\rm o}$}\label{appendix:neii:v34}

The measured intensity of $\lambda$3542.85 (3d~$^4$P$_{5/2}$ -- 3p~$^4$S$^{\rm
o}_{3/2}$) and is 0.031, which agrees with the measurement
given by LLB01 (0.0325). Fig.\,\ref{3540-3595} shows that $\lambda$3565.82
(3d~$^4$P$_{3/2}$ -- 3p~$^4$S$^{\rm o}_{3/2}$) is partially resolved from
Ne~{\sc ii} M9 3p$^{\prime}$~$^2$F$^{\rm o}_{7/2}$ -- 3s$^{\prime}$~$^2$D$_{5}$
$\lambda$3568.50, which is about 7 times stronger. The fitted intensity of
$\lambda$3565.82 is 0.025, with a relatively large uncertainty due to the
weakness. This measurement also agrees with that of LLB01 (0.0243). The
intensity ratio of these two M34 lines is thus 1.24\,:\,1.0, which differs
from the {\it LS}\,coupling value 1.5\,:\,1. The other line $\lambda$3594.16
is not observed.

\subsection{Multiplet 9, 3p$^{\prime}$\,$^2$F$^{\rm o}$ --
3s$^{\prime}$\,$^2$D}\label{appendix:neii:v9}

See Section\,\ref{neii_orls:v9}.

\subsection{4f -- 3d transitions}\label{appendix:neii:4f-3d}

\subsubsection{Multiplet 55e,
4f\,2[5]$^{\rm o}$ -- 3d\,$^4$F}\label{appendix:neii:v55e}

See Section\,\ref{neii_orls:4f-3d}.

\subsubsection{Multiplet 52a,
4f\,2[4]$^{\rm o}$ -- 3d\,$^4$D}\label{appendix:neii:v52a}

$\lambda$4219.75 (4f~2[4]$^{\rm o}_{9/2}$ -- 3d~$^4$D$_{7/2}$) blends with
Ne~{\sc ii} M52a 4f~2[4]$^{\rm o}_{7/2}$ -- 3d~$^4$D$_{7/2}$ $\lambda$4219.37
and Ne~{\sc ii} M52d 4f~2[2]$^{\rm o}_{5/2}$ -- 3d~$^4$D$_{5/2}$
$\lambda$4220.89 (Fig.\,\ref{4176-4260}). The intensity contribution of
$\lambda$4219.37 (a few per cent) could be negligible. The fitted intensity of
$\lambda$4219.75 is 0.064, which agrees with the measurements
given by LLB01: 0.0674 (ESO 1.52~m), 0.0699 (WHT 1996) and 0.0616 (WHT 1997).
This measurement is higher than the predicted value
(Table\,\ref{relative:neii_4f-3d}). The measured intensity of the other line
$\lambda$4233.85 agrees well with LLB01, but is higher than the predicted
value. This measurement is probably unreliable due to weakness of the
$\lambda$4233.85 line.

\subsubsection{Multiplet 52b,
4f\,2[3]$^{\rm o}$ -- 3d\,$^4$D}\label{appendix:neii:v52b}

The measured intensity of $\lambda$4231.64 (4f~2[3]$^{\rm o}_{7/2}$ --
3d~$^4$D$_{5/2}$) is 0.027, which agrees with the measurements
of LLB01: 0.0244 (ESO 1.52m), 0.0328 (WHT 1996) and 0.0237 (WHT 1997). Here the
contribution of $\lambda$4231.53 (4f~2[3]$^{\rm o}_{5/2}$ -- 3d~$^4$D$_{5/2}$)
is included. The intensity of another M52b line $\lambda$4250.65 is 0.016, and
it also agrees with those of LLB01: 0.0169 (ESO 1.52~m) and 0.0154 (WHT 1996),
and 0.0121 (WHT 1997). Both measurements are higher than the predicted values
(Table\,\ref{relative:neii_4f-3d}), thus $\lambda\lambda$4231.64 and 4250.65
are not used for abundance determinations. The other two lines
$\lambda\lambda$4217.17 and 4217.15 are too faint to measure accurately.

%

\subsubsection{Multiplet 57b,
4f\,1[4]$^{\rm o}$ -- 3d\,$^4$F}\label{appendix:neii:v57b}

Only $\lambda$4397.99 (4f~1[4]$^{\rm o}_{7/2}$ -- 3d~$^4$F$_{5/2}$) is
observed (Fig.\,\ref{4380-4445}). Its fitted intensity is 0.024$\pm$0.004.
This intensity value is slightly
higher than those given by LLB01: 0.0219 (ESO 1.52~m), 0.0184 (WHT 1996) and
0.0185 (WHT 1997), but agrees with the predicted value within errors
(Table\,\ref{relative:neii_4f-3d}). The other M57b lines are not observed.

\subsubsection{Multiplet 60c,
4f\,1[3]$^{\rm o}$ -- 3d\,$^2$F}\label{appendix:neii:v60c}

Only $\lambda$4428.64 (4f~1[3]$^{\rm o}_{7/2}$ -- 3d~$^2$F$_{5/2}$) is
observed, and its fitted intensity is 0.044, which agrees with
the measurements given by LLB01: 0.0483 (ESO 1.52~m), 0.0451 (WHT 1996) and
0.0479 (WHT 1997). Here the contribution from Ne~{\sc ii} M61b 4f~2[3]$^{\rm
o}_{7/2}$ -- 3d~$^2$D$_{5/2}$ $\lambda$4428.52 is included, but the other two
blended lines Ne~{\sc ii} M60c 4f~1[3]$^{\rm o}_{5/2}$ -- 3d~$^2$F$_{5/2}$
$\lambda$4428.52 and Ne~{\sc ii} M61b 4f~2[3]$^{\rm o}_{5/2}$ --
3d~$^2$D$_{5/2}$ $\lambda$4428.41 were assumed to be negligible.

\subsubsection{Multiplet 61a,
4f\,2[4]$^{\rm o}$ -- 3d\,$^2$D}\label{appendix:neii:v61a}

The measured intensity of $\lambda$4430.94 (4f~2[4]$^{\rm o}_{7/2}$ --
3d~$^2$D$_{5/2}$) is 0.040, which is higher than the
observations of LLB01: 0.0331 (ESO 1.52~m), 0.0340 (WHT 1996) and 0.0317 (WHT
1997). This measurement is of large uncertainty due to blending. The
contribution from Ne~{\sc ii} M57a 4f~1[2]$^{\rm o}_{5/2}$ -- 3d~$^4$F$_{3/2}$
$\lambda$4430.90 and Ne~{\sc ii} M55a 4f~2[4]$^{\rm o}_{9/2}$ --
3d~$^4$F$_{7/2}$ $\lambda$4430.06 are included, but Ne~{\sc ii} M57a
4f~1[2]$^{\rm o}_{3/2}$ -- 3d~$^4$F$_{3/2}$ $\lambda$4431.11 was assumed to be
negligible.

%

\subsubsection{Multiplet 61d,
4f\,2[2]$^{\rm o}$ -- 3d\,$^2$D}\label{appendix:neii:v61d}

$\lambda$4457.05 (4f~2[2]$^{\rm o}_{5/2}$ -- 3d~$^2$D$_{3/2}$) is partially
resolved from O~{\sc iii} M49a 5g~F[4]$^{\rm o}_{3}$ -- 4f~D[3]$_{2}$
$\lambda$4458.55 (Fig.\,\ref{4444-4504}). The fitted intensity of
$\lambda$4457.05 is 0.026, with an uncertainty of about 20
per cent. Here the contribution from Ne~{\sc ii} M66c 4f~1[3]$^{\rm o}_{7/2}$
-- 3d~$^4$P$_{5/2}$ $\lambda$4457.36 is included, but the other two blended
lines Ne~{\sc ii} M61d 4f~2[2]$^{\rm o}_{3/2}$ -- 3d~$^2$D$_{3/2}$
$\lambda$4457.24 and Ne~{\sc ii} M66c 4f~1[3]$^{\rm o}_{5/2}$ --
3d~$^4$P$_{5/2}$ $\lambda$4457.24 were assumed to be negligible. This
measurement agrees with those of LLB01: 0.0250 (ESO 1.52~m), 0.0247 (WHT 1996)
and 0.0211 (WHT 1997), but is much higher than the predicted value
(Table\,\ref{relative:neii_4f-3d}). The other M61d lines are not observed.

\subsubsection{Multiplet 65,
4f\,0[3]$^{\rm o}$ -- 3d\,$^4$P}\label{appendix:neii:v65}

$\lambda$4413.22 (4f~0[3]$^{\rm o}_{7/2}$ -- 3d~$^4$P$_{5/2}$) blends with
O~{\sc ii} M5 3p~$^2$D$^{\rm o}_{5/2}$ -- 3s~$^2$P$_{3/2}$ $\lambda$4414.90
(Fig.\,\ref{4380-4445}). The fitted intensity of $\lambda$4413.22 is 0.028,
which agrees with the measurements of LLB01: 0.0296 (ESO
1.52~m), 0.0243 (WHT 1996) and 0.0271 (WHT 1997), but is higher than the
predicted value (Table\,\ref{relative:neii_4f-3d}). Here the contribution from
Ne~{\sc ii} M65 4f~0[3]$^{\rm o}_{5/2}$ -- 3d~$^4$P$_{5/2}$ $\lambda$4413.11
was assumed to be negligible, but Ne~{\sc ii} M57c 4f~1[3]$^{\rm o}_{5/2}$ --
3d~$^4$F$_{3/2}$ $\lambda$4413.11 is included. The other line $\lambda$4377.98
blends with N~{\sc iii} M18 $\lambda$4379.

\subsubsection{Multiplet 66c,
4f\,1[3]$^{\rm o}$ -- 3d\,$^4$P}\label{appendix:neii:v66c}

$\lambda$4421.39 (4f~1[3]$^{\rm o}_{5/2}$ -- 3d~$^4$P$_{3/2}$) is observed in
Fig.\,\ref{4380-4445}. Its fitted intensity is 0.0084. The
uncertainty of this measurement could be large due to weakness of this line.
The other two M66c lines $\lambda\lambda$4457.24 and 4457.36 blend with
Ne~{\sc ii} M61d 4f~2[2]$^{\rm o}_{5/2}$ -- 3d~$^2$D$_{3/2}$ $\lambda$4457.05.

%

\section{The N~{\sc iii} permitted lines}\label{appendix:e}

\subsection{Multiplet 1,
3p\,$^2$P$^{\rm o}$ -- 3s\,$^2$S}\label{appendix:niii:v1}

$\lambda$4097.33 is partially resolved from H~{\sc i} $\lambda$4101
(Fig.\,\ref{4060-4115}). Its fitted intensity is 2.246$\pm$0.112. Several
O~{\sc ii} and N~{\sc ii} ORLs
are blended with $\lambda$4097.33, and they contribute about 15--20 per cent
to the total intensity. The other line $\lambda$4103.39 blends with H~{\sc i}
$\lambda$4101, and contributes less than 1 per cent to the total intensity.

\subsection{Multiplet 2,
3d\,$^2$D -- 3p\,$^2$P$^{\rm o}$}\label{appendix:niii:v2}

Measurements of this multiplet are presented in Section\,\ref{oii_orls:v1}.
Here the ratio of the two lines $\lambda\lambda$4634.14 and 4641.85 were
assumed to be as in {\it LS}\,coupling. The resultant intensities of the
$\lambda\lambda$4634.14, 4640.64 and 4641.85 lines are 1.133, 2.262 and 0.226,
respectively, and the uncertainties are all less than 10 per
cent. Thus the intensity ratio of $\lambda$4640.64 to $\lambda$4641.81 is
10\,:\,1, which differs from the pure {\it LS}\,coupling ratio, i.e. 9\,:\,1.

\subsection{Multiplet 3, 3p$^{\prime}$\,$^4$D --
3s$^{\prime}$\,$^4$P$^{\rm o}$}\label{appendix:niii:v3}

See Section\,\ref{niii_orls:v3}.

\subsection{Multiplet 6, 3p$^{\prime}$\,$^2$D --
3s$^{\prime}$\,$^2$P$^{\rm o}$}\label{appendix:niii:v6}

$\lambda$4195.76 (3p$^{\prime}$~$^2$D$_{3/2}$ -- 3s$^{\prime}$~$^2$P$^{\rm
o}_{1/2}$) is observed in Fig.\,\ref{4176-4260}. Its measured intensity is
0.046$\pm$0.008. Here the
intensity contribution from N~{\sc ii} M49a 4f~2[4]$_{3}$ -- 3d~$^3$D$^{\rm
o}_{2}$ $\lambda$4195.97, which is about 8 per cent, has been corrected for.
Another line $\lambda$4215.77 (3p$^{\prime}$~$^2$D$_{3/2}$ --
3s$^{\prime}$~$^2$P$^{\rm o}_{3/2}$) is partially resolved from Ne~{\sc ii}
M52b 4f~2[3]$^{\rm o}_{7/2}$ -- 3d~$^4$D$_{7/2}$ $\lambda$4217.17. Two
Gaussian profiles were used to fit the complex, and the fitted intensity of
$\lambda$4215.77 is 0.0086, with an uncertainty of more than
20 per cent. Thus the ratio $\lambda$4195.76.$\lambda$4215.77 is 0.185,
slightly higher than the {\it LS}\,coupling value (0.20). This is interesting
because these two lines decay from the same upper level, thus their intensity
ratio depends only on the coupling scheme. The other line $\lambda$4200.10
(3p$^{\prime}$~$^2$D$_{5/2}$ -- 3s$^{\prime}$~$^2$P$^{\rm o}_{3/2}$) blends
with He~{\sc ii} 11g~$^2$G -- 4f~$^2$F$^{\rm o}$ $\lambda$4199.83 which is
probably about 4 times stronger.

\subsection{Multiplet 9, 3d$^{\prime}$\,$^4$F$^{\rm o}$ --
3p$^{\prime}$\,$^4$D}\label{appendix:niii:v9}

Only $\lambda$4881.81 (3d$^{\prime}$~$^4$F$^{\rm o}_{3/2}$ --
3p$^{\prime}$~$^4$D$_{5/2}$) and $\lambda$4884.13 (3d$^{\prime}$~$^4$F$^{\rm
o}_{7/2}$ -- 3p$^{\prime}$~$^4$D$_{7/2}$) are observed (Fig.\,\ref{4875-4935}).
$\lambda$4881.81 blends with [Fe~{\sc iii}] $\lambda$4881.00, whose intensity
contribution is unknown. Measurements of $\lambda$4884.13 are of large
uncertainty due to weakness of the line.

\subsection{Multiplet 12,
5s\,$^2$S -- 4p\,$^2$P$^{\rm o}$}\label{appendix:niii:v12}

Measurements of $\lambda$4544.85 (5s~$^2$S$_{1/2}$ -- 4p~$^2$P$^{\rm o}_{3/2}$)
could be overestimated due to the strength of He~{\sc ii} 9g~$^2$G --
4f~$^2$F$^{\rm o}$ $\lambda$4541.59 (Fig.\,\ref{4505-4555}), which is more than
10 times stronger. The other line $\lambda$4539.71 is blended with He~{\sc ii}
$\lambda$4541.59.

\subsection{Multiplet 17,
5f\,$^2$F$^{\rm o}$ -- 4d\,$^2$D}\label{appendix:niii:v17}

$\lambda$3998.63 (5f~$^2$F$^{\rm o}_{5/2}$ -- 4d~$^2$D$_{3/2}$) is detected in
Fig.\,\ref{3940-4006}. Its fitted intensity is 0.025$\pm$0.005. The other two
lines $\lambda\lambda$4003.58 and 4003.72 are blending together. The measured
intensity ratio ($\lambda$4003.58\,+\,$\lambda$4003.58)/$\lambda$3998.63 is
1.54, which agrees with the value in {\it LS}\,coupling, i.e. 1.50.

\subsection{Multiplet 18,
5g\,$^2$G -- 4f\,$^2$F$^{\rm o}$}\label{appendix:niii:v18}

See Section\,\ref{niii_orls:v18}.

\section{The O~{\sc iii} permitted lines}\label{appendix:f}

\subsection{Multiplet 2,
3p\,$^3$D -- 3s\,$^3$P$^{\rm o}$}\label{appendix:oiii:v2}

\begin{figure}
\begin{center}
\includegraphics[width=7.5cm,angle=-90]{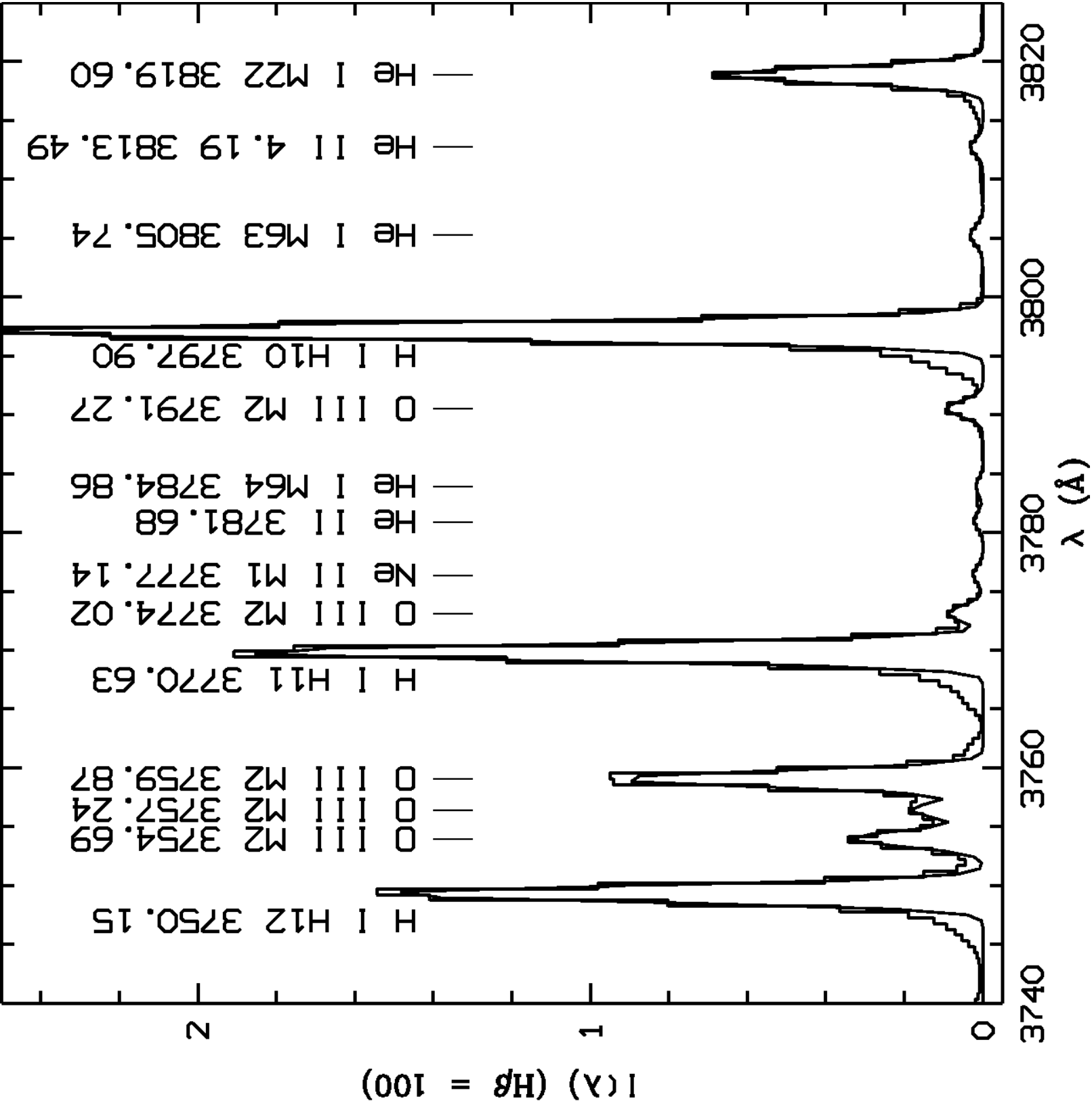}
\caption{Spectrum of NGC\,7009 from 3740 to 3825\,{\AA} showing the O~{\sc iii}
M2 lines and other emission features. The continuous curve is the sum of
Gaussian profile fits. Continuum has been subtracted and the spectrum has been
normalized such that H$\beta$ has an integrated flux of 100. Extinction has
not been corrected for.}
\label{3740-3825}
\end{center}
\end{figure}

Table\,\ref{relative:oiii_v2} presents the observed and predicted relative
intensities of the O~{\sc iii} M2 lines detected in NGC\,7009.
Fig.\,\ref{3740-3825} shows the observed M2 lines. $\lambda$3757.24 is
partially resolved from $\lambda$3754.69 and $\lambda$3759.87. Three Gaussian
profile fits were used to fit the complex, and resultant relative intensities
deviate from {\it LS}\,coupling. The uncertainty of the $\lambda$3757.24
intensity is the largest among the three (over 30 per cent), because it also
blends with
He~{\sc i} M66 14d~$^{1}$D$_{2}$ -- 2p~$^{1}$P$^{\rm o}_{1}$ $\lambda$3756.10,
whose contribution is unknown, and He~{\sc ii} 23g~$^2$G -- 4f~$^2$F$^{\rm o}$
$\lambda$3758.15, which contributes less than 10 per cent to the total
intensity (an estimate based on the hydrogenic theory of Storey \& Hummer
\citealt{sh1995}). The uncertainties of $\lambda\lambda$3754.69 and 3759.87
are both less than 5 per cent. Measurements of $\lambda$3774.02 is probably of
large error due to the strength of H~{\sc i} H11 $\lambda$3770.63, which is
more than 20 times stronger. Another line $\lambda$3791.27 is free of blend,
but its fitted intensity is 45 per cent lower than the value in pure {\it
LS}\,coupling. The other line $\lambda$3810.99 is not observed. Both of the
line ratios $\lambda$3791.27/$\lambda$3754.69 and
$\lambda$3774.02/$\lambda$3757.24,
where the two lines of each of the ratios decay from the same upper levels,
differ from the {\it LS}\,coupling assumption, and agree with the measurements
given by Liu et al. \cite{ld1993a} for NGC\,7009. Discussion of these line
ratios is in Section\,\ref{oiii_orls:fluorescence}.

\begin{table}
\centering
\caption{Comparison of the observed and predicted (in the {\it LS}\,coupling
assumption) relative intensities of O~{\sc iii} M2 lines in NGC\,7009.}
\label{relative:oiii_v2}
\begin{tabular}{cclll}
\hline
Line & $J_2-J_1$ & $I_{\rm LS}$ & $I_{\rm obs}$ & $\frac{I_{\rm obs}}{I_{\rm
LS}}$\\
\hline
$\lambda$3754.69 &  2--1  &  0.536  &  0.361   & 0.673\\
$\lambda$3757.24 &  1--0  &  0.238  &  0.196   & 0.823\\
$\lambda$3759.87 &  3--2  &  1.000  &  1.000   & 1.000\\
$\lambda$3774.02 &  1--1  &  0.179  &  0.092   & 0.516\\
$\lambda$3791.27 &  2--2  &  0.179  &  0.098   & 0.548\\
\hline
\end{tabular}
\end{table}

\subsection{Multiplet 3,
3p\,$^3$S -- 3s\,$^3$P$^{\rm o}$}\label{appendix:oiii:v3}

\begin{figure}
\begin{center}
\includegraphics[width=7.5cm,angle=-90]{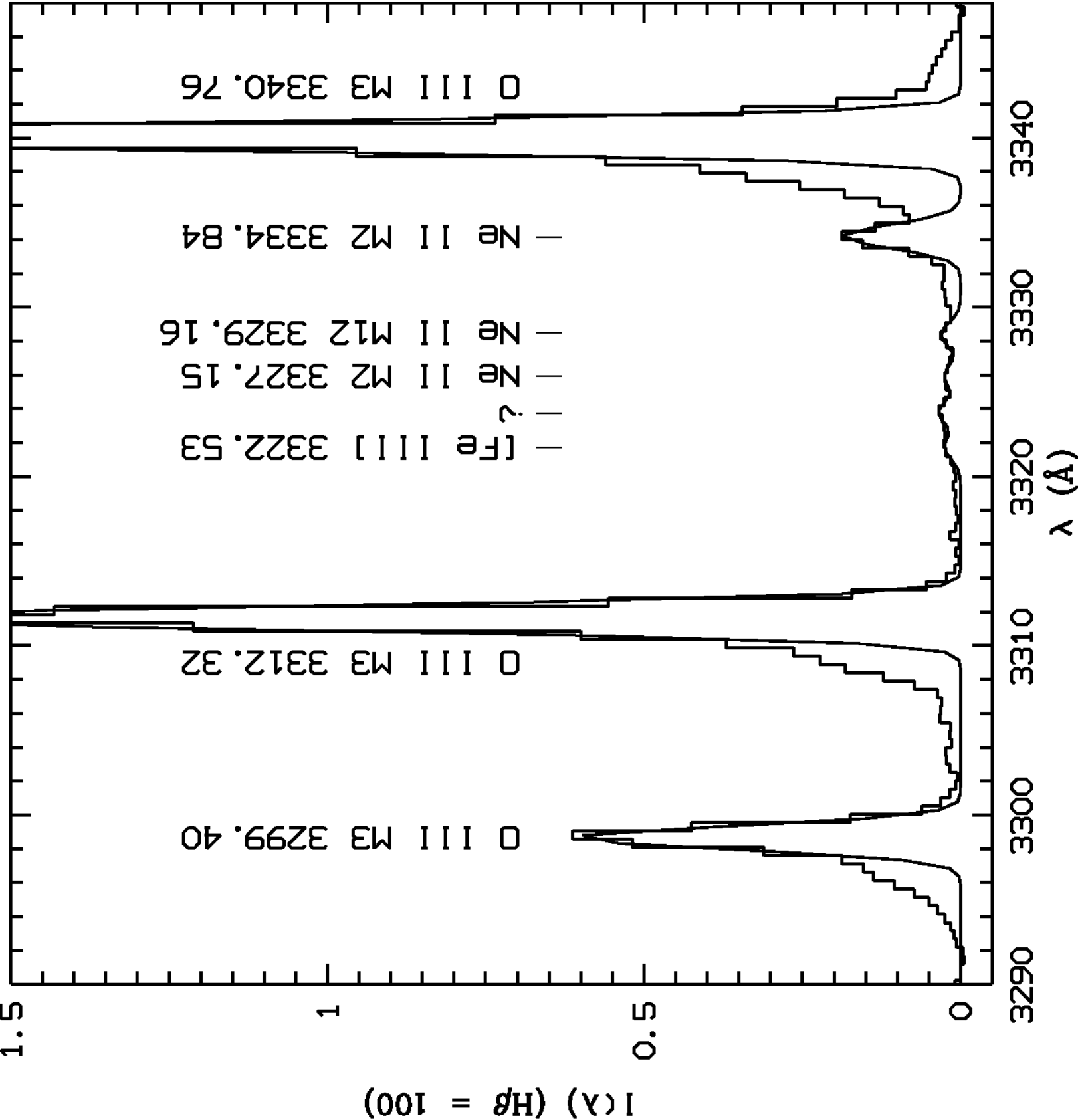}
\caption{Spectrum of NGC\,7009 from 3290 to 3348\,{\AA} showing the O~{\sc iii}
M3 lines and other emission features. The continuous curve is the sum of
Gaussian profile fits. Continuum has been subtracted and the spectrum has been
normalized such that H$\beta$ has an integrated flux of 100. Extinction has
not been corrected for.}
\label{3290-3348}
\end{center}
\end{figure}

Fig.\,\ref{3290-3348} shows the observed O~{\sc iii} M3 lines in NGC\,7009.
$\lambda$3340.76 blends with the [Ne~{\sc iii}] auroral line 2p$^4$~$^1$S$_{0}$
-- 2p$^4$~$^1$D$_{2}$ $\lambda$3342.50 and Ne~{\sc ii} M2 3p~$^4$D$^{\rm
o}_{1/2}$ -- 3s~$^4$P$_{1/2}$ $\lambda$3344.40. The intensity of [Ne~{\sc iii}]
$\lambda$3342.50 was estimated from the theoretical nebular-to-auroral line
ratio ($\lambda\lambda$3868.76\,+\,3967.47)/$\lambda$3342.50 which was
calculated by solving the population equations for a five-level atomic model
at a particular $N_\mathrm{e}$ and $T_\mathrm{e}$. We estimated that [Ne~{\sc
iii}] $\lambda$3342.50 contributes about 12 per cent to the total intensity of
the blend at $\lambda$3341. The contribution from Ne~{\sc ii} M2
$\lambda$3344.40 is negligible. The resultant $\lambda$3299.40/$\lambda$3340.76
and $\lambda$3312.32/$\lambda$3340.76 ratios are discussed in
Section\,\ref{oiii_orls:fluorescence}.


\subsection{Multiplet 4,
3p\,$^3$P -- 3s\,$^3$P$^{\rm o}$}\label{appendix:oiii:v4}

Only $\lambda$3047.12 and $\lambda$3059.30 are observed in the near-UV, but
their measurements are quite unreliable due to poor S/N.


\subsection{Multiplet 5,
3p\,$^1$P -- 3s\,$^1$P$^{\rm o}$}\label{appendix:oiii:v5}

\begin{figure}
\begin{center}
\includegraphics[width=7.5cm,angle=-90]{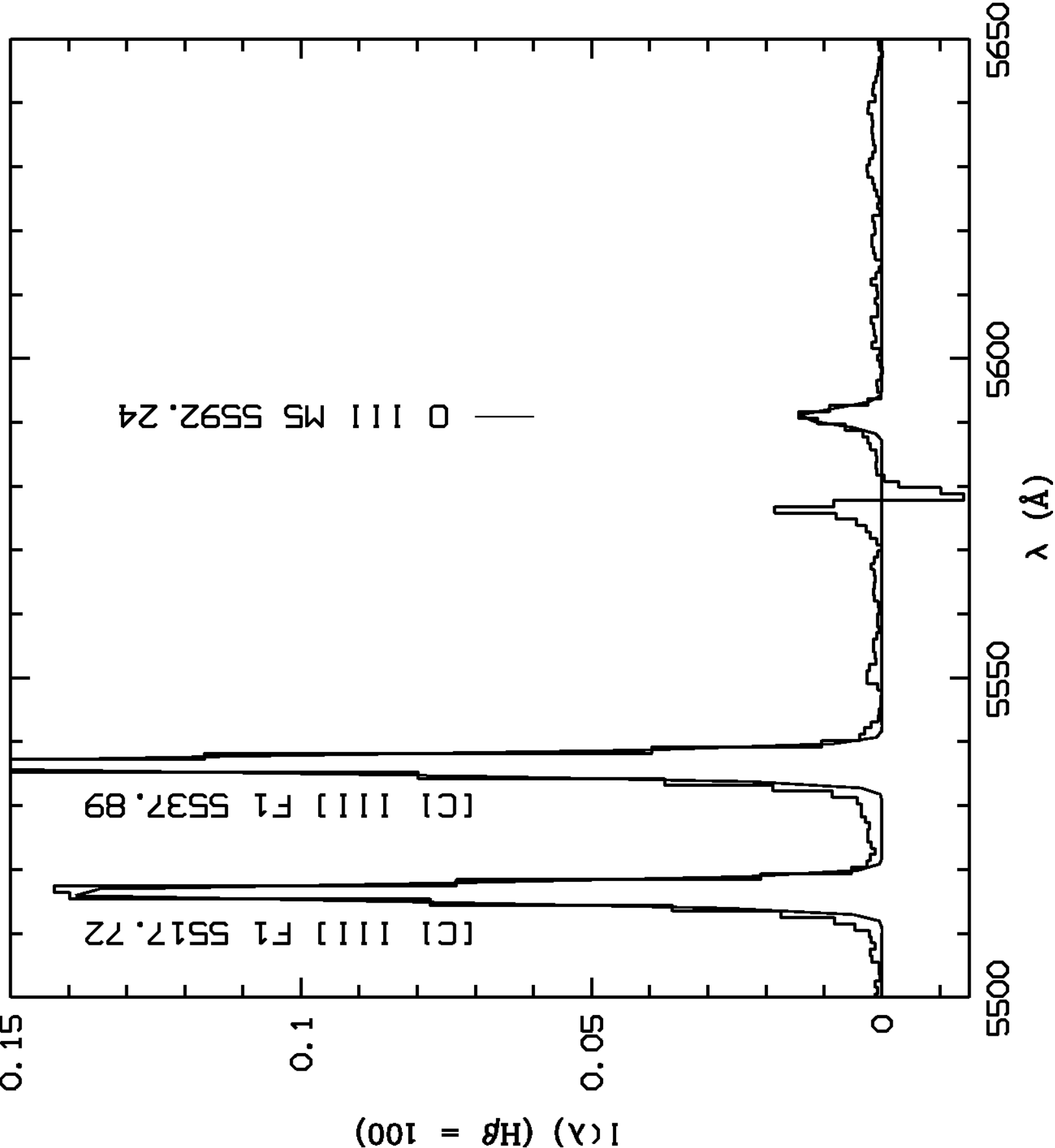}
\caption{Spectrum of NGC\,7009 from 5500 to 5650\,{\AA} showing the O~{\sc
iii} M5 $\lambda$5592.24 line. The [Cl~{\sc iii}] $\lambda$5517,\,5537 doublet
lines are also detected. Note the residual of [O~{\sc i}] $\lambda$5577
due to poor sky subtraction. The continuous curve is the sum of Gaussian
profile fits. Continuum has been subtracted and the spectrum has been
normalized such that H$\beta$ has an integrated flux of 100. Extinction has
not been corrected for.}
\label{5500-5650}
\end{center}
\end{figure}

$\lambda$5592.24 is observed in Fig.\,\ref{5500-5650}. Its fitted intensity is
0.047, which is slightly lower than the observation of Liu \&
Danziger \cite{ld1993a}: 0.0510$\pm$0.0031. The uncertainty of the current
measurement is about 10 per cent.

\subsection{Multiplet 8,
3d\,$^3$F$^{\rm o}$ -- 3p\,$^3$D}\label{appendix:oiii:v8}

See Section\,\ref{oiii_orls:v8}.

\subsection{Multiplet 12,
3d\,$^3$P$^{\rm o}$ -- 3p\,$^3$S}\label{appendix:oiii:v12}

\begin{figure}
\begin{center}
\includegraphics[width=7.5cm,angle=-90]{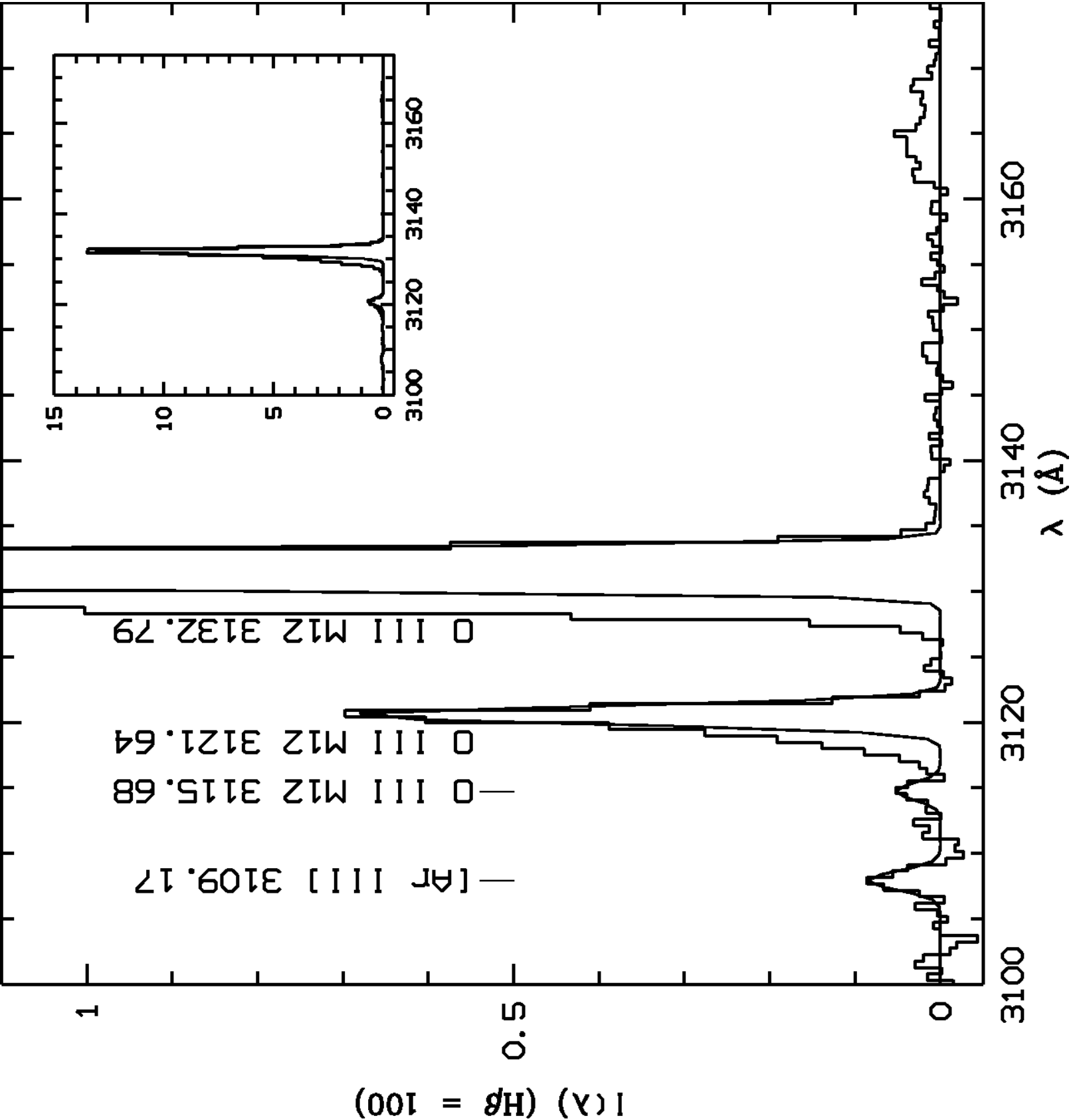}
\caption{Spectrum of NGC\,7009 from 3100 to 3175\,{\AA} showing the O~{\sc iii}
M12 lines. The continuous curve is the sum of Gaussian profile fit. Continuum
has been subtracted and the spectrum has been normalized such that H$\beta$ has
an integrated flux of 100. Extinction has not been corrected for.}
\label{3100-3175}
\end{center}
\end{figure}

The O~{\sc ii} M12 lines are shown in Fig.\,\ref{3100-3175}. The fitted
intensity of $\lambda$3132.79 is 37.81, with an uncertainty of about 7 per
cent. Our measurement is lower than that of Liu \& Danziger
\cite{ld1993a}: 43.4$\pm$4.3. 
The measured intensity ratio of the three lines
$\lambda$3115.68\,:\,$\lambda$3121.64\,:\,$\lambda$3132.79 is 1\,:\,13\,:\,343,
which differs significantly from the ratio in the pure {\it LS}\,coupling
assumption, i.e. 1\,:\,3\,:\,5. That is because this multiplet is mainly
excited by the Bowen fluorescence mechanism.

\subsection{Multiplet 14,
3d\,$^3$D$^{\rm o}$ -- 3p\,$^3$P}\label{appendix:oiii:v14}

\begin{figure}
\begin{center}
\includegraphics[width=7.5cm,angle=-90]{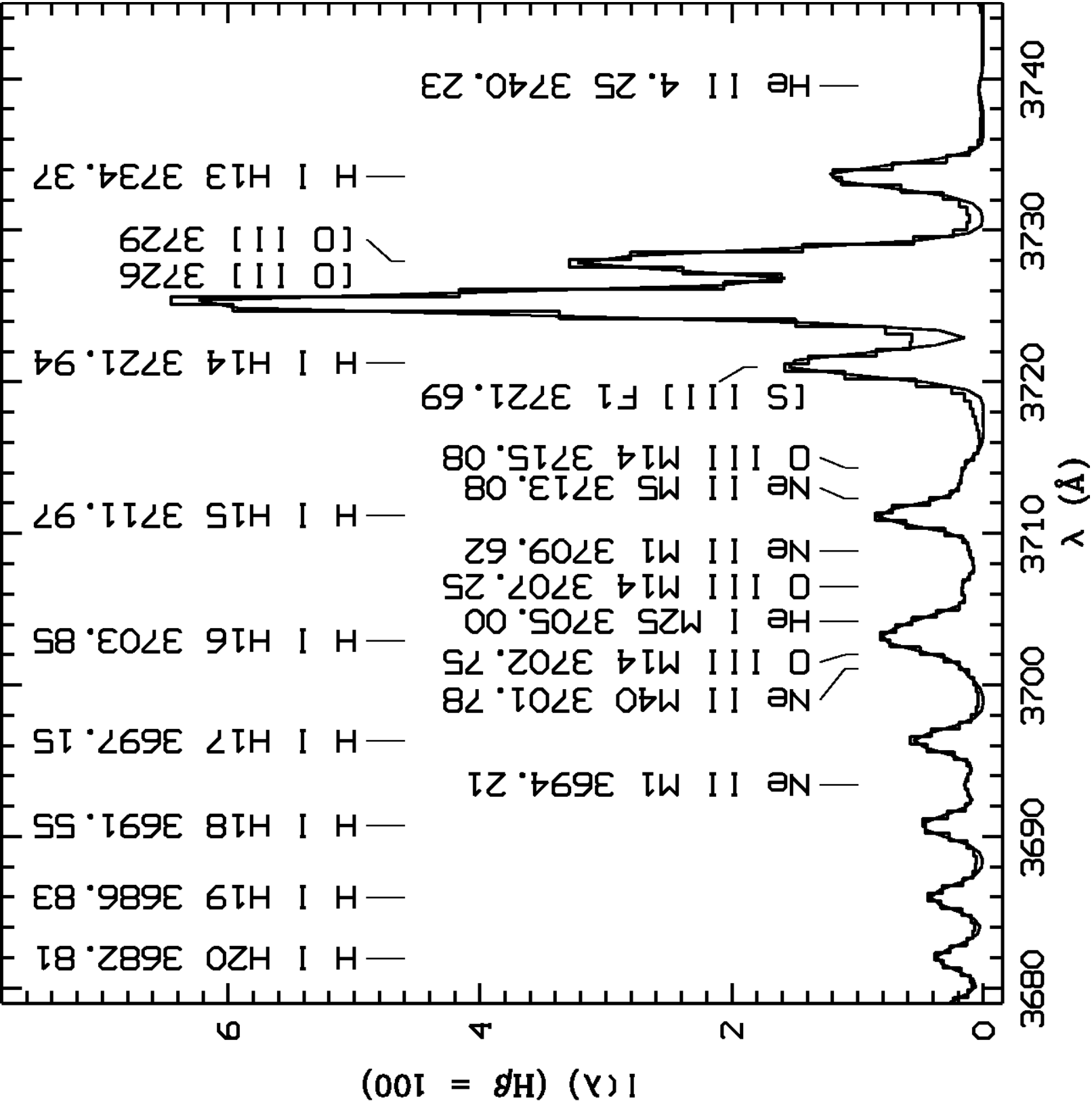}
\caption{Spectrum of NGC\,7009 from 3680 to 3745\,{\AA} showing the O~{\sc
iii} M14 lines and other emission lines. The H~{\sc i} Balmer series are
marked on top of the figure. The Ne~{\sc ii} M1 line $\lambda$3694.21
(3p\,$^{4}$P$^{\rm o}_{5/2}$\,--\,3s\,$^{4}$P$_{5/2}$) is also detected. The
continuous curve is the sum of Gaussian profile fits. Continuum has been
subtracted and the spectrum has been normalized such that H$\beta$ has an
integrated flux of 100. Extinction has not been corrected for.}
\label{3680-3745}
\end{center}
\end{figure}

$\lambda$3715.08 (3d~$^3$D$^{\rm o}_{3}$ -- 3p~$^3$P$_{2}$) is partially
blended with H~{\sc i} H15 $\lambda$3711.97 (Fig.\,\ref{3680-3745}). Also
blended here are $\lambda$3714.03 (3d~$^3$D$^{\rm o}_{1}$ -- 3p~$^3$P$_{1}$),
Ne~{\sc ii} M5 3p~$^2$D$^{\rm o}_{5/2}$ -- 3s~$^2$P$_{3/2}$ $\lambda$3713.08
and He~{\sc ii} 29g~$^2$G -- 4f~$^2$F$^{\rm o}$ $\lambda$3715.16. Reliable
measurements of the two M14 lines are difficult. Accurate measurements of
another line $\lambda$3707.25 are also difficult due to weakness. The other
lines are not observed.

\subsection{Multiplet 15,
3d\,$^3$P$^{\rm o}$ -- 3p\,$^3$P}\label{appendix:oiii:v15}

See Section\,\ref{oiii_orls:v15}.

\end{document}